\documentclass[11pt]{article}

\usepackage{graphicx}
\usepackage{amsmath,amssymb,amsfonts,dsfont,slashed}
\usepackage{cite}
\usepackage{booktabs}

\allowdisplaybreaks[1]

\def\Xint#1{\mathchoice
   {\XXint\displaystyle\textstyle{#1}}%
   {\XXint\textstyle\scriptstyle{#1}}%
   {\XXint\scriptstyle\scriptscriptstyle{#1}}%
   {\XXint\scriptscriptstyle\scriptscriptstyle{#1}}%
   \!\int}
\def\XXint#1#2#3{{\setbox0=\hbox{$#1{#2#3}{\int}$}
     \vcenter{\hbox{$#2#3$}}\kern-0.5\wd0}}
\providecommand{\dashint}[1][0pt]{\Xint{\hspace{#1}-}}

\providecommand{\ec}{\;,}
\providecommand{\ep}{\;.}
\providecommand{\nt}{\notag}

\newcommand{\diff}{\text{d}}

\newcommand{\mpp}{m_\text{p}}
\newcommand{\mpi}{M_{\pi}}
\newcommand{\mK}{M_K}
\newcommand{\tpi}{t_\pi}
\newcommand{\tK}{t_K}
\newcommand{\tN}{t_N}
\newcommand{\tm}{t_\text{m}}
\newcommand{\sm}{s_\text{m}}
\newcommand{\ta}{t_\text{a}}
\newcommand{\sa}{s_\text{a}}
\newcommand{\Wa}{W_\text{a}}

\DeclareMathOperator{\cosec}{cosec}
\DeclareMathOperator{\Res}{\text{Res}}
\providecommand{\Disc}{\text{Disc}\,}
\renewcommand{\Im}{\text{Im}\,}
\renewcommand{\Re}{\text{Re}\,}

\providecommand{\Ord}{\mathcal{O}}
\providecommand{\Amp}{\mathcal{A}}
\providecommand{\Pol}{\mathcal{P}}
\providecommand{\unity}{\mathds{1}}
\providecommand{\tdelta}{\tilde{\delta}}
\providecommand{\tgamma}{\tilde{\gamma}}

\providecommand{\MeV}{\,\text{MeV}}
\providecommand{\GeV}{\,\text{GeV}}

\providecommand{\ste}[1]{\left[#1\right]_{(0,0)}}
\providecommand{\asym}[1]{\left.#1\right|_{\text{asym}}}

\providecommand{\unsub}[1]{#1\big|^{0\text{-sub}}}
\providecommand{\onesub}[1]{#1\big|^{1\text{-sub}}}
\providecommand{\twosub}[1]{#1\big|^{2\text{-sub}}}
\providecommand{\nsub}[1]{#1\big|^{n\text{-sub}}}
\providecommand{\nsubasym}[1]{\left.#1\right|^{n\text{-sub}}_{\text{asym}}}

\providecommand{\redonesub}{\overset{1\text{-sub}}{\longrightarrow}}
\providecommand{\redunsub}{\overset{0\text{-sub}}{\longrightarrow}}

\begin{document}

\setlength{\unitlength}{1mm}

\numberwithin{equation}{section}

\author{C.~Ditsche$^a$, M.~Hoferichter$^a$, B.~Kubis$^a$, U.-G. Mei{\ss}ner$^{a,b}$}

\title{\bf Roy--Steiner equations for pion--nucleon scattering}

\date{}

\maketitle

\begin{center}
{\small
$^a${\it Helmholtz--Institut f\"ur Strahlen- und Kernphysik (Theorie) 
   and Bethe Center for Theoretical Physics, Universit\"at Bonn, D-53115 Bonn, Germany}

\bigskip

$^b${\it Institut f\"ur Kernphysik, Institute for Advanced Simulation, 
   and J\"ulich Center for Hadron Physics, Forschungszentrum J\"ulich, D-52425  J\"ulich, Germany}
}
\end{center}

\bigskip

\begin{abstract}
Starting from hyperbolic dispersion relations, we derive a closed system of Roy--Steiner equations for pion--nucleon scattering that respects analyticity, unitarity, and crossing symmetry. 
We work out analytically all kernel functions and unitarity relations required for the lowest partial waves. 
In order to suppress the dependence on the high-energy regime we also consider once- and twice-subtracted versions of the equations, where we identify the subtraction constants with subthreshold parameters. 
Assuming Mandelstam analyticity we determine the maximal range of validity of these equations. 
As a first step towards the solution of the full system we cast the equations for the $\pi\pi\to\bar NN$ partial waves into the form of a Muskhelishvili--Omn\`es problem with finite matching point, which we solve numerically in the single-channel approximation. 
We investigate in detail the role of individual contributions to our solutions and discuss some consequences for the spectral functions of the nucleon electromagnetic form factors.
\end{abstract}


\medskip

\tableofcontents

\newpage

\section{Introduction}
\label{sec:introduction}

Pion--nucleon scattering is one of the most basic and fundamental processes in 
strong-interaction physics. Even though a large data basis exists and numerous
investigations based on a cornucopia of methods (dispersion relations, quark models,
resonance models, chiral perturbation theory, just to name a few)
have been performed for many decades, the pion--nucleon ($\pi N$) scattering amplitude is still not known to
sufficient precision in the low-energy region.\footnote{The exceptions are the
$S$-wave scattering lengths, which can be extracted with high precision from the
beautiful data on pionic hydrogen and pionic deuterium, see~\cite{PSI:hydrogen,PSI:deuterium,piNcoupling:short,piNcoupling:long}.} 
This becomes most obvious in the scalar-isoscalar sector, which features the
so-called pion--nucleon $\sigma$ term $\sigma_{\pi N}$, i.e.\ the scalar form factor of
the nucleon at zero momentum transfer. Its value is
a measure of the light quark contribution to the nucleon mass (and it can also
be related to its strange quark contribution), 
see e.g.\ the classical paper~\cite{Gasser:sigma1980}. The $\sigma$ term has gained renewed interest 
as it parameterizes the spin-independent cross section for possible dark matter 
candidates scattering off nuclei~\cite{darkmatter:Bottino,darkmatter:Ellis} (for
a recent review cf.~\cite{Olive2012}). In principle, lattice QCD would
be the method of choice to pin down the $\sigma$ term---however, a direct
computation of the scalar form factor necessarily involves disconnected
diagrams, which is not yet under sufficient control. Similarly, the
indirect extraction of $\sigma_{\pi N}$ from the derivative of the nucleon
mass is still hampered with systematic uncertainties related to the chiral
extrapolations utilized, see e.g.~\cite{WalkerLoud2011}. Therefore,
in this paper we follow a different path, namely setting up the powerful
machinery of Roy--Steiner (RS) equations that will ultimately allow for a
precise determination of the pion--nucleon scattering amplitude at low energies.

More specifically, RS equations are based on hyperbolic dispersion relations (HDRs), 
a particular kind of dispersion relations along hyperbolae in the Mandelstam plane. 
Dispersion relations are a widely used tool
that is built upon very general principles, such as Lorentz invariance,
unitarity, crossing symmetry, and analyticity. There are multiple uses of
dispersion relations---they can be used to stabilize extrapolation of
experimental data to threshold and allow for a continuation into unphysical 
regions, as it is e.g.\ required for the extrapolation of the pion--nucleon 
scattering amplitude to the so-called Cheng--Dashen point~\cite{ChengDashen}, which is
crucial for the extraction of the $\sigma$ term.
We notice that unitarity constraints can most
conveniently be formulated in terms of partial-wave amplitudes.
The resulting partial-wave dispersion relations (PWDRs) together with 
unitarity constraints allow to study processes at low energies with high
precision. We just mention a few examples. The most prominent example
is of course pion--pion ($\pi\pi$) scattering, which is intimately linked to the
spontaneous and explicit chiral symmetry breaking in QCD. The Roy equations~\cite{Roy} 
are the appropriate PWDRs, which have been extensively studied
in the last years~\cite{ACGL,DFGS02,GKPY,CCL:Regge,CCL:PWA,Moussallam2011},
leading to a determination of the fundamental $\pi\pi$ scattering amplitude
with unprecedented precision. The pion--pion system, however, is special as all
channels are identical. This is different for the simplest scattering 
process in QCD involving strange quarks, namely pion--kaon ($\pi K$) scattering, which 
has been investigated in~\cite{piK:comparison,piK:RS}. 
As far as crossing symmetry and isospin quantum numbers are concerned, the pion--kaon
system is similar to the pion--nucleon case considered here.
Crossing symmetry relates the $s$-/$u$-channel ($\pi N\to\pi N$) and the 
$t$-channel ($\pi\pi\to\bar NN$) amplitudes, with the $s$-channel amplitudes
relevant e.g.\ for $\sigma$-term physics, while the $t$-channel amplitudes
feature prominently in the dispersive analysis of the nucleon form factors.
The final aim of solving the full (subtracted) RS system for $\pi N$ scattering
is a precise determination of the lowest partial-wave amplitudes in the
low-energy (physical and unphysical) region as well as the
pertinent low-energy parameters, such as the $\pi N$ coupling constant and the so-called
subthreshold parameters, and to provide reliable theoretical
errors for the fundamental pion--nucleon scattering amplitude for the first time.

In the low-energy region, the pion--nucleon amplitude is well represented
by its $S$- and $P$-wave projections. Due to the spin of the nucleon, one has
in total six partial waves in the $s$- and $u$-channel, commonly denoted as
$f_{0+}^\pm$, $f_{1+}^\pm$, $f_{1-}^\pm$, where the superscript $I=\pm$ refers to the
isospin, $l\in\{0,1\}$ in the subscript to the 
orbital angular momentum, and the $\pm$ to the total angular momentum
$j = l\pm 1/2$. Similarly, there are three $t$-channel $S$- and $P$-waves, called $f^0_+$, $f_\pm^1$,
where the superscript refers to total angular momentum $J$ 
and the $+/-$ to parallel/antiparallel antinucleon--nucleon helicities,
such that there is one wave with even and two with 
odd isospin (due to Bose symmetry). It was pointed out in~\cite{BecherLeutwyler}
how to generalize the Roy equations for $\pi\pi$ scattering to the $\pi N$
system based on fixed-$t$ dispersion relations. These amount to coupled integral equations for
the nine partial waves, where the effect of the higher partial waves is
encoded in the respective kernels of these integral equations.
Here, we follow a somewhat different path by utilizing 
hyperbolic dispersion relations as pioneered by Hite and Steiner a long time ago~\cite{HiteSteiner}. 
The main advantage of HDRs is that they combine the $s$- 
and the $t$-channel (i.e.\ all three) physical regions, which is
obviously not true for e.g.\ usual fixed-$t$ dispersion relations. It is known that a
reliable continuation to the subthreshold region in dispersion theory can only
be made by using input information also from the $t$-channel, 
cf.\ e.g.~\cite{Koch:piNsigmaterm,Hoehler:1999:sigmaterm,Stahov:1999,Stahov:2002}.
Furthermore, the knowledge of the absorptive parts in the dispersion relations is needed only
in regions where the corresponding partial-wave expansions converge, and HDRs are
considered the best choice fulfilling these requirements that yields still manageable angular
kernels~\cite{HiteSteiner}. In addition, the underlying hyperbolic relation
$(s-a)(u-a)=b$ (with $a$, $b$ real-valued parameters) also respects 
$s\leftrightarrow u$ crossing symmetry of the
$\pi N$ amplitude. Due to the tunable parameters
$a$, $b$, better convergence properties can be achieved with HDRs and they are
found to be especially powerful for determining the $\sigma$ term~\cite{Koch:piNsigmaterm}.
The derivation of the RS equations for the $\pi N$ system is given by a series of steps:
first, one expands the $s$-/$t$-channel absorptive parts of
the HDRs in $s$-/$t$-channel partial waves, respectively. Second, one
projects the full, partial-wave-expanded HDRs onto both $s$- and $t$-channel partial waves, 
resulting in what we will refer to as the $s$- and $t$-channel 
part of the RS system in the following. The resulting system of
equations exhibits the following general structure: it features the
nucleon-pole-term contributions, integrals over the imaginary parts of the
$s$-(and $u$-)channel as well as integrals over $t$-channel absorptive parts,
both from the corresponding threshold to infinity. The generic properties of the equations 
are then determined by the integral kernels. In the equation for each partial 
wave, the corresponding kernels consist of the self-coupling, singular Cauchy kernel 
and an analytic remainder that in addition involves the coupling to all other partial waves. 
In particular, these kernel functions automatically incorporate the analytic properties
expected for a given partial wave: the Cauchy kernel corresponds to the right-hand cut, 
while the remainder contains all left-hand-cut contributions.

Another important issue is the possibility to subtract dispersion relations.
This can be advantageous for various reasons:
first, in some cases the asymptotic behavior of the integrand is such that subtractions have to be 
performed to ensure convergence of the dispersive integral. Similarly,
if the high-energy behavior is not known, it can be subsumed in subtraction 
constants, which are a priori unknown. In some cases, these subtraction
constants can be related to phenomenology or the parameters of a low-energy 
effective field theory like e.g.\ chiral perturbation theory (ChPT). Second,
one can even introduce subtractions that are not necessarily required by the asymptotic behavior 
in order to lessen the dependence on high-energy input, however, at the expense of 
introducing the corresponding subtraction polynomials. Third, subtracting
the dispersion relations is especially useful in the $\pi N$ case, since subtracting at the so-called 
subthreshold point allows for a relation to the subthreshold expansion 
and is convenient for the continuation to the Cheng--Dashen point. In
addition, such subtractions are well suited for the $t$-channel problem to be
discussed later. In what follows, we will consider unsubtracted as well as
subtracted versions of the RS equations.

\begin{figure}[!t]
\centering
\includegraphics[width=0.8\textwidth]{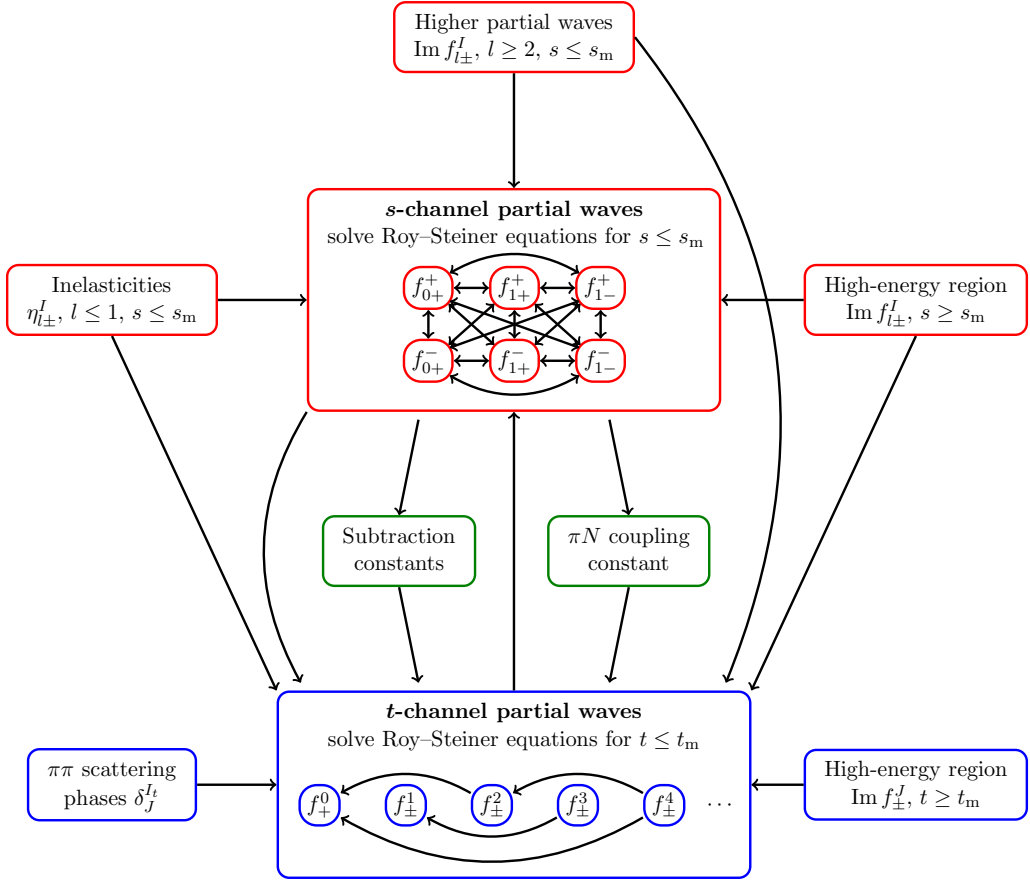}
\caption{Flowchart of the solution strategy for the Roy--Steiner system for $\pi N$ scattering.}
\label{fig:flowchart}
\end{figure} 

Next, we will outline the strategy to solve the RS equations, as depicted in Fig.~\ref{fig:flowchart}:
first, one solves the $t$-channel part of the
RS system, which takes the form of a Muskhelishvili--Omn\`es (MO) 
problem~\cite{Muskhelishvili,Omnes} (using rather well known $s$-channel partial waves
and $\pi\pi$ phase shifts as input). Then, one uses the $t$-channel MO solutions to solve the
$s$-channel part, and finally the procedure is repeated (iterated) until
self-consistency of the partial waves and parameters is reached and the results 
have converged, cf.\ Fig.~\ref{fig:flowchart}. In both the $s$- and $t$-channel part of the system one
actually solves the equations in the low-energy region and for the lowest partial waves, 
while the amplitudes in the high-energy region as well as higher partial waves are
needed as input. The separation between both energy regions occurs at the so-called matching points $\sm$ and $\tm$
in the $s$- and $t$-channel, respectively. Due to the complexity 
of the full problem, we will not yet solve the whole set of RS equations in this article,
but concentrate on the $t$-channel part of the system as a first step. The solution of this $t$-channel subproblem is
interesting by itself, as it features in the dispersive analysis of the
nucleon electromagnetic form factors as well as the scalar form factor, which is, in turn, essential for the extraction
of the $\sigma$ term. At present, in the unphysical region
only the KH80 solution~\cite{KH80,Hoehler} has been used. It is, however,
well-known that this solution does not include more recent and precise data
and that the $\pi N$ coupling constant used there differs significantly
from more modern determinations. Furthermore, no analysis of the theoretical uncertainties is performed 
(apart from an iteration uncertainty, cf.\ Sect.~\ref{subsubsec:piNMO:input:expwa}), 
which is an absolute requirement for any modern theoretical investigation.
Therefore, a new $t$-channel solution is needed as a first step for solving the 
full system. Finally, a consistent set of partial-wave amplitudes for all channels is especially 
important as far as the $\sigma$-term extraction is concerned, and it has been pointed out that
the KH80 solution seems to suffer from internal inconsistencies~\cite{Koch:piNsigmaterm,Stahov:1997,Stahov:2002}, 
which emphasizes the necessity of a full system of PWDRs.

The original Roy equations for $\pi\pi$ scattering~\cite{Roy} were solely based on fixed-$t$ dispersion relations. 
This approach fails for processes involving non-identical particles, since crossing symmetry intertwines different physical processes. 
For this reason, a combination of fixed-$t$ and hyperbolic dispersion relations was used in~\cite{piK:comparison,piK:RS} to construct integral equations for $\pi K$ scattering for the partial waves of both $s$- and $t$-channel, which are therefore referred to as Roy--Steiner equations. 
In this work, we solely consider HDRs, a path that has already proven useful in the construction of RS equations for $\gamma\gamma\to\pi\pi$~\cite{ggpipi}. 
Our solution strategy for the $t$-channel MO equations follows~\cite{piK:RS}, however, there is a major difference between $\pi\pi\to\bar KK$ and $\pi\pi\to\bar NN$ as far as inelasticities in the unitarity relation are concerned, since the pseudophysical region in the $\pi N$ case is much larger due to the large nucleon mass. 
In both cases, the first non-negligible contribution besides $\pi\pi$ intermediate states originate from $\bar KK$, which play an important role for the $S$-wave in view of the occurrence of the $f_0(980)$ resonance. 
For $\pi\pi\to\bar KK$ the inelasticities can simply be accounted for by using phase-shift solutions for the corresponding partial waves, while physical input for $\pi\pi\to\bar NN$ is only available above the two-nucleon threshold. 
Once the $t$-channel problem is solved, the remaining equations take the form of the conventional $\pi\pi$ Roy equations, such that known results concerning existence and uniqueness of solutions~\cite{GasserWanders,Wanders} may be transferred to the $s$-channel RS equations as well. 

This work is organized as follows: 
in Sect.~\ref{sec:preliminaries} we specify our conventions and review HDRs for the invariant amplitudes of $\pi N$ scattering. 
In Sects.~\ref{sec:RS} and \ref{sec:subRS} we derive a closed system of RS equations as well as a once- and twice-subtracted version, and show how the $t$-channel equations can be cast into the form of a MO problem. 
Sect.~\ref{sec:piNMO} is devoted to the explicit solution of the $t$-channel MO problem: 
we first review the MO problem with a finite matching point and state the explicit solution for the $\pi N$ $t$-channel amplitudes. 
Then we collect the necessary input and discuss the numerical results. 
Finally, we briefly discuss the application to nucleon form factors before concluding in Sect.~\ref{sec:conclusion}. 
The explicit derivation of the $s$- and $t$-channel RS equations is described in full detail in Appendices~\ref{sec:spwp} and \ref{sec:tpwp}, respectively. 
In Appendix~\ref{sec:convergence} we determine the range of convergence of our equations, while Appendix~\ref{sec:regge} contains a discussion of the asymptotic regions in the dispersion integrals.

\section{Preliminaries}
\label{sec:preliminaries}

\subsection{Kinematics}
\label{subsec:preliminaries:kinematics}

We take the $s$-channel reaction of $\pi N$ scattering to be $\pi(q)+N(p)\to\pi(q')+N(p')$ and the $t$-channel reaction to be $\pi(q)+\pi(-q')\to\bar{N}(-p)+N(p')$ with the usual Mandelstam variables
\begin{equation}
s=(p+q)^2\ec\qquad
t=(p-p')^2\ec\qquad
u=(p-q')^2\ec
\end{equation}
which fulfill
\begin{equation}
s+t+u=2m^2+2\mpi^2=\Sigma\ec
\end{equation}
where $m$ and $\mpi$ denote the nucleon and pion mass, respectively. We will use the masses of~\cite{PDG}, with the isospin limit defined by the charged particles, i.e.\ $\mpi\equiv M_{\pi^\pm}$ and $m\equiv\mpp$ (later also $\mK\equiv M_{K^\pm}$ for the kaon mass).
Unless stated otherwise, $u$ is always to be understood as a function of $s$ and $t$
\begin{equation}
u(s,t)=\Sigma-s-t\ep
\end{equation}
We define the generic kinematical K\"all\'en function
\begin{equation}
\lambda_x^{PQ}=\lambda\big(x,M_P^2,M_Q^2\big)=\big[x-(M_P-M_Q)^2\big]\big[x-(M_P+M_Q)^2\big]\ec
\end{equation}
and for the equal-mass case
\begin{equation}
\sigma^P_x=\sigma\big(x,M_P^2\big)=\frac{\sqrt{\lambda_x^{PP}}}{x}=\sqrt{1-\frac{4M_P^2}{x}}\ep
\end{equation}
Furthermore, we introduce the general definitions\footnote{For more on $\pi N$ kinematics and for $\pi N$ conventions in general we refer to~\cite{Hoehler}. Note that the convention for $\nu$ therein and which we have adopted here differs from the choice $\nu=s-u$ of e.g.~\cite{HiteSteiner}.}
\begin{align}
\Sigma&=2s_0\ec & \nu(s,t)&=\frac{s-u}{4m}=\frac{2s+t-\Sigma}{4m}=\frac{2(s-s_0)+t}{4m}\ec\nt\\
W^2&=s\ec & \nu_B(t)&=-\frac{s+u-2m^2}{4m}=\frac{t-2\mpi^2}{4m}=\nu(s=m^2,t)\ec
\end{align}
with $W$ as the total center-of-mass-system (CMS) energy, as well as the abbreviation
\begin{equation}
\lambda_x=\lambda_x^{\pi N}=\lambda\big(x,m^2,\mpi^2\big)=\big[x-s_-\big]\big[x-s_+\big]\ec \qquad s_\pm=W_\pm^2=(m\pm\mpi)^2\ec
\end{equation}
where $W_-$ and $W_+$ denote the ($s$-channel) pseudothreshold and threshold energies, respectively.
Additional related useful definitions and relations are
\begin{equation}
\Sigma_\pm=m^2\pm\mpi^2\ec \qquad \Sigma_+=s_0\ec \qquad \Sigma_-=W_+W_-\ec \qquad \Sigma_-^2=s_+s_-\ec \qquad \Sigma=s_++s_-\ep
\end{equation}

The CMS kinematics of the elastic $s$-channel reaction $\pi N\to\pi N$ above threshold (i.e.\ for $s\geq s_+$) with CMS momentum $q=|\mathbf{q}|$, nucleon energy $E$, and scattering angle $z_s=\cos\theta_s$ are then given by
\begin{align}
q(s)&=\sqrt{\frac{\lambda_s}{4s}}\ec & E(\pm W)&=\pm\sqrt{m^2+q^2}=\frac{s+\Sigma_-}{2(\pm W)}=\pm E(W)\ec\nt\\
z_s(s,t)&=1-\frac{s+u-\Sigma}{2q^2}=1+\frac{t}{2q^2}\ec & 4q^2&=s-\Sigma+\frac{\Sigma_-^2}{s}\ep
\end{align}

For the $t$-channel reaction $\pi\pi\to\bar{N}N$ with CMS momenta $q_t$ for the pions and $p_t$ for the nucleons and scattering angle $z_t=\cos\theta_t$, the CMS kinematics above threshold (i.e.\ for $t\geq4m^2$) read
\begin{align}
q_t(t)&=\sqrt{\frac{t}{4}-\mpi^2}=\frac{\sqrt{t}}{2}\sigma^\pi_t=+iq_-\ec & p_t(t)&=\sqrt{\frac{t}{4}-m^2}=\frac{\sqrt{t}}{2}\sigma^N_t=+ip_-\nt\\
z_t(s,t)&=\frac{s-u}{4p_tq_t}=\frac{2s+t-\Sigma}{4p_tq_t}=\frac{m\nu}{p_tq_t}\ec
\end{align}
where below the corresponding two-particle thresholds $\tpi$ and $\tN$ one has to use the quantities
\begin{equation}
\label{pmqm}
q_-(t)=\sqrt{\mpi^2-\frac{t}{4}}\geq0 \quad \forall\,t\leq \tpi=4\mpi^2\ec \qquad
p_-(t)=\sqrt{m^2-\frac{t}{4}}\geq0 \quad \forall\,t\leq \tN=4m^2\ec
\end{equation}
whose phases are constrained in general to $p_tq_t=-p_-q_-$ and fixed here by convention.
Relations valid in all kinematical ranges can be written down by relying on the quantities
\begin{equation}
\label{pt2qt2}
q_t^2(t)=\frac{t-\tpi}{4}=-q_-^2(t)\ec \qquad p_t^2(t)=\frac{t-\tN}{4}=-p_-^2(t)\ec
\end{equation}
from which roots in the corresponding regimes may be taken.\footnote{We use the non-cyclic convention $^a_b\times^c_d$ for a reaction $a+b\to c+d$ in order to stick to the usual $\pi N$ conventions of~\cite{Hoehler}, rather than the cyclic convention $^a_b\times^d_c$ that leads to symmetric kinematical relations for the $s$-, $t$-, and $u$-channel and is therefore sometimes used in the literature. While the cyclic convention is especially favorable when all four particles are identical like e.g.\ in the case of $\pi\pi$ scattering, it leads to different sign conventions for the CMS scattering angles and also to different isospin crossing matrices (cf.\ Sect.~\ref{subsec:preliminaries:isospin}). The non-cyclic convention, however, is well-suited for $s\leftrightarrow u$ crossing symmetric situations like e.g.\ $\pi N$ scattering, with $t=0$ corresponding to an undeflected pion (i.e.\ forward scattering) in both the $s$- and $u$-channel and thus $z_s(t=0)=1=z_u(t=0)$ rather than $z_s(t=0)=1=-z_u(t=0)$ for the cyclic convention.}

\begin{figure}
\centering
\includegraphics[scale=0.8]{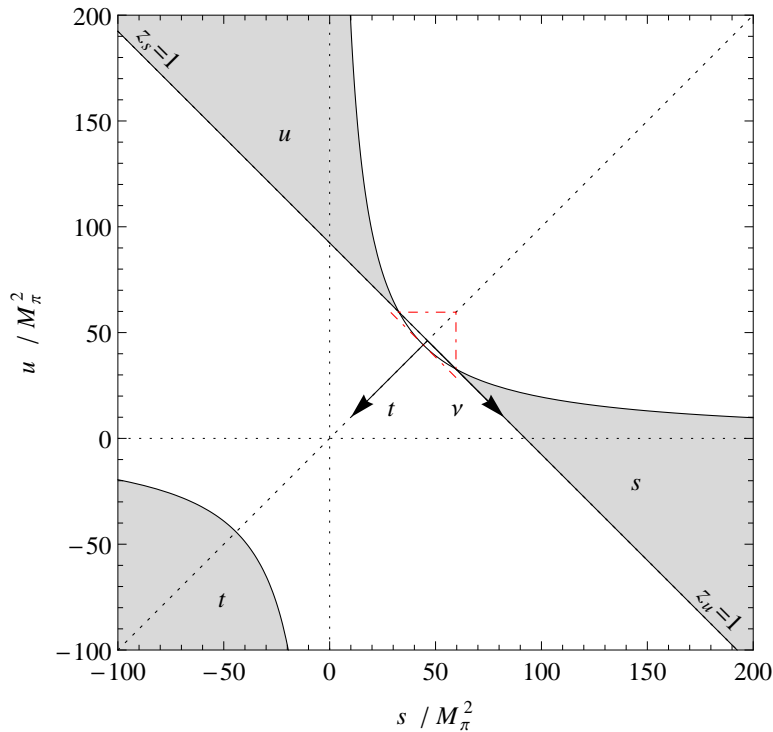}
\caption[Physical regions for $\pi N$ scattering.]{Physical regions for $s$-, $t$-, and $u$-channel reactions of $\pi N$ scattering (shaded) and the subthreshold triangle (dot-dashed) enclosing the subthreshold lens.}
\label{fig:physbounds}
\end{figure}

The physical regions for the $s$-, $t$-, and $u$-channel reactions are restricted to kinematical regions where the Kibble function
$\Phi$~\cite{Kibble} is non-negative.
For $\pi N$ scattering we have
\begin{equation}
\label{piNKibble}
\frac{\Phi}{t}=su-\Sigma_-^2=4sq^2(1+z_s)=4p_t^2q_t^2(1-z_t^2)\ec
\end{equation}
such that the boundaries are given by
\begin{align}
\Phi&=-s\big[u-(\Sigma-s)\big]\bigg[u-\frac{\Sigma_-^2}{s}\bigg]\nt\\
&=\frac{t}{4}\Big[t-\Big(\Sigma-2\sqrt{(2m\nu)^2-\Sigma_-^2}\Big)\Big]\Big[t-\Big(\Sigma+2\sqrt{(2m\nu)^2-\Sigma_-^2}\Big)\Big]=0\ec
\end{align}
and the corresponding physical regions are shown in Fig.~\ref{fig:physbounds}.

$\pi N$ scattering in the isospin limit can be described by the four Lorentz-invariant amplitudes $A^\pm(s,t)$ and $B^\pm(s,t)$, as well as the related amplitudes $D^\pm(s,t)$ convenient for low-energy theorems (all to be defined in Sect.~\ref{subsec:preliminaries:isospin}).
These amplitudes are real inside the Mandelstam subthreshold triangle defined by the lines $s=s_+$, $u=s_+$, and $t=\tpi$, i.e.\ below the thresholds for the physical $s$- and $u$-channel reactions and below the $\pi\pi$ threshold,\footnote{Note that $t\leq0$ is necessary for both the $s$- and $u$-channel reaction to be physical.} including in particular the small on-shell but unphysical lens-shaped low-energy region (subthreshold lens) close to $(\nu=0,t=0)$ depicted in Fig.~\ref{fig:physbounds}.

The analytic structure of the invariant amplitudes governs the analytic structure of both the $s$- and $t$-channel partial-wave amplitudes, for details we refer to~\cite{Hoehler} and references therein.
Here, we only mention the different analytic structures of the $s$-channel $\pi N$ scattering invariant amplitudes (in the complex $s$-plane)
\begin{itemize}
 \item right-hand cut (RHC): physical $s$-channel cut along $s\geq s_+$,
 \item nucleon pole: at $s=m^2$ from the $s$-channel nucleon-exchange pole term $1/(s-m^2)$,
 \item crossed cut: along $s\leq s_-$ as combination of the $u$-channel cut $s\leq s_-$ and the $t$-channel cut $s\leq-\Sigma_-$,
 \item left-hand cut (LHC): collective name for all cuts in the unphysical region, i.e.\ for $\Re\{s\}<s_+$.
\end{itemize}
In addition, the mapping between the complex $s$- and $q^2$-planes involves a circular cut in the complex $s$-plane at $|s|=\Sigma_-=s_{(+)}s_{(-)}$, where $s_{(+)}(q^2)$ and $s_{(-)}(q^2)$ are the two solutions
\begin{equation}
 s_{(\pm)}(q^2)=2q^2+\Sigma_+\pm2\sqrt{(q^2+m^2)(q^2+\mpi^2)}
\end{equation}
for a given $q^2$ (note the cut for $-m^2\leq q^2\leq-\mpi^2$) with $s_{(+)}(0)=s_+$ and $s_{(-)}(0)=s_-$. This circular cut becomes relevant once amplitudes are considered as functions of $q^2$ rather than $s$, e.g.\ for the partial waves. 
The additional analytic structures of the $s$-channel partial-wave amplitudes due to the partial-wave projection are
\begin{itemize}
 \item kinematical cuts: for $s\leq0$ from terms depending on $W=\sqrt{s}$ in the partial-wave projection formula,
 \item short nucleon cut:\footnote{Actually, there are two short nucleon cuts as discussed in the appendix of~\cite{Doering}. The second one, however, is situated on an unphysical sheet.} along $\Sigma_-^2/m^2\leq s\leq m^2+2\mpi^2$ from evaluating the $u$-channel nucleon-exchange pole term $1/(u(s,z_s)-m^2)$ for $z_s=\pm1$,
 \item circular-cut contributions: from $t$-channel exchange of particles with mass $m_t\geq2\mpi$, i.e.\ evaluating $1/(t(s,z_s)-m_t^2)$ for $z_s=\pm1$ and $m_t^2=\tpi$,
 \item crossed-cut contributions for $s\leq0$ and singularities at $s=0$: from partial-wave projection of the aforementioned $u$- and $t$-channel exchanges.
\end{itemize}
Finally, we mention some kinematical points of specific interest (cf.\ e.g.~\cite{KaufmannHite,Hoehler}):
the Cheng--Dashen point at $(s=u=m^2,t=2\mpi^2)=(\nu=0,\nu_B=0)$ is pivotal for $\pi N$ $\sigma$-term physics, since the Born-term-subtracted amplitude $\bar D^+(\nu=0,t=2\mpi^2)=A^+(\nu=0,t=2\mpi^2)-g^2/m$ is related to the $\sigma$ term by a low-energy theorem~\cite{ChengDashen,Brown,BKM96,BecherLeutwyler,HiteKaufmannJacob}.\footnote{Note that since $z_s^\text{CD}=z_s(m^2,2\mpi^2)=-\mpi^2/(4m^2-\mpi^2)\approx-5.56\times10^{-3}$ is close to zero, the amplitudes at the CD point are dominated by the ($s$-channel) $S$-wave.}
The subthreshold point at $(s=u=s_0,t=0)=(\nu=0,\nu_B=-\mpi^2/(2m))$ serves as expansion point for the subthreshold expansion, while the ($s$-channel) threshold point $(s=s_+,t=0,u=s_-)=(\nu=\mpi,\nu_B=-\mpi^2/(2m))$ is relevant for the threshold expansion/parameters (e.g.\ scattering lengths).

\subsection{Isospin structure}
\label{subsec:preliminaries:isospin}

The most general Lorentz-invariant and parity-conserving $T$-matrix element for the process $\pi^a(q)+N(p)\to\pi^b(q')+N(p')$ with isospin indices $a$ and $b$ is given in terms of Lorentz-invariant amplitudes $A$, $B$, and $D$ according to
\begin{align}
\label{genlorentzinvamps}
T_{fi}^{ba}(s,t)&=\frac{1}{2}\{\tau^b,\tau^a\}T_{fi}^+(s,t)+\frac{1}{2}[\tau^b,\tau^a]T_{fi}^-(s,t)
=\delta^{ba}T_{fi}^+(s,t)+i\epsilon_{bac}\tau^cT_{fi}^-(s,t)\ec\nt\\
T_{fi}^I(s,t)&=\bar{u}_f(p')\bigg\{A^I(s,t)+\frac{\slashed q'+\slashed q}{2}B^I(s,t)\bigg\}u_i(p)
=\bar{u}_f(p')\bigg\{D^I(s,t)-\frac{[\slashed q',\slashed q]}{4m}B^I(s,t)\bigg\}u_i(p)\ec\nt\\
D^I(s,t)&=A^I(s,t)+\nu(s,t)B^I(s,t)\ec \qquad I\in\{+,-\}\ec
\end{align}
where we have introduced the isospin index $I=+/-$ for the part that is even/odd under interchange of $a$ and $b$.
Furthermore, the $\pi N$ scattering amplitudes $\Amp$ have definite crossing properties under interchange of $s$ and $u$ for fixed $t$, i.e.~under change of sign of $\nu$, such that one can work with amplitudes
\begin{equation}
\tilde{\Amp}(\nu^2,t)=\begin{cases}\Amp(\nu,t)\qquad\text{if }\Amp(\nu,t)=+\Amp(-\nu,t)\ec\\\frac{\Amp(\nu,t)}{\nu}\qquad\,\,\;\text{if }\Amp(\nu,t)=-\Amp(-\nu,t)\ec\end{cases}
\end{equation}
which are even functions of $\nu$ and thus free of kinematical square root branch cuts in the complex $t$-plane originating from $p_t$ or $q_t$.
Explicitly, the above amplitudes fulfill
\begin{equation}
\label{crossingamps}
A^\pm(\nu,t)=\pm A^\pm(-\nu,t)\ec \qquad B^\pm(\nu,t)=\mp B^\pm(-\nu,t)\ep
\end{equation}

The amplitudes of all ten $\pi N$ scattering reactions can be written in terms of only two independent matrix elements with total $s$-channel isospin index $I_s\in\{1/2,3/2\}$.
In agreement with~\cite{Hoehler} (i.e.~using the usual Condon--Shortley phase convention for the Clebsch--Gordan coefficients~\cite{PDG}, but the non-cyclic kinematical convention according to Sect.~\ref{subsec:preliminaries:kinematics}) we assign the isospin-doublets of both the nucleons and antinucleons according to the fundamental representation of the Lie-algebra of $SU(2)$
\begin{equation}
|p\rangle=\bigg|\frac{1}{2},\frac{1}{2}\bigg\rangle\ec \qquad |n\rangle=\bigg|\frac{1}{2},-\frac{1}{2}\bigg\rangle\ec \qquad 
|\bar{n}\rangle=\bigg|\frac{1}{2},\frac{1}{2}\bigg\rangle\ec \qquad |\bar{p}\rangle=\bigg|\frac{1}{2},-\frac{1}{2}\bigg\rangle\ec
\end{equation}
and the isospin-triplet of the pions according to
\begin{equation}
|\pi^+\rangle=|1,1\rangle\ec \qquad |\pi^0\rangle=|1,0\rangle\ec \qquad |\pi^-\rangle=|1,-1\rangle\ec
\end{equation}
which leads to the following properties under charge conjugation $C$
\begin{equation}
\label{chargeconjugation}
C|p\rangle=|\bar{p}\rangle\ec \qquad C|n\rangle=-|\bar{n}\rangle\ec \qquad C|\pi^\pm\rangle=-|\pi^\mp\rangle\ec \qquad C|\pi^0\rangle=|\pi^0\rangle\ep
\end{equation}
Thus, the relations between the spherical and the Cartesian components of the pion-multiplet are
\begin{equation}
|\pi^\pm\rangle=\mp\frac{1}{\sqrt{2}}(|\pi_1\rangle\pm i|\pi_2\rangle)\ec \qquad |\pi^0\rangle=|\pi_3\rangle\ep
\end{equation}
By decomposing the initial and final isospin states of the $\pi N$ system into linear combinations of $s$-channel isospin eigenstates, e.g.\ 
\begin{equation}
|\pi^+p\rangle=\bigg|\frac{3}{2},\frac{3}{2}\bigg\rangle\ec \qquad 
|\pi^-p\rangle=\sqrt{\frac{1}{3}}\bigg|\frac{3}{2},-\frac{1}{2}\bigg\rangle-\sqrt{\frac{2}{3}}\bigg|\frac{1}{2},-\frac{1}{2}\bigg\rangle\ec \qquad 
|\pi^0n\rangle=\sqrt{\frac{2}{3}}\bigg|\frac{3}{2},-\frac{1}{2}\bigg\rangle+\sqrt{\frac{1}{3}}\bigg|\frac{1}{2},-\frac{1}{2}\bigg\rangle\ec
\end{equation}
we can readily obtain the relations between the $\pi N$ isospin amplitudes
\begin{alignat}{4}
\label{sampisorels}
\Amp_+&= \Amp(\pi^+ p\to\pi^+ p)    &&=\Amp(\pi^-n\to\pi^-n) &&=\Amp^+-\Amp^-        &&=\Amp^{3/2}\ec\nt\\
\Amp_-&= \Amp(\pi^- p\to\pi^- p)    &&=\Amp(\pi^+n\to\pi^+n) &&=\Amp^++\Amp^-        &&=\frac{1}{3}(2\Amp^{1/2}+\Amp^{3/2})\ec\nt\\
\Amp_0&= \Amp(\pi^-p\to\pi^0n)      &&=\Amp(\pi^+n\to\pi^0p) &&=-\sqrt{2}\Amp^-      &&=-\frac{\sqrt{2}}{3}(\Amp^{1/2}-\Amp^{3/2})\ec\nt\\
      &\quad\;\Amp(\pi^0p\to\pi^0p) &&=\Amp(\pi^0n\to\pi^0n) &&=\Amp^+               &&=\frac{1}{3}(\Amp^{1/2}+2\Amp^{3/2})\ec\nt\\
      &                             &&                       &&\quad\;\Amp^++2\Amp^- &&=\Amp^{1/2}\ep
\end{alignat}
From these we can infer the so-called isospin triangle relation
\begin{equation}
\Amp_+-\Amp_-=\sqrt{2}\Amp_0\ec
\end{equation}
and the relations for the isospin even/odd amplitudes with $I=+/-$ and the amplitudes in the $s$-channel isospin basis $I_s\in\{1/2,3/2\}$ can be summarized in matrix notation as
\begin{equation}
\label{schannelcrossing}
\begin{pmatrix}\Amp^+\\\Amp^-\end{pmatrix}=C_{\nu s}\begin{pmatrix}\Amp^{1/2}\\\Amp^{3/2}\end{pmatrix}\ec \qquad \begin{pmatrix}\Amp^{1/2}\\\Amp^{3/2}\end{pmatrix}=C_{s\nu}\begin{pmatrix}\Amp^+\\\Amp^-\end{pmatrix}\ec \qquad C_{\nu s}=\frac{1}{3}C_{s\nu}=\frac{1}{3}\begin{pmatrix}1&2\\1&-1\end{pmatrix}\ep
\end{equation}
The $s$-channel isospin amplitudes with $I_s\in\{1/2,3/2\}$ and the corresponding $u$-channel isospin amplitudes with $I_u\in\{1/2=N,3/2=\Delta\}$ can be shown to obey the $s\leftrightarrow u$ crossing isospin relations
\begin{equation}
\label{uchannelcrossing}
\begin{pmatrix}\Amp^{1/2}\\\Amp^{3/2}\end{pmatrix}=C_{su}\begin{pmatrix}\Amp^N\\\Amp^\Delta\end{pmatrix}\ec \qquad \begin{pmatrix}\Amp^N\\\Amp^\Delta\end{pmatrix}=C_{us}\begin{pmatrix}\Amp^{1/2}\\\Amp^{3/2}\end{pmatrix}\ec \qquad C_{su}=C_{us}=\frac{1}{3}\begin{pmatrix}-1&4\\2&1\end{pmatrix}\ec
\end{equation}
and combining this with~\eqref{schannelcrossing} yields
\begin{equation}
\label{sanduchannelcrossing}
\begin{pmatrix}\Amp^+\\\Amp^-\end{pmatrix}=C_{\nu u}\begin{pmatrix}\Amp^N\\\Amp^\Delta\end{pmatrix}\ec \qquad 
C_{\nu u}=C_{\nu s}C_{su}=\frac{1}{3}\begin{pmatrix}1&2\\-1&1\end{pmatrix}\ec \qquad C_{u \nu}=C_{\nu u}^{-1}=\begin{pmatrix}1&-2\\1&1\end{pmatrix}\ep
\end{equation}

For the $t$-channel reactions, the $|\bar{N}N\rangle$ isospin states are superpositions of the states $|I_t=1,(I_t)_3\rangle$ and $|I_t=0,0\rangle$
\begin{equation}
\label{barNNisospinstates1}
|\bar{n}p\rangle=|1,1\rangle\ec \qquad 
|\bar{n}n\rangle=\frac{1}{\sqrt{2}}\big(|1,0\rangle+|0,0\rangle\big)\ec \qquad 
|\bar{p}p\rangle=\frac{1}{\sqrt{2}}\big(|1,0\rangle-|0,0\rangle\big)\ec \qquad 
|\bar{p}n\rangle=|1,-1\rangle\ec
\end{equation}
from which we can deduce\footnote{Note that~\eqref{barNNisospinstates1} and~\eqref{barNNisospinstates2} are in perfect agreement with the usual Clebsch--Gordan coefficients~\cite{PDG}, but differ from~\cite{Hoehler} wherein different conventions are used in these and corresponding equations. In particular, the analog of~\eqref{tampisorels} in~\cite{Hoehler} seems to (exceptionally) follow the cyclic kinematical convention. Nevertheless, all other relations, especially the crossing matrix~\eqref{tchannelcrossing} and the important relations~\eqref{ampspmtoampsIt01}, are identical.}
\begin{equation}
\label{barNNisospinstates2}
|1,0\rangle=\frac{1}{\sqrt{2}}(|\bar{n}n\rangle+|\bar{p}p\rangle)\ec \qquad 
|0,0\rangle=\frac{1}{\sqrt{2}}(|\bar{n}n\rangle-|\bar{p}p\rangle)\ec
\end{equation}
whereas the decomposition of the $|\pi\pi\rangle$ isospin states reads
\begin{align}
|\pi^+\pi^0\rangle&=\frac{1}{\sqrt{2}}(|2,1\rangle+|1,1\rangle)\ec & 
|\pi^+\pi^-\rangle&=\frac{1}{\sqrt{6}}|2,0\rangle+\frac{1}{\sqrt{2}}|1,0\rangle+\frac{1}{\sqrt{3}}|0,0\rangle\ec\nt\\
|\pi^-\pi^0\rangle&=\frac{1}{\sqrt{2}}(|2,-1\rangle-|1,-1\rangle)\ec & 
|\pi^0\pi^0\rangle&=\sqrt{\frac{2}{3}}|2,0\rangle-\frac{1}{\sqrt{3}}|0,0\rangle\ep
\end{align}
By strictly using the non-cyclic kinematical convention together with the properties under charge conjugation~\eqref{chargeconjugation} we can obtain the $t$-channel amplitudes from the $s$-channel ones via crossing
\begin{equation}
\Amp_\pm=-\Amp(\bar{p}p\to\pi^\pm\pi^\mp)\ec \qquad \Amp_0=\Amp(\pi^+n\to\pi^0p)=-\Amp(\bar{n}p\to\pi^+\pi^0)=\Amp(\bar{n}p\to\pi^0\pi^+)\ec
\end{equation}
which together with the $s$-channel isospin relations~\eqref{sampisorels} on the one hand and the $t$-channel isospin decompositions above on the other hand yields the following relations for the reactions with a proton as target particle
\begin{alignat}{4}
\label{tampisorels}
\Amp(\bar{p}p\to\pi^+\pi^-)&=-\Amp_+                    &&=-\Amp^++\Amp^- &&=-\Amp^{3/2}                               &&=-\frac{1}{\sqrt{6}}\Amp^0+\frac{1}{2}\Amp^1\ec\nt\\
\Amp(\bar{p}p\to\pi^-\pi^+)&=-\Amp_-                    &&=-\Amp^+-\Amp^- &&=-\frac{1}{3}(2\Amp^{1/2}+\Amp^{3/2})      &&=-\frac{1}{\sqrt{6}}\Amp^0-\frac{1}{2}\Amp^1\ec\nt\\
\Amp(\bar{n}p\to\pi^+\pi^0)&=-\Amp_0                    &&=\sqrt{2}\Amp^- &&=\frac{\sqrt{2}}{3}(\Amp^{1/2}-\Amp^{3/2}) &&=\frac{1}{\sqrt{2}}\Amp^1\ec\nt\\
\Amp(\bar{p}p\to\pi^0\pi^0)&=\frac{1}{2}(\Amp_++\Amp_-) &&=\Amp^+         &&=\frac{1}{3}(\Amp^{1/2}+2\Amp^{3/2})       &&=\frac{1}{\sqrt{6}}\Amp^0\ep
\end{alignat}
Thereby we can easily deduce the $s\leftrightarrow t$ crossing isospin relations
\begin{equation}
\label{tchannelcrossing}
\begin{pmatrix}\Amp^{1/2}\\\Amp^{3/2}\end{pmatrix}=C_{st}\begin{pmatrix}\Amp^0\\\Amp^1\end{pmatrix}\ec \quad C_{st}=\begin{pmatrix}\frac{1}{\sqrt{6}}&1\\\frac{1}{\sqrt{6}}&-\frac{1}{2}\end{pmatrix}\ec \qquad \begin{pmatrix}\Amp^0\\\Amp^1\end{pmatrix}=C_{ts}\begin{pmatrix}\Amp^{1/2}\\\Amp^{3/2}\end{pmatrix}\ec \quad C_{ts}=\frac{2}{3}\begin{pmatrix}\sqrt{\frac{3}{2}}&\sqrt{6}\\1&-1\end{pmatrix}\ec
\end{equation}
and the fact that $\Amp^+$ and $\Amp^-$ have well-defined quantum number $I_t=0$ and $I_t=1$, respectively,
\begin{equation}
\label{ampspmtoampsIt01}
\begin{pmatrix}\Amp^+\\\Amp^-\end{pmatrix}=C_{\nu t}\begin{pmatrix}\Amp^0\\\Amp^1\end{pmatrix}\ec \qquad C_{\nu t}=C_{\nu s}C_{st}=\begin{pmatrix}\frac{1}{\sqrt{6}}&0\\0&\frac{1}{2}\end{pmatrix}\ep
\end{equation}
Since
\begin{equation}
G|\pi\rangle=-|\pi\rangle \quad\Rightarrow\quad G|\pi\pi\rangle=|\pi\pi\rangle\ec
\end{equation}
the antinucleon--nucleon initial state in the reaction $\bar{N}N\to\pi\pi$ has to be an eigenstate of $G$-parity with eigenvalue $+1$, i.e.\ it can only couple to states with an even number of pions.
The result for charge conjugation of an antifermion--fermion or antiboson--boson pair
\begin{equation}
C|\bar ff\rangle=(-1)^{L+S}|\bar ff\rangle\ec \qquad C|\bar bb\rangle=(-1)^L|\bar bb\rangle\ec
\end{equation}
yields
\begin{equation}
G|\bar{N}N\rangle=(-1)^{J+I_t}|\bar{N}N\rangle\ec
\end{equation}
from which we can conclude that for reactions with a two-pion final state (i.e.\ $G=+1$) only the combinations ($J$ even, $I_t=0$) and ($J$ odd, $I_t=1$) are allowed. The same combinations arise from the symmetry properties of the symmetric isosinglet for $I_t=0$ and the antisymmetric isotriplet for $I_t=1$ due to the fact that the exchange of two pions in an orbital state with total angular momentum $J=L$ yields a factor of $(-1)^J$. According to~\eqref{ampspmtoampsIt01} this leads to the following selection rules for the partial-wave decomposition of the $t$-channel amplitudes: the partial-wave expansion of the amplitudes $\Amp^{I=+/-}$ or $\Amp^{I_t=0/1}$ contains only partial waves with even/odd $J$, respectively, and the transition between the two sets of amplitudes involves the isospin crossing coefficients $c_J$ with
\begin{equation}
\label{crossingcoefficients}
c_J=\begin{cases}\frac{1}{\sqrt{6}}\qquad\text{if }J\text{ is even}\ec\\\frac{1}{2}\,\,\,\,\qquad\text{if }J\text{ is odd}\ep\end{cases}
\end{equation}

\subsection{Hyperbolic dispersion relations}
\label{subsec:preliminaries:hdrs}

In~\cite{HiteSteiner} it was shown how to construct HDRs for the $\pi N$ scattering amplitudes, using hyperbolae in the Mandelstam plane of the form
\begin{equation}
(s-a)(u-a)=b\ec
\end{equation}
with hyperbola parameter $b$ and asymptotes $s=a$ and $u=a$.  They obey the relation
\begin{equation}
t=-\frac{b}{s-a}+\Sigma-s-a\ec
\end{equation}
and
\begin{align}
\label{sunuoftab}
s(t;a,b)&=\frac{1}{2}\big(\Sigma-t+4m\nu(t;a,b)\big)\ec & 4m\nu(t;a,b)&=\sqrt{(t-\Sigma+2a)^2-4b}\ec\nt\\
u(t;a,b)&=\frac{1}{2}\big(\Sigma-t-4m\nu(t;a,b)\big)\ec & t_{(\pm)}(\nu;a,b)&=\Sigma-2a\pm2\sqrt{(2m\nu)^2+b}\ep
\end{align}
In the following $b$ is considered as a function of $s$ and $t$ for a given value of $a$,
\begin{equation}
\label{bofsta}
b(s,t;a)=(s-a)(\Sigma-s-t-a)\ec
\end{equation}
and hence for given $s$ and $a$ one considers a family of hyperbolae wherein all members are uniquely defined by $t$.
Under the assumption that no subtractions are necessary (cf.\ Appendix~\ref{sec:regge}), the HDRs for the $\pi N$ scattering Lorentz-invariant amplitudes can be written as
\begin{alignat}{2}
\label{hdr}
A^+(s,t;a)&=\frac{1}{\pi}\int\limits_{s_+}^\infty\diff s'\left[\frac{1}{s'-s}+\frac{1}{s'-u}-\frac{1}{s'-a}\right]\Im A^+(s',t')
 +\frac{1}{\pi}\int\limits_{\tpi}^\infty\diff t'\;\frac{\Im A^+(s',t')}{t'-t}\ec\\
A^-(s,t;a)&=\frac{1}{\pi}\int\limits_{s_+}^\infty\diff s'\left[\frac{1}{s'-s}-\frac{1}{s'-u}\right]\Im A^-(s',t')
 +\frac{1}{\pi}\int\limits_{\tpi}^\infty\diff t'\;\frac{\nu}{\nu'}\frac{\Im A^-(s',t')}{t'-t}\ec\nt\\
B^+(s,t;a)&=N^+(s,t)+\frac{1}{\pi}\int\limits_{s_+}^\infty\diff s'\left[\frac{1}{s'-s}-\frac{1}{s'-u}\right]\Im B^+(s',t')
 +\frac{1}{\pi}\int\limits_{\tpi}^\infty\diff t'\;\frac{\nu}{\nu'}\frac{\Im B^+(s',t')}{t'-t}\ec\nt\\
B^-(s,t;a)&=N^-(s,t;a)+\frac{1}{\pi}\int\limits_{s_+}^\infty\diff s'\left[\frac{1}{s'-s}+\frac{1}{s'-u}-\frac{1}{s'-a}\right]\Im B^-(s',t')
 +\frac{1}{\pi}\int\limits_{\tpi}^\infty\diff t'\;\frac{\Im B^-(s',t')}{t'-t}\ec\nt
\end{alignat}
where we have defined the abbreviation
\begin{equation}
\nu'(s',t')=\nu(s',t')=\frac{2s'+t'-\Sigma}{4m}\ec
\end{equation}
and under the integrals one has to use
\begin{equation}
t'(s',s,t;a)=-\frac{b(s,t;a)}{s'-a}+\Sigma-s'-a\ec \qquad
 s'(t',s,t;a)=\frac{1}{2}\left[\Sigma-t'+\sqrt{(t'-\Sigma+2a)^2-4b(s,t;a)}\right]\ec
\end{equation}
since the {\em external kinematics} ($s,t,u$) and the {\em internal kinematics} ($s',t',u'$) are related by
\begin{equation}
\label{extbint}
(s-a)(u-a)=b=(s'-a)(u'-a)\ec \qquad s+t+u=\Sigma=s'+t'+u'\ep
\end{equation}
Only the amplitudes $B^\pm$ contain the Born-term contributions $N^\pm$ due to the nucleon poles given by (cf.~\cite{Hoehler} for $\bar N^\pm$)
\begin{align}
\label{hdrnucleonpoleterms}
N^+(s,t)&=\bar N^+(s,t)\ec &
 \bar N^+(s,t)&=g^2\left[\frac{1}{m^2-s}-\frac{1}{m^2-u}\right]=\frac{g^2}{m}\frac{\nu}{\nu_B^2-\nu^2}\ec\nt\\
N^-(s,t;a)&=\bar N^-(s,t)-\frac{g^2}{m^2-a}\ec &
 \bar N^-(s,t)&=g^2\left[\frac{1}{m^2-s}+\frac{1}{m^2-u}\right]=\frac{g^2}{m}\frac{\nu_B}{\nu_B^2-\nu^2}\ec
\end{align}
where the usual pseudoscalar $\pi N$ coupling constant $g$ and the pseudovector $\pi N$ coupling constant $f$ are given by\footnote{Note that~\cite{Hoehler} quotes a value of $14.28$ based on~\cite{Bugg}. For more information on conventions as well as the current value see~\cite{Nijmegen:1997,GWU:2006,piNcoupling:short,piNcoupling:long}.}
\begin{equation}
\label{piNcouplingvalue}
\frac{g^2}{4\pi}=\frac{4m^2f^2}{\mpi^2}\approx13.7\ep
\end{equation}
In order to express the integrands in terms of the corresponding CMS scattering angles according to
\begin{equation}
X(s',z_s')=X(s',t')\Big|_{t'=t'(s',z_s')}\ec \qquad X(t',z_t')=X(s',t')\Big|_{s'=s'(t',z_t')}\ec \qquad X\in\left\{A^\pm,B^\pm\right\}\ec
\end{equation}
we define
\begin{align}
\label{primedkinematics}
z_s'(s',t')&=z_s(s',t')=1+\frac{t'}{2q'^2}\ec \qquad q'(s')=q(s')\ec\nt\\
z_t'(s',t')&=z_t(s',t')=\frac{m\nu'}{p_t'q_t'}\ec \qquad p_t'(t')=p_t(t')=ip_-'(t')\ec \qquad q_t'(t')=q_t(t')=iq_-'(t')\ec
\end{align}
which leads to the relations
\begin{alignat}{2}
\label{internalkinematicsofsta}
t'(s',z_s')&=-2q'^2(1-z_s')\ec \qquad
 &&z_s'(s',s,t;a)=1-\frac{1}{2q'^2}\left[s'-\Sigma+a+\frac{b(s,t;a)}{s'-a}\right]\ec\nt\\
s'(t',z_t')&=\frac{1}{2}(\Sigma-t'+4p_t'q_t'z_t')\ec \qquad
 &&z_t'(t',s,t;a)=\frac{1}{4p_t'q_t'}\sqrt{(t'-\Sigma+2a)^2-4b(s,t;a)}\ep
\end{alignat}
Note that $b$ is linearly related to $z_s'$ for the $s$-channel, but only to $z_t'^2$ for the $t$-channel, which will have important consequences in Appendix~\ref{sec:convergence}, where it will be shown that the HDRs~\eqref{hdr} incorporate contributions from the direct as well as from the crossed channels, but not from double-spectral regions, provided the parameters are chosen appropriately.
Furthermore, one can check explicitly that $A^+$ and $B^-$ are indeed functions of $\nu^2$, while $A^-$ and $B^+$ are proportional to $\nu$.
Since moreover $4m\nu'=4p_t'q_t'z_t'=s'-u'$ is always real, one may also write the above HDRs~\eqref{hdr} in terms of reduced amplitudes $A^-/\nu$ and $B^+/\nu$, respectively.
This fact will be used in Sect.~\ref{sec:subRS} and Appendix~\ref{subsec:regge:asymptotics}.

In contrast, for usual fixed-$t$ dispersion relations external and internal kinematics are related by
\begin{equation}
t=t'\ec \qquad s+t+u=\Sigma=s'+t'+u'\ep
\end{equation}
It is remarkable that the HDRs have the simple form of~\eqref{hdr} and~\eqref{hdrnucleonpoleterms}, which by neglecting the terms depending on $a$ (or equivalently for $|a|\to\infty$) reduce to fixed-$t$ dispersion relations, provided, however, that the $t$-channel integrals are discarded.
Moreover, the hyperbolae then reduce to fixed-$t$ lines, and thus we will refer to the limit $|a|\to\infty$ as ``fixed-$t$ limit'' in the following.\footnote{As explained in Appendix~\ref{subsec:convergence:lehmannellipse}, only the limit $a\to-\infty$ is compatible with range-of-convergence considerations.}

\section{Roy--Steiner system for pion--nucleon scattering}
\label{sec:RS}

In this section, we first collect the results for the partial-wave hyperbolic dispersion relations (PWHDRs) that follow from the HDRs~\eqref{hdr} via partial-wave expansion in and projection onto both $s$- and $t$-channel partial waves as explained in detail in Appendices~\ref{sec:spwp} and~\ref{sec:tpwp}, in order to state the closed system  of RS equations for $\pi N$ scattering.
Then, we elaborate on the corresponding partial-wave unitarity relations for the $s$- and especially the $t$-channel.
Finally, we use the threshold behavior of the $t$-channel partial waves $f^J_\pm(t)$ in order to cast the $t$-channel part of the RS system in the form of a MO problem, whose solution will be the subject of Sect.~\ref{sec:piNMO}.

\subsection{Partial-wave hyperbolic dispersion relations}
\label{subsec:RS:pwhdr}

The $s$-channel partial-wave amplitudes are conventionally denoted by $f^I_{l\pm}(W)$ with isospin (i.e.\ crossing) index $I\in\{+,-\}$ and total angular momentum $j=l\pm1/2=l\pm$, where the orbital angular momentum can take the values $l\geq0$ for $j=l+$ and $l\geq1$ for $j=l-$.
Using a shorthand notation for the $z_s$-projections of the invariant amplitudes
\begin{equation}
\label{sprojinvampl}
X^I_l(s)=\int\limits_{-1}^1\diff z_s\;P_l(z_s)X^I(s,t)\Big|_{t=t(s,z_s)=-2q^2(1-z_s)} \qquad \text{for }X\in\{A,B\}\ec
\end{equation}
the well-known $s$-channel partial-wave projection formula reads~\cite{FrazerFulco:PWDR}
\begin{align}
\label{sprojform}
f^I_{l\pm}(W)&=\frac{1}{16\pi W}\Big\{(E+m)\big[A^I_l(s)+(W-m)B^I_l(s)\big]+(E-m)\big[-A^I_{l\pm1}(s)+(W+m)B^I_{l\pm1}(s)\big]\Big\}\ep
\end{align}
By construction, the $f^I_{l\pm}(W)$ obey the MacDowell symmetry relation~\cite{MacDowell} in the complex $W$-plane
\begin{equation}
\label{macdowell}
f^I_{l+}(W)=-f^I_{(l+1)-}(-W) \qquad \forall\;l\geq0\ec
\end{equation}
due to which only half of the complex $W$-plane is actually needed. Alternatively, this relation can be used the other way around to derive the partial waves with $j=l-$ from the ones with $j=l+$.
Expanding the absorptive parts of the HDRs~\eqref{hdr} into $s$-channel and $t$-channel partial waves, respectively, and subsequently projecting the full HDRs onto the $s$-channel partial waves $f^I_{l\pm}(W)$ yields the $s$-channel PWHDRs of~\cite{HiteSteiner}
\begin{align}
\label{spwhdr}
f^I_{l+}(W)&=N^I_{l+}(W)
+\frac{1}{\pi}\int\limits^\infty_{W_+}\diff W'\;\sum\limits_{l'=0}^\infty
 \Big\{K^I_{ll'}(W,W')\,\Im f^I_{l'+}(W')+K^I_{ll'}(W,-W')\,\Im f^I_{(l'+1)-}(W')\Big\}\nt\\
&\quad+\frac{1}{\pi}\int\limits^\infty_{\tpi}\diff t'\;\sum\limits_J
 \Big\{G_{lJ}(W,t')\,\Im f^J_+(t')+H_{lJ}(W,t')\,\Im f^J_-(t')\Big\}\nt\\
&=-f^I_{(l+1)-}(-W) \qquad \forall\;l\geq0\ec
\end{align}
which constitutes the $s$-channel part of the full RS system.
Here, $N^I_{l\pm}(W)$ represent the contributions due to the nucleon pole terms in the amplitudes $B^\pm$ as given in~\eqref{hdr}.
Each $s$-channel partial wave $f^I_{l\pm}(W)$ is coupled to the absorptive parts of all other $s$-channel partial waves via the kernels $K^I_{ll'}(W,W')$, which contain the usual Cauchy kernel responsible for the physical cut and an analytically known remainder (denoted by dots below) containing only left-hand cut contributions
\begin{equation}
K^I_{ll'}(W,W')=\frac{\delta_{ll'}}{W'-W}+\dots \quad\forall\;l,l'\geq0\ec
\end{equation}
as well as to the absorptive parts of the $t$-channel partial waves $f^J_\pm(t)$ via the kernels $G_{lJ}(W,t')$ and $H_{lJ}(W,t')$, where the lower index $\pm$ denotes parallel($+$) or antiparallel($-$) antinucleon--nucleon helicities and the total ($t$-channel) angular momentum  $J$ can take the values $J\geq0$ or $J\geq1$, respectively.
Due to Bose statistics (i.e.~crossing symmetry), the summations over $J$ in~\eqref{spwhdr} run over even/odd values of $J$ for the crossing even/odd partial waves (upper index $I=+/-$), respectively, as explained in Sect.~\ref{subsec:preliminaries:isospin}.
For the sake of completeness and convenience, in Appendix~\ref{sec:spwp} the different contributions to~\eqref{spwhdr} will be discussed along the lines of~\cite{HiteSteiner,BaackeSteiner,Steiner:piNpotential,Steiner:PWCR} (correcting several typographical errors, adjusting the conventions, and partially extending the presentation therein at the same time).

For the $t$-channel partial-wave projection, by virtue of  $s\leftrightarrow u$ crossing symmetry it is possible to use only half the interval in the cosine of the $t$-channel CMS scattering angle and thus the projection can be written as~\cite{FrazerFulco:tPW}
\begin{align}
\label{tprojform}
f^J_+(t)&=-\frac{1}{4\pi}\int\limits^1_0\diff z_t\;P_J(z_t)\bigg\{\frac{p_t^2}{(p_tq_t)^J}A^I(s,t)\Big|_{s=s(t,z_t)}-\frac{m}{(p_tq_t)^{J-1}}z_tB^I(s,t)\Big|_{s=s(t,z_t)}\bigg\}
 & &\forall\;J\geq0\ec\nt\\
f^J_-(t)&=\frac{1}{4\pi}\frac{\sqrt{J(J+1)}}{2J+1}\frac{1}{(p_tq_t)^{J-1}}\int\limits^1_0\diff z_t\Big[P_{J-1}(z_t)-P_{J+1}(z_t)\Big]B^I(s,t)\Big|_{s=s(t,z_t)}
 & &\forall\;J\geq1\ec
\end{align}
where again $I=+/-$ if $J$ is even/odd, such that the integrands are always functions of the squared angle $z_t^2$.
These formulae are valid literally only for $t\geq\tN$, but can actually be used for all kinematical cases, cf.\ the discussion following~\eqref{texpform}.
For a closed system of RS equations we need to derive the analog of~\eqref{spwhdr} for the $t$-channel partial waves $f^J_\pm(t)$, cf.~\cite{Roy,piK:RS,ggpipi}.
The result takes the form
\begin{align}
\label{tpwhdr}
f^J_+(t)&=\tilde N^J_+(t)+\frac{1}{\pi}\int\limits^{\infty}_{W_+}\diff W'\sum\limits^\infty_{l=0}\Big\{
\tilde G_{J l}(t,W')\,\Im f^I_{l+}(W')+\tilde G_{J l}(t,-W')\,\Im f^I_{(l+1)-}(W')\Big\}\nt\\
&\qquad+\frac{1}{\pi}\int\limits^{\infty}_{\tpi}\diff t'\sum\limits_{J'}\Big\{
\tilde K^1_{J J'}(t,t')\,\Im f^{J'}_+(t')+\tilde K^2_{J J'}(t,t')\,\Im f^{J'}_-(t')\Big\} \qquad \forall\;J\geq0\ec\nt\\
f^J_-(t)&=\tilde N^J_-(t)+\frac{1}{\pi}\int\limits^{\infty}_{W_+}\diff W'\sum\limits^\infty_{l=0}\Big\{
\tilde H_{J l}(t,W')\,\Im f^I_{l+}(W')+\tilde H_{J l}(t,-W')\,\Im f^I_{(l+1)-}(W')\Big\}\nt\\
&\qquad+\frac{1}{\pi}\int\limits^{\infty}_{\tpi}\diff t'\sum\limits_{J'}\tilde K^3_{J J'}(t,t')\,\Im f^{J'}_-(t') \qquad \forall\;J\geq1\ec
\end{align}
where again $I=+/-$ if $J$ is even/odd and the sum over $J'$ runs over even/odd values of $J'$ if $J$ is even/odd (cf.\ Sect.~\ref{subsec:preliminaries:isospin}).
As for the $s$-channel case, the kernels for the corresponding $t$-channel partial waves can be split into the Cauchy kernel and well-defined remainders
\begin{equation}
\tilde K^{1}_{JJ'}(t,t')=\frac{\delta_{JJ'}}{t'-t}+\dots \quad\forall\;J,J'\geq0\ec \qquad 
\tilde K^{3}_{JJ'}(t,t')=\frac{\delta_{JJ'}}{t'-t}+\dots \quad\forall\;J,J'\geq1\ec
\end{equation}
but, in contrast to the $s$-channel case, only higher $t$-channel partial waves can couple to lower ones, since $\tilde K^{1,2,3}_{JJ'}(t,t')=0$ for all $J'<J$, which will be a key ingredient in reducing the $t$-channel part~\eqref{tpwhdr} of the RS system to a MO problem in Sect.~\ref{subsec:RS:thrMOfJpm}.
The technical details of the derivation of the different contributions to~\eqref{tpwhdr} are relegated to Appendix~\ref{sec:tpwp}.

There are three aspects of convergence in the RS system of PWHDRs constructed in Appendices~\ref{sec:spwp} and~\ref{sec:tpwp}:
first, the question of convergence of the integrals in the high-energy regime is linked to the number of necessary subtractions of the dispersion relations, which will be discussed in Sect.~\ref{sec:subRS}.
Moreover, for the full system of RS equations to be valid, the convergence of
both the partial-wave expansion of the imaginary parts inside the integrals and the $s$- and $t$-channel partial-wave projection of the full HDR equations needs to be shown. Analyzing these two constraints yields the ranges of convergence in $s$ and $t$ for~\eqref{spwhdr} and~\eqref{tpwhdr}, respectively.
As explained in detail in Appendix~\ref{sec:convergence}, the hyperbola parameter $a$ can actually be tuned in order to obtain the largest possible domain of validity.
For the $s$-channel part~\eqref{spwhdr} of the RS system the combined analysis of $s$- and $t$-channel constraints leads to an optimal value of $a$ and a corresponding range of convergence in $s$ of (cf.\ Appendix~\ref{subsec:convergence:sprojection})
\begin{equation}
a=-23.19\,\mpi^2 \quad\Rightarrow\quad
s\in\big[s_+=(m+\mpi)^2,97.30\,\mpi^2\big] \quad\Leftrightarrow\quad
W\in[W_+=1.08\GeV,1.38\GeV]\ec
\end{equation}
where $s_+=59.64\,\mpi^2$, while for the $t$-channel part~\eqref{tpwhdr} we find (cf.\ Appendix~\ref{subsec:convergence:tprojection})
\begin{equation}
a=-2.71\,\mpi^2 \quad\Rightarrow\quad
t\in[\tpi=4\mpi^2,205.45\,\mpi^2] \quad\Leftrightarrow\quad
\sqrt{t}\in[\sqrt{\tpi}=0.28\GeV,2.00\GeV]\ep
\end{equation}
Note that different choices of $a$ for the $s$- and $t$-channel partial-wave projections are perfectly justified, as we may start from different sets of HDRs.
However, the choice of $a$ is not only crucial for the ranges of convergence, but also influences the high-energy behavior of the imaginary parts, whose estimation via Regge asymptotics is discussed in Appendix~\ref{sec:regge}.
For this purpose one splits the corresponding integration ranges $s_+\leq s'\leq\infty$ and $\tpi\leq t'\leq\infty$ of the HDRs~\eqref{hdr} at some appropriate values $\sa=\Wa^2$ and $\ta$, respectively, in order to describe the asymptotic $s$- and $t$-channel contributions to the invariant amplitudes in terms of Regge amplitudes.
The remaining non-asymptotic parts are then given by the corresponding integrals over $s_+\leq s'\leq\sa$ and $\tpi\leq t'\leq\ta$, respectively, plus the nucleon pole terms $N^I(s,t)$ for the amplitudes $B^I(s,t)$.
However, eventually the high-energy region is of only little practical relevance, in particular if subtractions are performed in order to suppress the dependence on higher energies (cf.\ Sect.~\ref{sec:subRS}). 

In order to use partial-wave unitarity relations that are diagonal in the $s$-channel partial waves, we have to work in the  $s$-channel isospin basis $I_s\in\{1/2,3/2\}$ rather than in the isospin even/odd basis $I=+/-$ (as will be explained in Sect.~\ref{subsec:RS:unirel}), and therefore in analogy to~\eqref{schannelcrossing} we define
\begin{equation}
\begin{pmatrix}X^{1/2}\\X^{3/2}\end{pmatrix}=C_{s\nu}\begin{pmatrix}X^+\\X^-\end{pmatrix}\ec \qquad
\begin{pmatrix}X^+\\X^-\end{pmatrix}=C_{\nu s}\begin{pmatrix}X^{1/2}\\X^{3/2}\end{pmatrix}\ec \qquad
\text{for}\;X\in\{f_{l\pm},N_{l\pm},K_{ll'}\}\ec
\end{equation}
and the abbreviation
\begin{equation}
K_{ll'}^{1/2+3/2}(W,W')=K_{ll'}^{1/2}(W,W')+K_{ll'}^{3/2}(W,W')=2K_{ll'}^+(W,W')+K_{ll'}^-(W,W')\ep
\end{equation}
The full closed RS system of PWDRs for both $s$- and $t$-channel partial waves in the corresponding isospin bases $I_s\in\{1/2,3/2\}$ and $I_t\in\{0,1\}$ that follows from rewriting~\eqref{spwhdr} and~\eqref{tpwhdr} reads\footnote{All sums run over both even and odd values, and the formulae for the $f^{I}_{(l+1)-}$ are given explicitly for convenience.}
\begin{align}
\label{sRSpwhdr}
f^{1/2}_{l+}(W)&=N^{1/2}_{l+}(W)+\frac{1}{\pi}\int\limits^{\infty}_{W_+}\diff W'\sum\limits^\infty_{l'=0}\frac{1}{3}
\Big\{K_{ll'}^{1/2}(W,W')\,\Im f^{1/2}_{l'+}(W')+2K_{ll'}^{3/2}(W,W')\,\Im f^{3/2}_{l'+}(W')\nt\\
&\qquad+K_{ll'}^{1/2}(W,-W')\,\Im f^{1/2}_{(l'+1)-}(W')+2K_{ll'}^{3/2}(W,-W')\,\Im f^{3/2}_{(l'+1)-}(W')\Big\}\nt\\
&\quad+\frac{1}{\pi}\int\limits^{\infty}_{\tpi}\diff t'\sum\limits_{J=0}^{\infty}\frac{\big(3-(-1)^J\big)}{2}
\Big\{G_{lJ}(W,t')\,\Im f^J_+(t')+H_{lJ}(W,t')\,\Im f^J_-(t')\Big\}\ec\nt\\
f^{3/2}_{l+}(W)&=N^{3/2}_{l+}(W)+\frac{1}{\pi}\int\limits^{\infty}_{W_+}\diff W'\sum\limits^\infty_{l'=0}\frac{1}{3}
\Big\{K_{ll'}^{3/2}(W,W')\,\Im f^{1/2}_{l'+}(W')+K_{ll'}^{1/2+3/2}(W,W')\,\Im f^{3/2}_{l'+}(W')\nt\\
&\qquad+K_{ll'}^{3/2}(W,-W')\,\Im f^{1/2}_{(l'+1)-}(W')+K_{ll'}^{1/2+3/2}(W,-W')\,\Im f^{3/2}_{(l'+1)-}(W')\Big\}\nt\\
&\quad+\frac{1}{\pi}\int\limits^{\infty}_{\tpi}\diff t'\sum\limits_{J=0}^{\infty}(-1)^J
\Big\{G_{lJ}(W,t')\,\Im f^J_+(t')+H_{lJ}(W,t')\,\Im f^J_-(t')\Big\}\ec\nt\\
f^{1/2}_{(l+1)-}(W)&=N^{1/2}_{(l+1)-}(W)\nt\\
&\quad-\frac{1}{\pi}\int\limits^{\infty}_{W_+}\diff W'\sum\limits^\infty_{l'=0}\frac{1}{3}
\Big\{K_{ll'}^{1/2}(-W,W')\,\Im f^{1/2}_{l'+}(W')+2K_{ll'}^{3/2}(-W,W')\,\Im f^{3/2}_{l'+}(W')\nt\\
&\qquad+K_{ll'}^{1/2}(-W,-W')\,\Im f^{1/2}_{(l'+1)-}(W')+2K_{ll'}^{3/2}(-W,-W')\,\Im f^{3/2}_{(l'+1)-}(W')\Big\}\nt\\
&\quad-\frac{1}{\pi}\int\limits^{\infty}_{\tpi}\diff t'\sum\limits_{J=0}^{\infty}\frac{\big(3-(-1)^J\big)}{2}
\Big\{G_{lJ}(-W,t')\,\Im f^J_+(t')+H_{lJ}(-W,t')\,\Im f^J_-(t')\Big\}\ec\nt\\
f^{3/2}_{(l+1)-}(W)&=N^{3/2}_{(l+1)-}(W)\nt\\
&\quad-\frac{1}{\pi}\int\limits^{\infty}_{W_+}\diff W'\sum\limits^\infty_{l'=0}\frac{1}{3}
\Big\{K_{ll'}^{3/2}(-W,W')\,\Im f^{1/2}_{l'+}(W')+K_{ll'}^{1/2+3/2}(-W,W')\,\Im f^{3/2}_{l'+}(W')\nt\\
&\qquad+K_{ll'}^{3/2}(-W,-W')\,\Im f^{1/2}_{(l'+1)-}(W')+K_{ll'}^{1/2+3/2}(-W,-W')\,\Im f^{3/2}_{(l'+1)-}(W')\Big\}\nt\\
&\quad-\frac{1}{\pi}\int\limits^{\infty}_{\tpi}\diff t'\sum\limits_{J=0}^{\infty}(-1)^J
\Big\{G_{lJ}(-W,t')\,\Im f^J_+(t')+H_{lJ}(-W,t')\,\Im f^J_-(t')\Big\}\ec
\end{align}
together with
\begin{align}
\label{tRSpwhdr}
f^J_+(t)&=\tilde N^J_+(t)+\frac{1}{\pi}\int\limits^{\infty}_{W_+}\diff W'\sum\limits^\infty_{l=0}\frac{1}{3}
\Big\{\tilde G_{J l}(t,W')\Big[\Im f^{1/2}_{l+}(W')+\frac{1+3(-1)^J}{2}\Im f^{3/2}_{l+}(W')\Big]\nt\\
&\qquad+\tilde G_{J l}(t,-W')\Big[\Im f^{1/2}_{(l+1)-}(W')+\frac{1+3(-1)^J}{2}\Im f^{3/2}_{(l+1)-}(W')\Big]\Big\}\nt\\
&\quad+\frac{1}{\pi}\int\limits^{\infty}_{\tpi}\diff t'\sum\limits^\infty_{J'=J}\frac{1+(-1)^{J+J'}}{2}
\Big\{\tilde K^1_{J J'}(t,t')\,\Im f^{J'}_+(t')+\tilde K^2_{J J'}(t,t')\,\Im f^{J'}_-(t')\Big\} \qquad \forall\;J\geq0\ec\nt\\
f^J_-(t)&=\tilde N^J_-(t)+\frac{1}{\pi}\int\limits^{\infty}_{W_+}\diff W'\sum\limits^\infty_{l=0}\frac{1}{3}
\Big\{\tilde H_{J l}(t,W')\Big[\Im f^{1/2}_{l+}(W')+\frac{1+3(-1)^J}{2}\Im f^{3/2}_{l+}(W')\Big]\nt\\
&\qquad+\tilde H_{J l}(t,-W')\Big[\Im f^{1/2}_{(l+1)-}(W')+\frac{1+3(-1)^J}{2}\Im f^{3/2}_{(l+1)-}(W')\Big]\Big\}\nt\\
&\quad+\frac{1}{\pi}\int\limits^{\infty}_{\tpi}\diff t'\sum\limits^\infty_{J'=J}\frac{1+(-1)^{J+J'}}{2}
\tilde K^3_{J J'}(t,t')\,\Im f^{J'}_-(t') \qquad \forall\;J\geq1\ep
\end{align}
Note that in the above $t$-channel part~\eqref{tRSpwhdr} the sums over $J'$ are limited to $J'\geq J$ due to~\eqref{vanishingtildeK}.

\subsection{Partial-wave unitarity relations}
\label{subsec:RS:unirel}

From the unitarity of the $S$-matrix $S=\unity+i\,T$ one can easily obtain the general unitarity relation by taking matrix elements and inserting a complete set of intermediate states
\begin{equation}
\label{genunitarityrel}
\langle f|T|i\rangle-\langle f|T^\dagger|i\rangle=i\sum\limits_{\{j\}}\int\diff\Pi^{(j)}_{n_j}\langle f|T^\dagger|j\rangle\langle j|T|i\rangle\ec
\end{equation}
where $\diff\Pi^{(j)}_{n_j}$ denotes the $n_j$-particle Lorentz-invariant phase space (LIPS) for intermediate state $j$, which in the case of $n_j$ identical intermediate particles implicitly includes an additional symmetry factor $1/S^{(j)}_{n_j}=1/n_j!$ in order to avoid multiple counting in the phase space integral.
Imposing overall 4-momentum conservation $\delta^{(4)}(\Sigma p_f-\Sigma p_i)$ and using time-reversal invariance of the strong interactions immediately yields the generalized optical theorem for the dimensionless invariant amplitudes $T_{fi}$
\begin{equation}
\label{genopttheorem}
\Im T_{fi}=\frac{1}{2}\sum\limits_{\{j\}}\int\diff\Pi^{(j)}_{n_j}(2\pi)^4\delta^{(4)}(\Sigma p_j-\Sigma p_i)T_{fj}^*T_{ji}\ep
\end{equation}
Under the additional assumption of hermitian analyticity of the $S$-matrix (i.e.\ the amplitudes $T_{fi}$ obey the Schwarz reflection principle $T_{fi}^*(s)=T_{fi}(s^*)$ and are real on part of the real axis) it follows
\begin{equation}
\label{discontinuity}
\Disc T_{fi}(s)=\lim_{\epsilon\to0}\big[T_{fi}(s+i\epsilon)-T_{fi}(s-i\epsilon)\big]
=2i\lim_{\epsilon\to0}\Im T_{fi}(s+i\epsilon)\ec
\end{equation}
for the physical limit corresponding to the $s$-channel process, and hence~\eqref{genopttheorem} may also be proven in the framework of perturbation theory to all orders.
By normalizing the 4-momentum states according to $\langle p'|p\rangle=2E_{\mathbf{p}}(2\pi)^3\delta^{(3)}(\mathbf{p}'-\mathbf{p})$ for both bosons and fermions,
for generic two-by-two scattering $ab\to cd$ with one particular intermediate 2-particle state $j=j_1j_2$ (with CMS 3-momentum modulus $p_j$) and after partial integration of the 2-particle LIPS the optical theorem~\eqref{genopttheorem} takes the form
\begin{equation}
\label{2ptrevinvopttheorem}
\Im T_{fi}=\frac{1}{S^{(j)}_2}\frac{1}{16\pi}\frac{2p_j}{\sqrt{s}}\int\frac{\diff\Omega_j}{4\pi}T_{fj}^*T_{ji}\ec
\qquad
p_j=\sqrt{\frac{\lambda_s^{j_1j_2}}{4s}}\ec
\end{equation}
leading to the usual form of the differential cross section (with $p_f$ and $p_i$ in analogy to $p_j$)
\begin{equation}
\label{cms2by2dcs}
\frac{\diff\sigma_{fi}}{\diff\Omega}=
\frac{p_fp_i}{\pi}\frac{\diff\sigma_{fi}}{\diff t}=
\frac{p_f}{p_i}\bigg|\frac{T_{fi}}{8\pi\sqrt{s}}\bigg|^2\ep
\end{equation}

A partial-wave decomposition of the invariant amplitudes $T_{fi}$ allows for a reduction of the unitarity constraint~\eqref{2ptrevinvopttheorem} to unitarity relations for each partial wave separately.
In the presence of spin the $T$ operator for two-by-two scattering can be diagonalized by using the eigenstates of total angular momentum $J$ as basis, which can be achieved most easily in the CMS via the helicity formalism~\cite{JacobWick}.
With $\lambda_P$ denoting the helicity of the corresponding particle, one can take the $T$-matrix elements in the basis of single particle momenta and helicities and by applying the respective phase space integration in the CMS, the corresponding invariant helicity amplitudes $T^{\lambda_c,\lambda_d;\lambda_a,\lambda_b}_{fi}$ can be written in terms of states of relative motion for both incoming and outgoing particle pairs. Thereby, the differential cross section for a reaction with a given set of helicities can be derived in full analogy to~\eqref{cms2by2dcs}.
With the usual angular conventions of~\cite{JacobWick,Hoehler} and the azimuthal angle $\varphi$ set to zero, the partial-wave expansion of these helicity amplitudes in the helicity basis then reads
\begin{equation}
\label{diffmodJWpwe}
T^{\lambda_c,\lambda_d;\lambda_a,\lambda_b}_{fi}(s,t)=\sqrt{S_fS_i}16\pi\sum\limits_J(2J+1)T^J_{\lambda_c,\lambda_d;\lambda_a,\lambda_b}(s)d^J_{\lambda_a-\lambda_b,\lambda_c-\lambda_d}(\theta)\ec
\end{equation}
where $d^j_{mm'}(\theta)$ are the Wigner $d$-functions\footnote{A comprehensive review on Wigner functions, in particular a comparison of different angular conventions used in the literature, is given in~\cite{VarshalovichMK}.} and the sum runs over integer/half-integer values of $J$ for an even/odd number of half-integer spins present in the initial or final state.
In the case of spinless particles with $d^J_{00}(\theta)=P_J(\cos\theta)$ and $J=l$ the expansion simplifies to
\begin{equation}
\label{diffmodJWpwenospin}
T_{fi}(s,t)=\sqrt{S_fS_i}16\pi\sum\limits_{J=0}^\infty(2J+1)T_{fi}^J(s)P_J(\cos\theta)\ep
\end{equation}
Note that we have added here explicit symmetry factors $S_i$ and $S_f$ to the partial-wave expansion of~\cite{JacobWick} in order to take care of identical particles in the initial and final state in a symmetric fashion.
This normalization reproduces the standard normalizations for spinless processes as well as for $\pi N\to\pi N$, and furthermore ensures that no symmetry factors occur in the elastic unitarity relations for the partial waves, since they always cancel with the symmetry factor implicitly included in the LIPS (cf.~\eqref{2ptrevinvopttheorem}).
We will explicitly demonstrate the effect of this convention for the symmetry factors on the extended unitarity relation for $\pi\pi\to\bar NN$ partial waves by considering $\pi\pi\to\pi\pi$ with $\bar KK$ and $\bar NN$ intermediate states below.

Due to the invariance of strong interactions under time reversal and parity, the helicity partial waves obey the symmetry properties
\begin{equation}
T^J_{\lambda_a,\lambda_b;\lambda_c,\lambda_d}(s)=T^J_{\lambda_c,\lambda_d;\lambda_a,\lambda_b}(s)=T^J_{-\lambda_c,-\lambda_d;-\lambda_a,-\lambda_b}(s)\ep
\end{equation}
If the particles are spinless or if the matrix $T^J(s)$ in helicity space is diagonal in some appropriate basis (as it is e.g.\ for $\pi N\to\pi N$ in the $s$-channel isospin basis $I_s\in\{1/2,3/2\}$), the unitarity relation~\eqref{2ptrevinvopttheorem} for partial waves of generic elastic scattering $ab\to ab$ (i.e.\ $f=j=i$) reads
\begin{equation}
\label{elpwunitrel}
\Im T^J_{fi}(s)=\frac{2p}{\sqrt{s}}\big|T^J_{fi}(s)\big|^2\ec
\end{equation}
which is solved by a parameterization of $T^J_{fi}(s)$ via the real phase shift $\delta^J_{fi}(s)$
\begin{equation}
\label{elpwunitnorm}
T^J_{fi}(s)=\frac{\sqrt{s}}{2p}\sin\delta^J_{fi}(s)e^{i\delta^J_{fi}(s)}\ec
\end{equation}
where~\eqref{elpwunitrel} and~\eqref{elpwunitnorm} are valid for each diagonal element $T^J_{fi}(s)$ of $T^J(s)$.
For $s$ above the lowest inelastic threshold $s_\text{inel}$ these equations have to be modified by introducing real inelasticities $0\leq\eta^J_{fi}(s)\leq1$ according to
\begin{equation}
\label{inelpwunitarity}
T_{fi}^J(s)=\frac{\sqrt{s}}{2p}\frac{\eta^J_{fi}(s)e^{2i\delta^J_{fi}(s)}-1}{2i}\ec \qquad
\Im T^J_{fi}(s)=\frac{2p}{\sqrt{s}}\big|T^J_{fi}(s)\big|^2+\frac{\sqrt{s}}{8p}\Big[1-\big(\eta^J_{fi}(s)\big)^2\Big]\ec
\end{equation}
with $\eta^J_{fi}(s)<1$ for $s>s_\text{inel}$ due to additional intermediate states contributing in~\eqref{genunitarityrel}.
These partial waves are then related to the diagonal elements of the corresponding $S$-matrix via
\begin{equation}
S^J_{fi}(s)=\eta^J_{fi}(s)e^{2i\delta^J_{fi}(s)}=1+i\frac{4p}{\sqrt{s}}T^J_{fi}(s)\ep
\end{equation}

After these general remarks, we now turn to $\pi N$ scattering:
the reduced $s$-channel partial-wave amplitudes $f^I_{l\pm}(W)$ in the $s$-channel isospin basis $I_s\in\{1/2,3/2\}$ are conventionally normalized according to (cf.~\eqref{inelpwunitarity} and e.g.~\cite{Hoehler,BecherLeutwyler})
\begin{equation}
f^{I_s}_{l\pm}(W)=\frac{1}{q}\frac{\big[S^{I_s}_{l\pm}(W)\big]_{\pi N\to\pi N}-1}{2i}=\frac{1}{q}\frac{\eta^{I_s}_{l\pm}(W)e^{2i\delta^{I_s}_{l\pm}(W)}-1}{2i}
\overset{W<W_\text{inel}}{=}\frac{\sin\delta^{I_s}_{l\pm}(W)}{q}e^{i\delta^{I_s}_{l\pm}(W)}\ec
\end{equation}
where for the elastic form we have used the fact that the lowest inelastic intermediate state is $\pi\pi N$ and thus $\eta^{I_s}_{l\pm}(W)=1$ below the inelastic threshold $W_\text{inel}=W_++\mpi$.
The $s$-channel partial-wave unitarity relation corresponding to the normalization given above reads
\begin{equation}
\label{selunitrel}
\Im f^{I_s}_{l\pm}(W)=q\big|f^{I_s}_{l\pm}(W)\big|^2\,\theta\big(W-W_+\big)+\frac{1-\big(\eta^{I_s}_{l\pm}(W)\big)^2}{4q}\,\theta\big(W-W_\text{inel}\big)\ec
\end{equation}
leading to the branch cut for $W>W_+$.

For the (necessarily inelastic) $t$-channel partial-wave unitarity relations one needs the dimensionless partial-wave amplitudes $t^{I_t}_J(t)$ of elastic $\pi\pi$ scattering.
They are conventionally defined from the dimensionless isospin amplitudes of $\pi\pi\to\pi\pi$ via (with $t$-channel isospin $I_t\in\{0,1,2\}$, total angular momentum $J=l$, and symmetry factors $\sqrt{S_fS_i}=2$ for identical pions, cf.~\eqref{diffmodJWpwenospin} and~\cite{ACGL,BecherLeutwyler})
\begin{equation}
T^{I_t}(s,t)=32\pi\sum\limits_{J=0}^\infty(2J+1)t^{I_t}_J(t)P_J(\cos\theta^{\pi\pi})\ec
\end{equation}
that are normalized according to
\begin{equation}
\frac{\diff\sigma^{I_t}_{\pi\pi\to\pi\pi}}{\diff\Omega}=\left|\frac{T^{I_t}(s,t)}{8\pi\sqrt{t}}\right|^2\ep
\end{equation}
The corresponding elastic unitarity relation then takes the form
\begin{equation}
\label{pipielunitrel}
\Im t^{I_t}_J(t)=\sigma^\pi_t\big|t^{I_t}_J(t)\big|^2\,\theta\big(t-\tpi\big)\ec \qquad \sigma^\pi_t=\frac{2q_t}{\sqrt{t}}=\sqrt{1-\frac{\tpi}{t}}\ec
\end{equation}
and hence the partial waves can be parameterized as
\begin{equation}
\label{pipipwa}
t^{I_t}_J(t)=\frac{1}{\sigma^\pi_t}\frac{\big[S^{I_t}_J(t)\big]_{\pi\pi\to\pi\pi}-1}{2i}=\frac{1}{\sigma^\pi_t}\frac{\eta^{I_t}_J(t)e^{2i\delta^{I_t}_J(t)}-1}{2i}\overset{\eta^{I_t}_J(t)=1}{=}\frac{\sin\delta^{I_t}_J(t)}{\sigma^\pi_t}e^{i\delta^{I_t}_J(t)}\ep
\end{equation}

The reduced $t$-channel $\pi N$ partial-wave amplitudes $f^J_\pm(t)$ are related to $\pi N$ helicity amplitudes $F_{\bar\lambda\lambda}(s,t)$ and dimensionless partial waves $F^J_\pm(t)$ via (cf.~\cite{Hoehler,FrazerFulco:tPW})
\begin{align}
\label{barNNtopipiJWpwe}
F_{++}(s,t)&=\;\;\; F_{--}(s,t)=\frac{4\pi\sqrt{t}}{q_t}\sum\limits_{J=0}^\infty(2J+1)F^J_+(t)P_J(\cos\theta_t)\ec &
 F_+^J(t)&=\frac{q_t}{p_t}(p_tq_t)^J\frac{2}{\sqrt{t}}f_+^J(t)\ec\nt\\
F_{+-}(s,t)&=-F_{-+}(s,t)=\frac{4\pi\sqrt{t}}{q_t}\sum\limits_{J=1}^\infty\frac{2J+1}{\sqrt{J(J+1)}}F^J_-(t)\sin\theta_tP_J'(\cos\theta_t)\ec &
 F_-^J(t)&=\frac{q_t}{p_t}(p_tq_t)^Jf_-^J(t)\ec
\end{align}
and they are normalized according to
\begin{equation}
\label{barNNtopipidiffcrosssect}
\frac{\diff\bar\sigma_{\pi\pi\to\bar NN}}{\diff\Omega}=
\frac{p_t}{q_t}\sum\limits_{\bar\lambda,\lambda}\left|\frac{F_{\bar\lambda\lambda}(s,t)}{8\pi\sqrt{t}}\right|^2=
\frac{2p_t}{q_t}\left\{\left|\frac{F_{++}(s,t)}{8\pi\sqrt{t}}\right|^2+\left|\frac{F_{+-}(s,t)}{8\pi\sqrt{t}}\right|^2\right\}=
\frac{4p_t^2}{q_t^2}\frac{\diff\bar\sigma_{\bar NN\to\pi\pi}}{\diff\Omega}\ep
\end{equation}
The general formulae~\eqref{barNNtopipiJWpwe} and~\eqref{barNNtopipidiffcrosssect} are also valid for isospin even/odd parts $F^I_{\bar\lambda\lambda}(s,t)$ with crossing index $I=+/-$ and $J$ even/odd, accordingly.
Note that when referring to the $t$-channel isospin basis $I_t\in\{0,1\}$ as in the following, the isospin crossing coefficients $c_J$ of~\eqref{crossingcoefficients} need to be included.
In general, the $t$-channel partial waves may be parameterized as
\begin{equation}
\label{barNNtopipipwa}
f^J_\pm(t)=\big|f^J_{\pm}(t)\big|e^{i\varphi^J(t)}=\Re f^J_{\pm}(t)+i\,\Im f^J_{\pm}(t)\ep
\end{equation}
By considering only $\pi\pi$ intermediate states in the region $t<(4\mpi)^2$ (which is elastic with respect to $\pi\pi$ scattering, but unphysical with respect to the $\pi N$ $t$-channel) in the general unitarity relation~\eqref{genunitarityrel} for $\bar NN\to\pi\pi$, the $f^J_\pm(t)$ can be shown to obey the ``elastic'' $t$-channel unitarity relation
\begin{equation}
\label{tunitrel}
\Im f^J_{\pm}(t)=\sigma^\pi_t\big(t^{I_t}_J(t)\big)^*f^J_{\pm}(t)\,\theta\big(t-\tpi\big) \qquad \forall\;t\in[\tpi,16\mpi^2)
\end{equation}
(where the coefficients $c_J$ cancel), which leads to the branch cut for $t>\tpi$.
Since the imaginary part $\Im f^J_{\pm}(t)$ itself must be real, from~\eqref{tunitrel} together with~\eqref{pipipwa} and~\eqref{barNNtopipipwa} one can immediately infer 
\begin{equation}
\label{watson}
f^J_{\pm}(t)=\big|f^J_{\pm}(t)\big|e^{i\delta^{I_t}_J(t)} \qquad \forall\;t\in[\tpi,16\mpi^2)\ec
\end{equation}
i.e.\ the phases of the $t$-channel partial waves $f^J_{\pm}(t)$ are given by the phases of the $\pi\pi$ partial waves $t^{I_t}_J(t)$ modulo $\pi$ (by convention we choose the phases to coincide exactly), which is also known as Watson's final state interaction theorem~\cite{Watson}.
It is common practice to assume that the contributions due to $4\pi$ and other intermediate states can safely be ignored for $t\lesssim40\mpi^2\approx0.78\GeV^2$ (see e.g.~\cite{Hoehler,HiteKaufmannJacob}). However, as demonstrated in~\cite{ACCGL} in the context of the scalar pion form factor, this is certainly only true in the $S$-wave below the threshold $\tK=4\mK^2\approx0.97\GeV^2$ for the production of $\bar KK$ intermediate states, while in the $P$-wave inelasticities effectively start to set in around the $\pi\omega$ threshold at $0.85\GeV^2$.

It is crucial to note that~\eqref{tunitrel} is invariant under rescaling of $f^J_{\pm}(t)$ with real factors, whereas elastic unitarity relations as~\eqref{selunitrel} (for $W<W_\text{inel}$) and~\eqref{pipielunitrel} are always nonlinear in the corresponding partial wave.
Hence, fixing the normalization of all different partial waves that are needed in extended $t$-channel unitarity relations (i.e.\ allowing for additional intermediate states) in a consistent manner can only be done resorting to the corresponding elastic reactions, as we will now demonstrate for a system of coupled-channel equations with $\pi$, $K$, and $N$ degrees of freedom.
Writing $T_{11}=T_{\pi\pi\to\pi\pi}$, $T_{12}=T_{\bar KK\to\pi\pi}$, $T_{13}=T_{\bar NN\to\pi\pi}$ etc.\ for the $T$-matrix elements and using the invariance of strong interactions under time reversal, the general unitarity relation reads in terms of matrix elements
\begin{equation}
S_{fj}^{*}S_{ji}=\delta_{fi}\ec \qquad S_{fi}=\delta_{fi}+iT_{fi}=\delta_{if}+iT_{if}=S_{if}\ep
\end{equation}
In particular, one can read off the extended elastic unitarity relation for $\pi\pi\to\pi\pi$ and the extended unitarity relation for $\bar NN\to\pi\pi$ with $\pi\pi$, $\bar KK$, and $\bar NN$ intermediate states
\begin{equation}
\label{smatrixelemunitrel}
\delta_{11}=1=|S_{11}|^2+|S_{12}|^2+|S_{13}|^2\ec \qquad \delta_{13}=0=S_{11}^{*}S_{13}+S_{12}^{*}S_{23}+S_{13}^{*}S_{33}\ec
\end{equation}
and thus, by dropping the $\bar NN$ intermediate states in the second relation (since we are finally interested in the extended $t$-channel unitarity relation of $\pi N$ scattering in the region below the $\bar NN$ threshold), we obtain
\begin{equation}
\label{tmatrixelemunitrel}
2\,\Im T_{11}=|T_{11}|^2+|T_{12}|^2+|T_{13}|^2\ec \qquad 2\,\Im T_{13}=T_{11}^{*}T_{13}+T_{12}^{*}T_{23}\ep
\end{equation}
Introducing now the reduced $t$-channel partial waves $g^{I_t}_J(t)$ of $\pi K$ scattering (with isospin $I_t=0/1$ corresponding to $J=l$ even/odd due to Bose symmetry in $\pi\pi$, symmetry factors $\sqrt{S_fS_i}=\sqrt{2}$ and the partial waves defined from dimensionless isospin amplitudes, cf.~\eqref{diffmodJWpwenospin} and~\cite{piK:RS})
\begin{equation}
G^{I_t}(s,t)=16\pi\sqrt{2}\sum\limits_{J=0}^\infty(2J+1)(k_tq_t)^Jg^{I_t}_J(t)P_J(\cos\theta_t^{\pi K})\ec \qquad k_t=\sqrt{\frac{t}{4}-\mK^2}=\frac{\sqrt{t}}{2}\sigma^K_t\ec
\end{equation}
the first relation of~\eqref{tmatrixelemunitrel} may be decomposed into partial waves, and performing the angular integrations of the phase space integrals leads to the partial-wave unitarity relation for $\pi\pi$ scattering with $\pi\pi$, $\bar KK$, and $\bar NN$ intermediate states
\begin{equation}
\label{pipiextelunitrel}
\Im t^{I_t}_J(t)=\sigma^\pi_t\Big|t_J^{I_t}(t)\Big|^2\,\theta\big(t-\tpi\big)+(k_tq_t)^{2J}\sigma^K_t\Big|g_J^{I_t}(t)\Big|^2\,\theta\big(t-\tK\big)
+\frac{t}{16q_t^2}\frac{\sigma^N_t}{c_J^2}\bigg\{\Big|F_+^J(t)\Big|^2+\Big|F_-^J(t)\Big|^2\bigg\}\theta\big(t-\tN\big)\ep
\end{equation}
For $t<\tK$ (or if $I_t+J$ equals an odd number) this reproduces the elastic unitarity relation for $\pi\pi$ scattering~\eqref{pipielunitrel}, which corresponds to the relation (cf.~\eqref{pipipwa})
\begin{equation}
\big[S^{I_t}_J(t)\big]_{\pi\pi\to\pi\pi}=1+i\frac{4q_t}{\sqrt{t}}t^{I_t}_J(t)\,\theta\big(t-\tpi\big)\ep
\end{equation}
Comparing~\eqref{pipiextelunitrel} with the elastic unitarity relation for the partial waves (cf.~\eqref{smatrixelemunitrel})
\begin{equation}
\label{smatrixpwaunitrel}
\Big|\big[S^{I_t}_J(t)\big]_{\pi\pi\to\pi\pi}\Big|^2+\Big|\big[S^{I_t}_J(t)\big]_{\pi\pi\to\bar KK}\Big|^2
+2\left\{\Big|\big[S^J_+(t)\big]^{I_t}_{\pi\pi\to\bar NN}\Big|^2+\Big|\big[S^J_-(t)\big]^{I_t}_{\pi\pi\to\bar NN}\Big|^2\right\}=1
\end{equation}
for the cases $t<\tN$ and $t\geq\tN$ successively then allows to fix the normalization of the partial-wave $S$-matrix elements of the inelastic channels (both in the natural $t$-channel isospin basis $I_t\in\{0,1\}$) to\footnote{Note that our symmetric normalization of the helicity partial waves~\eqref{diffmodJWpwe} together with~\eqref{barNNtopipiJWpwe} and~\eqref{barNNtopipidiffcrosssect} leads to an additional factor of $1/\sqrt{2}$ to $\big[S^J_\pm(t)\big]^{I_t}_{\pi\pi\to\bar NN}$ in comparison with~\cite{Hoehler,FrazerFulco:tPW}, where one should read in addition $\big[S^J_\pm(t)\big]_{\pi\pi\to\bar NN}\equiv c_J\big[S^J_\pm(t)\big]^{I_t}_{\pi\pi\to\bar NN}$.}
\begin{equation}
\label{ppKKppNNnorm}
\big[S^{I_t}_J(t)\big]_{\pi\pi\to\bar KK}=i\frac{4(k_tq_t)^{J+\frac{1}{2}}}{\sqrt{t}}g^{I_t}_J(t)\,\theta\big(t-\tK\big)\ec \qquad
\big[S^J_\pm(t)\big]^{I_t}_{\pi\pi\to\bar NN}=\frac{i}{c_J\sqrt{2}}\sqrt{\frac{p_t}{q_t}}F^J_\pm(t)\,\theta\big(t-\tN\big)\ep
\end{equation}
These $S$-matrix elements indeed reproduce the correctly normalized differential cross sections
\begin{equation}
\frac{\diff\sigma^{I_t}_{\pi\pi\to\bar KK}}{\diff\Omega}=\frac{k_t}{q_t}\left|\frac{G^{I_t}(s,t)}{8\pi\sqrt{t}}\right|^2 
\end{equation}
and~\eqref{barNNtopipidiffcrosssect}, respectively.
Furthermore, from the unitarity bound of the $t$-channel partial-wave $S$ matrix of $\pi N$ scattering only (cf.~\eqref{smatrixpwaunitrel}) together with its explicit form~\eqref{ppKKppNNnorm} and the relations~\eqref{barNNtopipiJWpwe} to the corresponding partial waves $f^J_\pm$ we can deduce that the partial waves fall off asymptotically at least as fast as (cf.~\cite{Hoehler})
\begin{equation}
\label{fJpmasymptotics}
f^J_+(t)\sim t^{-J+\frac{1}{2}}\ec \qquad f^J_-(t)\sim t^{-J}\ec \qquad \text{for }\;t\to\infty\ec
\end{equation}
i.e.\ $f^J_\pm(t)\to0$ for $t\to\infty$ by unitarity at least for all $J>0$\;; this asymptotic vanishing is usually assumed to hold for the $S$-wave as well.
By virtue of similar considerations, the normalization of the remaining partial waves in the second relation of~\eqref{tmatrixelemunitrel} can be fixed.
We may introduce the reduced $t$-channel partial waves $h^J_\pm(t)$ of $KN$ scattering in analogy to the $\pi N$ case via dimensionless helicity amplitudes (cf.~\eqref{barNNtopipiJWpwe} and~\cite{MHammerD})
\begin{align}
H_{++}(s,t)&=
\frac{4\pi\sqrt{t}}{k_t}\sum\limits_{J=0}^\infty(2J+1)H^J_+(t)P_J(\cos\theta_t^{KN})\ec &
 H_+^J(t)&=\frac{k_t}{p_t}(p_tk_t)^J\frac{2}{\sqrt{t}}h_+^J(t)\ec\nt\\
H_{+-}(s,t)&=
\frac{4\pi\sqrt{t}}{k_t}\sum\limits_{J=1}^\infty\frac{2J+1}{\sqrt{J(J+1)}}H^J_-(t)\sin\theta_t^{KN}P_J'(\cos\theta_t^{KN})\ec &
 H_-^J(t)&=\frac{k_t}{p_t}(p_tk_t)^Jh_-^J(t)\ec
\end{align}
where it is important to note that, in contrast to $\pi N$ scattering, also the combinations $I_t=0$ with odd $J$ and $I_t=1$ with even $J$ are allowed due to lack of Bose symmetry in $\bar KK$.
In order not to bloat the notation, we refrain from using an additional index for $I_t$, and in the following
e.g.\ $h_\pm^{J=\text{even/odd}}$ is always to be understood as $h_\pm^{(J=\text{even/odd},I_t=0/1)}$, respectively, and not $h_\pm^{(J=\text{odd/even},I_t=0/1)}$, since only the former can couple to the $t$-channel process $\pi\pi\to\bar NN$.
The normalization is fixed by
\begin{equation}
\frac{\diff\bar\sigma_{K\bar K\to\bar NN}}{\diff\Omega}=
\frac{p_t}{k_t}\sum\limits_{\bar\lambda,\lambda}\left|\frac{H_{\bar\lambda\lambda}(s,t)}{8\pi\sqrt{t}}\right|^2=
\frac{2p_t}{k_t}\left\{\left|\frac{H_{++}(s,t)}{8\pi\sqrt{t}}\right|^2+\left|\frac{H_{+-}(s,t)}{8\pi\sqrt{t}}\right|^2\right\}=
\frac{4p_t^2}{k_t^2}\frac{\diff\bar\sigma_{\bar NN\to K\bar K}}{\diff\Omega}\ec
\end{equation}
so that the dimensionless partial-wave amplitudes $H_\pm^J(t)$ are related to the diagonal elements of the corresponding $S$-matrix according to
\begin{equation}
\label{KKNNnorm}
\big[S^J_\pm(t)\big]^{I_t}_{K\bar K\to\bar NN}=\frac{i}{c_J^{KN}}\sqrt{\frac{p_t}{k_t}}H^J_\pm(t)\,\theta\big(t-\tN\big)\ec \qquad c_J^{KN}=\frac{1}{2} \quad \forall\;J\ep
\end{equation}
Plugging the partial-wave matrix elements into the partial-wave projection of the second relation of~\eqref{tmatrixelemunitrel} for either parallel or antiparallel antinucleon--nucleon helicities yields the extended $t$-channel unitarity relation for the $t$-channel partial waves $f^J_{\pm}(t)$ (extending~\eqref{tunitrel} for $\bar KK$ intermediate states)
\begin{equation}
\label{exttunitrel}
\Im f^J_{\pm}(t)=\sigma^\pi_t\big(t^{I_t}_J(t)\big)^*f^J_{\pm}(t)\,\theta\big(t-\tpi\big)
+2c_J\sqrt{2}\,k_t^{2J}\sigma^K_t\big(g^{I_t}_J(t)\big)^*h^J_\pm(t)\,\theta\big(t-\tK\big)\ep
\end{equation}

Finally, we can use~\eqref{smatrixpwaunitrel} to derive the inelasticities $\eta_J^{I_t}(t)$ of the $\pi\pi$ scattering amplitude that are consistent with~\eqref{exttunitrel}. Below the $\bar NN$ threshold, inserting~\eqref{pipipwa} and~\eqref{ppKKppNNnorm} into~\eqref{smatrixpwaunitrel} leads to
\begin{equation}
\eta_J^{I_t}(t)=\sqrt{1-4\sigma_t^\pi\sigma_t^K(k_tq_t)^{2J}\big|g^{I_t}_J(t)\big|^2\,\theta\big(t-\tK\big)}\ep
\end{equation}

\subsection{From Roy--Steiner equations to the Muskhelishvili--Omn\`es problem}
\label{subsec:RS:thrMOfJpm}

\subsubsection[Threshold behavior of the $t$-channel partial waves]{Threshold behavior of the \boldmath{$t$}-channel partial waves}
\label{subsubsec:RS:thrfJpm}

The asymptotic behavior of $f^J_\pm(t)$ for $p_t\to0$ and $q_t\to0$ (which is equivalent to $t\to\tN=4m^2$ and $t\to\tpi=4\mpi^2$, respectively) can be derived directly from the partial-wave projection~\eqref{tprojform}.
Since $A^I(t,z_t)$ and $B^I(t,z_t)$ have definite symmetry properties under $s\leftrightarrow u$ and since $s-u=4m\nu=4p_tq_tz_t$, we can write down the expansions
\begin{equation}
A^I(t,z_t)=\sum\limits_{J'}(p_tq_t)^{J'}P_{J'}(z_t)a_{J'}(t)\ec \qquad B^I(t,z_t)=\sum\limits_{J'}(p_tq_t)^{J'}P_{J'}(z_t)b_{J'}(t)\ec
\end{equation}
where only even/odd values of $J'$ contribute according to the symmetry properties of $A^I$ and $B^I$ (i.e.\ even $J'$ for $A^+$, $B^-$ and odd $J'$ for $A^-$, $B^+$).
Let us first consider the limit $p_t\to0$, i.e.\ the behavior of $f^J_\pm(t)$ at the $t$-channel threshold $\tN$.
As far as the leading asymptotic behavior is concerned, the functions $a_{J'}(t)$ and $b_{J'}(t)$ can be evaluated at $t=\tN$ and will thus be considered as constant coefficients in the following.
Inserting these expansions into~\eqref{tprojform} (where $J$ even/odd corresponds to $I=+/-$), we find for $J=0$ that
\begin{equation}
\label{thresholdf0p}
f^0_+(t\to\tN)=\Ord(p_t^2)
\end{equation}
at the physical threshold, while for $J\geq1$ we obtain
\begin{align}
f^J_+(t\to\tN)&=\frac{m b_{J-1}}{8\pi}\int\limits^1_{-1}\diff z_tP_J(z_t)z_tP_{J-1}(z_t)+\Ord(p_t^2)=
\frac{m b_{J-1}}{8\pi}\frac{J}{2J+1}\frac{2}{2J-1}+\Ord(p_t^2)\ec\\
f^J_-(t\to\tN)&=\frac{b_{J-1}}{8\pi}\frac{\sqrt{J(J+1)}}{2J+1}\int\limits^1_{-1}\diff z_tP_{J-1}(z_t)P_{J-1}(z_t)+\Ord(p_t^2)=
\frac{b_{J-1}}{8\pi}\frac{\sqrt{J(J+1)}}{2J+1}\frac{2}{2J-1}+\Ord(p_t^2)\ec\nt
\end{align}
such that
\begin{equation}
f^J_+(t\to\tN)=\Ord(1)\ec \qquad f^J_-(t\to\tN)=\Ord(1)\ec \qquad \forall\;J\geq1\ep
\end{equation}
However, the linear combination
\begin{equation}
\label{Gammadef}
\Gamma^J(t)=m\sqrt{\frac{J}{J+1}}f^J_-(t)-f^J_+(t)\qquad\forall\;J\geq1
\end{equation}
 vanishes at threshold (cf.~\cite{FrazerFulco:NFF,Pietarinen,Hoehler})
\begin{equation}
\label{thresholdfJpm}
\Gamma^J(t\to\tN)=\Ord(p_t^2)\qquad\forall\;J\geq1\ep
\end{equation}
The same reasoning may be applied to the limit $q_t\to0$ as well, but as $A^I$ contributes at the same order as $B^I$ in the expansion of $f^J_+(t)$, no relation between the threshold values of different amplitudes may be inferred for $q_t\to0$
\begin{equation}
f^J_+(t\to\tpi)=\Ord(1)\qquad\forall\;J\geq0\ec \qquad f^J_-(t\to\tpi)=\Ord(1)\qquad\forall\;J\geq1\ep
\end{equation}

In fact, the properties of $f^J_\pm(t)$ at the $t$-channel threshold are crucial to ensure convergence in the RS equations.
From the partial-wave expansion~\eqref{texpform} we can easily derive the leading contributions to the invariant amplitudes (given explicitly for $J\leq2$)
\begin{align}
\frac{A^+(\nu,t)}{4\pi}&=-\frac{f^0_+(t)}{p_t^2}+\frac{15}{2}m^2\nu^2\frac{\Gamma^2(t)}{p_t^2}+\frac{5}{2}q_t^2f^2_+(t)+\dots\ec&
\frac{A^-(\nu,t)}{4\pi}&=3m\nu\frac{\Gamma^1(t)}{p_t^2}+\dots\ec\nt\\
\frac{B^+(\nu,t)}{4\pi}&=\frac{15}{\sqrt{6}}m\nu f^2_-(t)+\dots\ec&
\frac{B^-(\nu,t)}{4\pi}&=\frac{3}{\sqrt{2}}f^1_-(t)+\dots\ec
\end{align}
demonstrating how the threshold behavior~\eqref{thresholdf0p} and~\eqref{thresholdfJpm} ensures that the partial-wave expansion does not introduce spurious kinematical poles at $p_t\to0$ into the expansion of the invariant amplitudes and thereby into the HDRs~\eqref{hdr}.
To illustrate the consequences of this point, we briefly comment on the several places in our RS system~\eqref{spwhdr} and~\eqref{tpwhdr} where the threshold behavior of $f^J_\pm(t)$ features:
\begin{enumerate}
\item
Although $G_{lJ}(W,t')$ and $H_{lJ}(W,t')$ diverge as $p_t'^{-2}$ for $t'\to\tN$ according to~\eqref{GHasymptotics}, the relation
\begin{equation}
\Res\Big[H_{lJ}(W,t'),t'=\tN\Big]=-m\sqrt{\frac{J}{J+1}}\Res\Big[G_{lJ}(W,t'),t'=\tN\Big]
\end{equation}
together with~\eqref{thresholdf0p} and~\eqref{thresholdfJpm} ensures that the corresponding integrals in~\eqref{spwhdr} are well defined. We have checked that the explicit expressions in~\eqref{GH_explicit} fulfill this equation.
\item
The $p_t'^{-2}$ divergence~\eqref{tildeKasymptotics} of $\tilde K^1_{JJ'}(t,t')$ and $\tilde K^2_{JJ'}(t,t')$ for $t'\to\tN$ cancels in~\eqref{tpwhdr} provided that
\begin{equation}
\label{threshold_tildeK12}
\Res\Big[\tilde K^2_{JJ'}(t,t'),t'=\tN\Big]=-m\sqrt{\frac{J'}{J'+1}}\Res\Big[\tilde K^1_{JJ'}(t,t'),t'=\tN\Big]\ep
\end{equation}
This relation can easily be verified for the kernels given in~\eqref{tildeKJJ} and~\eqref{tildeK02}, cf.~\eqref{tildeK12rel}.
\item
Based on the asymptotic forms~\eqref{tchannel_pole_asym} of the pole-term projections $\tilde N^J_\pm(t)$, one may check their threshold behavior to be analogous to~\eqref{thresholdf0p} and~\eqref{thresholdfJpm}. Note that in this special case the relations hold for $q_t\to0$ as well, since $A^I$ does not contribute to the pole terms:
\begin{equation}
\label{threshold_tildeN}
\tilde N^0_+(p_tq_t\to0)=\Ord(p_t^2q_t^2)\ec\qquad
m\sqrt{\frac{J}{J+1}}\tilde N^J_-(p_tq_t\to0)-\tilde N^J_+(p_tq_t\to0)=\Ord(p_t^2q_t^2)\qquad\forall\;J\geq1\ep
\end{equation}
\end{enumerate}

\subsubsection[Muskhelishvili--Omn\`es problem for the $t$-channel partial waves]{Muskhelishvili--Omn\`es problem for the \boldmath{$t$}-channel partial waves}
\label{subsubsec:RS:MOfJpm}

Using the properties of the kernel functions for $t$-channel exchange as given in Appendix~\ref{subsec:tpwp:t} together with the threshold behavior of the partial waves as discussed in Sect.~\ref{subsubsec:RS:thrfJpm}, we can rewrite the $t$-channel part~\eqref{tpwhdr} of the (unsubtracted) RS system as
\begin{align}
\label{MOsub0}
f^0_+(t)&=\Delta^0_+(t)
-\frac{1}{\pi}\int\limits^\infty_{\tpi}\diff t'\frac{\Im f^0_+(t')}{t'-\tN}
+\frac{1}{\pi}\int\limits^\infty_{\tpi}\diff t'\frac{\Im f^0_+(t')}{t'-t}\ec\nt\\
f^J_+(t)&=\Delta^J_+(t)
+\frac{1}{\pi}\int\limits^{\infty}_{\tpi}\diff t'\frac{m\sqrt{\frac{J}{J+1}}\Im f^J_-(t')-\Im f^J_+(t')}{t'-\tN}
+\frac{1}{\pi}\int\limits^\infty_{\tpi}\diff t'\frac{\Im f^J_+(t')}{t'-t}\qquad\forall\;J\geq1\ec\nt\\
f^J_-(t)&=\Delta^J_-(t)
+\frac{1}{\pi}\int\limits^\infty_{\tpi}\diff t'\frac{\Im f^J_-(t')}{t'-t}\qquad\forall\;J\geq1\ec
\end{align}
where we have defined the abbreviations
\begin{align}
\label{Deltadef}
\Delta^J_\pm(t)&=\tilde N^J_\pm(t)+\bar\Delta^J_\pm(t)\ec\nt\\
\bar\Delta^J_+(t)&=\frac{1}{\pi}\int\limits^{\infty}_{W_+}\diff W'\sum\limits^\infty_{l=0}
\Big\{\tilde G_{Jl}(t,W')\,\Im f^I_{l+}(W')+\tilde G_{Jl}(t,-W')\,\Im f^I_{(l+1)-}(W')\Big\}\nt\\
&\qquad+\frac{1}{\pi}\int\limits^{\infty}_{\tpi}\diff t'\sum^{\infty}\limits_{J'=J+2}\frac{1+(-1)^{J+J'}}{2}
\Big\{\tilde K^1_{JJ'}(t,t')\,\Im f^{J'}_+(t')+\tilde K^2_{JJ'}(t,t')\,\Im f^{J'}_-(t')\Big\}\qquad\forall\;J\geq0\ec\nt\\
\bar\Delta^J_-(t)&=\frac{1}{\pi}\int\limits^{\infty}_{W_+}\diff W'\sum\limits^\infty_{l=0}
\Big\{\tilde H_{Jl}(t,W')\,\Im f^I_{l+}(W')+\tilde H_{Jl}(t,-W')\,\Im f^I_{(l+1)-}(W')\Big\}\nt\\
&\qquad+\frac{1}{\pi}\int\limits^{\infty}_{\tpi}\diff t'\sum^{\infty}\limits_{J'=J+2}\frac{1+(-1)^{J+J'}}{2}
\tilde K^3_{JJ'}(t,t')\,\Im f^{J'}_-(t')\qquad\forall\;J\geq1\ec
\end{align}
for the inhomogeneities $\Delta^J_\pm(t)$, which besides the $t$-channel projections $\tilde N^J_\pm(t)$ of the nucleon pole terms contain the coupling to all $s$-channel partial waves as well as to the higher $t$-channel partial waves.
Note that $\Delta^J_\pm(t)$ only contains the left-hand cut and therefore is real for all $t\geq\tpi$\;.
By virtue of~\eqref{Gammadef} and the analogous definition
\begin{equation}
\Delta_\Gamma^J(t)=m\sqrt{\frac{J}{J+1}}\Delta^J_-(t)-\Delta^J_+(t)\ec
\end{equation}
the equations~\eqref{MOsub0} can be cast into the form of a MO problem for $f^0_+(t)$, $f^J_-(t)$, and the linear combinations $\Gamma^J(t)$ 
\begin{align}
\label{MOGamsub0}
f^0_+(t)&=\Delta^0_+(t)
+\frac{t-\tN}{\pi}\int\limits^\infty_{\tpi}\diff t'\frac{\Im f^0_+(t')}{(t'-\tN)(t'-t)}\ec\nt\\
\Gamma^J(t)&=\Delta_\Gamma^J(t)
+\frac{t-\tN}{\pi}\int\limits^\infty_{\tpi}\diff t'\frac{\Im\Gamma^J(t')}{(t'-\tN)(t'-t)}\qquad\forall\;J\geq1\ec\nt\\
f^J_-(t)&=\Delta^J_-(t)
+\frac{1}{\pi}\int\limits^\infty_{\tpi}\diff t'\frac{\Im f^J_-(t')}{t'-t}\qquad\forall\;J\geq1\ec
\end{align}
where for $f^0_+(t)$ and $\Gamma^J(t)$ combining the integrals effectively yields one subtraction at the threshold $\tN$ and the additional roots at $t'=\tN$ in the denominators are canceled by the threshold behavior of the numerators.
The solution for $f^J_+(t)$ can then easily be recovered via~\eqref{Gammadef}.

How these equations (or their subtracted analogs derived in Sect.~\ref{subsec:subRS:MO}) can be used to determine $f^0_+(t)$, $f^1_\pm(t)$, and $f^2_\pm(t)$ with the help of MO techniques will be described in the following sections.
Note that such an easy rewriting scheme is not possible for the $s$-channel part~\eqref{spwhdr} of the RS PWHDRs, since in the corresponding $s$-channel integrals also $0\leq l'\leq l$ contribute.

\section{Subtracted Roy--Steiner system for pion--nucleon scattering}
\label{sec:subRS}

The Froissart--Martin bound~\cite{Froissart,Martin:Froissart} limits the number of subtractions necessary for the convergence of the integrals in the high-energy regime to 2, since the total cross section does not increase faster than $\log^2s$ for $s\to\infty$.\footnote{While the original Froissart bound assumes validity of the Mandelstam representation for the scattering amplitude, the result by Martin is based on somewhat less restrictive assumptions.}
The influence of the high-energy contributions to dispersion integrals may be reduced by means of suitable subtractions for the trade-off of introducing corresponding subtraction polynomials with subtraction constants that are a priori unknown.
For the MO integrals in~\eqref{MOGamsub0} subtracting in $t$ at subtraction points below $\tpi$ with the additional constraint $s=u$ in order to preserve crossing symmetry is favorable.
A particularly useful choice is the subthreshold expansion, which amounts to subtracting in $t$ at zero:
first, it is very convenient for extrapolation to the Cheng--Dashen point in order to elaborate on the $\pi N$ $\sigma$ term (cf.\ Sect.~\ref{subsec:preliminaries:kinematics}); second, subtracting at the subthreshold point facilitates matching to chiral perturbation theory, which is expected to work best in the subthreshold region.\footnote{For the application of heavy-baryon ChPT to $\pi N$ scattering in the subthreshold region see~\cite{piNintriangle}. Conversely, analyticity and unitarity are used in~\cite{GasparyanLutz} to stabilize the extrapolation of $\pi N$ partial waves derived from ChPT amplitudes in the subthreshold region into the physical region, thus enabling the determination of the chiral parameters by matching to experimental information in terms of $s$-channel phase shifts.}
To this end, we first briefly review the subthreshold expansion of the scattering amplitudes and then discuss its application in order to write down both the once- and twice-subtracted form of the HDRs~\eqref{hdr}.

\subsection{Subthreshold expansion}
\label{subsec:subRS:ste}

The subthreshold expansion refers to the expansion of Born-subtracted amplitudes around the subthreshold point $(s=u=s_0,t=0)=(\nu=0,t=0)$ (cf.\ Sect.~\ref{subsec:preliminaries:kinematics}), where the nucleon pole terms are subtracted since they are rapidly varying in this kinematical region.
Subtracting the {\it pseudovector} Born terms (indicated by bars) yields
\begin{align}
\bar A^+(s,t)&=A^+(s,t)-\frac{g^2}{m}\ec & 
\bar B^+(s,t)&=B^+(s,t)-g^2\left[\frac{1}{m^2-s}-\frac{1}{m^2-u}\right]\ec\nt\\
\bar A^-(s,t)&=A^-(s,t)\ec & 
\bar B^-(s,t)&=B^-(s,t)-g^2\left[\frac{1}{m^2-s}+\frac{1}{m^2-u}\right]+\frac{g^2}{2m^2}\ec
\end{align}
while for the {\it pseudoscalar} Born-subtracted (indicated by tildes) amplitudes $\tilde A^\pm$ and $\tilde B^\pm$ the terms $-g^2/m$ and $+g^2/2m^2$ need to be dropped (cf.~\ref{hdrnucleonpoleterms}).
Due to the crossing symmetry of the amplitudes~\eqref{crossingamps} (similarly for HDRs~\eqref{hdr}) one can write the subthreshold expansion generically for crossing-even amplitudes as (cf.~\cite{Hoehler})
\begin{equation}
X(\nu,t)=\sum\limits_{m,n}x_{mn}\big(\nu^2\big)^mt^n\ec \qquad
X\in\left\{\bar{A}^+,\tilde{A}^+,\frac{\bar{A}^-}{\nu},\frac{\tilde{A}^-}{\nu},\frac{\bar{B}^+}{\nu},\frac{\tilde{B}^+}{\nu},\bar{B}^-,\tilde{B}^-,\bar{D}^+,\tilde{D}^+,\frac{\bar{D}^-}{\nu},\frac{\tilde{D}^-}{\nu}\right\}\ec
\end{equation}
and thus explicitly for the pseudovector Born-subtracted amplitudes as
\begin{align}
\bar A^+(\nu,t)&=\sum\limits_{m,n=0}^\infty a_{mn}^+\nu^{2m}t^{n}\ec & 
\bar B^+(\nu,t)&=\sum\limits_{m,n=0}^\infty b_{mn}^+\nu^{2m+1}t^{n}\ec\nt\\
\bar A^-(\nu,t)&=\sum\limits_{m,n=0}^\infty a_{mn}^-\nu^{2m+1}t^{n}\ec & 
\bar B^-(\nu,t)&=\sum\limits_{m,n=0}^\infty b_{mn}^-\nu^{2m}t^{n}\ec
\end{align}
where the corresponding subthreshold parameters of the amplitudes $\bar D^\pm=\bar A^\pm+\nu\bar B^\pm$ are related by
\begin{equation}
d_{mn}^+=a_{mn}^++b_{m-1,n}^+\ec \qquad d_{mn}^-=a_{mn}^-+b_{mn}^-\ep
\end{equation}
Note that due to $b_{-1,n}^+=0$ in particular
\begin{equation}
d_{0n}^+=a_{0n}^+\ep
\end{equation}
From the expansions
\begin{equation}
\label{subthrexpkernelfracs}
\frac{1}{s'-s}-\frac{1}{s'-u}=\frac{4m\nu}{(s'-s_0)^2}+\Ord\big(\nu^3,\nu t\big)\ec \qquad
\frac{1}{s'-s}+\frac{1}{s'-u}=\frac{2}{s'-s_0}-\frac{t}{(s'-s_0)^2}+\Ord\big(\nu^2,\nu^2t,t^2\big)\ec
\end{equation}
one then can read off the subthreshold expansions of the Born-unsubtracted amplitudes up to and including first order
\begin{align}
\label{subthrexp}
A^+(\nu,t)&=\frac{g^2}{m}+d_{00}^++d_{01}^+t+\Ord\big(\nu^2,\nu^2t,t^2\big)\ec\nt\\ 
A^-(\nu,t)&=\nu a_{00}^-+\Ord\big(\nu^3,\nu t\big)\ec \qquad
 B^+(\nu,t)=g^2\frac{4m\nu}{(m^2-s_0)^2}+\nu b_{00}^++\Ord\big(\nu^3,\nu t\big)\ec\nt\\
B^-(\nu,t)&=g^2\left[\frac{2}{m^2-s_0}-\frac{t}{(m^2-s_0)^2}\right]-\frac{g^2}{2m^2}+b_{00}^-+b_{01}^-t+\Ord\big(\nu^2,\nu^2t,t^2\big)\ep
\end{align}

\subsection{Sum rules for subthreshold parameters}
\label{subsec:subRS:sumrules}

Subtracting simultaneously at $s_0=\Sigma/2<s_+$ and $t_0=0<\tpi$ corresponds to the subthreshold expansion around $(\nu=0,t=0)$ and thus allows for the determination of sum rules for the subthreshold parameters.
Matching the expansions~\eqref{subthrexp} to the corresponding expansions of the HDRs~\eqref{hdr} by equating the coefficients (where it is crucial to keep track of all implicit dependencies in the expansions) together with introducing the abbreviation
\begin{equation}
h_0(s')=\frac{2}{s'-s_0}-\frac{1}{s'-a}
\end{equation}
then yields the following sum rules for the lowest subthreshold parameters
{\allowdisplaybreaks
\begin{align}
\label{sumrules}
d_{00}^+&=-\frac{g^2}{m}+\frac{1}{\pi}\int\limits_{s_+}^\infty\diff s'\,h_0(s')\ste{\Im A^+(s',z_s')}
+\frac{1}{\pi}\int\limits_{\tpi}^\infty\frac{\diff t'}{t'}\ste{\Im A^+(t',z_t')}\ec\nt\\
b_{00}^-&=\frac{g^2}{2m^2}-\frac{g^2}{m^2-a}+\frac{1}{\pi}\int\limits_{s_+}^\infty\diff s'\,h_0(s')\ste{\Im B^-(s',z_s')}
+\frac{1}{\pi}\int\limits_{\tpi}^\infty\frac{\diff t'}{t'}\ste{\Im B^-(t',z_t')}\ec\nt\\
d_{01}^+&=\frac{1}{\pi}\int\limits_{s_+}^\infty\diff s'\Bigg\{h_0(s')\ste{\partial_t\Im A^+(s',z_s')}-\frac{\ste{\Im A^+(s',z_s')}}{(s'-s_0)^2}\Bigg\}\nt\\
&\quad+\frac{1}{\pi}\int\limits_{\tpi}^\infty\frac{\diff t'}{t'}\Bigg\{\ste{\partial_t\Im A^+(t',z_t')}+\frac{1}{t'}\ste{\Im A^+(t',z_t')}\Bigg\}\ec\nt\\
b_{01}^-&=\frac{1}{\pi}\int\limits_{s_+}^\infty\diff s'\Bigg\{h_0(s')\ste{\partial_t\Im B^-(s',z_s')}-\frac{\ste{\Im B^-(s',z_s')}}{(s'-s_0)^2}\Bigg\}\nt\\
&\quad+\frac{1}{\pi}\int\limits_{\tpi}^\infty\frac{\diff t'}{t'}\Bigg\{\ste{\partial_t\Im B^-(t',z_t')}+\frac{1}{t'}\ste{\Im B^-(t',z_t')}\Bigg\}\ec\nt\\
\frac{a_{00}^-}{4m}&=\frac{1}{\pi}\int\limits_{s_+}^\infty\diff s'\,\frac{\ste{\Im A^-(s',z_s')}}{(s'-s_0)^2}
+\frac{1}{\pi}\int\limits_{\tpi}^\infty\frac{\diff t'}{t'}\ste{\frac{\Im A^-(t',z_t')}{4p_t'q_t'z_t'}}\ec\nt\\
\frac{b_{00}^+}{4m}&=\frac{1}{\pi}\int\limits_{s_+}^\infty\diff s'\,\frac{\ste{\Im B^+(s',z_s')}}{(s'-s_0)^2}
+\frac{1}{\pi}\int\limits_{\tpi}^\infty\frac{\diff t'}{t'}\ste{\frac{\Im B^+(t',z_t')}{4p_t'q_t'z_t'}}\ep
\end{align}}\noindent
The subscript $(0,0)$ indicates that $z_s'$ and $z_t'$ in the $s$- and $t$-channel integrals, respectively, are to be evaluated at $(\nu=0,\,t=0)$, which according to~\eqref{internalkinematicsofsta} and~\eqref{bofsta} amounts to using
\begin{align}
\label{zsdeltzs}
\ste{z_s'}&=1-\frac{(s'-s_0)^2}{2q'^2(s'-a)}\ec & 
\ste{\partial_tz_s'}&=\frac{s_0-a}{2q'^2(s'-a)}\ec\nt\\
\ste{z_t'^2}&=\frac{t'(t'-4(s_0-a))}{16p_t'^2q_t'^2}=1+\frac{t'4a-\tN\tpi}{16p_t'^2q_t'^2}\ec & 
\ste{\partial_tz_t'^2}&=\frac{s_0-a}{4p_t'^2q_t'^2}\ec
\end{align}
where again we have used the fact that the $t$-channel integrands depend on the squared angle $z_t'^2$ only.
Note that these sum rules as such are valid independent of the choice of $a$, but in practice one will incur an $a$-dependence once approximations are made (such as truncation of the partial-wave expansion, approximation of the high-energy region by Regge theory, etc.).

\subsection{Subtracted hyperbolic dispersion relations}
\label{subsec:subRS:HDRs}

A single subtraction at $(\nu=0,t=0)$ only affects $A^+(\nu,t)$ and $B^-(\nu,t)$, since both $A^-(\nu,t)$ and $B^+(\nu,t)$ are proportional to $\nu$.
Based on the unsubtracted HDRs~\eqref{hdr}, the explicit subthreshold expansions~\eqref{subthrexp}, and the corresponding sum rules~\eqref{sumrules}, we obtain the once-subtracted HDRs
\begin{align}
\label{hdr1sub}
A^+(s,t;a)&=\frac{g^2}{m}+d_{00}^+
+\frac{1}{\pi}\int\limits_{\tpi}^\infty\diff t'
\Bigg\{\frac{\Im A^+(t',z_t')}{t'-t}-\frac{\ste{\Im A^+(t',z_t')}}{t'}\Bigg\}\\
&\quad+\frac{1}{\pi}\int\limits_{s_+}^\infty\diff s'
\Bigg\{\bigg[\frac{1}{s'-s}+\frac{1}{s'-u}-\frac{1}{s'-a}\bigg]\Im A^+(s',z_s')-h_0(s')\,\ste{\Im A^+(s',z_s')}\Bigg\}\ec\nt\\
B^-(s,t;a)&=g^2\bigg[\frac{1}{m^2-s}+\frac{1}{m^2-u}\bigg]-\frac{g^2}{2m^2}+b_{00}^-
+\frac{1}{\pi}\int\limits_{\tpi}^\infty\diff t'
\Bigg\{\frac{\Im B^-(t',z_t')}{t'-t}-\frac{\ste{\Im B^-(t',z_t')}}{t'}\Bigg\}\nt\\
&\quad+\frac{1}{\pi}\int\limits_{s_+}^\infty\diff s'
\Bigg\{\bigg[\frac{1}{s'-s}+\frac{1}{s'-u}-\frac{1}{s'-a}\bigg]\Im B^-(s',z_s')-h_0(s')\,\ste{\Im B^-(s',z_s')}\Bigg\}\ec\nt
\end{align}
together with the unaltered equations~\eqref{hdr} for $A^-$ and $B^+$.
Note that the dependence on $a$ of the Born-term contribution $N^-$ is canceled by the sum rule~\eqref{sumrules} for $b_{00}^-$, which is why the subtraction constants are formally included in the subtracted nucleon pole terms in the following for convenience (i.e.\ preserving the generic form of the HDRs~\eqref{hdr}).

Similarly, a second subtraction at $(\nu=0,t=0)$ yields the twice-subtracted HDRs
\begin{align}
\label{hdr2sub}
A^+(s,t;a)&=\frac{g^2}{m}+d_{00}^++d_{01}^+t\nt\\
&\quad+\frac{1}{\pi}\int\limits_{\tpi}^\infty\diff t'
\Bigg\{\frac{\Im A^+(t',z_t')}{t'-t}-\bigg(\frac{1}{t'}+\frac{t}{t'^2}\bigg)\ste{\Im A^+(t',z_t')}-\frac{t}{t'}\ste{\partial_t\Im A^+(t',z_t')}\Bigg\}\nt\\
&\quad+\frac{1}{\pi}\int\limits_{s_+}^\infty\diff s'
\Bigg\{\bigg[\frac{1}{s'-s}+\frac{1}{s'-u}-\frac{1}{s'-a}\bigg]\Im A^+(s',z_s')-h_0(s')\,t\ste{\partial_t\Im A^+(s',z_s')}\nt\\
&\qquad-\bigg(h_0(s')-\frac{t}{(s'-s_0)^2}\bigg)\ste{\Im A^+(s',z_s')}\Bigg\}\ec\nt\\
B^-(s,t;a)&=g^2\bigg[\frac{1}{m^2-s}+\frac{1}{m^2-u}\bigg]-\frac{g^2}{2m^2}+b_{00}^-+b_{01}^-t\nt\\
&\quad+\frac{1}{\pi}\int\limits_{\tpi}^\infty\diff t'
\Bigg\{\frac{\Im B^-(t',z_t')}{t'-t}-\bigg(\frac{1}{t'}+\frac{t}{t'^2}\bigg)\ste{\Im B^-(t',z_t')}-\frac{t}{t'}\ste{\partial_t\Im B^-(t',z_t')}\Bigg\}\nt\\
&\quad+\frac{1}{\pi}\int\limits_{s_+}^\infty\diff s'
\Bigg\{\bigg[\frac{1}{s'-s}+\frac{1}{s'-u}-\frac{1}{s'-a}\bigg]\Im B^-(s',z_s')-h_0(s')\,t\ste{\partial_t\Im B^-(s',z_s')}\nt\\
&\qquad-\bigg(h_0(s')-\frac{t}{(s'-s_0)^2}\bigg)\ste{\Im B^-(s',z_s')}\Bigg\}\ec\nt\\
A^-(s,t;a)&=a_{00}^-\nu
+\frac{\nu}{\pi}\int\limits_{\tpi}^\infty\diff t'
\Bigg\{\frac{\Im A^-(t',z_t')}{\nu'(t'-t)}-\frac{\ste{\Im A^-(t',z_t')/\nu'}}{t'}\Bigg\}\nt\\
&\quad+\frac{1}{\pi}\int\limits_{s_+}^\infty\diff s'
\Bigg\{\bigg[\frac{1}{s'-s}-\frac{1}{s'-u}\bigg]\Im A^-(s',z_s')-\frac{4m\nu\ste{\Im A^-(s',z_s')}}{(s'-s_0)^2}\Bigg\}\ec\nt\\
B^+(s,t;a)&=g^2\bigg[\frac{1}{m^2-s}-\frac{1}{m^2-u}\bigg]+b_{00}^+\nu
+\frac{\nu}{\pi}\int\limits_{\tpi}^\infty\diff t'
\Bigg\{\frac{\Im B^+(t',z_t')}{\nu'(t'-t)}-\frac{\ste{\Im B^+(t',z_t')/\nu'}}{t'}\Bigg\}\nt\\
&\quad+\frac{1}{\pi}\int\limits_{s_+}^\infty\diff s'
\Bigg\{\bigg[\frac{1}{s'-s}-\frac{1}{s'-u}\bigg]\Im B^+(s',z_s')-\frac{4m\nu\ste{\Im B^+(s',z_s')}}{(s'-s_0)^2}\Bigg\}\ec
\end{align}
where $A^-$ and $B^+$ can also be written as
\begin{align}
\label{hdrAmBp2sub}
\frac{A^-(s,t;a)}{4m\nu}&=\frac{a_{00}^-}{4m}
+\frac{1}{\pi}\int\limits_{\tpi}^\infty\diff t'
\Bigg\{\frac{\Im A^-(t',z_t')}{4p_t'q_t'z_t'(t'-t)}-\frac{1}{t'}\ste{\frac{\Im A^-(t',z_t')}{4p_t'q_t'z_t'}}\Bigg\}\nt\\
&\quad+\frac{1}{\pi}\int\limits_{s_+}^\infty\diff s'
\Bigg\{\frac{\Im A^-(s',z_s')}{(s'-s)(s'-u)}-\frac{\ste{\Im A^-(s',z_s')}}{(s'-s_0)^2}\Bigg\}\ec\nt\\
\frac{B^+(s,t;a)}{4m\nu}&=\frac{g^2}{(m^2-s)(m^2-u)}+\frac{b_{00}^+}{4m}
+\frac{1}{\pi}\int\limits_{\tpi}^\infty\diff t'
\Bigg\{\frac{\Im B^+(t',z_t')}{4p_t'q_t'z_t'(t'-t)}-\frac{1}{t'}\ste{\frac{\Im B^+(t',z_t')}{4p_t'q_t'z_t'}}\Bigg\}\nt\\
&\quad+\frac{1}{\pi}\int\limits_{s_+}^\infty\diff s'
\Bigg\{\frac{\Im B^+(s',z_s')}{(s'-s)(s'-u)}-\frac{\ste{\Im B^+(s',z_s')}}{(s'-s_0)^2}\Bigg\}\ep
\end{align}
These subtractions require a modification of the nucleon-pole-term projections and the kernel functions for both the $s$- and $t$-channel contributions  calculated in Appendices~\ref{sec:spwp} and~\ref{sec:tpwp} as well as the asymptotic contributions given in Appendix~\ref{sec:regge}.
The differences on the right-hand side of the once-/twice-subtracted HDRs~\eqref{hdr1sub}/\eqref{hdr2sub} compared to the unsubtracted HDRs~\eqref{hdr} are the sources for the necessary modifications which are derived in Appendices~\ref{subsec:spwp:subtractions},
\ref{subsec:tpwp:subtractions}, and \ref{subsec:regge:asymptotics} for the $s$-channel kernels and pole terms, their $t$-channel analogs, and the asymptotic contributions, respectively.

\subsection[Subtracted $t$-channel Muskhelishvili--Omn\`es problem]{Subtracted \boldmath{$t$}-channel Muskhelishvili--Omn\`es problem}
\label{subsec:subRS:MO}

Using the subtracted kernels and pole terms as derived in Appendix~\ref{subsec:tpwp:subtractions} leads to the subtracted analogs of the unsubtracted $t$-channel MO problem~\eqref{MOsub0}, which we will state explicitly in the following for $J\leq2$ (the equations for $J\geq3$ are unaltered for up to two subtractions).
For one subtraction we may write
\begin{align}
\label{MOsub1}
f^0_+(t)&=\onesub{\Delta^0_+}(t)-\frac{t}{\pi}\int\limits_{\tpi}^\infty\diff t'\frac{\Im f^0_+(t')}{t'(t'-\tN)}
+\frac{t}{\pi}\int\limits_{\tpi}^\infty\diff t'\frac{\Im f^0_+(t')}{t'(t'-t)}\ec\nt\\
f^1_+(t)&=\onesub{\Delta^1_+}(t)+\frac{\tN}{\pi}\int\limits_{\tpi}^\infty\diff t'\frac{\frac{m}{\sqrt{2}}\Im f^1_-(t')-\Im f^1_+(t')}{t'(t'-\tN)}
+\frac{t}{\pi}\int\limits_{\tpi}^\infty\diff t'\frac{\Im f^1_+(t')}{t'(t'-t)}\ec\nt\\
f^1_-(t)&=\onesub{\Delta^1_-}(t)+\frac{t}{\pi}\int\limits_{\tpi}^\infty\diff t'\frac{\Im f^1_-(t')}{t'(t'-t)}\ec\nt\\
f^2_+(t)&=\onesub{\Delta^2_+}(t)+\frac{1}{\pi}\int\limits_{\tpi}^\infty\diff t'\frac{m\sqrt{\frac{2}{3}}\Im f^2_-(t')-\Im f^2_+(t')}{t'-\tN}
+\frac{1}{\pi}\int\limits_{\tpi}^\infty\diff t'\frac{\Im f^2_+(t')}{t'-t}\ec\nt\\
f^2_-(t)&=\onesub{\Delta^2_-}(t)+\frac{1}{\pi}\int\limits_{\tpi}^\infty\diff t'\frac{\Im f^2_-(t')}{t'-t}\ec
\end{align}
while two subtractions yield
\begin{align}
\label{MOsub2}
f^0_+(t)&=\twosub{\Delta^0_+}(t)-\frac{t^2}{\pi}\int\limits_{\tpi}^\infty\diff t'\frac{\Im f^0_+(t')}{t'^2(t'-\tN)}
+\frac{t^2}{\pi}\int\limits_{\tpi}^\infty\diff t'\frac{\Im f^0_+(t')}{t'^2(t'-t)}\ec\nt\\
f^1_+(t)&=\twosub{\Delta^1_+}(t)+\frac{\tN t}{\pi}\int\limits_{\tpi}^\infty\diff t'\frac{\frac{m}{\sqrt{2}}\Im f^1_-(t')-\Im f^1_+(t')}{t'^2(t'-\tN)}
+\frac{t^2}{\pi}\int\limits_{\tpi}^\infty\diff t'\frac{\Im f^1_+(t')}{t'^2(t'-t)}\ec\nt\\
f^1_-(t)&=\twosub{\Delta^1_-}(t)+\frac{t^2}{\pi}\int\limits_{\tpi}^\infty\diff t'\frac{\Im f^1_-(t')}{t'^2(t'-t)}\ec\nt\\
f^2_+(t)&=\twosub{\Delta^2_+}(t)+\frac{\tN}{\pi}\int\limits_{\tpi}^\infty\diff t'\frac{m\sqrt{\frac{2}{3}}\Im f^2_-(t')-\Im f^2_+(t')}{t'(t'-\tN)}
+\frac{t}{\pi}\int\limits_{\tpi}^\infty\diff t'\frac{\Im f^2_+(t')}{t'(t'-t)}\ec\nt\\
f^2_-(t)&=\twosub{\Delta^2_-}(t)+\frac{t}{\pi}\int\limits_{\tpi}^\infty\diff t'\frac{\Im f^2_-(t')}{t'(t'-t)}\ep
\end{align}
It is important to note that $S$- and $D$-waves are coupled, as $\Delta^0_+$ contains contributions from $J=2$ according to~\eqref{Deltadef}.
While the integrands containing the Cauchy kernel in~\eqref{MOsub1} and~\eqref{MOsub2} for $J=0$ and $J=1$ clearly show the corresponding number of subtractions at $t_0=0$, for $J=2$ there is always one subtraction less or no subtraction at all.
Note that the integrands containing linear combinations of the partial waves are proportional to $\tN/t'$ (if affected by the subtractions at all), which results in a suppressed internal high-energy dependence inside the integral due to division by $t'$ without an increased external high-energy dependence due to multiplication with $\tN$ rather than $t$ as for a usual subtraction at zero.

The un-~\eqref{MOsub0}, once-~\eqref{MOsub1}, and twice-subtracted~\eqref{MOsub2} equations are of the original form of the (subtracted) MO problem with integrals of the absorptive parts times the Cauchy kernel, if the remaining $t$-independent integrals (which may, however, come with $t$-dependent prefactors) are absorbed into a redefinition of the inhomogeneities $\Delta^J_\pm(t)$.
This problem is well defined due to the threshold behavior of the partial waves at $t=\tN$.
However, the price for taking advantage of the convergence properties of the integrals this way is that reasonable approximations for the starting values for the partial waves are needed as input, since the solutions can only be found iteratively.

Therefore, we prefer to utilize the threshold behavior of the partial waves and use the linear combinations $\Gamma^J(t)$ in order to rewrite the equations in analogy to~\eqref{MOGamsub0}, i.e.\ to modify the original form of the (subtracted) MO problem in a well-defined manner.
The general $n$-times subtracted (with $n\in\{0,1,2\}$) versions of the MO equations~\eqref{MOGamsub0} for all $J$ then read
\begin{align}
\label{MOGamsubn}
f^0_+(t)&=\nsub{\Delta^0_+}(t)
+\frac{t^n(t-\tN)}{\pi}\int\limits^\infty_{\tpi}\diff t'\frac{\Im f^0_+(t')}{t'^n(t'-\tN)(t'-t)}\ec\nt\\
\Gamma^J(t)&=\nsub{\Delta_\Gamma^J}(t)
+\frac{t^{(n-J)\theta(n-J)}(t-\tN)}{\pi}\int\limits^\infty_{\tpi}\diff t'\frac{\Im\Gamma^J(t')}{t'^{(n-J)\theta(n-J)}(t'-\tN)(t'-t)}\qquad\forall\;J\geq1\ec\nt\\
f^J_-(t)&=\nsub{\Delta^J_-}(t)
+\frac{t^{(n-J+1)\theta(n-J)}}{\pi}\int\limits^\infty_{\tpi}\diff t'\frac{\Im f^J_-(t')}{t'^{(n-J+1)\theta(n-J)}(t'-t)}\qquad\forall\;J\geq1\ec
\end{align}
where the Heaviside step function is to be understood in its right-continuous form, i.e.\ $\theta(0)=1$.
Again, the equations for $f^0_+(t)$ and $\Gamma^J(t)$ exhibit one additional subtraction at $\tN$, such that the combined number of subtractions for all $J\geq0$ can be given as $(n-J+1)\theta(n-J)$.
For convenience, we explicitly show the terms in $\Delta^0_+(t)$ that couple the $D$- to the $S$-waves
\begin{align}
\label{MODwavecouplingsubn}
\unsub{\Delta^0_+}(t)&=-\frac{5}{16}\frac{t-\tN}{\pi}\int\limits^{\infty}_{\tpi}\diff t'
\bigg\{\big[t'+t-(\tN+\tpi)+6a\big]\frac{\Im\Gamma^2(t')}{t'-\tN}+\frac{m}{\sqrt{6}}\Im f^2_-(t')\bigg\}+\dots\ec\\
\onesub{\Delta^0_+}(t)&=-\frac{5}{16}\frac{t-\tN}{\pi}\int\limits^{\infty}_{\tpi}\frac{\diff t'}{t'}
\bigg\{\bigg[t't+\frac{\tN\tpi}{2}\bigg]\frac{\Im\Gamma^2(t')}{t'-\tN}+\tpi\frac{m}{\sqrt{6}}\Im f^2_-(t')\bigg\}+\dots\ec\nt\\
\twosub{\Delta^0_+}(t)&=-\frac{5}{16}\frac{t-\tN}{\pi}\int\limits^{\infty}_{\tpi}\frac{\diff t'}{t'^2}
\bigg\{\frac{1}{2}\big[(t'+t)\tN\tpi-t't(\tN+\tpi)\big]\frac{\Im\Gamma^2(t')}{t'-\tN}+t\tpi\frac{m}{\sqrt{6}}\Im f^2_-(t')\bigg\}+\dots\ec\nt
\end{align}
which converge for $t'\to\tN$ due to the threshold behavior of $\Gamma^2$ and vanish for $t\to\tN$ due to the exceptional behavior of the ($n$-times subtracted) kernel $\tilde K^2_{02}$ (cf.~\eqref{tilde2K02asymptotics} and~\eqref{tilde2K02asymptoticsnsub});
the respective remainder denoted by dots above is then given by
\begin{align}
\dots&=\nsub{\tilde N^0_+}(t)+\frac{1}{\pi}\int\limits^{\infty}_{W_+}\diff W'\sum\limits^\infty_{l=0}
\Big\{\nsub{\tilde G_{0l}}(t,W')\,\Im f^+_{l+}(W')+\nsub{\tilde G_{0l}}(t,-W')\,\Im f^+_{(l+1)-}(W')\Big\}\nt\\
&\qquad+\frac{1}{\pi}\int\limits^{\infty}_{\tpi}\diff t'\sum^{\infty}\limits_{J'=4}\frac{1+(-1)^{J'}}{2}
\Big\{\nsub{\tilde K^1_{0J'}}(t,t')\,\Im f^{J'}_+(t')+\nsub{\tilde K^2_{0J'}}(t,t')\,\Im f^{J'}_-(t')\Big\}\ep
\end{align}
Note that this $D$- to $S$-wave coupling becomes independent of $a$ by subtracting once or twice, while the corresponding $F$- to $P$-wave coupling also depends on $a$ in the once-subtracted case (cf.~\eqref{subtractedtildeKkernels}).

\section{Solving the \boldmath{$t$}-channel Muskhelishvili--Omn\`es problem}
\label{sec:piNMO}

In this section the solution of the MO problem for the lowest $t$-channel partial waves $f^J_\pm(t)$ with $J\in\{0,1,2\}$ will be discussed.
First, the explicit analytical solutions will be stated.
Then, the numerical input needed will be collected.
Finally, the numerical results will be discussed.

\subsection{Muskhelishvili--Omn\`es problem with finite matching point}
\label{subsec:piNMO:MO}

We assume to know the imaginary part of the $t$-channel partial waves $f_\pm^J(t)$ above the finite matching point $\tm$ as well as the scattering phases $\delta^{I_t}_J(t)$ of the $\pi\pi$ partial waves $t^{I_t}_J(t)$ for $4\mpi^2=\tpi\leq t\leq\tm$, which in the elastic region are also the phases of the $f_\pm^J(t)$ due to Watson's final state theorem, cf.~\eqref{watson}.
All inelastic contributions will be neglected.
Under these assumptions, we have to solve equations of the MO type~\cite{Muskhelishvili,Omnes}
\begin{equation}
\label{mogeneral}
f(t)=\Delta(t)+\frac{1}{\pi}\int\limits_{\tpi}^{\tm}\diff t'\frac{T(t')^*f(t')}{t'-t}+\frac{1}{\pi}\int\limits_{\tm}^\infty\diff t'\frac{\Im f(t')}{t'-t}
\end{equation}
for $f(t)$ in the range $\tpi\leq t\leq \tm$ with finite $\tm$~\cite{piK:RS}, where the physical values of the integrals are obtained in the limit $t\rightarrow t+i\epsilon$ and the discontinuity of $f(t)$ across the right-hand cut is given by unitarity (cf.\ hermitian analyticity~\eqref{discontinuity} and the elastic $t$-channel unitarity relation~\eqref{tunitrel})
\begin{equation}
\label{moimpart}
\frac{\Disc f(t)}{2i}=\Im f(t)=T(t)^*f(t)\,\theta\big(t-\tpi\big)\ec
\end{equation}
where the inhomogeneity $\Delta(t)$ contains potential left-hand cut contributions to $f(t)$ (i.e.\ it is real for $\tpi\leq t$) and the elastic amplitude $T(t)$ is given by
\begin{equation}
T(t)=\sin\delta(t)e^{i\delta(t)}\ep
\end{equation}
We briefly review the result of~\cite{Omnes,piK:RS} in the following sections.

\subsubsection{General solution}

To begin with, we consider the homogeneous problem for a function $f_0(t)$ with non-vanishing imaginary part only for $\tpi\leq t\leq\tm$. The solution can then be written as
\begin{equation}
f_0(t)=\Omega(t)\Sigma_0(t)\ec
\end{equation}
with the Omn\`es function\footnote{Note that for a finite matching point it is not mandatory to work with a subtracted Omn\`es function. However, subtracting once at $t=0$ ensures the usual normalization $\Omega(0)=1$.}
\begin{align}
\label{moomnesfunction}
\Omega(t)&=\exp\Bigg\{\frac{t}{\pi}\int\limits_{\tpi}^{\tm}\frac{\diff t'}{t'}\frac{\delta(t')}{t'-t}\Bigg\}
 =|\Omega(t)|\exp\Big\{i\delta(t)\theta\big(t-\tpi\big)\theta\big(\tm-t\big)\Big\}\ec
 \qquad \Omega(0)=1\ec\nt\\
|\Omega(t)|&=\exp\Bigg\{\frac{t}{\pi}\;\dashint\limits_{\tpi}^{\tm}\frac{\diff t'}{t'}\frac{\delta(t')}{t'-t}\Bigg\}
 =|\bar\Omega(t)|\,|\tm-t|^{x(t)}\ec
 \qquad x(t)=\frac{\delta(t)}{\pi}\ec\nt\\
|\bar\Omega(t)|&=\bigg|\frac{\tm}{\tpi}(t-\tpi)\bigg|^{-x(t)}\exp\Bigg\{\frac{t}{\pi}\int\limits_{\tpi}^{\tm}\frac{\diff t'}{t'}\frac{\delta(t')-\delta(t)}{t'-t}\Bigg\}\ec
\end{align}
where we have analytically separated the endpoint singularities of the principal value integral.
By assuming the reasonable asymptotic behavior $f_0(t)\to0$ for $t\to\infty$ (cf.~\eqref{fJpmasymptotics}), the only analytic structures of $\Sigma_0(t)$ allowed by $f_0(t)$ and $\Omega(t)$ are poles at the endpoints $t=\tpi$ and $t=\tm$.
Since $\Omega(t)$ is regular at $\tpi$ due to $\delta(\tpi)=0$, the regularity of $f_0(t)$ excludes poles at $t=\tpi$ and restricts the order of the poles at $t=\tm$ to
\begin{equation}
\label{mondef}
n=\left\lfloor x\right\rfloor\ec \qquad x=\frac{\delta(\tm)}{\pi}
\end{equation}
($\lfloor x\rfloor$ denotes the largest integer $\leq x$). In this way, we find
\begin{equation}
\Sigma_0(t)=\frac{\Pol_{n-1}(t)}{(\tm-t)^n}\ec
\end{equation}
where $\Pol_{n-1}(t)$ is an arbitrary real polynomial of degree $n-1$ that introduces $n$ free parameters to the Omn\`es problem.
For $n=0$ the homogeneous solution vanishes according to $\Pol_{-1}(t)=0$ and no free parameter enters the problem.

The general solution reads
\begin{equation}
\label{moresult}
f(t)=\Delta(t)+\Omega(t)\Bigg\{\frac{\Pol_{n-1}(t)}{(\tm-t)^n}+\frac{1}{\pi}\int\limits_{\tpi}^{\tm}\diff t'\frac{\Delta(t')\sin\delta(t')}{|\Omega(t')|(t'-t)}
 +\frac{1}{\pi}\int\limits_{\tm}^\infty\diff t'\frac{\Im f(t')}{|\Omega(t')|(t'-t)}\Bigg\}\ec
\end{equation}
which may also be written in terms of a principal value integral as
\begin{equation}
\label{mosolution}
f(t)=\Bigg[\Delta(t)\cos\delta(t)+|\Omega(t)|\Bigg\{\frac{\Pol_{n-1}(t)}{(\tm-t)^n}
 +\frac{1}{\pi}\dashint\limits_{\tpi}^{\tm}\diff t'\frac{\Delta(t')\sin\delta(t')}{|\Omega(t')|(t'-t)}
 +\frac{1}{\pi}\int\limits_{\tm}^\infty\diff t'\frac{\Im f(t')}{|\Omega(t')|(t'-t)}\Bigg\}\Bigg]e^{i\delta(t)}\ec
\end{equation}
in accordance with~\cite{Omnes} for $\tm\rightarrow \infty$.
Note that due to Watson's theorem~\eqref{watson} the prefactor in square brackets can be identified with the modulus $|f(t)|$, and since the phase $\delta(t)$ is known, we only need to solve the MO problem for this modulus for $\tpi\leq t\leq\tm$.

\subsubsection{Subtractions}

If $x>1$, suitable subtractions need to be performed in~\eqref{mosolution} to ensure integrability for $t'\to\tm$.
Let us begin with the case $1<x<2$, i.e.\ $n=1$. We may write
\begin{equation}
\int\limits_{\tm}^\infty\diff t'\frac{\Im f(t')}{|\Omega(t')|(t'-t)}=\frac{1}{\tm-t}
\Bigg\{\tm \int\limits_{\tm}^\infty\frac{\diff t'}{t'}\frac{\Im f(t')}{|\Omega(t')|}
+t\int\limits_{\tm}^\infty\frac{\diff t'}{t'}\frac{\tm-t'}{|\Omega(t')|}\frac{\Im f(t')}{t'-t}
\Bigg\}\ec
\end{equation}
where the second integral is now convergent.
The first integral is still divergent, of course, but it does not depend on $t$ any more and can thus be absorbed into a redefinition of the (constant) polynomial $\Pol_0$ in~\eqref{mosolution} due to the common prefactor $(\tm-t)^{-1}$.
For higher values of $x$ this subtraction and redefinition prescription needs to be iterated, whereby all $n$ parameters contained in the polynomial receive corresponding contributions.
Applying this reasoning to both integrals of~\eqref{mosolution} for general $x$ and using the highest number of subtractions allowed by the degree of the polynomial, the result is given by
\begin{align}
\label{moftgen}
|f(t)|&=\Delta(t)\cos\delta(t)+\frac{|\Omega(t)|}{(\tm-t)^n}\Bigg\{\Pol_{n-1}(t)
 +\frac{t^n}{\pi}\dashint\limits_{\tpi}^{\tm}\frac{\diff t'}{t'^n}\frac{(\tm-t')^n}{|\Omega(t')|}\frac{\Delta(t')\sin\delta(t')}{t'-t}\nt\\
&\qquad+\frac{t^n}{\pi}\int\limits_{\tm}^\infty\frac{\diff t'}{t'^n}\frac{(\tm-t')^n}{|\Omega(t')|}\frac{\Im f(t')}{t'-t}\Bigg\}\ep
\end{align}

In order to reduce the influence of the high-energy contributions on the Omn\`es integrals, subtractions may also be introduced already right from the beginning~\eqref{mogeneral}.
With $l$ such subtractions, the analog of~\eqref{moftgen} becomes
\begin{align}
\label{moftfinal}
|f(t)|&=\Delta(t)\cos\delta(t)+\frac{t^l|\Omega(t)|}{(\tm-t)^n}\Bigg\{\Pol_{n-1}(t)
 +\frac{t^n}{\pi}\dashint\limits_{\tpi}^{\tm}\frac{\diff t'}{t'^{n+l}}\frac{(\tm-t')^n}{|\Omega(t')|}\frac{\Delta(t')\sin\delta(t')}{t'-t}\nt\\
&\qquad+\frac{t^n}{\pi}\int\limits_{\tm}^\infty\frac{\diff t'}{t'^{n+l}}\frac{(\tm-t')^n}{|\Omega(t')|}\frac{\Im f(t')}{t'-t}\Bigg\}\ep
\end{align}
This constitutes the final general result, which for $l=1$ and $n\in\{0,1\}$ reduces to the results quoted in~\cite{piK:RS}.

\subsubsection{Numerical treatment}

The asymptotic behavior of the Omn\`es function $|\Omega(t)|$ for $t\to \tm$ requires some care in the numerical evaluation of the integrals in~\eqref{moftfinal}.
Although by construction the singularities for $t'\to\tm$ are integrable, the corresponding cusps generate large contributions to the integral and a fully numerical treatment would require a very careful distribution of mesh points in order to catch the effect.
In the following, we will demonstrate how these endpoint singularities can be separated analytically (cf.\ the appendix of~\cite{piK:RS}).
For the sake of simplicity, we discuss here the case of $n=l=0$, which already displays all relevant features; the generalization is then straightforward.
To this end, we split the integrals close to the matching point $\tm$ and approximate $|\Omega(t)|$ by its asymptotic form in the proximity of $\tm$
\begin{equation}
|\Omega(t\approx\tm)|\approx|\bar\Omega(\tm)|\,|\tm-t|^x\ep
\end{equation}
For $\tau\to 0^+$, we may thus rewrite the integrals above the matching point as
\begin{align}
\label{mosplitintabove}
\int\limits_{\tm}^\infty\diff t'\frac{\Im f(t')}{|\Omega(t')|(t'-t)}&=\int\limits_{\tm+\tau}^\infty\diff t'\frac{\Im f(t')}{|\Omega(t')|(t'-t)}
+\frac{\Im f(\tm)}{|\bar\Omega(\tm)|}\int\limits_{\tm}^{\tm +\tau}\frac{\diff t'}{|\tm-t'|^x(t'-t)}\nt\\
&=\int\limits_{\tm+\tau}^\infty\diff t'\frac{\Im f(t')}{|\Omega(t')|(t'-t)}
+\frac{\Im f(\tm)}{|\bar\Omega(\tm)|(\tm-t)^x}I_+(t)\ec
\end{align}
and similarly below the matching point either for $\tpi\leq t<\tm-\tau$ 
\begin{align}
\label{mosplitintbelow}
\dashint\limits_{\tpi}^{\tm}\diff t'\frac{\Delta(t')\sin \delta(t')}{|\Omega(t')|(t'-t)}&=
\int\limits_{\tpi}^{\tm-\tau}\frac{\diff t'}{t'-t}
\bigg(\frac{\Delta(t')\sin \delta(t')}{|\Omega(t')|}-\frac{\Delta(t)\sin\delta(t)}{|\Omega(t)|}\bigg)
+\frac{\Delta(t)\sin \delta(t)}{|\Omega(t)|}\log\frac{\tm-\tau-t}{t-\tpi}\nt\\
&\qquad+\frac{\Delta(\tm)\sin\delta(\tm)}{|\bar\Omega(\tm)|(\tm-t)^x}I_-(t)\ec
\end{align}
or for $\tm-\tau\leq t\leq\tm$
\begin{equation}
\label{mosplitintbelowasym}
\dashint\limits_{\tpi}^{\tm}\diff t'\frac{\Delta(t')\sin \delta(t')}{|\Omega(t')|(t'-t)}=
\int\limits_{\tpi}^{\tm-\tau}\frac{\diff t'}{t'-t}\frac{\Delta(t')\sin \delta(t')}{|\Omega(t')|}
+\frac{\Delta(\tm)\sin\delta(\tm)}{|\bar\Omega(\tm)|(\tm-t)^x}\tilde{I}_-(t)\ep
\end{equation}
The substitution $v(t')=(t'-\tm)/(\tm-t)$ leads to the integrals (with $x\in(0,1)$)
\begin{align}
\label{mointegrals}
I_\pm(t)&=\int\limits_0^{\tilde\tau(t)}\frac{\diff v}{v^x(1\pm v)}=
\frac{\tilde\tau(t)^{1-x}}{1-x}\mp\int\limits_0^{\tilde\tau(t)}\diff v\frac{v^{1-x}}{1\pm v}\ec \qquad \tilde\tau(t)=\frac{\tau}{\tm-t}\ec\nt\\
\tilde{I}_-(t)&=\dashint[3.5pt]\limits_0^{\tilde\tau(t)}\frac{\diff v}{v^x(1-v)}=
-\log|\tilde\tau(t)-1|+\frac{\tilde\tau(t)^{1-x}}{1-x}+\int\limits_0^{\tilde\tau(t)}\diff v\frac{v^{1-x}-1}{1-v}\ep
\end{align}
Separating the singularities as shown above, the remaining integrals can be solved by using standard integration routines. For sufficiently small $\tau$ (i.e.\ if $\tau$ is of the same order of magnitude as the discretization error of the integration routine), the above approximations are well justified and this procedure allows for a stable numerical evaluation of the Omn\`es integrals.

\subsubsection{Continuity at the matching point}

The continuity of the Omn\`es solution $f(t)$ at the matching point $\tm$  is analytically ensured by the asymptotic form of the corresponding integrals of~\eqref{mointegrals} (with $0<x<1$, cf.\ the appendix of~\cite{piK:RS}) 
\begin{align}
I_+(\tm)&=\int\limits_0^{\infty}\frac{\diff v}{v^x(1+v)}=\pi\cosec\pi x\ec\quad
\tilde{I}_-(\tm)=\dashint\limits_0^{\infty}\frac{\diff v}{v^x(1-v)}=-\pi\cot\pi x\ep
\end{align}
Taking equation~\eqref{mosolution} in the limit $t\to\tm$ from below, plugging in these asymptotic expressions for the integrals, and using \eqref{mondef} indeed reduces the square bracket to $|f(\tm)|$.
This analytical equality may also be used as a check of the numerical evaluation.

However, the continuity of the first and higher derivatives is not ensured in a similar manner.
Since the solution must not depend on the value of the matching point, an unphysical cusp or non-smooth behavior of the modulus of the solution at the matching point only indicates that the input in terms of the absorptive part is not precise enough; moreover, the physical condition of a smooth behavior at the matching point ensures the uniqueness of the solution~\cite{GasserWanders}.
Physically consistent input given, this smoothness constraint may actually be used in order to tune/estimate/fit the subtraction constants (cf.~\cite{piK:RS}).

\subsection[Explicit solution of the $t$-channel Muskhelishvili--Omn\`es problem]{Explicit solution of the \boldmath{$t$}-channel Muskhelishvili--Omn\`es problem}
\label{subsec:piNMO:explsol}

Here, we will give the explicit solutions for the $n$-times subtracted $t$-channel MO problem~\eqref{MOGamsubn} using the general results of Sect.~\ref{subsec:piNMO:MO}.
The crucial ingredient for the following discussion is Watson's theorem~\eqref{watson}, which states that below the onset of inelasticities the phases $\varphi^J_\pm(t)$ of the $t$-channel partial waves $f^J_\pm(t)$ are given by the corresponding $\pi\pi$ scattering phases $\delta^{I_t}_J(t)$ with $I_t\in\{0,1\}$, i.e.\ explicitly for $J\in\{0,1,2\}$
\begin{equation}
\varphi^0_+(t)=\delta^0_0(t)=\delta_0(t)\ec \qquad \varphi^1_\pm(t)=\delta^1_1(t)=\delta_1(t)\ec \qquad \varphi^2_\pm(t)=\delta^0_2(t)=\delta_2(t)\ep
\end{equation}
These identities enter the solutions at two places:
first, in this kinematical region we can use the same Omn\`es function $\Omega_J$ for both $f^J_\pm$ and thus also for the linear combination $\Gamma^J$.
Second, in this range of $t$ the linear relation~\eqref{Gammadef} is also valid for the moduli such that after solving for $\big|\Gamma^J\big|$ we can recover
\begin{equation}
\label{fJpfromfJmandGammaJ}
\big|f^J_+(t)\big|=m\sqrt{\frac{J}{J+1}}\big|f^J_-(t)\big|-\big|\Gamma^J(t)\big|\ep
\end{equation}
Using once-subtracted Omn\`es functions in the convention (cf.~\eqref{moomnesfunction}),
\begin{align}
\label{MOOmegaJ}
\Omega_J(t)&=\exp\Bigg\{\frac{t}{\pi}\int\limits_{\tpi}^{\tm}\frac{\diff t'}{t'}\frac{\delta_J(t')}{t'-t}\Bigg\}=
\big|\Omega_J(t)\big|\exp\Big\{i\delta_J(t)\theta\big(t-\tpi\big)\theta\big(\tm-t\big)\Big\}\ec
 \qquad \Omega_J(0)=1\ec\nt\\
\big|\Omega_J(t)\big|&=\bigg|1-\frac{t}{\tm}\bigg|^{x_J(t)}\,\bigg|\frac{t}{\tpi}-1\bigg|^{-x_J(t)}\exp\Bigg\{\frac{t}{\pi}\int\limits_{\tpi}^{\tm}\frac{\diff t'}{t'}
\frac{\delta_J(t')-\delta_J(t)}{t'-t}\Bigg\}\ec
 \qquad x_J(t)=\frac{\delta_J(t)}{\pi}\ec
\end{align}
the general $n$-times subtracted ($n\in\{0,1,2\}$) solutions of~\eqref{MOGamsubn} for $t\in[\tpi,\tm]$ and $\lfloor\delta_J(\tm)/\pi\rfloor=0$ read
\begin{align}
\label{MOGamsubnDeltasol}
f^0_+(t)&=\nsub{\Delta^0_+}(t)
+\Omega_0(t)\frac{t^n(t-\tN)}{\pi}\Bigg\{\int\limits_{\tpi}^{\tm}\diff t'\frac{\nsub{\Delta^0_+}(t')\sin\delta_0(t')}{t'^n(t'-\tN)|\Omega_0(t')|(t'-t)}\nt\\
&\qquad+\int\limits_{\tm}^{\infty}\diff t'\frac{\Im f_+^0(t')}{t'^n(t'-\tN)|\Omega_0(t')|(t'-t)}\Bigg\}\ec\nt\\
\Gamma^J(t)&=\nsub{\Delta_\Gamma^J}(t)
+\Omega_J(t)\frac{t^{(n-J)\theta(n-J)}(t-\tN)}{\pi}\Bigg\{\int\limits_{\tpi}^{\tm}\diff t'\frac{\nsub{\Delta_\Gamma^J}(t')\sin\delta_J(t')}{t'^{(n-J)\theta(n-J)}(t'-\tN)|\Omega_J(t')|(t'-t)}\nt\\
&\qquad+\int\limits_{\tm}^{\infty}\diff t'\frac{\Im\Gamma^J(t')}{t'^{(n-J)\theta(n-J)}(t'-\tN)|\Omega_J(t')|(t'-t)}\Bigg\}\qquad\forall\;J\geq1\ec\nt\\
f^J_-(t)&=\nsub{\Delta^J_-}(t)
+\Omega_J(t)\frac{t^{(n-J+1)\theta(n-J)}}{\pi}\Bigg\{\int\limits_{\tpi}^{\tm}\diff t'\frac{\nsub{\Delta^J_-}(t')\sin\delta_J(t')}{t'^{(n-J+1)\theta(n-J)}|\Omega_J(t')|(t'-t)}\nt\\
&\qquad+\int\limits_{\tm}^{\infty}\diff t'\frac{\Im f^J_-(t')}{t'^{(n-J+1)\theta(n-J)}|\Omega_J(t')|(t'-t)}\Bigg\}\qquad\forall\;J\geq1\ep
\end{align}
Now, we can use the spectral representations of the inverse of the Omn\`es functions in the un-, once-, and twice-subtracted form
\begin{align}
\label{omnesspecrep}
\Omega_J^{-1}(t)&=\frac{1}{\pi}\int\limits_{\tpi}^{\tm}\diff t'\frac{\Im\Omega_J^{-1}(t')}{t'-t}
=-\frac{1}{\pi}\int\limits_{\tpi}^{\tm}\diff t'\frac{\sin\delta_J(t')}{|\Omega_J(t')|(t'-t)}\nt\\
&=1-\frac{t}{\pi}\int\limits_{\tpi}^{\tm}\diff t'\frac{\sin\delta_J(t')}{t'|\Omega_J(t')|(t'-t)}
=1-t\,\dot\Omega_J(0)-\frac{t^2}{\pi}\int\limits_{\tpi}^{\tm}\diff t'\frac{\sin\delta_J(t')}{t'^2|\Omega_J(t')|(t'-t)}\ec
\end{align}
with the derivative of the Omn\`es function\footnote{Note that for $\tm\to\infty$ (and neglecting inelasticities in the single-channel approximation) this quantity is closely related to the pion vector radius for $J=1$\,: $\lim_{\tm\to\infty}\dot\Omega_1=\tfrac{1}{6}\langle r^2\rangle_\pi^V$\,.}
\begin{equation}
\dot\Omega_J(0)=\frac{\diff}{\diff t}\Omega_J(t)\bigg|_{t=0}=\frac{1}{\pi}\int\limits_{\tpi}^{\tm}\diff t'\frac{\delta_J(t')}{t'^2}\ec
\end{equation}
in order to explicitly perform the integrals over terms that are either constant or come with appropriate factors of $t'$ or $p_t'^2$, i.e.\ all terms involving the subthreshold parameters as well as the term proportional to $\delta_{J1}/(m^2-a)$ for the unsubtracted case.
For this purpose we define $\tilde\Delta^J_\pm(t)$ via removing all constant or subthreshold-parameter contributions from the inhomogeneities $\Delta^J_\pm(t)$ (cf.~\eqref{Deltadef} and~\eqref{DeltahatNJpmsubn})
\begin{equation}
\label{tildeDeltadef}
\nsub{\tilde\Delta^J_\pm}(t)=\nsub{\Delta^J_\pm}(t)-\nsub{\Delta\hat N^J_\pm}(t)=\hat N^J_\pm(t)+\nsub{\bar\Delta^J_\pm}(t)\ec
\end{equation}
and thereby we obtain
\begin{align}
\label{MOGamsubntildeDeltasol}
f^0_+(t)&=\nsub{\tilde\Delta^0_+}(t)
+\Omega_0(t)\frac{t-\tN}{\pi}\Bigg\{\nsub{\chi^0_+}(t)\nt\\
&\qquad+t^n\Bigg[
\int\limits_{\tpi}^{\tm}\diff t'\frac{\nsub{\tilde\Delta^0_+}(t')\sin\delta_0(t')}{t'^n(t'-\tN)|\Omega_0(t')|(t'-t)}
+\int\limits_{\tm}^{\infty}\diff t'\frac{\Im f^0_+(t')}{t'^n(t'-\tN)|\Omega_0(t')|(t'-t)}
\Bigg]\Bigg\}\ec\nt\\
\Gamma^J(t)&=\nsub{\tilde\Delta_\Gamma^J}(t)
+\Omega_J(t)\frac{t-\tN}{\pi}\Bigg\{\nsub{\chi^J_\Gamma}(t)\nt\\
&\qquad+t^{(n-J)\theta(n-J)}\Bigg[
\int\limits_{\tpi}^{\tm}\diff t'\frac{\nsub{\tilde\Delta_\Gamma^J}(t')\sin\delta_J(t')}{t'^{(n-J)\theta(n-J)}(t'-\tN)|\Omega_J(t')|(t'-t)}\nt\\
&\qquad\quad+\int\limits_{\tm}^{\infty}\diff t'\frac{\Im\Gamma^J(t')}{t'^{(n-J)\theta(n-J)}(t'-\tN)|\Omega_J(t')|(t'-t)}
\Bigg]\Bigg\}\qquad\forall\;J\geq1\ec\nt\\
f^J_-(t)&=\nsub{\tilde\Delta^J_-}(t)
+\Omega_J(t)\frac{1}{\pi}\Bigg\{\nsub{\chi^J_-}(t)\nt\\
&\qquad+t^{(n-J+1)\theta(n-J)}\Bigg[
\int\limits_{\tpi}^{\tm}\diff t'\frac{\nsub{\tilde\Delta^J_-}(t')\sin\delta_J(t')}{t'^{(n-J+1)\theta(n-J)}|\Omega_J(t')|(t'-t)}\nt\\
&\qquad\quad+\int\limits_{\tm}^{\infty}\diff t'\frac{\Im f^J_-(t')}{t'^{(n-J+1)\theta(n-J)}|\Omega_J(t')|(t'-t)}
\Bigg]\Bigg\}\qquad\forall\;J\geq1\ec
\end{align}
with
\begin{align}
\twosub{\chi^0_+}(t)&=-\frac{1}{16}\bigg\{\bigg[\frac{g^2}{m}+d_{00}^++\tpi\frac{b_{00}^+}{12}\bigg]\big(1-t\,\dot\Omega_0(0)\big)+\bigg[d_{01}^+ -\frac{b_{00}^+}{12}\bigg]t\bigg\}
\redonesub-\frac{1}{16}\bigg[\frac{g^2}{m}+d_{00}^+\bigg]
\redunsub0\ec\nt\\
\twosub{\chi^J_\Gamma}(t)&=\frac{1}{12}\frac{a_{00}^-}{4m}\delta_{J1}
\redonesub0\redunsub0\ec\nt\\
\twosub{\chi^J_-}(t)&=\frac{\sqrt{2}}{12}\bigg\{\bigg[-\frac{g^2}{2m^2}+b_{00}^-\bigg]\big(1-t\,\dot\Omega_1(0)\big)+b_{01}^-t\bigg\}\delta_{J1}+\frac{\sqrt{6}}{15}\frac{b_{00}^+}{4m}\delta_{J2}\nt\\
&\redonesub\frac{\sqrt{2}}{12}\bigg[-\frac{g^2}{2m^2}+b_{00}^-\bigg]\delta_{J1}
\redunsub0\ep
\end{align}
Note that also in the unsubtracted case the explicit dependence on $a$ cancels.\footnote{Actually, this has to be the case: e.g.\ the constant term proportional to $(m^2-a)^{-1}$ in the nucleon pole terms~\eqref{hdrnucleonpoleterms}, which was introduced to the dispersion relations via the hyperbolic kinematical relations and which can be thought of as a contribution of the contour integral from the circle with infinite radius, leads to constant pole-term contributions to the partial waves (cf.~\eqref{hatNJpm} and~\eqref{DeltahatNJpmunsub}). These (unphysical) contributions do not vanish asymptotically, generate an unphysical behavior on $a$, and thus they must cancel in any (physical) solution. Hence, the dispersion integrals for the unsubtracted case both for the Omn\`es solution and the spectral representation of the Omn\`es function are strictly speaking not correct: there should be contributions from the contour at infinity. However, this problem can be solved most easily by removing all ``dangerous'' parts of the inhomogeneities via~\eqref{omnesspecrep}, which ensures that all these potential contributions from the contour at infinity cancel.}

Finally, due to Watson's theorem~\eqref{watson} we can separate the unknown moduli from the known $\pi\pi$ phases and solve the MO problem for the moduli directly
\begin{align}
\label{MOGamsubnabsvalsol}
\big|f^0_+(t)\big|&=\nsub{\tilde\Delta^0_+}(t)\cos\delta_0(t)
+(t-\tN)\frac{|\Omega_0(t)|}{\pi}\Bigg\{\nsub{\chi^J_0}(t)\nt\\
&\qquad+t^n\Bigg[
\;\dashint\limits_{\tpi}^{\tm}\diff t'\frac{\nsub{\tilde\Delta^0_+}(t')\sin\delta_0(t')}{t'^n(t'-\tN)|\Omega_0(t')|(t'-t)}
+\int\limits_{\tm}^{\infty}\diff t'\frac{\Im f^0_+(t')}{t'^n(t'-\tN)|\Omega_0(t')|(t'-t)}
\Bigg]\Bigg\}\ec\nt\\
\big|\Gamma^J(t)\big|&=\nsub{\tilde\Delta_\Gamma^J}(t)\cos\delta_J(t)
+(t-\tN)\frac{|\Omega_J(t)|}{\pi}\Bigg\{\nsub{\chi^J_\Gamma}(t)\nt\\
&\qquad+t^{(n-J)\theta(n-J)}\Bigg[
\;\dashint\limits_{\tpi}^{\tm}\diff t'\frac{\nsub{\tilde\Delta_\Gamma^J}(t')\sin\delta_J(t')}{t'^{(n-J)\theta(n-J)}(t'-\tN)|\Omega_J(t')|(t'-t)}\nt\\
&\qquad\quad+\int\limits_{\tm}^{\infty}\diff t'\frac{\Im\Gamma^J(t')}{t'^{(n-J)\theta(n-J)}(t'-\tN)|\Omega_J(t')|(t'-t)}
\Bigg]\Bigg\}\qquad\forall\;J\geq1\ec\nt\\
\big|f^J_-(t)\big|&=\nsub{\tilde\Delta^J_-}(t)\cos\delta_J(t)
+\frac{|\Omega_J(t)|}{\pi}\Bigg\{\nsub{\chi^J_-}(t)\nt\\
&\qquad+t^{(n-J+1)\theta(n-J)}\Bigg[
\;\dashint\limits_{\tpi}^{\tm}\diff t'\frac{\nsub{\tilde\Delta^J_-}(t')\sin\delta_J(t')}{t'^{(n-J+1)\theta(n-J)}|\Omega_J(t')|(t'-t)}\nt\\
&\qquad\quad+\int\limits_{\tm}^{\infty}\diff t'\frac{\Im f^J_-(t')}{t'^{(n-J+1)\theta(n-J)}|\Omega_J(t')|(t'-t)}
\Bigg]\Bigg\}\qquad\forall\;J\geq1\ep
\end{align}
On the one hand, the subtraction-independent pole terms $\hat N^J_\pm$ are real for $t\geq\tpi-(\mpi^2/m)^2$ and grow rapidly with $J$ for $t$ in the vicinity of $\tpi$, as discussed in Appendix~\ref{subsec:tpwp:n}.
On the other hand, in the elastic region the phases $\delta_J$ are given by the corresponding $\pi\pi$ scattering phases such that $\delta_J(\tpi)=0$ and thus $\Im f^J_\pm(\tpi)=0$.
Since furthermore phenomenologically the $\pi\pi$ phases grow slower for higher $J$, we thus expect the partial waves $f^J_\pm$ (and thereby also their moduli $|f^J_\pm|$) to be increasingly dominated by the pole terms for increasing $J$ and $t\to\tpi$.
However, we do not solve for $f^J_+$ directly but for the linear combination $\Gamma^J$, for which, in turn, the pole-term contributions $\hat N^J_\pm$ cancel at $\tpi$, cf.\ Appendix~\ref{subsec:tpwp:subtractions}.
The pole-term domination of $|f^J_+|$ enters when calculating these parallel helicity moduli from the solutions for the $|\Gamma^J|$ and the (pole-term dominated) antiparallel helicity moduli $|f^J_-|$ via~\eqref{fJpfromfJmandGammaJ}, where in addition the relative importance of the latter increases with $J$ due to the factor $\sqrt{J/(J+1)}$.
Both the pole-term domination and the dependence of $|f^J_+|$ on $|f^J_-|$ will be explicitly demonstrated in Sect.~\ref{subsec:piNMO:results}.

\subsection{Input}
\label{subsec:piNMO:input}

In this section we will discuss all input that is needed to solve the (elastic) $t$-channel MO problem~\eqref{MOGamsubnabsvalsol} as given in Sect.~\ref{subsec:piNMO:explsol}.

\subsubsection{Pion--pion phases and Omn\`es functions}
\label{subsubsec:piNMO:input:phases}

We use $\pi\pi$ scattering phase shifts $\delta^{I_t}_J(t)$ of~\cite{CCL:Regge,CCL:PWA} for $J\in\{0,1,2\}$ with $I_t\in\{0,1\}$ ($I_t$ even/odd for $J$ even/odd) which are constructed for $\sqrt{t}\in[2\mpi,1.15\GeV]$.\footnote{For $\pi\pi$ scattering the validity of the Roy equations can be shown rigorously for $\tpi\leq t\leq60\mpi^2$ based on axiomatic field theory~\cite{Roy}. Assuming Mandelstam analyticity, this range can be extended to $\tpi\leq t\leq68\mpi^2$~\cite{Basdevant}, which corresponds to $2\mpi\leq\sqrt{t}\leq1.15\GeV$, by reasoning along the lines of Appendix~\ref{sec:convergence} for $\pi\pi$ scattering.}
Schenk-like parameterizations~\cite{Schenk,ACGL}
\begin{equation}
\label{schenk}
\tan\delta^{I_t}_J(t)=\sigma^\pi_tq_t^{2J}\Big\{A^{I_t}_J+B^{I_t}_Jq_t^2+C^{I_t}_Jq_t^4+\dots\Big\}\frac{\tpi-r^{I_t}_J}{t-r^{I_t}_J}\ec
\end{equation}
where the parameter $r^{I_t}_J$ denotes the point where the corresponding phase shift passes through $\pi/2$  and the Schenk parameters $A^{I_t}_J$ etc.\ may be related to the coefficients of the threshold expansion, ensure both the vanishing at threshold $\delta^{I_t}_J(\tpi)=0$ as well as the correct square-root-power behavior above threshold.
Thus, $\delta^0_0$ is linear, $\delta^1_1$ cubic, and $\delta^0_2$ quintic in $\sigma^\pi_t$.

\begin{figure}[t!]
\centering
\includegraphics{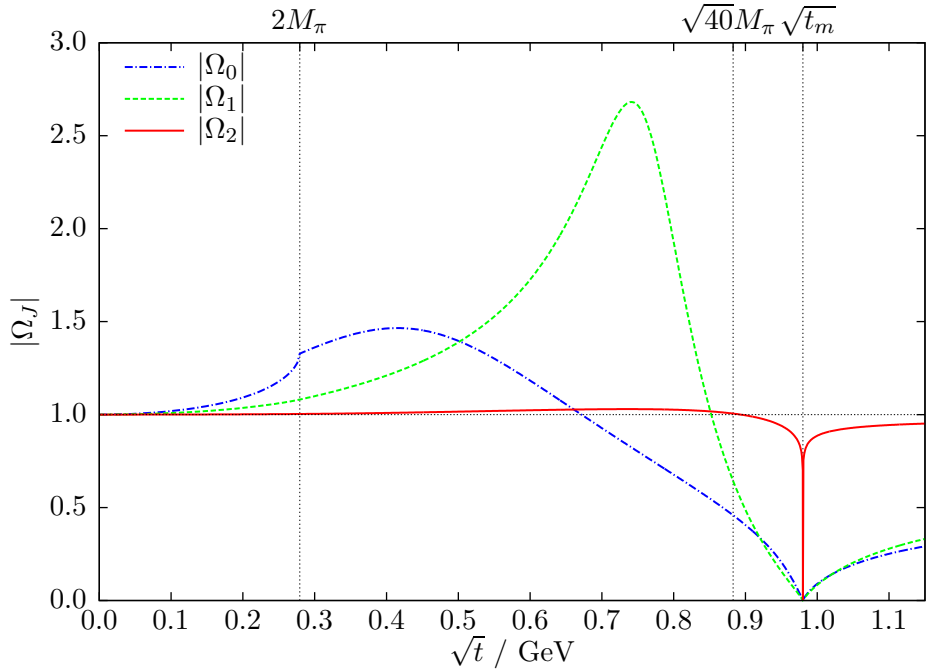}
\caption[Moduli of the lowest once-subtracted finite-matching-point Omn\`es functions.]{Moduli $|\Omega_J|$ of the lowest once-subtracted finite-matching-point Omn\`es functions for $\sqrt{\tm}=0.98\GeV$.}
\label{fig:Omega}
\end{figure}

In Fig.~\ref{fig:Omega} we show the moduli $|\Omega_J|$ of the resulting once-subtracted finite-matching-point Omn\`es functions according to~\eqref{MOOmegaJ} for $J\in\{0,1,2\}$, where the choice $\sqrt{\tm}=0.98\GeV$ ensures that $x_J(t)\in(0,1)$ and hence $n_J=\lfloor x_J(\tm)\rfloor=0$ for $t\in[\tpi,\tm]$.
Therefore, all functions are normalized to unity at $t=0$, finite for all $t$, and vanish at $t=\tm$ due to the finite-matching-point prefactor $|\tm-t|^{x_J(t)}$.
Furthermore, for $J=0$ the Omn\`es function exhibits a cusp (i.e.\ a discontinuity of the derivative) at the physical $\pi\pi$ threshold $\tpi$ and decreases approximately linearly over a wide range in $t$, for $J=1$ it is fully dominated by the $\rho(770)$ peak, and for $J=2$ it is almost flat (equaling one again roughly at the end of the KH80 energy range at $0.88\GeV$ and dropping rapidly above).

Using instead the parameterization of the $\pi\pi$ phases as given in~\cite{GKPY} for the numerical evaluations leads to deviations in these Omn\`es functions, and thereby the solutions of the MO problem~\eqref{MOGamsubnabsvalsol}, that, however, are much smaller than the effects of the alterations described in Sect.~\ref{subsec:piNMO:results}.\footnote{As stated in~\cite{Hoehler}, the Karlsruhe--Helsinki dispersive partial-wave analyses KH78 and KH80 (see Sect.~\ref{subsubsec:piNMO:input:expwa} for more details) use as input the $\pi\pi$ phase shifts of~\cite{FroggattPetersen77}, which are based on Roy-equation fits. In principle, the differences between these phase shifts and the recent results~\cite{CCL:Regge,CCL:PWA,GKPY} are sources of discrepancies between the KH80 results and the solutions of the MO problem. However, this point is of minor importance for the results discussed in Sect.~\ref{subsec:piNMO:results}.}

\subsubsection{General remarks on existing pion--nucleon partial-wave analyses}
\label{subsubsec:piNMO:input:expwa}

Before summarizing the input from $\pi N$ partial-wave analyses that will be used in the following, some general remarks are in order:
first of all, we will use the Karlsruhe--Helsinki dispersive partial-wave analysis KH80~\cite{KH80,Hoehler} both as input for $s$-channel partial waves as well as subthreshold parameters and as reference for our MO $t$-channel partial-wave solutions, since KH80 is still the only consistent analysis for all the partial waves and parameters entailed in our RS framework.
KH80 is based on $\pi N\to\pi N$ data only (and isospin invariance) and uses Pietarinen's expansion method~\cite{Pietarinen72} in combination with conformal mapping techniques, aided in particular by fixed-$t$ analyticity.
Its solutions for both channels are given as tables in~\cite{Hoehler}.\footnote{In~\cite{Hoehler} the results for the $t$-channel partial waves are quoted as KH78 solution, but according to~\cite{Hoehler:1999:sigmaterm} these tables are actually calculated from the KH80 $s$-channel solutions. Thus we will speak of the $t$-channel partial waves in~\cite{Hoehler} as KH80 solution as well. In general, KH80 is an update of KH78 including more recent data and improved fixed-$t$ analyticity constraints.}
Moreover, an iteration uncertainty of about $3\,\%$ for the iterative KH80 procedure is stated, which, however, cannot replace a thorough analysis of the systematic uncertainties.
The subsequent Karlsruhe analysis KA84~\cite{KA84} improves on KH80 especially for higher partial waves by using a modified PWDR framework and thereby smoothing KH80, but no consistent subthreshold parameters are derived in this framework.\footnote{For a comparison of KH80 with KA84 and also an improvement of the formalism outlined in~\cite{BaackeSteiner}, see~\cite{Koch:comparison}.}
The same holds true for the continuously updated VPI/GWU(SAID) $s$-channel analyses, see e.g.~\cite{SP98,GWU:2006,GWU:2008,SAID}, for which at most the $\pi N$ coupling constant and some of the necessary subthreshold parameters are determined.
For the $t$-channel partial waves in the unphysical region $t\in[\tpi,\tN]$, there also exists an unpublished solution~\cite{Pietarinen} extending the KH80 energy range $\sqrt{t}\in[2\mpi,\sqrt{40}\mpi=0.88\GeV]$ to roughly $1\GeV$.
While this solution is compatible with KH80 within the aforementioned range, it seems to suffer from internal inconsistencies for higher energies.\footnote{There are e.g.\ rather obvious outliers (corresponding to unphysical jumps) in the phases.}
For the $t$-channel partial waves in the physical region $t\geq\tN$, however, there exists a partial-wave analysis~\cite{Anisovich}, which at least in principle could be used as input.
Finally, a partial update of the KH80 analysis including new data and using more computational power was reported in~\cite{Sainio}, but so far only results for forward $\pi N$ scattering have been published~\cite{Metsae}.

\subsubsection[$s$-channel partial waves]{\boldmath{$s$}-channel partial waves}

\begin{figure}[t!]
\centering
\includegraphics[width=0.47\linewidth]{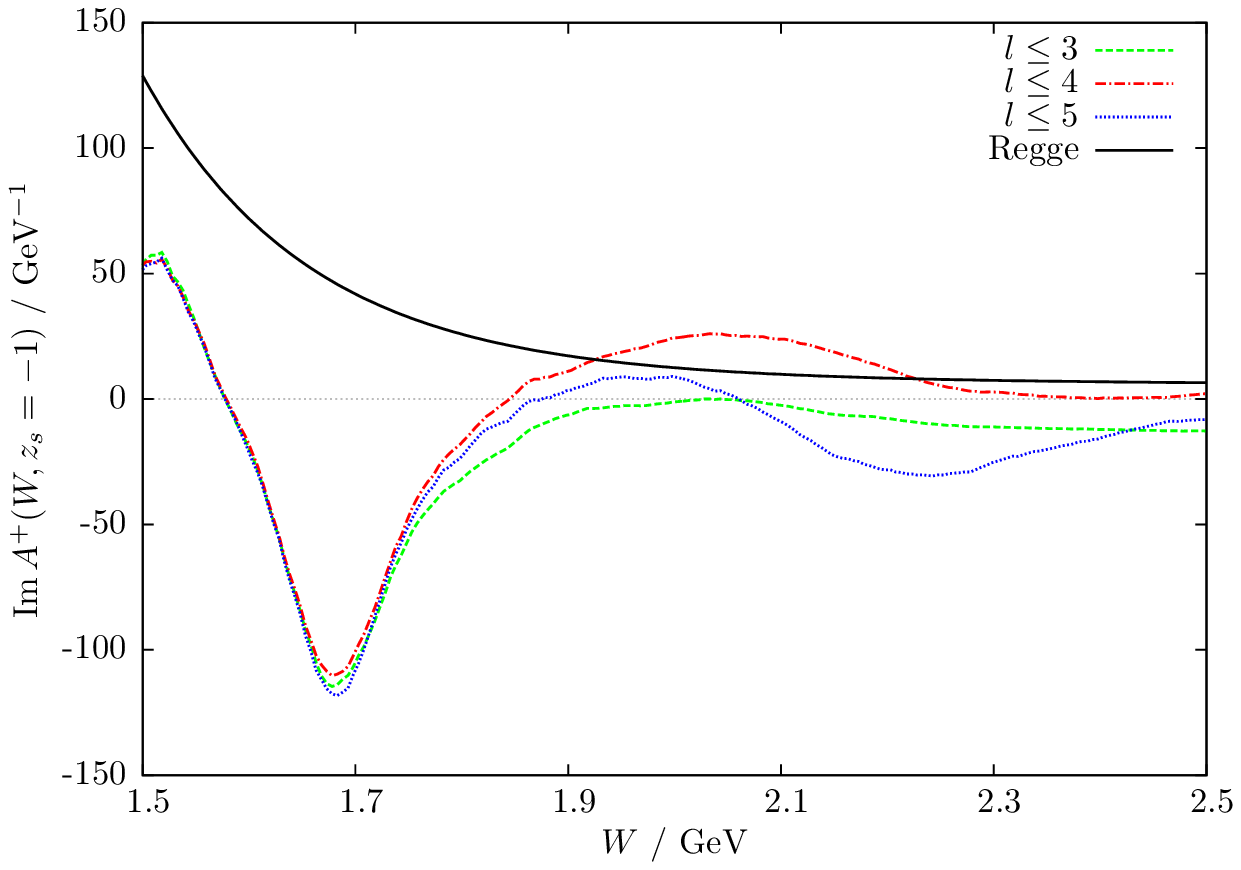}\hspace{0.05\linewidth}
\includegraphics[width=0.47\linewidth]{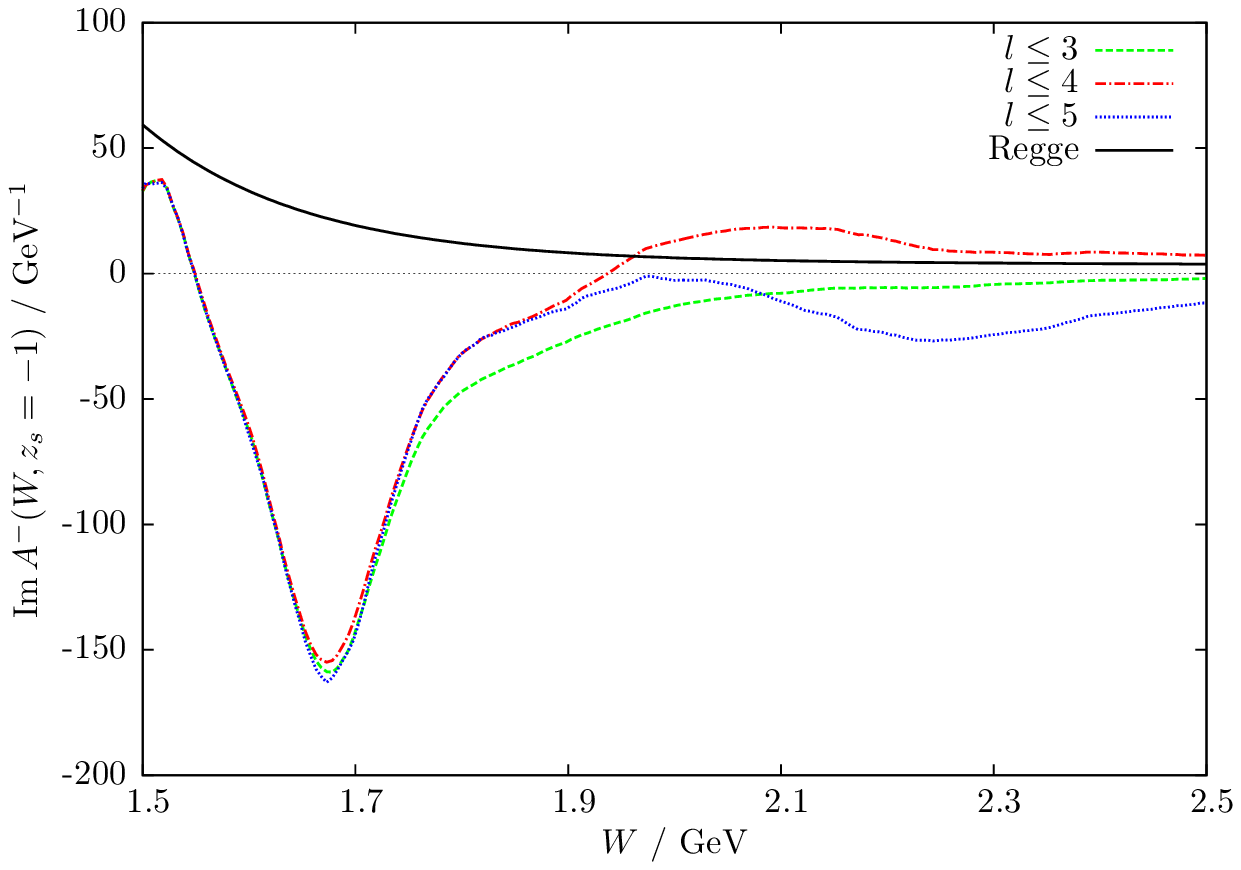}\\
\includegraphics[width=0.47\linewidth]{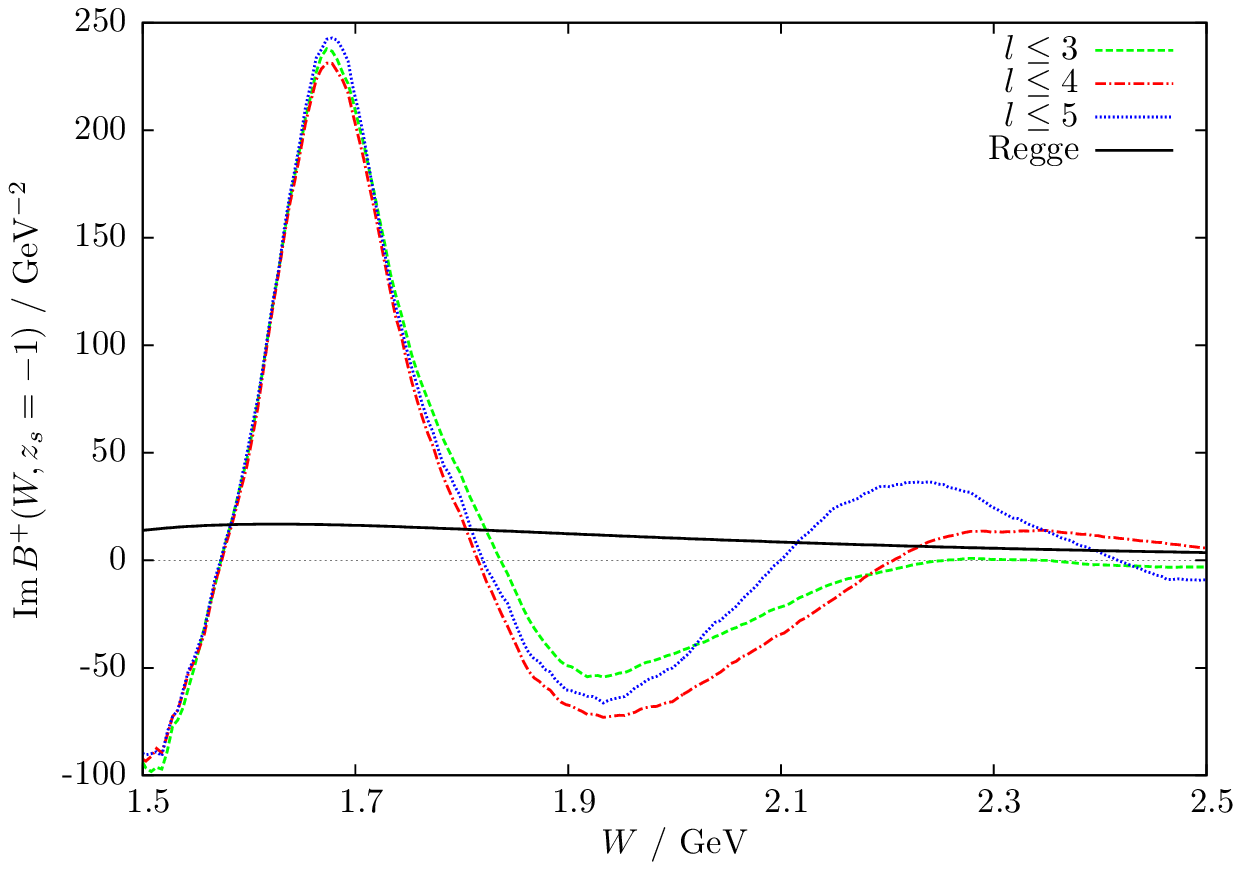}\hspace{0.05\linewidth}
\includegraphics[width=0.47\linewidth]{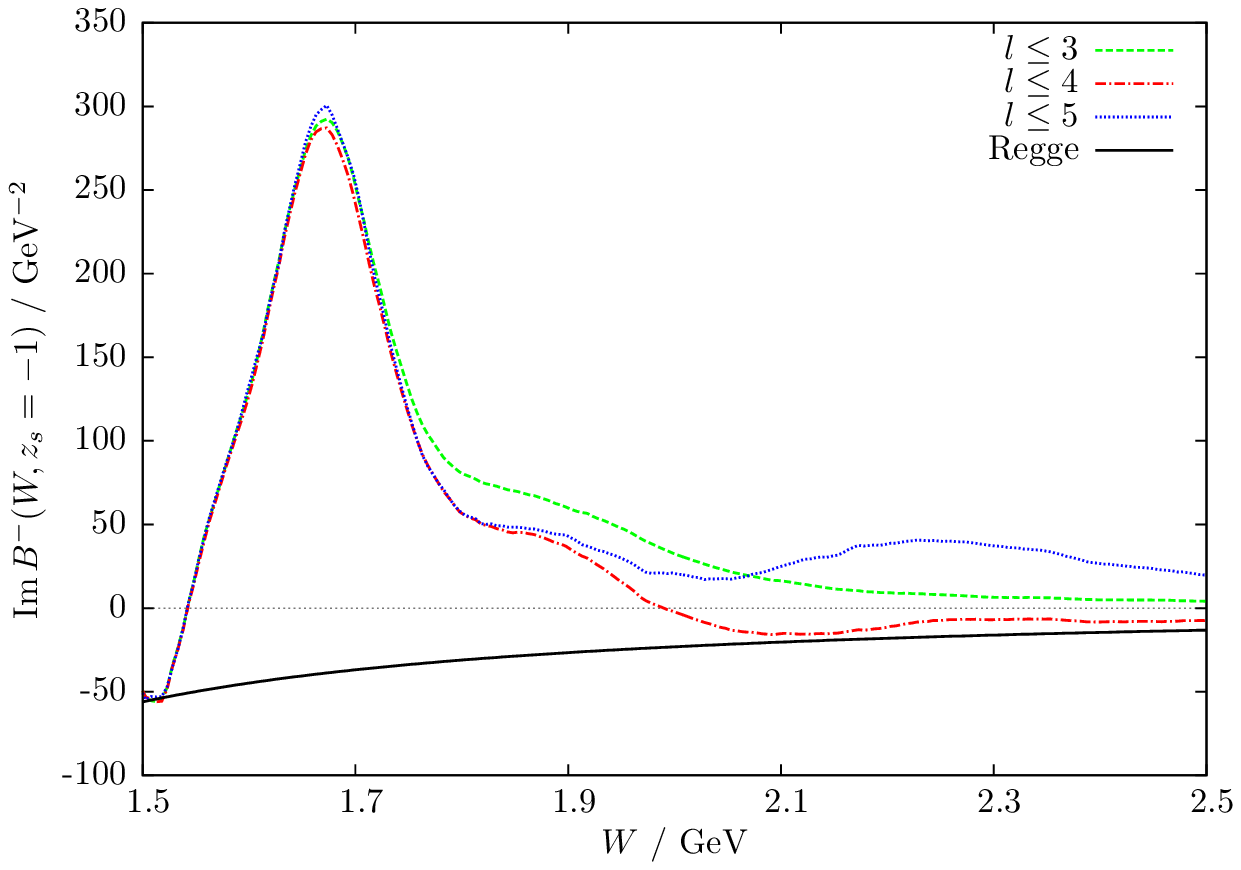}
\caption[Matching of $s$-channel absorptive parts between partial-wave contributions and Regge model.]{Matching of the $s$-channel absorptive parts between KH80 partial-wave contributions for $l\leq3$, $l\leq4$, and $l\leq5$ and the $\pi N$ backward scattering Regge model~\cite{piNuRegge}.}
\label{fig:ABpmtoRegge}
\end{figure}

We use the KH80 solution for the $s$-channel partial waves from~\cite{Hoehler} for $W_+\leq W\leq\Wa=2.5\GeV$.
On the one hand, this is roughly the same energy range as for the continuously updated GWU ``current solution''~\cite{SAID} such that we are able to compare between KH80 and GWU solutions as input.
However, the effect of taking the GWU solution (or the ``smoothed'' KH80 solution~\cite{SAID}) instead as input for the $t$-channel Omn\`es problem (i.e.\ the corresponding inhomogeneities $\tilde\Delta^J_\pm$) turns out to be much smaller than the effects discussed in Sect.~\ref{subsec:piNMO:results}.
On the other hand, at $\Wa=2.5\GeV$ a reasonable transition from the truncated sum of partial waves below $\Wa$ to the Regge model for the full invariant amplitudes above $\Wa$ can be achieved as we will demonstrate now.
Summing up all partial waves with $l\leq5$ would encompass all 4-star resonances of~\cite{PDG}, but of both $l=5$ 4-star resonances, $N(2220)$ as $H_{1,9}$ and $\Delta(2420)$ as $H_{3,11}$, especially the latter is mostly out of this energy range due to its broad width of roughly $700\MeV$.
Hence we expect the best agreement with the Regge model~\cite{piNuRegge}, which is based on differential cross section and polarization data for $\pi N$ backward scattering with $W\geq3\GeV$ as discussed in Appendix~\ref{subsec:regge:s}, for $l\leq4$ and a scattering angle of $z_s=-1$ corresponding to backward scattering.
Since deviations between summing up contributions for $l\leq3$, $l\leq4$, and $l\leq5$ start to show up around $1.5\GeV$ and we are interested in the matching to the Regge model at the end of the GWU range of validity around $2.5\GeV$, only this region is shown in Fig.~\ref{fig:ABpmtoRegge} (in the spirit of~\cite{piK:RS}).
Note that only $l\leq4$ yields the correct sign compared to the Regge contribution in all four cases.
Moreover, it turns out that for $l\leq5$ the agreement is even worse than for $l\leq3$.
Hence, in the following all higher partial waves with $l\geq5$ will be neglected below $\Wa$.

\subsubsection[$t$-channel partial waves]{\boldmath{$t$}-channel partial waves}
\label{subsubsec:piNMO:input:tpws}

The assumption of elastic unitarity breaks down in the $S$-wave as soon as the $\bar KK$ channel opens, which manifests itself in the appearance of the $f_0(980)$ resonance.
In principle, there are several ways how this phenomenon may be accommodated in a single-channel description.

First, inelastic contributions could be included directly in the solution of the MO equations along the lines discussed in~\cite{PhamTruong,Caprini} provided that the inelasticities are well known.
However, in the case of $f^0_+$ this would in particular require knowledge of the $\bar KK\to\bar NN$ $S$-wave, but it is unclear how reliable input for this partial wave could be obtained independently from the present approach.

Second, one could retain a rather low matching point $\tm$, but try to model the energy region above $\tm$ by means of a resonance description in order to establish a more meaningful matching condition.
This strategy proved quite successful in $\gamma\gamma\to\pi\pi$~\cite{ggpipi}, where the input above the matching point is dominated by the $f_2(1270)$.
However, in the case of the $f_0(980)$ this strategy is subject to several difficulties: its pole position is very close to the two-kaon threshold, such that the subtle interplay between the $\pi\pi$ and $\bar KK$ channels can certainly not be approximated by a simple Breit--Wigner description.
To circumvent this problem, one would be compelled to further decrease the matching point and include the $f_0(980)$ dynamics by hand using a Flatt\'e-like parameterization~\cite{Flatte}, which is a modified relativistic version of the Breit--Wigner differential mass distribution.
However, while the $f_0\pi\pi$ coupling constant has been thoroughly investigated~\cite{GKPY:f0} based on the recent dispersive analysis~\cite{GKPY} (which yields phases that are basically consistent with the phases of~\cite{CCL:Regge,CCL:PWA}), the $f_0NN$ coupling constant is only very poorly known, with different meson-exchange models disagreeing significantly on the strength of the coupling and the continuation to the physical pole~\cite{Nagels:1977,Maessen:1989,Stoks:1994,Rijken:2002}.
We conclude that including the $f_0(980)$ in our approach reliably as well as extending the energy range of our representation for $f^0_+$ beyond the two-kaon threshold will require a full solution of the underlying two-channel Omn\`es problem~\cite{twochannelOmnes}.
In this work we will content ourselves with the single-channel approximation.

Since therefore we can solve the single-channel MO problem in the elastic region only and furthermore iteration with the $s$-channel RS solutions (for which, in turn, accurate MO solutions are needed as input) as well as a consistent determination of the $\pi N$ coupling and the subthreshold parameters is necessary to finally arrive at precise quantitative results for the partial waves of both channels, here we will only give qualitative results for the $t$-channel partial waves by comparing with KH80.

Hence, in the following all $t$-channel absorptive parts above $\tm$ are set to zero.
Consequently, all $t$-channel Regge contributions are omitted (since $\tm<\ta$; cf.\ the discussion of the $t$-channel asymptotics in Appendix~\ref{subsec:regge:t}).
Note that otherwise one would have to avoid double counting of the asymptotic regions of the $t$-channel partial waves in the MO problem.
Finally, also all higher partial waves with $J\geq3$ are neglected.

\subsubsection{Subthreshold parameters}

To precisely determine the subthreshold parameters is not an easy task, since there simply is no experimental data available to analyze the $t$-dependence of the amplitudes close to $t=0$ and thus means of analytic continuation or extrapolation are needed.
Accordingly, in the literature there are only few determinations of all parameters that enter the subtracted RS system.
The KH80 results (cf.~\cite{Hoehler}, wherein the error estimates are quoted to be ``based on deviations from the internal consistency'' and the total uncertainty to be ``somewhat larger'') and all more recent dispersion theoretical analyses that we are aware of are collected in Table~\ref{tab:subthrpar} (cf.~\cite{BecherLeutwyler}).
\begin{table}
\renewcommand{\arraystretch}{1.3}
\centering
\begin{tabular}{c r r r r r r}
\toprule
 & KH80 & St(KA84) & St(SP98) & Oa(KH80) & Oa(SP98) & Fe(KA84) \\
\midrule
$d^+_{00}~\big[\mpi^{-1}\big]$\!\! & $-1.46\pm0.10$ & $-1.39\pm0.02$ & $-1.32\pm0.02$ & $-1.46\pm0.04$ & $-1.29\pm0.02$ & $-1.58$ \\
$d^+_{01}~\big[\mpi^{-3}\big]$\!\! & $ 1.14\pm0.02$ & $ 1.14\pm0.01$ & $ 1.15\pm0.02$ & $ 1.15\pm0.11$ & $ 1.23\pm0.04$ & $ 1.36$ \\
$a^-_{00}~\big[\mpi^{-2}\big]$\!\! & $-8.83\pm0.10$ & $-8.82\pm0.04$ & $-8.97\pm0.01$ & $-9.26\pm0.17$ & $-8.92\pm0.07$ & $-8.47$ \\
$b^+_{00}~\big[\mpi^{-3}\big]$\!\! & $-3.54\pm0.06$ & $-3.49\pm0.03$ & $-3.48\pm0.02$ & $-3.56\pm0.10$ & $-3.42\pm0.04$ & $-7.90$ \\
$b^-_{00}~\big[\mpi^{-2}\big]$\!\! & $10.36\pm0.10$ & $10.35\pm0.02$ & $10.45\pm0.01$ & $10.84\pm0.18$ & $10.37\pm0.08$ & $10.34$ \\
$b^-_{01}~\big[\mpi^{-4}\big]$\!\! & $ 0.24\pm0.01$ & $ 0.22\pm0.01$ & $ 0.24\pm0.01$ & $ 0.26\pm0.22$ & $ 0.26\pm0.10$ & $ 0.14$ \\
\bottomrule
\end{tabular}
\renewcommand{\arraystretch}{1.0}
\caption[Subthreshold parameter values.]{Subthreshold parameter values as given by KH80/H\"{o}hler~\cite{Hoehler}, Stahov~\cite{Stahov:1999}, Oades~\cite{Oades:1999,Oades:2002}, and Fettes (heavy-baryon ChPT)~\cite{Fettes:Diss}. See main text for details.}
\label{tab:subthrpar}
\end{table}
Note that there are several determinations of only some of these parameters, which are therefore not listed in Table~\ref{tab:subthrpar}.
In~\cite{Stahov:1999} the subthreshold parameters are determined  by means of interior dispersion relations together with fixed-$t$ dispersion relations, and by using as input the $s$-channel partial waves of both KA84 and VPI/SP98~\cite{SP98,SP98MENU} as well as the $t$-channel partial waves of KH80 (and those of~\cite{Pietarinen} in the consistent energy range).\footnote{These are most probably the ``new subthreshold parameters'' mentioned in~\cite{Hoehler:1999:phenomenology}, where no explicit reference is given.}
In contrast, finite-contour dispersion relations are used in~\cite{Oades:1999} to derive subthreshold parameter values---again for both KH80 and VPI/SP98 input (amongst others).\footnote{Note that some of the results of~\cite{Oades:1999} are corrected in~\cite{Oades:2002}, where also a modified version of the finite-contour dispersion relations together with conformal mapping techniques is applied (it is mentioned therein that the subthreshold parameters do not change substantially). Since the applied fitting procedure does not respect the exact analytic equality of the parameters $d^+_{0n}$ and $a^+_{0n}$, however, the (corrected) values agree only within the given errors, but not exactly.}

The subthreshold parameters are the standard expansion parameters for the Lorentz-invariant amplitudes, but neither these amplitudes nor the kinematical variables $\nu$ and $t$ are natural for heavy-baryon ChPT, and hence values obtained in analyses using this framework are not very satisfactory, cf.~\cite{Oades:2002}.
However, for comparison we also state the corresponding values for a third-order calculation~\cite{Fettes:piN1,Fettes:Diss} as given in~\cite{Fettes:Diss} (Fit 1 therein corresponding to KA84); note that according to~\cite{Fettes:piN2,Fettes:Diss} some of the parameter values even deteriorate when calculated up to fourth order.

As can be seen already from the deviations between the different determinations of subthreshold parameters in Table~\ref{tab:subthrpar}, the errors on the central values are in general unrealistically small (i.e.\ only statistical fit errors for specific input in a given framework, thus neglecting systematic errors).
Hence we can conclude that there is no precise and consistent determination of the subthreshold parameters including realistic errors.
Since we want to compare our MO results with the KH80 solutions, for consistency we use the KH80 subthreshold parameters as given in Table~\ref{tab:subthrpar} as well as the outdated KH80 $\pi N$ pseudoscalar coupling value of 14.28 instead of the new value of 13.7 as given in~\eqref{piNcouplingvalue}.\footnote{Note that the $\pi N$ coupling and the subthreshold parameters are related, as the difference $d^-_{00}-g^2/(2m)$ is given by an integral over a total cross section, cf.~\cite{BecherLeutwyler}.}

\subsection{Results}
\label{subsec:piNMO:results}

The numerical results that will be presented in this section are to be understood as a qualitative ``KH80 consistency check'' in order to show that the $t$-channel RS--MO machinery works, and as a first step towards a numerical analysis of the full RS system.
In particular, by variation of either the coupling or the subthreshold parameters we can alter the results significantly, since these variations produce the most sizable effects on the MO solutions compared to the other variation that will be discussed in the following.
However, it is by no means clear a priori what the parameter values or their errors are, and only a self-consistent determination of all parameters and partial waves in a second step will allow for reliable quantitative results. 
Therefore, the necessary first task in this program is to check our method and the internal consistency of the KH80 results by using KH80 input as described in Sect.~\ref{subsec:piNMO:input} and comparing our $t$-channel MO results with those of KH80.
Moreover, we will investigate different systematic effects on the (subtracted) MO solutions $|f^J_\pm|$, which should prove valuable for the solution of the full system:
after discussing exemplarily the importance of the different contributions to the MO inhomogeneities $\tilde\Delta^J_\pm(t)$, we will also discuss both the connection to the ``fixed-$t$ limit''\footnote{Accordingly, the $s$-channel integral of the HDRs reduces to the fixed-$t$ result, cf.\ Sect.~\ref{subsec:preliminaries:hdrs}. However, even in this limit the HDRs contain additional information as compared to fixed-$t$ dispersion relations, since those do not provide equations for the $t$-channel partial waves in the first place.} $a\to-\infty$ and the effect of changing the matching point $\tm$.
Except for the $a\to-\infty$ results, we will always use the optimal hyperbola parameter value of $a=-2.71\,\mpi^2$ as obtained in Appendix~\ref{subsec:convergence:tprojection}.

\clearpage
\subsubsection{Contributions to Muskhelishvili--Omn\`es inhomogeneities}

\begin{figure}[t!]
\centering
\includegraphics{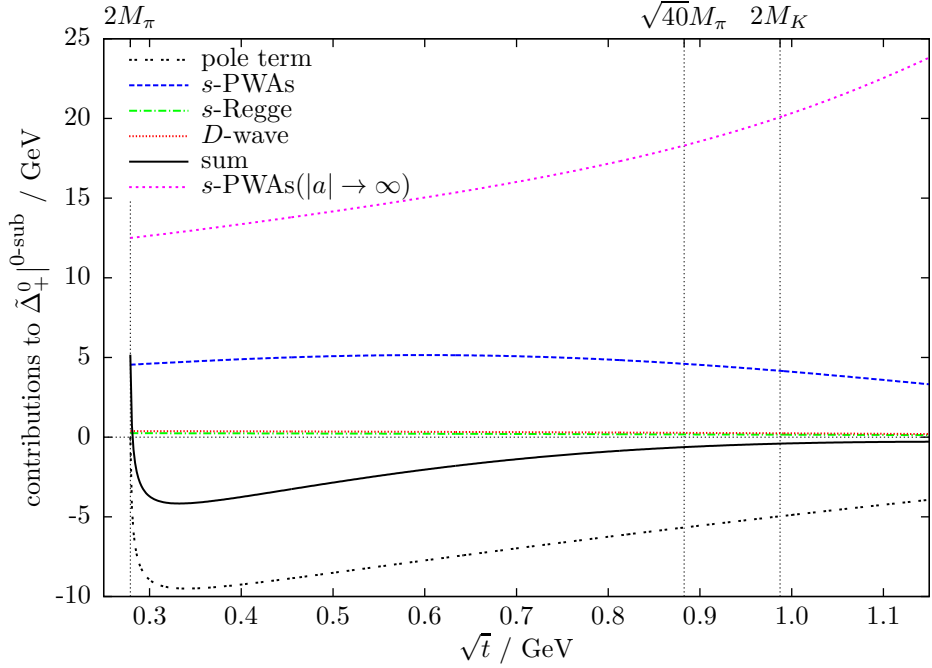}
\caption[Contributions to unsubtracted $S$-wave MO inhomogeneity.]{Contributions to the unsubtracted $S$-wave MO inhomogeneity $\unsub{\tilde\Delta^0_+}$. See text for details.}
\label{fig:Deltatil0psub0}
\end{figure}
\begin{figure}[t!]
\centering
\includegraphics{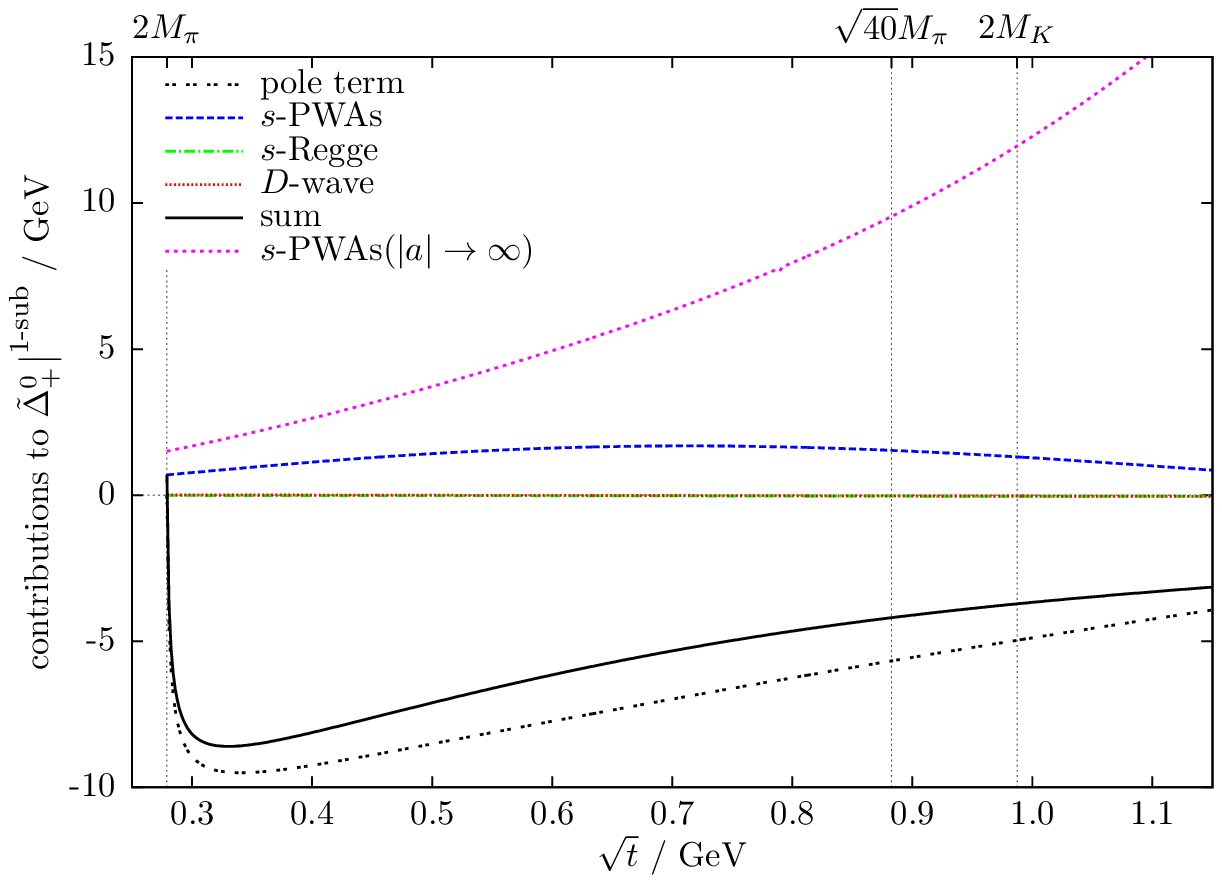}
\caption[Contributions to once-subtracted $S$-wave MO inhomogeneity.]{Contributions to the once-subtracted $S$-wave MO inhomogeneity $\onesub{\tilde\Delta^0_+}$.}
\label{fig:Deltatil0psub1}
\end{figure}
\begin{figure}[t!]
\centering
\includegraphics{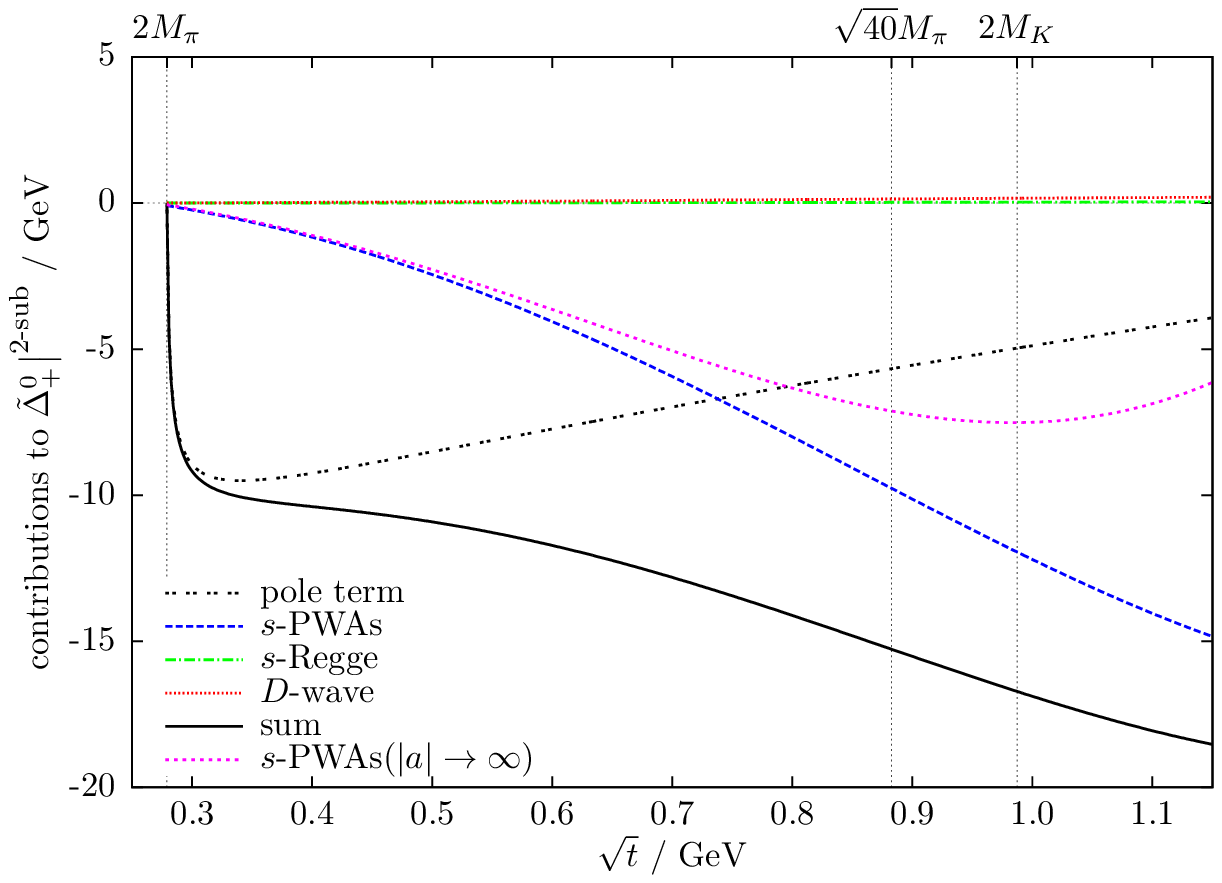}
\caption[Contributions to twice-subtracted $S$-wave MO inhomogeneity.]{Contributions to the twice-subtracted $S$-wave MO inhomogeneity $\twosub{\tilde\Delta^0_+}$.}
\label{fig:Deltatil0psub2}
\end{figure}

In Figs.~\ref{fig:Deltatil0psub0},~\ref{fig:Deltatil0psub1}, and~\ref{fig:Deltatil0psub2} we show the different contributions to the MO inhomogeneities $\tilde\Delta^J_\pm(t)$ exemplarily for $\tilde\Delta^0_+$ for the un-, once-, and twice-subtracted case, respectively, from the $\pi\pi$ threshold $\tpi$ up to $1.15\GeV$, also indicating the upper limit of the KH80 solution as well as the $\bar KK$ threshold $\tK$ as the uppermost limit of approximate elasticity for $J=0$.
We choose the $S$-wave for the following reasons: for $J=0$ the nucleon pole term is zero at $\tpi$ and does not dominate all other contributions as it does for the higher partial waves; in addition, for the $S$-wave we can also show the coupling of the $D$-wave as the leading example for the coupling of higher partial waves.
The pole term $\hat N^0_+$ is independent of both the number of subtractions and $a$ and thus serves as reference in all three plots (double-dashed).
The $s$-channel contributions are shown separately for the sum of all partial waves with $l\leq4$ in the range $W\in[W_+,\Wa]$ (dashed) and the Regge contributions of the full invariant amplitudes for $W>\Wa$ (dot-dashed). 
Even in the unsubtracted case both the $s$-channel Regge as well as the $t$-channel $D$-wave contributions (dotted) are very small and almost negligible in comparison to the other parts.
From this it is also clear that the coupling of higher $t$-channel partial waves (e.g.\ $F$-wave contributions to $P$-waves) can be completely omitted.
The solid line denotes the sum of all these contributions and we have checked for $J\in\{0,1,2\}$ and $n\in\{0,1,2\}$ that the expected threshold behavior according to~\eqref{thresholdf0p} and~\eqref{thresholdfJpm} (as for the corresponding partial waves) is indeed fulfilled.
While all results are given for the optimal value of $a$ unless stated otherwise, for comparison we also show the non-Regge $s$-channel contributions in the ``fixed-$t$ limit'' $a\to-\infty$.
Since this is a very drastic alteration (the RS system is not strictly valid in this case as will be explained below), the difference of this contributions for the two $a$ values gives a very ample bound on the dependence on $a$.
While the Regge contributions vanish for $a\to-\infty$ as discussed in Appendix~\ref{subsec:regge:s}, the $D$-wave coupling is not even well defined for $a\to-\infty$ in this framework as can be seen from the explicit $a$-dependence in the unsubtracted case leading to an infinite contribution, cf.~\eqref{MODwavecouplingsubn}.
By comparing the three plots it is clearly seen that in the once-subtracted case all contributions except the pole term are suppressed, while in the twice-subtracted case an additional $t$-dependence is introduced such that they are strongly suppressed at $\tpi$, but at least the $s$-channel partial-wave contributions are comparable to the pole term around $0.75\GeV$.
For small $t$, the differences between the two $a$ values are also suppressed by each subtraction as expected.

\subsubsection{Muskhelishvili--Omn\`es solutions: comparison with KH80}
\label{subsubsec:piNMO:results:MOKH80}

\begin{figure}[t!]
\centering
\includegraphics{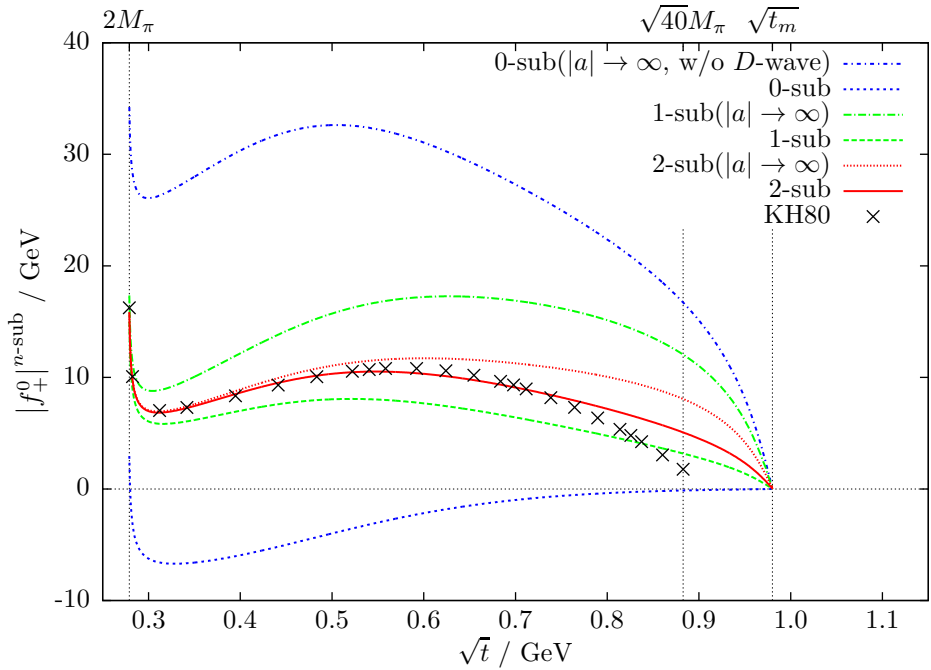}
\caption{MO solutions for the $S$-wave.}
\label{fig:modf0pDsubn}
\end{figure}
\begin{figure}[t!]
\centering
\includegraphics{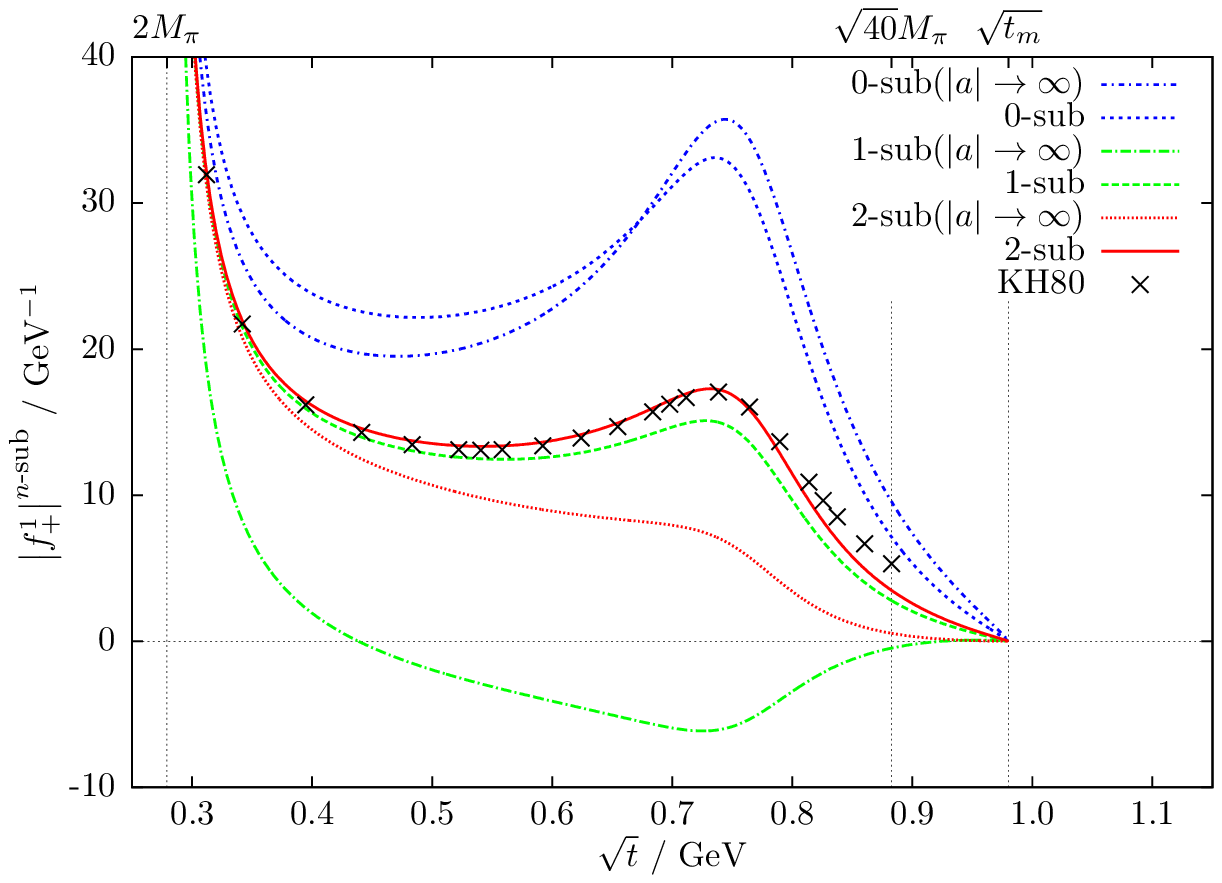}\\\includegraphics{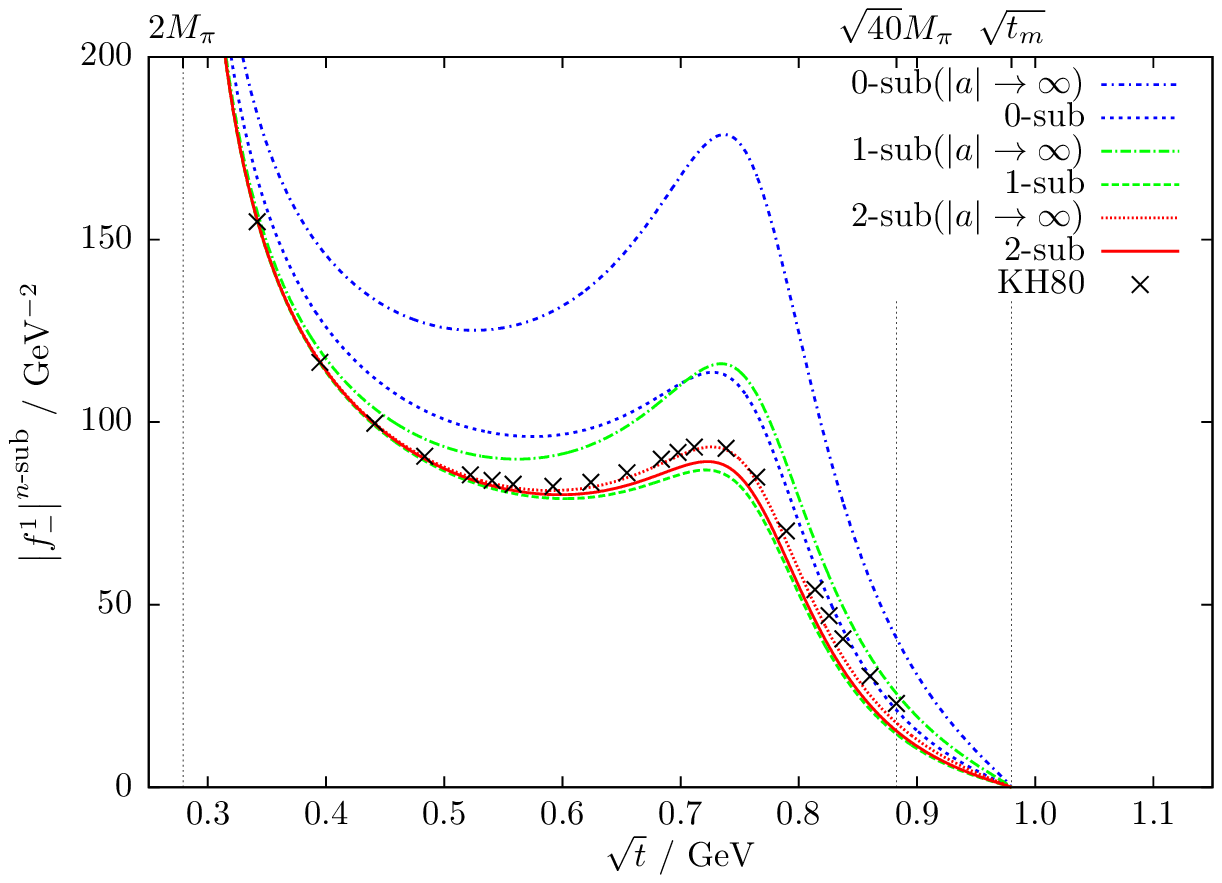}
\caption{MO solutions for the $P$-waves.}
\label{fig:modf1pmDsubn}
\end{figure}
\begin{figure}[t!]
\centering
\includegraphics{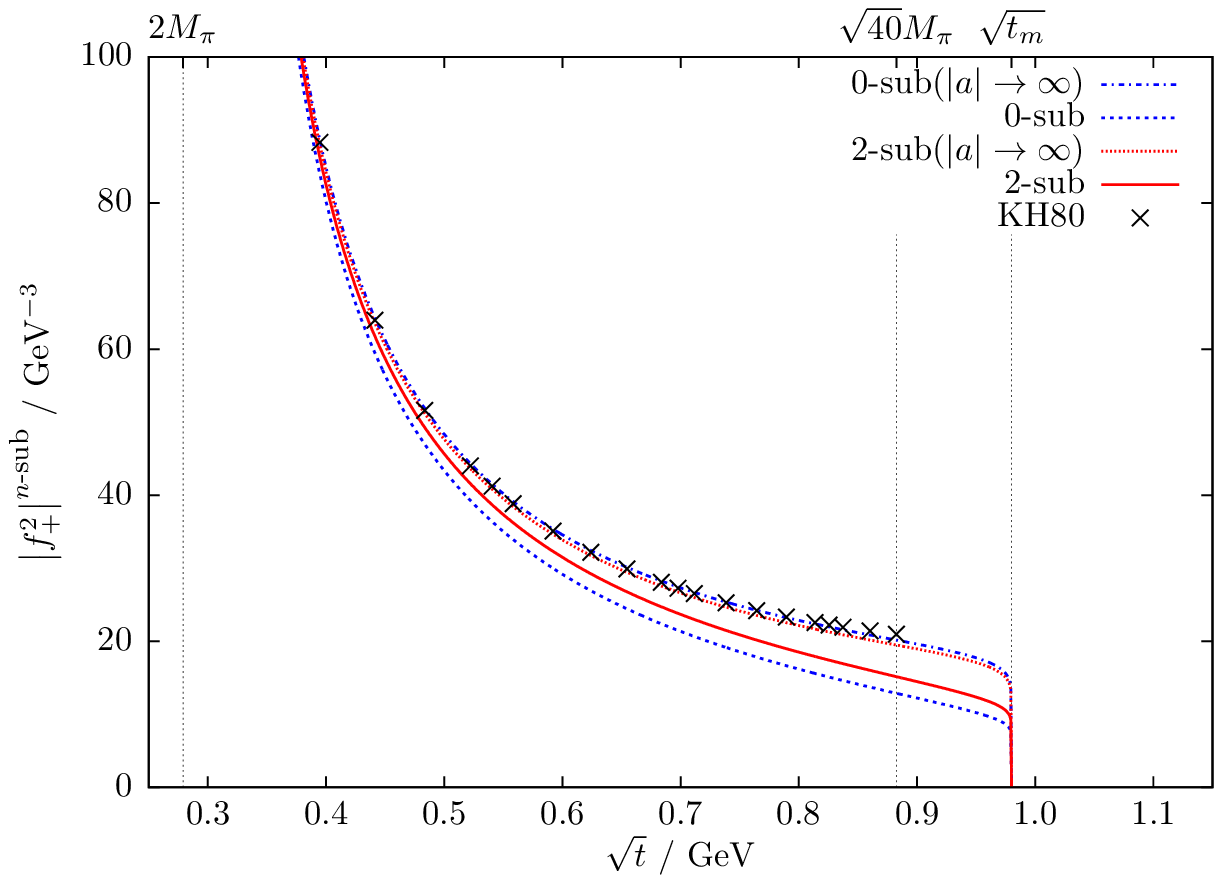}\\\includegraphics{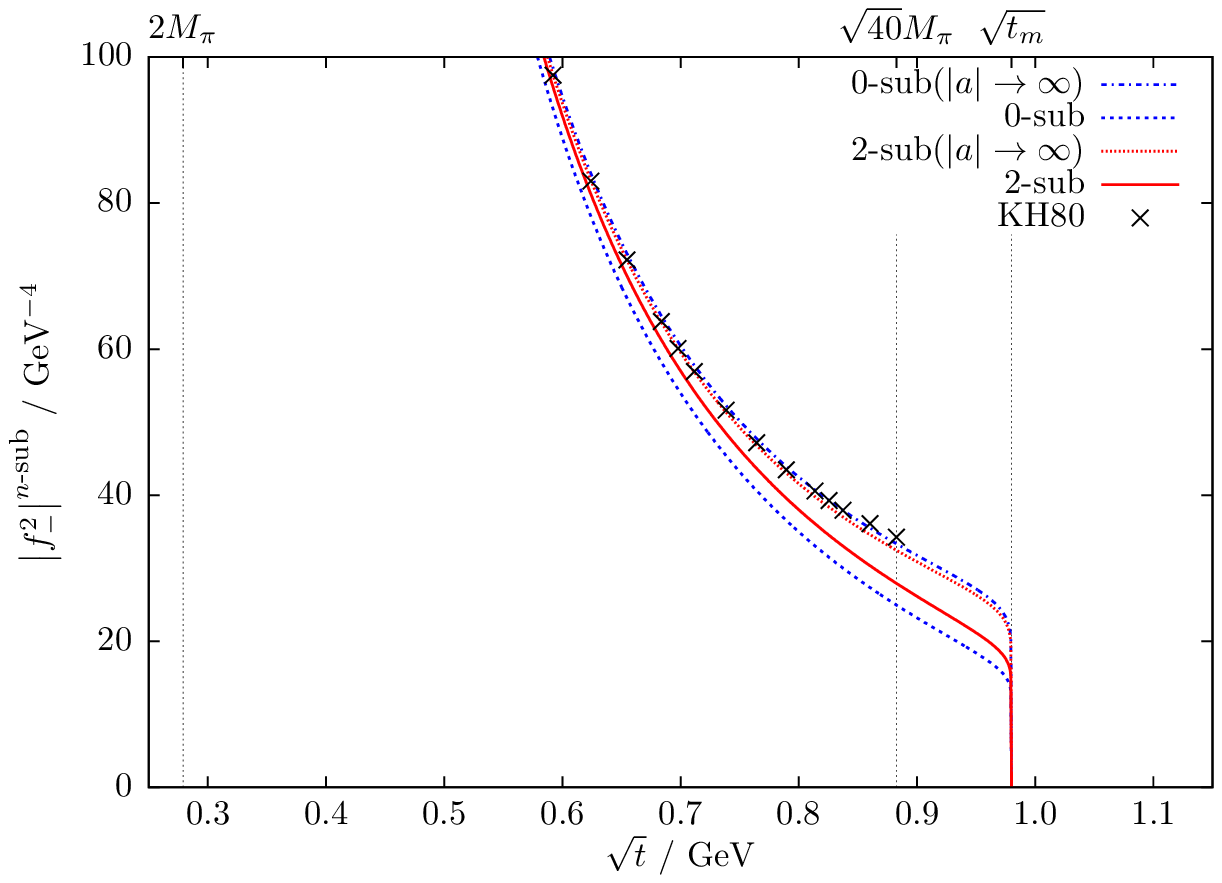}
\caption[MO solutions for the $D$-waves.]{MO solutions for the $D$-waves. For $J\geq2$ one subtraction has no effect.}
\label{fig:modf2pmDsubn}
\end{figure}
We will compare our un-, once-, and twice subtracted MO solutions for $|f^J_\pm(t)|$ with $J\in\{0,1,2\}$ for $t\in[\tpi,\tm]$ with the KH80 results given as tables in~\cite{Hoehler}.
Note that for $J\geq2$ the un- and once-subtracted solutions coincide.
The $a$-dependence (which is fully contained in $\tilde\Delta^J_\pm$) can be used as a crude measure for the systematic uncertainties due to neglecting $t$-channel input above $\tm$ (i.e.\ ``non-analytic'' input), since the physical result must be independent of $a$.
Thus, for the five lowest $t$-channel partial waves we show our ``KH80 consistency MO solution'' for the un-, once-, and twice-subtracted case, each for both the optimal value of $a$ and $a\to-\infty$ in Figs.~\ref{fig:modf0pDsubn},~\ref{fig:modf1pmDsubn}, and~\ref{fig:modf2pmDsubn}.
Here, we have chosen to use the same value of $\sqrt{\tm}=0.98\GeV$ for all considered partial waves, which in principle is not necessary (the effect of varying $\tm$ will be considered explicitly below).
As discussed in Sects.~\ref{subsubsec:piNMO:input:phases} and~\ref{subsubsec:piNMO:input:tpws}, this choice is mainly motivated by the $S$-wave phase, which is just below $\pi$ at this energy (reaching $\pi$ around the $\bar KK$ threshold $\sqrt{\tK}=2\mK=0.987\GeV$) such that no additional subtractions are necessary in the MO scheme.
In general, neglecting any input above the matching point enforces $|f^J_\pm(\tm)|=0$ on the MO solutions.
Nevertheless, even for the $S$-wave we expect reasonable agreement with KH80 for this choice of $\tm$, since both KH80 and~\cite{Pietarinen} suggest that the modulus $|f^0_+|$ has a minimum or even an approximate zero between $0.9\GeV$ and $\sqrt{\tK}$.

In general, the solutions are fixed on both ends of the solution interval $[\tpi,\tm]$: on the left due to the pole term and on the right due to the input above $\tm$.
In our case the solutions are forced to go to zero at the matching point since the input above $\tm$ is set to zero.
With increasing $J$ the pole term becomes larger and thus more dominating.
Therefore, the differences between the $n$-subtracted solutions and also the different $a$ values decrease close to $\tpi$.
As they furthermore agree very well with the KH80 solution in the respective pole-term-dominated regions, we only show the remaining regions.
Since the $D$-wave coupling for the unsubtracted case depends on $a$, this contribution is omitted for the $a\to-\infty$ limit (and thus the solutions for the two $a$ values do not coincide at $\tpi$).
Obviously, the occurrence of a negative modulus (i.e.\ the unsubtracted $|f^0_+|$ for optimal $a$ and the once-subtracted $|f^1_+|$ for $a\to-\infty$) only indicates that too much input information is missing in this particular case in order to yield a reasonable solution---a problem that can be cured by subtractions.
The general pattern is as expected: the effect of varying $a$ is suppressed by both the subtraction procedure and higher $J$.
Furthermore, the agreement with the KH80 solution is strongly aided by subtracting. This is clear since each subtraction power on the one hand suppresses the lacking input above $\tm$ and on the other hand introduces additional consistent information via the subthreshold parameters as subtraction constants.
Hence, the twice-subtracted solution for optimal $a$ is our central ``consistency result''.
The $S$-wave shows a nice convergence behavior in $n$, but around $0.8\GeV$ it starts to deviate from KH80, which is not surprising as the $f_0(980)$ is expected to have an important impact (cf.\ Sect.~\ref{subsubsec:piNMO:input:tpws}).
As far as the $P$-waves are concerned, the numerical results confirm the analytic expectation that $|f^1_+|$ is much less well determined or constrained than $|f^1_-|$:
basically, the MO equations for $|f^1_+|$ effectively contain one low-energy subtraction less.
Moreover, in the necessary intermediate step of solving the MO problem for $|\Gamma^1|$ the pole-term contributions $\hat N^J_\pm$ cancel at $\tpi$ (as discussed in Appendix~\ref{subsec:tpwp:subtractions}) as for the $S$-wave, and thus the solution for $|f^1_+|$ is less pole-term dominated and hence more sensitive to the values of the subthreshold parameters.
Furthermore, the uncertainties of $|f^1_-|$ propagate into $|f^1_+|$ when calculating the latter from $|\Gamma^1|$ via~\eqref{fJpfromfJmandGammaJ}.
All this leads to a rather slow convergence behavior in $n$ for fixed $a$ as well as the loss of the expected convergence pattern in $n$ of the differences between the two $a$ values (note especially the crossing of the unsubtracted solutions for different $a$ values and the negative once-subtracted modulus).
However, our central twice-subtracted result for $|f^1_+|$ agrees rather well with KH80 especially in the $\rho(770)$ peak, even though our result for $|f^1_-|$ (which enters $|f^1_+|$) seems to be systematically smaller than KH80.
Since this underestimation might be due to forcing the solution to go to zero at $\tm$, we will investigate the effect of using a higher value for $\tm$ below.
Nevertheless, the $a\to-\infty$ variant of the twice-subtracted solution for $|f^1_-|$ agrees well with KH80 in the $\rho$ peak (though the agreement with KH80 becomes worse for $|f^1_-|$ in this limit).
The $D$-wave results are systematically smaller than KH80.  The change from one (or equivalently zero) to two subtractions towards KH80 is roughly one third of this discrepancy and furthermore approximately of the same absolute size for both partial waves, which is probably due to calculating $|f^2_+|$ by using the result for $|f^2_-|$ together with the fact that $\chi^J_\Gamma=0$ for all $J\neq1$.
For both $|f^2_+|$ and $|f^2_-|$ the accordance with KH80 (which is based on fixed-$t$ dispersion relations) in the ``fixed-$t$ limit'' $a\to-\infty$ is striking, the effect of varying $a$ being much larger than the effect of subtractions.

\subsubsection{Muskhelishvili--Omn\`es solutions: variation of the matching point}
\label{subsubsec:piNMO:results:tm}

\begin{figure}[t!]
\centering
\includegraphics{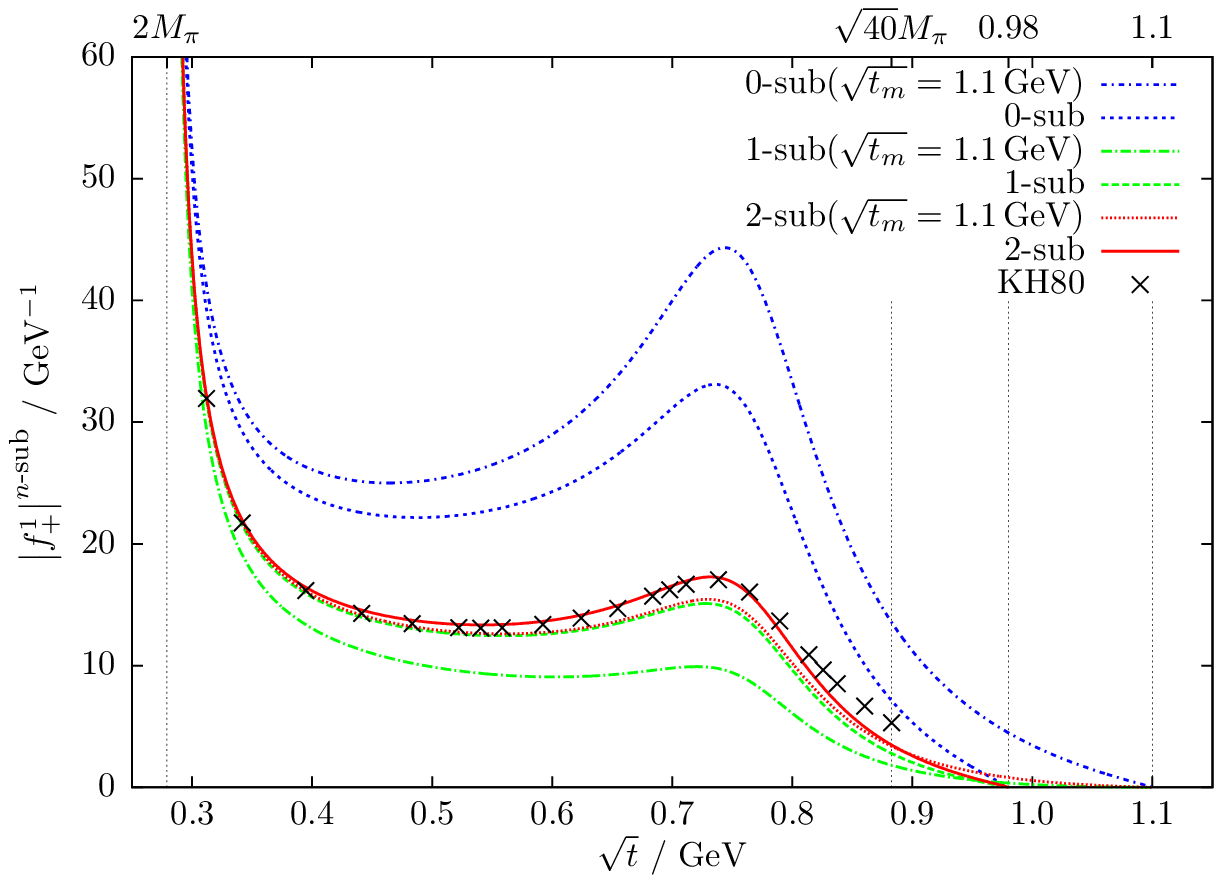}\\\includegraphics{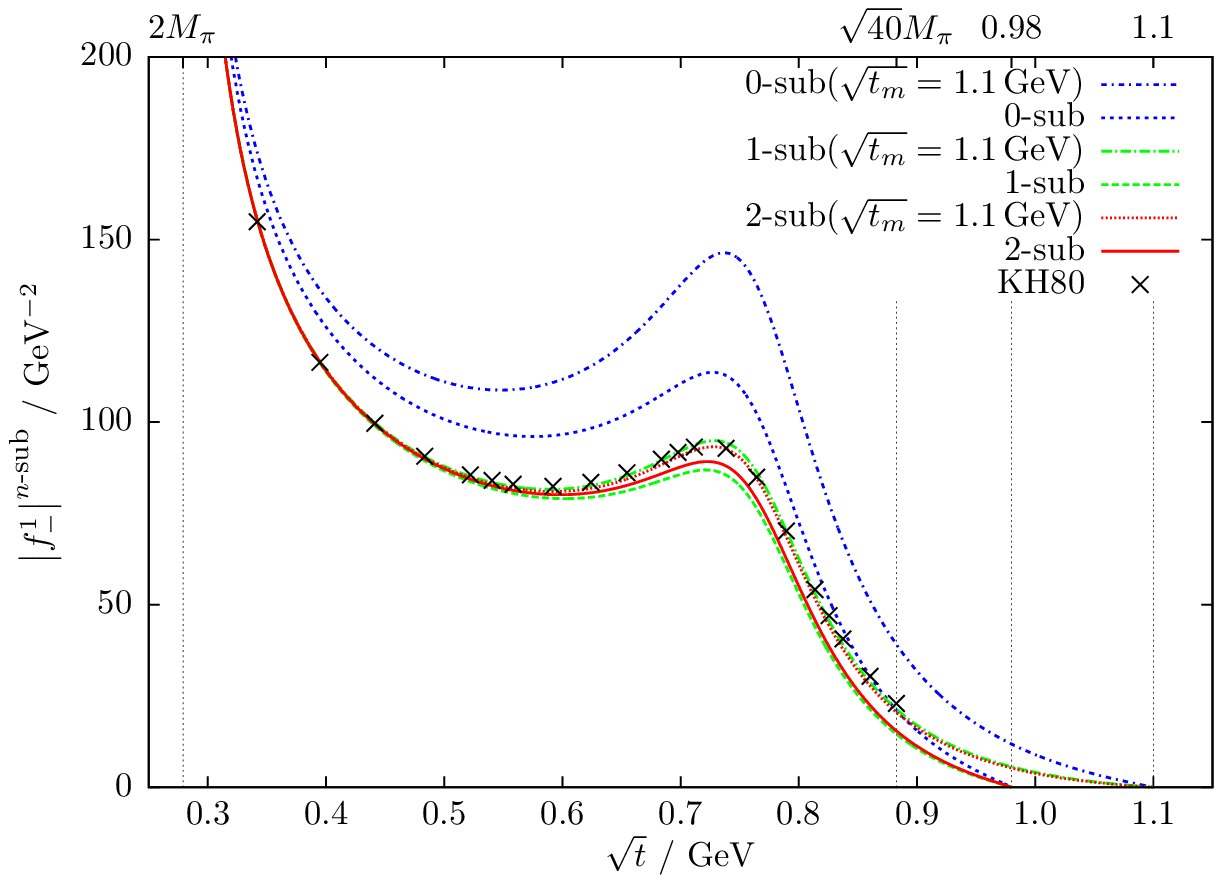}
\caption{MO solutions for the $P$-waves with $\sqrt{\tm}=1.1\GeV$.}
\label{fig:modf1pmDsubntm1100}
\end{figure}
\begin{figure}[t!]
\centering
\includegraphics{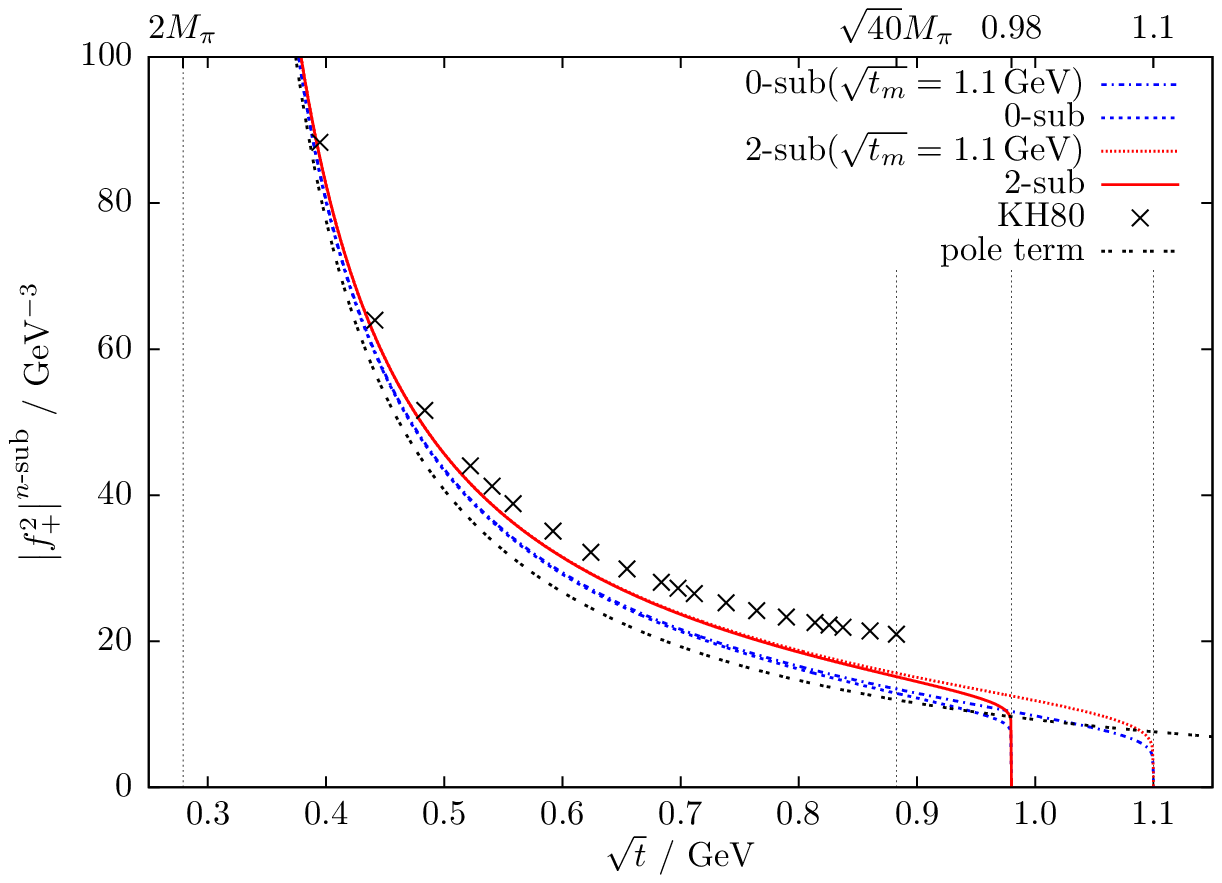}\\\includegraphics{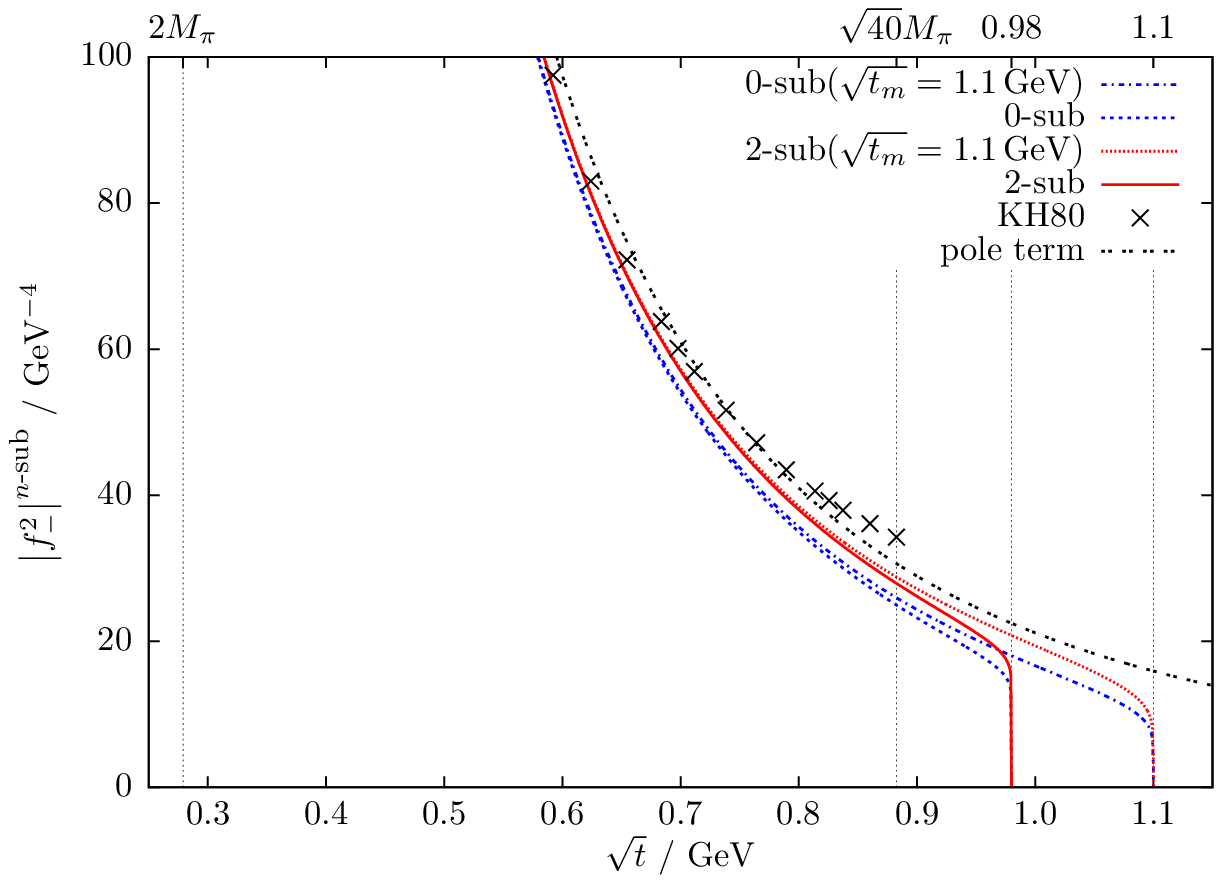}
\caption[MO solutions for the $D$-waves with $\sqrt{\tm}=1.1\GeV$.]{MO solutions for the $D$-waves with $\sqrt{\tm}=1.1\GeV$. For $J\geq2$ one subtraction has no effect.}
\label{fig:modf2pmDsubntm1100}
\end{figure}
Up to now we have used the $S$-wave-motivated value $\sqrt{\tm}=0.98\GeV$ for all considered partial waves.
The effect of changing $\sqrt{\tm}$ to e.g.\ $1.1\GeV$ is shown in Figs.~\ref{fig:modf1pmDsubntm1100} and~\ref{fig:modf2pmDsubntm1100} for the un-, once-, and twice-subtracted solutions for $J\in\{1,2\}$.
Again, for $J\geq2$ the un- and once-subtracted solutions coincide.
For $J=1$ it is generally assumed that $4\pi$ contributions can safely be neglected up to the $\pi\omega$ threshold around $0.92\GeV$; however, the $\pi\pi$ scattering $P$-wave inelasticity is small even above that energy and hence the impact of neglecting it (for both values of $\tm$) should be smaller than the effect of changing $\tm$.
For $J=2$ we do not expect sizable deviations from elasticity, since the $\pi\pi$ $D$-wave is essentially elastic in this energy range.
The $P$-wave solutions exhibit the expected behavior: the differences between the two matching-point values become smaller with each subtraction, but the convergence behavior in $n$ is again less good for $|f^1_+|$, where a higher value of $\tm$ does not lead to a better agreement with KH80, while for $|f^1_-|$ already one subtraction in combination with the higher matching point yields a description of the KH80 solution that is even better than the twice-subtracted version for $a\to-\infty$ discussed before. 
Therefore we conclude that on the one hand the KH80 solution for $|f^1_-|$ can be reproduced well with a higher matching point already in the once-subtracted case, but on the other hand the KH80 solution for $|f^1_+|$ calls for a second subtraction and is hard to be accommodated in our MO scheme for energies above roughly $0.8\GeV$.
The $D$-wave solutions, however, are hardly affected at all in the KH80 energy range by changing $\tm$.
As discussed in Sect.~\ref{subsec:piNMO:explsol}, they are expected to be dominated by the pole terms $\hat N^2_\pm$, which for comparison are also shown in Fig.~\ref{fig:modf2pmDsubntm1100}.
While for $|f^2_-|$ the KH80 solution indeed agrees rather well with the pole term itself throughout the whole KH80 energy range, for $|f^2_+|$ there are sizable (with respect to the scale) deviations between KH80 and the pole term in this region, which again fits the picture that the partial wave with parallel helicity is both analytically and numerically less well constrained.
Together with Fig.~\ref{fig:modf2pmDsubn} we can conclude that in the limit $a\to-\infty$ for $|f^2_-|$ the net effect of adding the dispersive integrals to the pole term is very small, while for $|f^2_+|$ the corresponding dispersive contributions (which thus are not mainly induced by $|f^2_-|$ in this limit) are crucial for the agreement with KH80.
For optimal $a$ (and independent from the choice of $\tm$), though, these contributions deteriorate the agreement with KH80 (with respect to the pole term) for $|f^2_-|$, whereas improving the agreement for $|f^2_+|$; in this case the corrections to $|f^2_+|$ are effectively due to $|f^2_-|$.

\subsubsection{Application to nucleon form factors}

\begin{figure}[t!]
\centering
\includegraphics{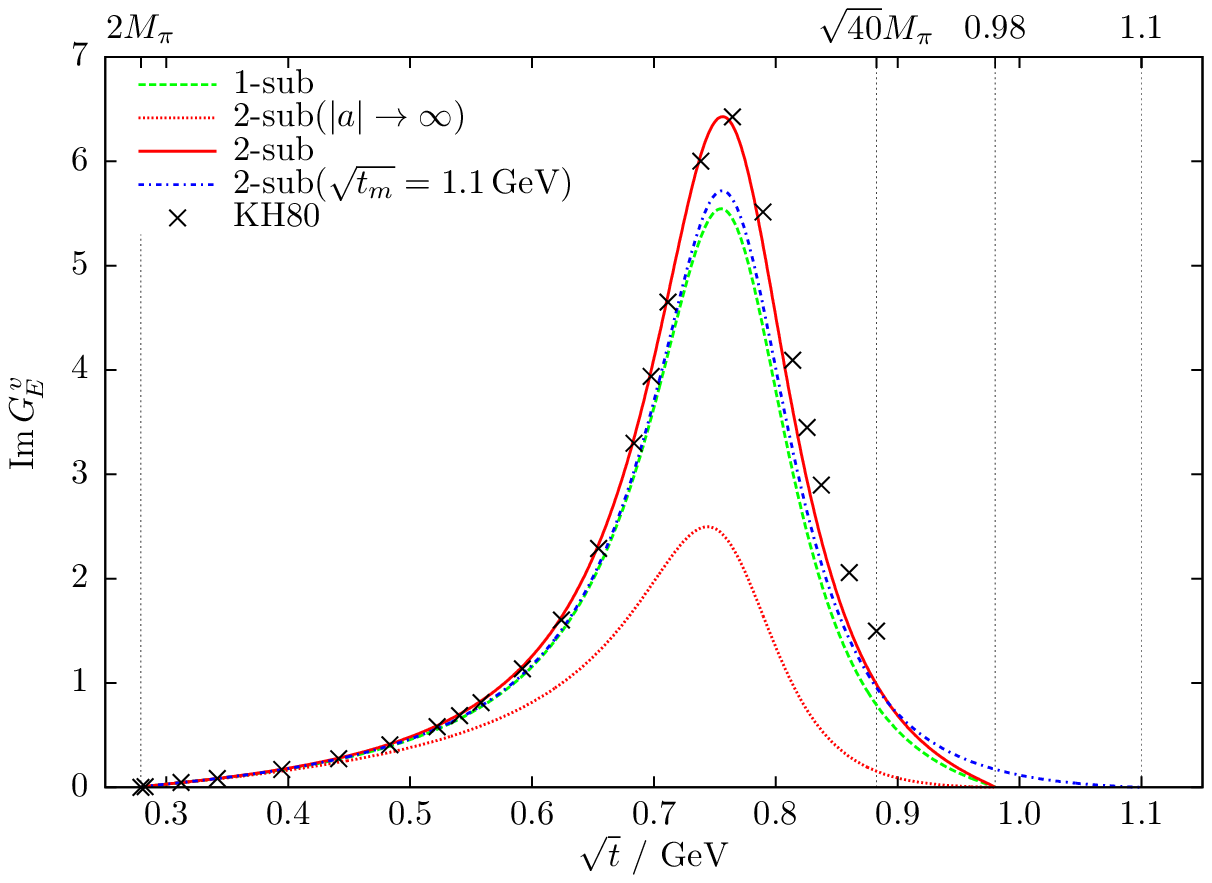}\\\includegraphics{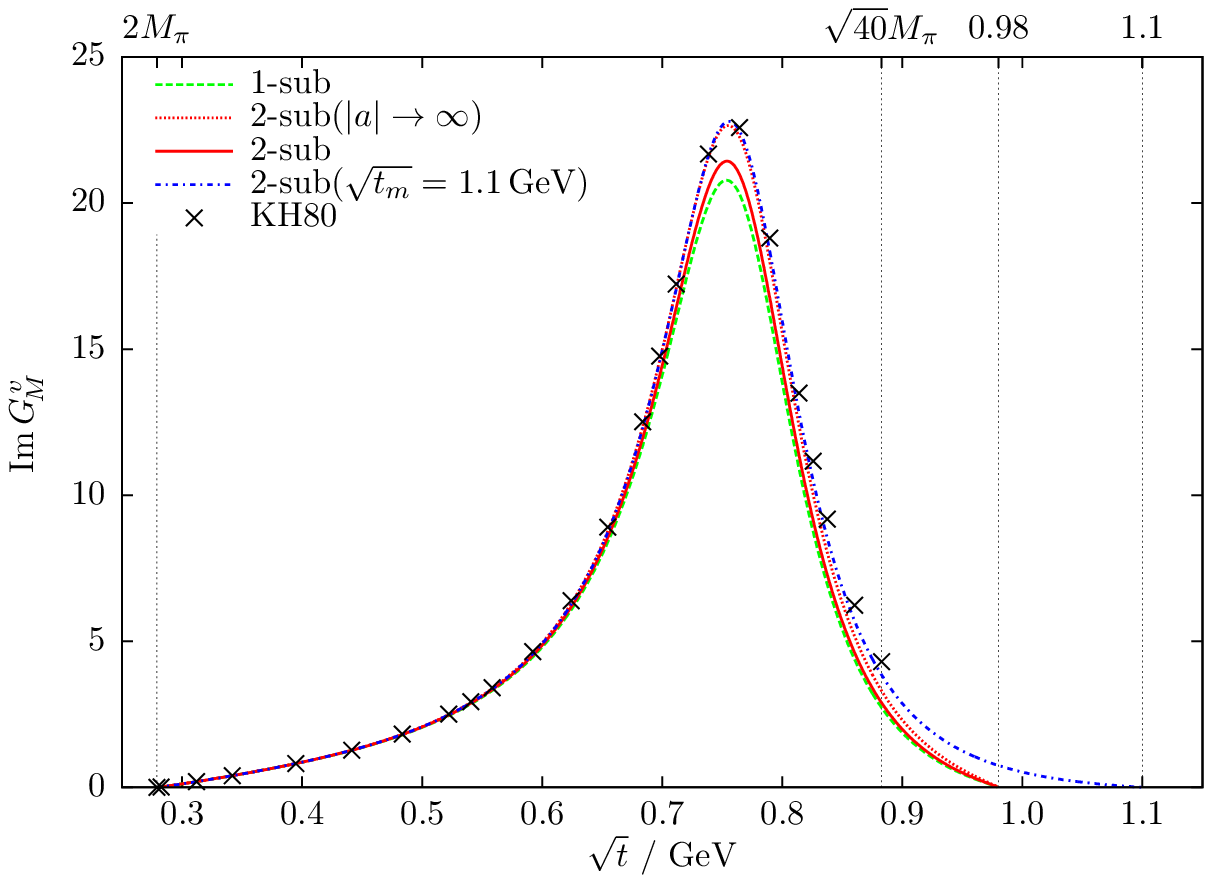}
\caption{Two-pion-continuum contribution to $\Im G_{\!E}^{\,v}(t)$ and $\Im G_{\!M}^{\,v}(t)$.}
\label{fig:ImGEMv}
\end{figure}
The $t$-channel partial waves considered in the previous sections are not only an integral part of any closed system of dispersion relations for $\pi N$ scattering fully consistent with crossing symmetry, but also an essential ingredient to dispersive analyses of nucleon form factors.
The contributions to the isovector spectral functions by two-pion intermediate states\footnote{$G$-parity dictates that intermediate states of an even (odd) number of pions only contribute to the isovector (isoscalar) spectral functions; cf.\ Sect.~\ref{subsec:preliminaries:isospin}.} in the case of the electromagnetic Sachs form factors read~\cite{FrazerFulco:NFF} (cf.~\cite{Hoehler:1974,Hoehler} for precise definitions and~\cite{BHM06} for a recent application)
\begin{equation}
\label{ImGEGMv}
\Im G_{\!E}^{\,v}(t)=\frac{q_t^3}{m\sqrt{t}}\big(F_\pi^V(t)\big)^*f^1_+(t)\,\theta\big(t-\tpi\big)\ec\qquad 
\Im G_{\!M}^{\,v}(t)=\frac{q_t^3}{\sqrt{2t}}\big(F_\pi^V(t)\big)^*f^1_-(t)\,\theta\big(t-\tpi\big)\ec
\end{equation}
while the imaginary part of the scalar form factor is determined by~\cite{GLS:FF}
\begin{equation}
\Im\sigma(t)=-\frac{3q_t}{4p_t^2\sqrt{t}}\big(F_\pi^S(t)\big)^*f^0_+(t)\,\theta\big(t-\tpi\big)\ec
\end{equation}
with the scalar and vector pion form factor $F_\pi^S(t)$ and $F_\pi^V(t)$, respectively.
In the case of the scalar form factors the approximation by $\pi\pi$ intermediate states breaks down as soon as the two-kaon threshold opens, and effects from $\bar KK$ intermediate states are known to be important for a dispersive description of $F_\pi^S(t)$~\cite{DGL,ACCGL}.
In contrast, the two-pion contribution dominates in the vector channel, where inelasticities set in more smoothly.
It is thus instructive to investigate the impact of our results for $|f^1_\pm(t)|$ on the spectral functions of the Sachs form factors.
To illustrate the corresponding effects we approximate the vector pion form factor by a simple twice-subtracted Omn\`es representation\footnote{This representation ensures that $F_\pi^V$ fulfills Watson's theorem, such that the phases in~\eqref{ImGEGMv} cancel. Strictly speaking, using any representation that goes beyond the two-pion approximation would be inconsistent unless the corresponding inelasticities are accounted for in the determination of $f^1_\pm$ and the unitarity relation~\eqref{ImGEGMv} as well, as exemplified by the breakdown of Watson's theorem and the spectral functions' becoming complex. Moreover, the precise value of $\langle r^2\rangle_\pi^V$ is immaterial in the present context, since we merely wish to convey how the uncertainties in $|f^1_\pm|$ propagate into the spectral functions. The present choice ensures a decent description of form-factor data, cf.~\cite{Colangelo:piVFF}.} (cf.~\cite{piVFF} and references therein)
\begin{equation}
F_\pi^V(t)=\exp\Bigg\{\frac{\langle r^2\rangle_\pi^V}{6}t+\frac{t^2}{\pi}\int\limits_{\tpi}^{\infty}\frac{\diff t'}{t'^2}\frac{\delta_1(t')}{t'-t}\Bigg\}
\end{equation}
using $\langle r^2\rangle_\pi^V=0.435\,{\rm fm}^2$.
The results for the once- and twice-subtracted versions of $|f^1_\pm|$ together with the comparison to KH80 are depicted in Fig.~\ref{fig:ImGEMv}.
As expected from the discussion in Sects.~\ref{subsubsec:piNMO:results:MOKH80} and~\ref{subsubsec:piNMO:results:tm}, the relative uncertainty in $\Im G_{\!E}^{\,v}$ is much larger than in $\Im G_{\!M}^{\,v}$, which is a result of the effectively lower number of subtractions in the calculation of $|f^1_+|$ and its enhanced subthreshold-parameter dependence.
However, since $\Im G_{\!M}^{\,v}$ is much larger than $\Im G_{\!E}^{\,v}$, the absolute deviations between the individual curves are actually of comparable size in both cases. 
We conclude that a new determination of the subthreshold parameters from a full solution of our RS system should lead to improved central values and associated uncertainties for the two-pion contribution to the spectral functions of both form factors.

\section{Conclusion}
\label{sec:conclusion}

In this article we have derived a closed system of Roy--Steiner equations for $\pi N$ scattering and analytically calculated the kernel functions for the lowest $s$- and $t$-channel partial waves. 
Furthermore, we have constructed the corresponding unitarity relations in detail, including inelastic contributions from $\bar KK$ intermediate states in the $t$-channel reaction. 
To pin down the optimal value of the free hyperbola parameter $a$, we have analyzed the domain of validity of the full system (assuming Mandelstam analyticity) and determined $a$ for both the $s$- and $t$-channel equations such that the range of convergence is maximized. 
We have introduced subtractions at the subthreshold point in order to suppress the dependence on the high-energy region and derived the corresponding once- and twice-subtracted versions of our Roy--Steiner system as well as sum rules for the subtraction constants and the necessary corrections to the kernel functions.

Casting the $t$-channel equations into the form of a Muskhelishvili--Omn\`es problem with finite matching point, we have then solved the equations for the $t$-channel numerically in the single-channel approximation. 
We have assessed the numerical importance of different input contributions for the Muskhelishvili--Omn\`es problem and its solutions and examined the behavior of the Muskhelishvili--Omn\`es solutions for the lowest $t$-channel partial waves ($J\in\{0,1,2\}$) with respect to varying  both the input and/or the framework in many ways, including their sensitivity to the $\pi\pi$ phase shifts, the number of subtractions ($n\in\{0,1,2\}$), variation of the matching point $\tm$, and taking the hyperbola parameter $a$ to $-\infty$. 
In general, we find consistency with the KH80 solutions. 
However, our analysis shows that the structure of the equations renders the $t$-channel partial waves $f^J_+$ systematically less well determined than their counterparts $f^J_-$ both due to an enhanced sensitivity to the subtraction constants and an effectively lower number of subtractions. 
Finally, we have briefly discussed some consequences for nucleon form factors, in particular our analysis gives a first indication where the largest uncertainties in the spectral functions are to be expected.

The next step in the solution of our system of Roy--Steiner equations will be the incorporation of $\bar KK$ intermediate states in a full two-channel Muskhelishvili--Omn\`es treatment of the $t$-channel $S$-wave, which will have immediate consequences for the extraction of the $\pi N$ $\sigma$ term via the scalar form factor of the nucleon~\cite{twochannelOmnes}. 
Having then solved the $t$-channel part of the system, the $s$-channel equations are solvable with techniques similar to those employed in the context of $\pi\pi$ Roy equations, and finally the iteration of the full system should determine the lowest partial waves as well as the subtraction parameters. 
We are confident that the framework proposed in this article will allow for a reliable extrapolation to the Cheng--Dashen point and thus for an accurate determination of the $\pi N$ $\sigma$ term.

\subsection*{Acknowledgments}

We are grateful to G.~Colangelo and B.~Moussallam for numerous helpful discussions, in particular for correspondence on the results of~\cite{CCL:Regge,CCL:PWA} and~\cite{piK:RS}.
We also acknowledge useful discussions and correspondence with 
M.~Albaladejo, S.~Descotes-Genon, H.-W.~Hammer, V.~A.~Nikonov, J.~A.~Oller, H.~Polinder, T.~Rijken, A.~V.~Sarantsev, and S.~P.~Schneider.
Partial financial support 
by the DFG (SFB/TR 16, ``Subnuclear Structure of Matter''),
by the project ``Study of Strongly Interacting Matter'' (HadronPhysics2, Grant Agreement No.~227431 and HadronPhysics3, Grant Agreement No.~283286) under the 7th Framework Program of the EU, 
and by the Bonn--Cologne Graduate School of Physics and Astronomy
is gratefully acknowledged.

\appendix

\section{Partial-wave projection for the \boldmath{$s$}-channel amplitudes}
\label{sec:spwp}

In this Appendix, the different contributions to the $s$-channel part~\eqref{spwhdr}of the RS system are discussed.

\subsection{Nucleon exchange}
\label{subsec:spwp:n}

The useful general definitions
\begin{equation}
\label{epsilonI}
\epsilon^I=\begin{cases}\epsilon^+=+1\ec\\\epsilon^-=-1\ec\end{cases} \quad
\tilde\epsilon_\pm=\frac{1\pm\epsilon^I}{2}\ec \qquad
\tilde\epsilon_+=\begin{cases}1\ec\\0\ec\end{cases} \quad
\tilde\epsilon_-=\begin{cases}0\ec\\1\ec\end{cases} \quad
\text{for }\begin{cases}I=+\text{ or $J$ even}\ec\\I=-\text{ or $J$ odd}\ec\end{cases}
\end{equation}
can be identified with
\begin{equation}
\epsilon^I\doteq(-1)^J\ec \qquad \tilde\epsilon_\pm\doteq\frac{1\pm(-1)^J}{2}\ec
\end{equation}
for the cases where the crossing-symmetry constraint applies (i.e.\ $J$ even/odd for $I=+/-$).
Projecting the HDR Born terms $N^I(s,t)$ of~\eqref{hdrnucleonpoleterms} onto $s$-channel partial waves via~\eqref{sprojform} leads to MacDowell-symmetric nucleon pole contributions
\begin{align}
\label{NIpm}
N^I_{l+}(W)&=\frac{g^2}{16\pi W}\bigg\{(E+m)(W-m)\bigg[\epsilon^I\frac{Q_l(y)}{q^2}+2\delta_{l0}\bigg(\frac{1}{m^2-s}-\frac{\tilde\epsilon_-}{m^2-a}\bigg)\bigg]\nt\\
&\qquad\qquad\!+(E-m)(W+m)\bigg[\epsilon^I\frac{Q_{l+1}(y)}{q^2}\bigg]\Bigg\}\nt\\
&=-N^I_{(l+1)-}(-W) \qquad \forall\;l\geq0\ec
\end{align}
which, by defining the abbreviation\footnote{Of course, also this form of the nucleon pole terms obeys the MacDowell symmetry relation~\eqref{macdowell}, since the term proportional to $\delta_{l+1,0}/\delta_{l0}$ vanishes for $\bar{N}^I_{l\pm}$ as a consequence of $l$ starting at $0/1$, respectively.}
\begin{equation}
\label{barNIpm}
\bar{N}^I_{l\pm}(W)=\frac{g^2}{16\pi W}\bigg\{(E+m)(W-m)\bigg[\epsilon^I\frac{Q_l(y)}{q^2}+\frac{2\delta_{l0}}{m^2-s}\bigg]
 +(E-m)(W+m)\bigg[\epsilon^I\frac{Q_{l\pm1}(y)}{q^2}+\frac{2\delta_{l\pm1,0}}{m^2-s}\bigg]\Bigg\}
\end{equation}
for later convenience, can also be written in the form
\begin{align}
\label{NIpmexpl}
N^+_{l+}(W)&=\bar{N}^+_{l+}, & N^-_{l+}(W)&=\bar{N}^-_{l+}-\frac{g^2}{4\pi}\frac{(E+m)(W-m)}{2W}\frac{\delta_{l0}}{m^2-a}, & &\forall\;l\geq0\ec\nt\\
N^+_{l-}(W)&=\bar{N}^+_{l-}, & N^-_{l-}(W)&=\bar{N}^-_{l-}-\frac{g^2}{4\pi}\frac{(E-m)(W+m)}{2W}\frac{\delta_{l1}}{m^2-a}, & &\forall\;l\geq1\ec
\end{align}
and where we have defined 
\begin{equation}
\label{ydef}
y(s)=1-\frac{s+m^2-\Sigma}{2q^2}=z_s(s,t(s,u=m^2))=x_s(s,s'=m^2)
\end{equation}
($x_s(s,s')$ will be introduced in~\eqref{xsdef}).
$Q_l(z)$ denotes the Legendre functions of the second kind.

The $Q_l(z)$ obey a recursion relation similar to the one for the usual Legendre polynomials $P_l(z)$ (for $l\geq0$)
\begin{align}
\label{legendrerecursion}
(l+1)P_{l+1}(z)+lP_{l-1}(z)&=(2l+1)zP_l(z)\ec\nt\\
(l+1)Q_{l+1}(z)+lQ_{l-1}(z)&=(2l+1)zQ_l(z)-\delta_{l0}\ec
\end{align}
which, together with $Q_l=P_l=0$ for $l<0$, leads in particular to (cf.~\eqref{wlminus1} for the general formula)
\begin{equation}
\label{ql123}
Q_1(z)=P_1(z)Q_0(z)-1\ec \qquad Q_2(z)=P_2(z)Q_0(z)-\frac{3}{2}z\ec \qquad Q_3(z)=P_3(z)Q_0(z)-\frac{5z^2}{2}+\frac{2}{3}\ep
\end{equation}
From the Neumann integral representation for general complex argument $z$~\cite{Bateman}
\begin{equation}
\label{qlneumann}
Q_l(z)=\frac{1}{2}\int\limits_{-1}^1\diff x\;\frac{P_l(x)}{z-x}=(-1)^{l+1}Q_l(-z)\ec
\end{equation}
one can read off the lowest function for general real argument $y$
\begin{equation}
\label{ql0}
Q_0(y\pm i\epsilon)=\frac{1}{2}\int\limits^1_{-1}\frac{\diff x}{y-x\pm i\epsilon}=\frac{1}{2}\log\bigg|\frac{1+y}{1-y}\bigg|\mp i\frac{\pi}{2}\,\theta(1-y^2)\ep
\end{equation}
We also need the analytic continuation for purely imaginary argument $z=iy$, e.g.\ for $y>1$
\begin{equation}
Q_0(iy)=\frac{1}{2}\log\frac{iy+1}{iy-1}=\frac{1}{2}\log\frac{1+iy}{1-iy}-i\frac{\pi}{2}=i\left(\arctan y-\frac{\pi}{2}\right)=-Q_0(-iy)\ep
\end{equation}
Functions with $l\geq1$ may then be obtained via either the recursion relation~\eqref{legendrerecursion} or the reduction formula~\eqref{wlminus1}.

\subsection[$s$- and $u$-channel exchange]{\boldmath{$s$}- and \boldmath{$u$}-channel exchange}
\label{subsec:spwp:s}

By introducing a convenient matrix notation via
\begin{equation}
\mathbf{A}^I=\begin{pmatrix}A^I\\B^I\end{pmatrix}, \qquad \mathbf{f}^I_l=\begin{pmatrix}f^I_{l+}\\f^I_{(l+1)-}\end{pmatrix},
\end{equation}
the crossing properties of the even and odd invariant amplitude combinations~\eqref{crossingamps} read
\begin{equation}
\label{matrixcrossingamps}
\mathbf{A}^I(\nu,t)=\epsilon^I\boldsymbol{\sigma}_3\;\mathbf{A}^I(-\nu,t)\ec \qquad \boldsymbol{\sigma}_3=\begin{pmatrix}1&0\\0&-1\end{pmatrix}.
\end{equation}
While the $s$-channel partial-wave projection~\eqref{sprojform} can be rewritten as
\begin{equation}
\label{sproj}
\mathbf{f}^I_l(W)=\int\limits_{-1}^1\diff z_s\;\mathbf{R}^l(W,z_s)\mathbf{A}^I(s,t)\Big|_{t=t(s,z_s)}\ec
\end{equation}
where the projection kernel matrix is given by
\begin{align}
\mathbf{R}^l(W,z_s)&=\begin{pmatrix}R^1_{l,l+1}&R^2_{l,l+1}\\R^1_{l+1,l}&R^2_{l+1,l}\end{pmatrix},\\
R^1_{kn}(W,z_s)&=\frac{1}{16\pi W}\Big\{(E+m)P_k(z_s)-(E-m)P_n(z_s)\Big\}=-R^1_{nk}(-W,z_s)\ec\nt\\
R^2_{kn}(W,z_s)&=\frac{1}{16\pi W}\Big\{(E+m)(W-m)P_k(z_s)+(E-m)(W+m)P_n(z_s)\Big\}=-R^2_{nk}(-W,z_s)\ec\nt
\end{align}
the $s$-channel partial-wave expansion, i.e.\ the inversion of~\eqref{sproj}, takes the form~\cite{CGLN}
\begin{equation}
\label{sexp}
\mathbf{A}^I(s,t)\Big|_{t=t(s,z_s)}=\sum\limits_{l=0}^\infty\mathbf{S}^l(W,z_s)\mathbf{f}^I_l(W)\ec
\end{equation}
with the expansion kernel matrix
\begin{align}
\label{sexpform}
\mathbf{S}^l(W,z_s)&=\begin{pmatrix}S^1_{l+1,l}&-S^1_{l,l+1}\\S^2_{l+1,l}&-S^2_{l,l+1}\end{pmatrix},\nt\\
S^1_{kn}(W,z_s)&=4\pi\bigg\{\frac{W+m}{E+m}P_k'(z_s)+\frac{W-m}{E-m}P_n'(z_s)\bigg\}=-S^1_{nk}(-W,z_s)\ec\nt\\
S^2_{kn}(W,z_s)&=4\pi\bigg\{\frac{1}{E+m}\;P_k'(z_s)-\frac{1}{E-m}\;P_n'(z_s)\bigg\}=-S^2_{nk}(-W,z_s)\ep
\end{align}
In accordance with the matrix form of the MacDowell symmetry relation~\eqref{macdowell}
\begin{equation}
\label{matrixmacdowell}
\mathbf{f}^I_l(W)=-\boldsymbol{\sigma}_1\mathbf{f}^I_l(-W)\ec \qquad \boldsymbol{\sigma}_1=\begin{pmatrix}0&1\\1&0\end{pmatrix},
\end{equation}
these kernels obey the symmetry relations
\begin{equation}
\label{sskernelmacdowell}
\mathbf{R}^l(W,z_s)=-\boldsymbol{\sigma}_1\mathbf{R}^l(-W,z_s)\ec \qquad
\mathbf{S}^l(W,z_s)=-\mathbf{S}^l(-W,z_s)\boldsymbol{\sigma}_1\ep
\end{equation}
With the definitions~\eqref{primedkinematics} the $s$- and $u$-channel terms of the HDRs~\eqref{hdr} thus can be cast into the matrix form
\begin{equation}
\label{sshdr}
\mathbf{A}^I(s,t)\Big|^{s+u}_{t=t(s,z_s)=-2q^2(1-z_s)}=\frac{1}{\pi}\int\limits_{s_+}^\infty\diff s'\;\mathbf{h}^I_s[s,s';z_s]\,\Im\mathbf{A}^I(s',t')\Big|_{t'=t'(s',z_s')=-2q'^2(1-z_s')}\ec
\end{equation}
where the HDR kernel matrix $\mathbf{h}^I_s$ is given by
\begin{align}
\label{xsdef}
\mathbf{h}^I_s(s,s';z_s)&=h_1\boldsymbol{\sigma}_0-\epsilon^Ih_2\boldsymbol{\sigma}_3\ec \qquad \boldsymbol{\sigma}_0=\unity_2=\begin{pmatrix}1&0\\0&1\end{pmatrix}\ec \qquad
x_s(s,s')=1-\frac{s+s'-\Sigma}{2q^2}\ec\nt\\
h_1(s,s')&=\frac{1}{s'-s}-\frac{1}{2}\frac{1}{s'-a}\ec \qquad
h_2(s,s';z_s)=\frac{1}{2q^2}\frac{1}{x_s-z_s}+\frac{1}{2}\frac{1}{s'-a}\ec
\end{align}
and $[s,s';z_s]$ indicates that the whole integrand is to be understood as a function of these variables, which can be achieved using
\begin{equation}
\label{alphabetadef}
z_s'(s,s';z_s)=\alpha z_s+\beta\ec \qquad \alpha(s,s')=\frac{q^2}{q'^2}\frac{s-a}{s'-a}\ec \qquad \beta(s,s')=1-\alpha-\frac{s'-s}{s'-a}\frac{s+s'-\Sigma}{2q'^2}\ep
\end{equation}
By expanding the absorptive part of the $s$- and $u$-channel HDR terms given in~\eqref{sshdr} into $s$-channel partial waves via~\eqref{sexp} and projecting out $s$-channel partial waves again by means of~\eqref{sproj}, we arrive at the partial-wave dispersion relations 
\begin{equation}
\label{sspwhdr}
\mathbf{f}^I_l(W)\Big|^{s+u}=\frac{1}{\pi}\int\limits_{W_+}^\infty\diff W'\;\sum\limits_{l'=0}^\infty\mathbf{K}^{ll',I}(W,W')\,\Im\mathbf{f}^I_{l'}(W')\ec
\end{equation}
where the $s$- and $u$-channel kernel matrix is defined by
\begin{equation}
\mathbf{K}^{ll',I}(W,W')=2W'\int\limits_{-1}^1\diff z_s\;\mathbf{R}^l(W,z_s)\mathbf{h}^I_s[W,W';z_s]\mathbf{S}^{l'}(W',z_s')\ep
\end{equation}
Due to the symmetry relations
\begin{equation}
-\boldsymbol{\sigma}_1\mathbf{K}^{ll',I}(-W,W')=\mathbf{K}^{ll',I}(W,W')=\mathbf{K}^{ll',I}(W,-W')\boldsymbol{\sigma}_1\ec
\end{equation}
which follow from the relations~\eqref{sskernelmacdowell}, the $s$- and $u$-channel kernel matrix can be written with only one kernel function according to
\begin{align}
\mathbf{K}^{ll',I}(W,W')&=\begin{pmatrix}K^I_{ll'}(W,W')&K^I_{ll'}(W,-W')\\-K^I_{ll'}(-W,W')&-K^I_{ll'}(-W,-W')\end{pmatrix},\nt\\
K^I_{ll'}(W,W')&=2W'\int\limits_{-1}^1\diff z_s\;\Big\{\mathbf{R}^l(W,z_s)\mathbf{h}^I_s[W,W';z_s]\mathbf{S}^{l'}(W',z_s')\Big\}_{1,1}\ec
\end{align}
where the subscript denotes the $1,1$-th element of the matrix in the brackets. The \mbox{PWDRs}~\eqref{sspwhdr} then take the form already stated in~\eqref{spwhdr}
\begin{align}
f^I_{l+}(W)\Big|^{s+u}&=\frac{1}{\pi}\int\limits^\infty_{W_+}\diff W'\;\sum\limits_{l'=0}^\infty\Big\{K^I_{ll'}(W,W')\,\Im f^I_{l'+}(W')+K^I_{ll'}(W,-W')\,\Im f^I_{(l'+1)-}(W')\Big\}\nt\\
&=-f^I_{(l+1)-}(-W)\Big|^{s+u}\ep
\end{align}

Defining the structure
\begin{align}
\varphi\big[a_{kn}\big|b(W,W')\big]&=\frac{W'}{W}\Big\{b(W,-W')a_{kn}+b(W,W')a_{k,n+1}\nt\\
 &\qquad\quad\;+b(-W,-W')a_{k+1,n}+b(-W,W')a_{k+1,n+1}\Big\}\ec
\end{align}
where $a_{kn}(s,s')$ is to be understood as a function invariant under sign changes in $W$ and $W'$, and introducing the kinematical abbreviations
\begin{align}
E'(W')&=E(W')\ec \qquad \delta(W,W')=\frac{E+m}{E'+m}\big[W'+W\big]\ec \qquad \varrho(W,W')=\frac{E+m}{E'+m}\big[W'-W+2m\big]\ec\nt\\
\varkappa^I(W,W')&=\frac{1}{2}\Big[\delta(W,W')+\epsilon^I\varrho(W,W')\Big]
=\frac{E+m}{E'+m}\Big[\tilde\epsilon_+(W'+m)+\tilde\epsilon_-(W-m)\Big]\ec
\end{align}
as well as the angular kernels
\begin{equation}
\label{angularkernels}
U_{ll'}(s,s')=\frac{1}{2}\int\limits_{-1}^1\diff z_s\;P_l(z_s)P_{l'}'(z_s')\ec \qquad
V_{ll'}(s,s')=\frac{1}{2}\int\limits_{-1}^1\diff z_s\;\frac{P_l(z_s)P_{l'}'(z_s')}{x_s-z_s}\ec
\end{equation}
the general $s$- and $u$-channel kernel function can be written as
\begin{align}
\label{sskernelfunction}
K^I_{ll'}(W,W')&=h_1\varphi\big[U_{ll'}\big|\delta(W,W')\big]-\frac{\epsilon^I}{2}\varphi\bigg[\frac{V_{ll'}}{q^2}+\frac{U_{ll'}}{s'-a}\bigg|\varrho(W,W')\bigg]\nt\\
&=\frac{\varphi\big[U_{ll'}\big|\delta(W,W')\big]}{s'-s}-\epsilon^I\frac{\varphi\big[V_{ll'}\big|\varrho(W,W')\big]}{2q^2}-\frac{\varphi\big[U_{ll'}\big|\varkappa^I(W,W')\big]}{s'-a}\ep
\end{align}
Since $\varphi[a_{kn}|b(W,W')]$ encodes the MacDowell symmetry~\eqref{macdowell} for both pairs $(k,W)$ and $(n,W')$, we can decompose it in two ways
\begin{align}
\varphi\big[a_{kn}\big|b(W,W')\big]&=\varphi_1\big[a_{kn}\big|b(W,W')\big]-\varphi_1\big[a_{k+1,n}\big|b(-W,W')\big]\ec\nt\\
&=\varphi_2\big[a_{kn}\big|b(W,W')\big]-\varphi_2\big[a_{k,n+1}\big|b(W,-W')\big]\ec\nt\\
\varphi_1\big[a_{kn}\big|b(W,W')\big]&=\frac{W'}{W}\Big\{b(W,-W')a_{kn}+b(W,W')a_{k,n+1}\Big\}\ec\nt\\
\varphi_2\big[a_{kn}\big|b(W,W')\big]&=\frac{W'}{W}\Big\{b(W,-W')a_{kn}+b(-W,-W')a_{k+1,n}\Big\}\ec
\end{align}
and with the definitions
\begin{equation}
K^{I,i}_{ll'}(W,W')=\frac{\varphi_i\big[U_{ll'}\big|\delta(W,W')\big]}{s'-s}-\epsilon^I\frac{\varphi_i\big[V_{ll'}\big|\varrho(W,W')\big]}{2q^2}-\frac{\varphi_i\big[U_{ll'}\big|\varkappa^I(W,W')\big]}{s'-a} \qquad i\in\{1,2\}\ec
\end{equation}
the kernels exhibit the following interrelations
\begin{equation}
K^I_{ll'}(W,W')=K^{I,1}_{ll'}(W,W')-K^{I,1}_{l+1,l'}(-W,W')=K^{I,2}_{ll'}(W,W')-K^{I,2}_{l,l'+1}(W,-W')
\end{equation}
that may be used to write down explicit expressions of the kernels in a compact form.
However, for numerical evaluations a different prescription is preferable. The part of~\eqref{sskernelfunction} that contains the $s$-channel cut can be decomposed according to
\begin{align}
\frac{\varphi\big[U_{ll'}\big|\delta(W,W')\big]}{s'-s}&=\frac{\gamma_{ll'}(W,W')}{W'-W}+\frac{1}{W'+W}\frac{W'}{W}\left\{\frac{E+m}{E'-m}U_{ll'}-\frac{E-m}{E'+m}U_{l+1,l'+1}\right\}\ec\nt\\
\gamma_{ll'}(W,W')&=\frac{W'}{W}\left\{\frac{E+m}{E'+m}U_{l,l'+1}-\frac{E-m}{E'-m}U_{l+1,l'}\right\}\ep
\end{align}
Using the identity
\begin{equation}
\int\limits_{-1}^1\diff z\;P_{l'}'(z)\big[P_{l\pm1}(z)-zP_l(z)\big]=\frac{2\delta_{ll'}}{2l+1}\begin{Bmatrix}-l\\l+1\end{Bmatrix}\ec
\end{equation}
we can easily calculate its residue at the pole $W'=W$ (where $\alpha=1$, $\beta=0$, and thus $z_s'=z_s$)
\begin{equation}
\Res\left[\frac{\gamma_{ll'}(W,W')}{W'-W},W'=W\right]=\gamma_{ll'}(W,W)=U_{l,l'+1}(s,s)-U_{l+1,l'}(s,s)=\delta_{ll'}\ec
\end{equation}
which together with the decompositions
\begin{align}
\label{Ubarandcpmdef}
U_{ll'}(W,W')&=U_{ll'}(W,W)+(W'-W)\bar{U}_{ll'}(W,W')\ec\nt\\
\frac{W'}{W}\frac{E\pm m}{E'\pm m}&=1+(W'-W)c_\pm\ec \qquad c_\pm(W,W')=\frac{(W'+W)\Sigma_-\pm2mWW'}{2W'(E'\pm m)s}\ec
\end{align}
leads us to the alternative form of the kernels $K^I_{ll'}(W,W')$
\begin{align}
\label{sskernelfunction2}
K^I_{ll'}(W,W')&=\frac{\delta_{ll'}}{W'-W}+\bar{K}^I_{ll'}(W,W')\ec\nt\\
\bar{K}^I_{ll'}(W,W')&=\bar{U}_{l,l'+1}(W,W')-\bar{U}_{l+1,l'}(W,W')+c_+U_{l,l'+1}-c_-U_{l+1,l'}+\frac{1}{W'+W}\frac{W'}{W}\bigg\{\frac{E+m}{E'-m}U_{ll'}\nt\\
&\qquad-\frac{E-m}{E'+m}U_{l+1,l'+1}\bigg\}-\epsilon^I\frac{\varphi\big[V_{ll'}\big|\varrho(W,W')\big]}{2q^2}-\frac{\varphi\big[U_{ll'}\big|\varkappa^I(W,W')\big]}{s'-a}\ec
\end{align}
where the first term is the usual Cauchy kernel for the $s$-channel cut (contributing only for $l=l'$) and the kernels $\bar{K}^I_{ll'}(W,W')$ contain only the left-hand cut.
In order to derive explicit expressions for the angular kernels $U_{ll'}(W=\sqrt{s},W'=\sqrt{s'})$ from~\eqref{angularkernels} and subsequently their regular parts $\bar{U}_{ll'}(W,W')$ from~\eqref{Ubarandcpmdef}, we use the following expansion~\cite{Bateman}
\begin{equation}
P_l(z_s)=\sum\limits_{\lambda=0}^l a_\lambda^l x^\lambda\ec \qquad a_\lambda^l=\frac{(-1)^\lambda(l+\lambda)!}{(\lambda!)^2(l-\lambda)!}\ec \qquad x=\frac{1-z_s}{2}\ec
\end{equation}
and hence
\begin{equation}
P_{l'}'(z_s')=-\frac{1}{2}\sum\limits_{\lambda'=0}^{l'-1}(\lambda'+1)a_{\lambda'+1}^{l'}x'^{\lambda'}\ec \qquad x'=\frac{1-z_s'}{2}=\omega+\alpha x\ec \qquad \omega(s,s')=\frac{1-(\alpha+\beta)}{2}\ec
\end{equation}
together with the binomial theorem and the Saalsch\"{u}tz identity~\cite{Bateman}
\begin{equation}
\label{saalschuetz}
\sum\limits_{\lambda=0}^{l}\frac{a_\lambda^l}{\mu+\lambda+1}=(-1)^l\frac{(\mu!)^2}{(\mu-l)!(\mu+l+1)!} \qquad (\mu\geq l)\ec
\end{equation}
to arrive at the general expression for the angular kernel $U_{ll'}$
\begin{equation}
\label{Udef}
U_{ll'}(s,s')=\frac{(-1)^{l+1}}{2}\sum\limits_{\lambda'=l}^{l'-1}(\lambda'+1)a_{\lambda'+1}^{l'}\sum\limits_{\mu=l}^{\lambda'}
\begin{pmatrix}\lambda'\\\mu\end{pmatrix}\frac{(\mu!)^2}{(\mu-l)!(\mu+l+1)!}\omega^{\lambda'-\mu}\alpha^\mu\ep
\end{equation}
These kernels show the following asymptotic behavior:
\begin{equation}
\label{Uasym}
U_{ll'}\sim q^{2l} \quad\text{for }\;q\to0\ec \qquad
U_{ll'}\sim q'^{-2l'+2} \quad\text{for }\;q'\to0\ec \qquad
U_{ll'}\sim q'^{-4l} \quad\text{for }\;q'\to\infty\ec
\end{equation}
and, in particular, the lowest kernels are given by (note that $U_{l0}=0$ and $U_{l1}=\delta_{l0}$)
\begin{align}
U_{ll'}&=0 \quad\text{for }\;l'\leq l\ec \qquad U_{l,l+1}=\alpha^l\ec \qquad U_{l,l+2}=(2l+3)\beta\alpha^l\ec\nt\\
U_{l,l+3}&=\frac{\alpha^l}{2}\Big\{(2l+5)\big[\alpha^2+(2l+3)\beta^2\big]-(2l+3)\Big\}\ep
\end{align}
From~\eqref{Udef} and~\eqref{angularkernels} we can easily deduce for $W'=W$
\begin{align}
U_{ll'}(W,W)=\sum\limits_{\lambda'=l}^{l'-1}u_{ll'}^{\lambda'}&=\begin{cases}0\quad\text{for }l'\leq l\text{ or }l'-l\text{ even}\ec\\1\quad\text{for }l'-l\text{ odd}\ec\end{cases}\nt\\
u_{ll'}^{\lambda'}&=\frac{(-1)^{l+\lambda'}(l'+\lambda'+1)!}{2(\lambda'+1)(l'-\lambda'-1)!(\lambda'-l)!(l+\lambda'+1)!}\ec
\end{align}
which again yields $U_{l,l'+1}(W,W)-U_{l+1,l'}(W,W)=\delta_{ll'}$. By defining
\begin{align}
\frac{q^2}{q'^2}&=1+(W'-W)d_1\ec \qquad d_1(W,W')=\frac{W'+W}{4q'^2}\bigg[\frac{\Sigma_-^2}{ss'}-1\bigg]\ec\nt\\
\frac{s-a}{s'-a}&=1+(W'-W)d_2\ec \qquad d_2(W,W')=-\frac{W'+W}{s'-a}\ec
\end{align}
we can rewrite the powers of $\alpha$ according to (note that $b_0=0$)
\begin{equation}
\label{bmudef}
\alpha^\mu=1+(W'-W)b_\mu\ec \qquad
b_\mu(W,W')=\sum\limits_{k=0}^{\mu-1}\begin{pmatrix}\mu\\k+1\end{pmatrix}(W'-W)^k\bigg\{d_1^{k+1}\left(\frac{s-a}{s'-a}\right)^\mu+d_2^{k+1}\bigg\}\ec
\end{equation}
which together with the definitions
\begin{align}
\omega&=(W'-W)\bar\omega\ec \qquad \bar\omega(W,W')=\frac{W'+W}{4q'^2}\frac{s+s'-\Sigma}{s'-a}\ec\nt\\
\tilde{U}_{ll'}(W,W')&=\frac{(-1)^{l+1}}{2}\sum\limits_{\lambda'=l}^{l'-1}(\lambda'+1)a_{\lambda'+1}^{l'}\sum\limits_{\mu=l}^{\lambda'-1}
\begin{pmatrix}\lambda'\\\mu\end{pmatrix}\frac{(\mu!)^2}{(\mu-l)!(\mu+l+1)!}\omega^{\lambda'-1-\mu}\alpha^\mu\ec
\end{align}
allows us to give the explicit form of the regular part $\bar{U}_{ll'}$ of the angular kernel $U_{ll'}$ as
\begin{align}
\label{Ubardef}
\bar{U}_{ll'}(W,W')=\sum\limits_{\lambda'=l}^{l'-1}u_{ll'}^{\lambda'}b_{\lambda'}+\bar\omega\tilde{U}_{ll'}\ec
\end{align}
from which we can easily obtain the lowest $\bar{U}_{ll'}$ (note that $\bar{U}_{l0}=0=\bar{U}_{l1}$)
\begin{align}
\bar{U}_{ll'}&=0 \quad\text{for }\;l'\leq l\ec \qquad \bar{U}_{l,l+1}=b_l\ec \qquad
\bar{U}_{l,l+2}=-(2l+3)\Big\{b_{l+1}-b_l+2\bar\omega\alpha^l\Big\}\ec\nt\\
\bar{U}_{l,l+3}&=(2l+5)(2l+3)\bigg\{(l+2)\bigg[\frac{b_{l+2}}{2l+3}+\frac{b_l}{2l+5}\bigg]-b_{l+1}-\bar\omega\alpha^l(1-\alpha+\beta)\bigg\}\ep
\end{align}
The angular kernels $V_{ll'}$ can be expressed by the kernels $U_{ll'}$ as follows: from the integral representation of $U_{ll'}$~\eqref{angularkernels} we can deduce that
\begin{equation}
P_{l'}'(z_s')=\sum\limits_{n=0}^{l'-1}(2n+1)U_{nl'}P_n(z_s)\ec
\end{equation}
and inserting this into the integral representation of $V_{ll'}$~\eqref{angularkernels} yields
\begin{equation}
V_{ll'}=\sum\limits_{n=0}^{l'-1}(2n+1)U_{nl'}\left\{\frac{1}{2}\int\limits_{-1}^1\diff z_s\;\frac{P_n(z_s)P_l(z_s)}{x_s-z_s}\right\}\ep
\end{equation}
By using the identity
\begin{equation}
\frac{1}{2}\int\limits_{-1}^1\diff x\;\frac{P_n(x)P_l(x)}{z-x}=P_n(z)Q_l(z) \qquad \text{for }\;n\leq l\ec
\end{equation}
we can write 
\begin{align}
V_{ll'}&=Q_l(x_s)P_{l'}'(x_s')-\sum\limits_{n=l+1}^{l'-1}(2n+1)U_{nl'}\Big\{P_n(x_s)Q_l(x_s)-P_l(x_s)Q_n(x_s)\Big\}\ec\nt\\
x_s'(s,s')&=\alpha x_s+\beta=1-\frac{s'+s-\Sigma}{2q'^2}=x_s(s',s)
\end{align}
(note that the sum vanishes for $l'\leq l+1$),
and with the aid of $W_{l-1}(z)$, which is a polynomial of degree $l-1$ in $z$ defined by~\cite{Bateman}
\begin{equation}
\label{wlminus1}
Q_l(z)=Q_0(z)P_l(z)-W_{l-1}(z)\ec \qquad W_{-1}=0\ec
\end{equation}
leading to the integral representation
\begin{equation}
\label{Wintrep}
W_{l-1}(z)=\frac{1}{2}\int\limits_{-1}^1\diff x\;\frac{P_l(z)-P_l(x)}{z-x}\ec
\end{equation}
the angular kernels $V_{ll'}$ take the general form
\begin{align}
\label{Vbardef}
V_{ll'}(s,s')&=Q_l(x_s)P_{l'}'(x_s')-\bar{V}_{ll'}\ec\nt\\
\bar{V}_{ll'}(s,s')&=\sum\limits_{n=l+1}^{l'-1}(2n+1)U_{nl'}\Big\{P_l(x_s)W_{n-1}(x_s)-P_n(x_s)W_{l-1}(x_s)\Big\}\ep
\end{align}
The $\bar{V}_{ll'}$ only contribute for $l'\geq l+2$
\begin{equation}
\bar{V}_{ll'}=0 \quad\text{for }\;l'\leq l+1\ec
\end{equation}
and we can immediately read off
\begin{equation}
V_{l0}=0\ec \qquad V_{l1}=Q_l(x_s)\ec \qquad V_{0l'}=Q_0(x_s)P_{l'}'(x_s')-\sum\limits_{n=1}^{l'-1}(2n+1)U_{nl'}W_{n-1}(x_s)\ec
\end{equation}
where the second equation can also be seen directly by comparing~\eqref{angularkernels} with~\eqref{qlneumann}. Furthermore, one easily obtains the asymptotic behavior
\begin{equation}
\label{Vasym}
V_{ll'}\sim q^{2l+2} \quad\text{for }\;q\to0\ec \qquad
V_{ll'}\sim q'^{-2l'+2} \quad\text{for }\;q'\to0\ec \qquad
V_{ll'}\sim q'^{-2l-2} \quad\text{for }\;q'\to\infty\ep
\end{equation}
From~\eqref{Wintrep} or from Christoffel's formula for $l\geq1$~\cite{Bateman}
\begin{equation}
W_{l-1}(z)=\sum\limits_{\lambda=0}^{\big\lfloor\frac{l-1}{2}\big\rfloor}\frac{2(l-\lambda)-(2\lambda+1)}{(l-\lambda)(2\lambda+1)}P_{l-(2\lambda+1)}(z)\ec \qquad
\bigg\lfloor\frac{l-1}{2}\bigg\rfloor=\begin{cases}\frac{l}{2}-1&\text{for $l\geq2$ even}\ec\\\frac{l-1}{2}&\text{for $l\geq1$ odd}\ec\end{cases}
\end{equation}
both yielding (also in agreement with~\eqref{ql123} and~\eqref{wlminus1}) besides $W_{-1}=0$
\begin{equation}
W_0=1\ec \qquad W_1(z)=\frac{3}{2}z\ec \qquad W_2(z)=\frac{5}{2}z^2-\frac{2}{3}\ec
\end{equation}
where it is useful to note that $W_l(z)$ like $P_l(z)$ contains only even/odd powers of $z$ for $l$ even/odd, respectively, 
we can immediately deduce the non-vanishing angular kernels $U_{ll'}$, $\bar{U}_{ll'}$, and $V_{ll'}$ for $l'\leq3$
\begin{align}
U_{l1}&=\delta_{l0}\ec & U_{l2}&=\alpha\delta_{l1}+3\beta\delta_{l0}\ec &
U_{l3}&=\alpha^2\delta_{l2}+5\alpha\beta\delta_{l1}+\frac{1}{2}\big\{5[\alpha^2+3\beta^2]-3\big\}\delta_{l0}\ec\nt\\
& & \bar{U}_{l2}&=b_1\delta_{l1}-3\big\{b_1+2\bar\omega\big\}\delta_{l0}\ec\nt\\
& & \bar{U}_{l3}&=b_2\delta_{l2}+5\big\{b_1-b_2-2\bar\omega\alpha\big\}&&\hspace{-0.85cm}\delta_{l1}-5\big\{3b_1-2b_2+3\bar\omega(1-\alpha+\beta)\big\}\delta_{l0}\ec\\
V_{l1}&=Q_l(x_s)\ec & V_{l2}&=3x_s'Q_l(x_s)-3\alpha\delta_{l0}\ec &
V_{l3}&=P_3'(x_s')Q_l(x_s)-\frac{5}{2}\alpha^2\delta_{l1}-\frac{15}{2}\alpha\big\{\alpha x_s+2\beta\big\}\delta_{l0}\ec\nt
\end{align}
that are needed for the kernels $K^I_{ll'}(W,W')$ for all combinations $(l\geq0,l'\leq2)$ according to~\eqref{sskernelfunction2}
{\allowdisplaybreaks
\begin{align}
\label{KIexplicit}
K_{l0}^I(W,W')&=\bigg\{\frac{1}{W'-W}+c_+-\frac{W'}{W}\frac{\varkappa^I(W,W')}{s'-a}\bigg\}\delta_{l0}\nt\\
&\quad-\frac{\epsilon^I}{2q^2}\frac{W'}{W}\bigg\{\varrho(W,W')Q_l(x_s)+\varrho(-W,W')Q_{l+1}(x_s)\bigg\}\ec\nt\\
K_{l1}^I(W,W')&=\bigg\{\frac{1}{W'-W}+b_1+\alpha c_+\bigg\}\delta_{l1}+\bigg\{-3\big[b_1+2\bar\omega-\beta c_+\big]+\frac{1}{W'+W}\frac{W'}{W}\bigg(\frac{E+m}{E'-m}\nt\\
&\qquad-\alpha\frac{E-m}{E'+m}\bigg)\bigg\}\delta_{l0}-\frac{\epsilon^I}{2q^2}\frac{W'}{W}\bigg\{-3\alpha\varrho(W,W')\delta_{l0}+\Big[3x_s'\varrho(W,W')+\varrho(W,-W')\Big]Q_l(x_s)\nt\\
&\qquad+\Big[3x_s'\varrho(-W,W')+\varrho(-W,-W')\Big]Q_{l+1}(x_s)\bigg\}\nt\\
&\quad-\frac{1}{s'-a}\frac{W'}{W}\Big\{\alpha\varkappa^I(W,W')\delta_{l1}+\Big[3\beta\varkappa^I(W,W')+\varkappa^I(W,-W')+\alpha\varkappa^I(-W,W')\Big]\delta_{l0}\Big\}\ec\nt\\
K_{l2}^I(W,W')&=\bigg\{\frac{1}{W'-W}+b_2+\alpha^2c_+\bigg\}\delta_{l2}+\bigg\{5\big[b_1-b_2-\alpha(2\bar\omega-\beta c_+)\big]+\frac{\alpha}{W'+W}\frac{W'}{W}\bigg(\frac{E+m}{E'-m}\nt\\
&\qquad-\alpha\frac{E-m}{E'+m}\bigg)\bigg\}\delta_{l1}+\bigg\{-2(8b_1-5b_2)-15\bar\omega(1-\alpha+\beta)+\frac{1}{2}(5[\alpha^2+3\beta^2]-3)c_+\nt\\
&\qquad-\alpha c_-+\frac{\beta}{W'+W}\frac{W'}{W}\bigg(3\frac{E+m}{E'-m}-5\alpha \frac{E-m}{E'+m}\bigg)\bigg\}\delta_{l0}\nt\\
&\quad-\frac{\epsilon^I}{2q^2}\frac{W'}{W}\bigg\{-\alpha\Big[15\Big\{\frac{\alpha}{2}x_s+\beta\Big\}\varrho(W,W')+3\varrho(W,-W')+\frac{5}{2}\alpha\varrho(-W,W')\Big]\delta_{l0}\nt\\
&\qquad-\frac{5}{2}\alpha^2\varrho(W,W')\delta_{l1}+\Big[P_3'(x_s')\varrho(W,W')+3x_s'\varrho(W,-W')\Big]Q_l(x_s)\nt\\
&\qquad+\Big[P_3'(x_s')\varrho(-W,W')+3x_s'\varrho(-W,-W')\Big]Q_{l+1}(x_s)\bigg\}\nt\\
&\quad-\frac{1}{s'-a}\frac{W'}{W}\bigg\{\alpha^2\varkappa^I(W,W')\delta_{l2}+\alpha\Big[5\beta\varkappa^I(W,W')+\varkappa^I(W,-W')+\alpha\varkappa^I(-W,W')\Big]\delta_{l1}\nt\\
&\qquad+\Big[\frac{1}{2}(5[\alpha^2+3\beta^2]-3)\varkappa^I(W,W')+3\beta\varkappa^I(W,-W')\nt\\
&\qquad\quad+5\alpha\beta\varkappa^I(-W,W')+\alpha\varkappa^I(-W,-W')\Big]\delta_{l0}\bigg\}\ep
\end{align}}\noindent
From $K^I_{l0}(W,W')$ in the form according to~\eqref{sskernelfunction}
\begin{align}
K^I_{l0}(W,W')&=\frac{1}{2W}\frac{W'}{E'+m}\Bigg\{(E+m)\bigg[(W+W')\frac{2\delta_{l0}}{s'-s}+\epsilon^I(W-W'-2m)\frac{Q_l(x_s)}{q^2}\bigg]\\
&+\epsilon^I(E-m)(W+W'+2m)\frac{Q_{l+1}(x_s)}{q^2}-(E+m)\Big[\tilde\epsilon_+(W'+m)+\tilde\epsilon_-(W-m)\Big]\frac{2\delta_{l0}}{s'-a}\Bigg\}\ec\nt
\end{align}
we can deduce that the nucleon pole terms~\eqref{NIpm} are reproduced by
\begin{equation}
\label{NIpmcheck}
N^I_{l+}(W)=-f^2K^I_{l0}(W,-W'=m)=-N^I_{(l+1)-}(-W) \qquad\forall\;l\geq0\ep
\end{equation}
The explicit formulae for the additional non-vanishing angular kernels $U_{ll'}$, $\bar{U}_{ll'}$, and $V_{ll'}$ for $(l\leq2,4\leq l'\leq6)$ needed for calculating the additional higher kernels $K^I_{ll'}$ for $(l\leq1,3\leq l'\leq5)$ via~\eqref{sskernelfunction2} are displayed in Appendix~\ref{subsec:spwp:higherkernels}.
Furthermore, we give the asymptotic behavior of the general kernel function $K^I_{ll'}(W,W')$, which can be inferred from the asymptotic behavior of the angular kernels~\eqref{Uasym} and~\eqref{Vasym},
\begin{align}
&\text{for }\;q\rightarrow 0 & K^I_{ll'}(W,W')&\sim q^{2l}\ec & K^I_{ll'}(-W,W')&\sim q^{2l+2}\ec\nt\\
&\text{for }\;q'\rightarrow 0 & K^I_{ll'}(W,W')&\sim q'^{-2l'}\ec & K^I_{ll'}(W,-W')&\sim q^{-2l'-2}\ec\nt\\
&\text{for }\;q'\rightarrow\infty & K^I_{ll'}(W,W')&\sim q'^{-2l-1}\ec
\end{align}
in agreement with the MacDowell symmetry relation~\eqref{macdowell}.
From~\eqref{NIpmcheck} we can then read off the asymptotic behavior of the nucleon pole terms
\begin{equation}
N^I_{l+}(W)\sim N^I_{(l+1)-}(-W)\sim q^{2l} \quad \text{for }q\to0\ep
\end{equation}

\subsection[$t$-channel exchange]{\boldmath{$t$}-channel exchange}
\label{subsec:spwp:t}

With definitions~\eqref{primedkinematics} and relations~\eqref{internalkinematicsofsta} the $t$-channel terms of the HDRs~\eqref{hdr} can be written as
\begin{equation}
\label{sthdr}
\mathbf{A}^I(s,t)\Big|^{t}_{t=t(s,z_s)}=\frac{1}{\pi}\int\limits_{\tpi}^\infty\diff t'\;\mathbf{h}^I_t[s,t';z_s]\,\Im\mathbf{A}^I(s',t')\Big|_{s'=s'(t',z_t')}\ec
\end{equation}
where the HDR kernel matrix $\mathbf{h}^I_t$ is given by
\begin{align}
\label{xtdef}
\mathbf{h}^I_t(s,t';z_s)&=\frac{1}{2q^2}\frac{1}{x_t-z_s}\begin{pmatrix}\lambda^I_1&0\\0&\lambda^I_2\end{pmatrix}, \qquad x_t(s,t')=1+\frac{t'}{2q^2}=z_s(s,t')\ec\nt\\
\lambda^I_n(s,t';z_s)&=\Big(\frac{\nu}{\nu'}\Big)^\frac{1+(-1)^n\epsilon^I}{2} \qquad (\text{with }\;x^0\equiv1\;\;\forall\;x)\ec
\end{align}
and the integrand is to be understood as a function of $[s,t';z_s]$ by using
\begin{align}
\label{ztprimetozs}
z_t'(s,t';z_s)=\frac{m\nu'}{p_t'q_t'}=\sqrt{\gamma z_s+\delta}\ec \qquad \gamma(s,t')&=\frac{q^2(s-a)}{2p_t'^2q_t'^2}\ec\nt\\ \delta(s,t')&=\frac{(t'-\Sigma+2a)^2-4(s-a)(2q^2+\Sigma-s-a)}{16p_t'^2q_t'^2}\ep
\end{align}
The $t$-channel partial-wave expansions of the invariant amplitudes, i.e.~the inversion of~\eqref{tprojform}, read~\cite{FrazerFulco:tPW}
\begin{align}
\label{texpform}
A^I(s',t')\Big|_{s'=s'(t',z_t')}&=-\frac{4\pi}{p_t'^2}\sum_J(2J+1)(p_t'q_t')^J\left\{P_J(z_t')f^J_+(t')-\frac{m}{\sqrt{J(J+1)}}z_t'P_J'(z_t')f^J_-(t')\right\}\ec\nt\\
B^I(s',t')\Big|_{s'=s'(t',z_t')}&=4\pi\sum_J\frac{2J+1}{\sqrt{J(J+1)}}(p_t'q_t')^{J-1}P_J'(z_t')f^J_-(t')\ec
\end{align}
where it is crucial that the sums only run over even $J$ for $I=+$ and odd $J$ for $I=-$ due to Bose symmetry.
Taken literally, the form~\eqref{texpform} of the partial-wave expansions is only valid for $t'\geq\tN$, since below the two-particle thresholds $\tN$ and $\tpi$ the CMS momenta $p_t'$ of the nucleons and $q_t'$ of the pions become purely imaginary and one has to use $p_-'$ and $q_-'$ instead, respectively (cf.~\eqref{pmqm} and~\cite{Hoehler}).
In particular, in the unphysical range $t'\in[\tpi,\tN)$ that we are interested in as the low-energy part of the integration range $t'\in[\tpi,\infty)$, we have $q_t'\in\mathbb{R}$ but $p_t',z_t'\in i\mathbb{R}$.
However, the squares $p_t'^2$ and $q_t'^2$ are always real (albeit not necessarily positive, cf.~\eqref{pt2qt2}) and since the combination $p_t'q_t'z_t'=m\nu'=m(2s'+t'-\Sigma)$ is always real as well, so is $z_t'^2$. 
Due to the fact that the Legendre polynomials and their derivatives have definite parity $P_J(-z)=(-1)^JP_J(z)$ and $P_J'(-z)=(-1)^{J-1}P_J'(z)$, a closer look at the expansions~\eqref{texpform} shows that in all cases only powers of the real combinations $p_t'q_t'z_t'$ and additional factors of powers of the likewise real squares $p_t'^2$ and $q_t'^2$ appear.
Therefore, we can symbolically use these formulae for all kinematical ranges and factor out powers of the real squared momenta whenever necessary in order to form explicitly real quantities.

By introducing the $t$-channel partial-wave amplitudes into the matrix notation via\footnote{In order to accommodate the fact that there is no $f^0_-$ to the matrix notation, we define $f^0_-\equiv0$ and in the following all corresponding quantities (e.g.\ integral kernels) are also understood to vanish.}
\begin{equation}
\mathbf{f}^J=\begin{pmatrix}f^J_+\\f^J_-\end{pmatrix},
\end{equation}
the expansions~\eqref{texpform} can be rewritten as
\begin{equation}
\label{texp}
\mathbf{A}^I(s',t')\Big|_{s'=s'(t',z_t')}=\sum_J\mathbf{T}^J(t',z_t')\mathbf{f}^J(t')\ec
\end{equation}
where the expansion kernel matrix is given by
\begin{align}
\mathbf{T}^J(t',z_t')&=\zeta_J \begin{pmatrix}u_J&v_J\\0&w_J\end{pmatrix}, \qquad \zeta_J(t')=4\pi(2J+1)(p_t'q_t')^{J-1}\ec\\
u_J(t',z_t')&=-\frac{q_t'}{p_t'}P_J(z_t')\ec \qquad v_J(t',z_t')=\frac{m}{\sqrt{J(J+1)}}\frac{q_t'}{p_t'}z_t'P_J'(z_t')\ec \qquad w_J(t',z_t')=\frac{1}{\sqrt{J(J+1)}}P_J'(z_t')\ep\nt
\end{align}
As the sum only runs over even $J$ for $I=+$ and odd $J$ for $I=-$ and thus the full information on the crossing properties is already contained in the index $J$, we can redefine
\begin{equation}
\lambda^I_n(s,t';z_s)=\lambda^J_n(s,t';z_s)=\Big(\frac{\nu}{\nu'}\Big)^\frac{1+(-1)^{n+J}}{2} \qquad (\text{with }\;x^0\equiv1\;\;\forall\;x)\ec
\end{equation}
and omit the index $I$ in favor of $J$ in the following. 
If we expand the imaginary part of the $t$-channel HDR terms in~\eqref{sthdr} into $t$-channel partial waves via~\eqref{texp} and project out $s$-channel partial waves again by use of~\eqref{sproj}, we can obtain the following PWDRs
\begin{equation}
\label{stpwhdr}
\mathbf{f}^I_l(W)\Big|^{t}=\frac{1}{\pi}\int\limits_{\tpi}^\infty\diff t'\;\sum_J\mathbf{G}^{lJ}(W,t')\,\Im\mathbf{f}^J(t')\ec
\end{equation}
where the $t$-channel kernel matrix is defined by
\begin{equation}
\mathbf{G}^{lJ}(W,t')=\int\limits_{-1}^1\diff z_s\;\mathbf{R}^l(W,z_s)\mathbf{h}^I_t[s,t';z_s]\mathbf{T}^J(t',z_t')\ep
\end{equation}
Due to the symmetry relation
\begin{equation}
\label{stkernelmacdowell}
\mathbf{G}^{lJ}(-W,t')=-\boldsymbol{\sigma}_1\mathbf{G}^{lJ}(W,t')\ec
\end{equation}
which follows from~\eqref{sskernelmacdowell} and is in accordance with the MacDowell symmetry~\eqref{matrixmacdowell}, the $t$-channel kernel matrix can be expressed by two kernel functions
\begin{equation}
\mathbf{G}^{lJ}(W,t')=\begin{pmatrix}G_{lJ}(W,t')&H_{lJ}(W,t')\\-G_{lJ}(-W,t')&-H_{lJ}(-W,t')\end{pmatrix},
\end{equation}
where in accordance with $f^0_-\equiv0$ for the matrix notation we set $H_{l0}\equiv0$,
and the PWDRs~\eqref{stpwhdr} take the form already given in~\eqref{spwhdr}
\begin{align}
f^I_{l+}(W)\Big|^{t}&=\frac{1}{\pi}\int\limits^\infty_{\tpi}\diff t'\;\sum\limits_J
 \Big\{G_{lJ}(W,t')\,\Im f^J_+(t')+H_{lJ}(W,t')\,\Im f^J_-(t')\Big\}\nt\\
&=-f^I_{(l+1)-}(-W)\Big|^{t}\ep
\end{align}

With the definitions
\begin{align}
\label{psidef}
\psi\big[a_{kn}\big|d(W)\big]=d(W)a_{kn}+d(-W)a_{k+1,n}\ec \qquad \eta_J(W,t')=\frac{2J+1}{4Wq^2}\frac{(p_t'q_t')^J}{p_t'^2}\ec
\end{align}
and by introducing the angular kernels
\begin{align}
A_{lJ}(s,t')&=\frac{1}{2}\int\limits_{-1}^1\diff z_s\;\lambda_1^J\frac{P_l(z_s)P_J(z_t')}{x_t-z_s}\ec \qquad 
B_{lJ}(s,t')=\frac{1}{2}\int\limits_{-1}^1\diff z_s\;\lambda_2^J\frac{P_l(z_s)P_J'(z_t')}{x_t-z_s}\ec\nt\\
C_{lJ}(s,t')&=\frac{1}{2}\int\limits_{-1}^1\diff z_s\;\lambda_1^J\frac{P_l(z_s)z_t'P_J'(z_t')}{x_t-z_s}=JA_{lJ}+B_{l,J-1}\ec
\end{align}
we can write the kernel functions as
\begin{align}
\label{stkernelfunctions}
G_{lJ}(W,t')&=-\eta_J\psi\big[A_{lJ}\big|E+m\big]
& &\forall\;J\geq0\ec\nt\\
H_{lJ}(W,t')&=\frac{\eta_J}{\sqrt{J(J+1)}}\left\{\frac{p_t'}{q_t'}\psi\big[B_{lJ}\big|(W-m)(E+m)\big]+m\psi\big[C_{lJ}\big|E+m\big]\right\}
& &\forall\;J\geq1\ep
\end{align}
If we use the decomposition
\begin{equation}
\frac{\nu}{\nu'}\frac{1}{x_t-z_s}=\frac{\mu_1}{z_t'}+\frac{\mu_2}{z_t'}\frac{1}{x_t-z_s}\ec
\qquad
\mu_1(s,t')=-\frac{q^2}{2p_t'q_t'}\ec \qquad \mu_2(s,t')=\frac{2s+t'-\Sigma}{4p_t'q_t'}\ec
\end{equation}
we find for the angular kernels for even $J$
\begin{align}
A_{lJ}(s,t')&=\frac{1}{2}\int\limits_{-1}^1\diff z_s\;\frac{P_l(z_s)P_J(z_t')}{x_t-z_s}\ec\nt\\
B_{lJ}(s,t')&=\frac{\mu_1}{2}\int\limits_{-1}^1\diff z_s\;P_l(z_s)\frac{P_J'(z_t')}{z_t'}+\frac{\mu_2}{2}\int\limits_{-1}^1\diff z_s\;\frac{P_l(z_s)P_J'(z_t')/z_t'}{x_t-z_s}\ec
\end{align}
and for odd $J$
\begin{align}
A_{lJ}(s,t')&=\frac{\mu_1}{2}\int\limits_{-1}^1\diff z_s\;P_l(z_s)\frac{P_J(z_t')}{z_t'}+\frac{\mu_2}{2}\int\limits_{-1}^1\diff z_s\;\frac{P_l(z_s)P_J(z_t')/z_t'}{x_t-z_s}\ec\nt\\
B_{lJ}(s,t')&=\frac{1}{2}\int\limits_{-1}^1\diff z_s\;\frac{P_l(z_s)P_J'(z_t')}{x_t-z_s}\ec
\end{align}
from which we can infer that only even powers of $z_t'$ occur and hence a square-root dependence on $z_s$ is avoided.
We now can work out the kernel functions explicitly, here given for all combinations $(l\geq0,J\leq2)$
{\allowdisplaybreaks
\begin{align}
\label{GH_explicit}
G_{l0}(W,t')&=-\frac{1}{4Wq^2p_t'^2}\Big\{(E+m)Q_l(x_t)-(E-m)Q_{l+1}(x_t)\Big\}\ec\nt\\
G_{l1}(W,t')&=\frac{3}{4}\bigg\{(2s+t'-\Sigma)G_{l0}(W,t')+\frac{E+m}{2Wp_t'^2}\delta_{l0}\bigg\}\ec\nt\\
H_{l1}(W,t')&=\frac{1}{\sqrt{2}}\bigg\{\frac{3}{4}Z_l(W,t')-mG_{l1}(W,t')\bigg\}\ec\nt\\
G_{l2}(W,t')&=\frac{5}{16}\bigg\{\Big[6s(s+t'-\Sigma)+(t'-\Sigma)^2+2\Sigma_-^2\Big]G_{l0}(W,t')+3\frac{(E+m)(s-a)}{Wp_t'^2}\delta_{l0}\bigg\}\ec\nt\\
H_{l2}(W,t')&=\frac{15}{16\sqrt{6}}\bigg\{(2s+t'-\Sigma)Z_l(W,t')-m\Big[4s(s+t'-\Sigma)+(t'-\Sigma)^2\Big]G_{l0}(W,t')\nt\\
&\qquad\qquad\quad-2\frac{E+m}{W}\bigg[\frac{m(s-a)}{p_t'^2}+W-m\bigg]\delta_{l0}\bigg\}\ec
\end{align}}\noindent
where we have defined
\begin{equation}
Z_l(W,t')=\frac{1}{Wq^2}\Big\{(E+m)(W-m)Q_l(x_t)+(E-m)(W+m)Q_{l+1}(x_t)\Big\}\ep
\end{equation}
From the expansion
\begin{equation}
Q_l(x_t)=\left(\frac{2q^2}{t'}\right)^{l+1}+\Ord\left(\left(\frac{2q^2}{t'}\right)^{l+2}\right) \quad \text{for }\;\frac{1}{x_t-1}=\frac{2q^2}{t'}\to0\ec
\end{equation}
we can finally deduce the asymptotic behavior of the non-vanishing general kernel functions~\eqref{stkernelfunctions}
\begin{align}
\label{GHasymptotics}
&\text{for }\;q\rightarrow 0 & G_{lJ}(W,t')&\sim H_{lJ}(W,t')\sim q^{2l}\ec & G_{lJ}(-W,t')&\sim H_{lJ}(-W,t')\sim q^{2l+2}\ec\nt\\
&\text{for }\;q_t'\rightarrow 0 & G_{lJ}(W,t')&\sim H_{lJ}(W,t')\sim 1\ec\nt\\
&\text{for }\;p_t'\rightarrow 0 & G_{lJ}(W,t')&\sim H_{lJ}(W,t')\sim p_t'^{-2}\ec\nt\\
&\text{for }\;t'\rightarrow\infty & G_{lJ}(W,t')&\sim H_{lJ}(W,t')\sim t'^{J-l-2}\ec
\end{align}
in accordance with the MacDowell symmetry relation~\eqref{macdowell}.

\subsection{Higher kernel functions}
\label{subsec:spwp:higherkernels}

Here, we display the explicit form of the additional angular kernels $U_{ll'}$, $\bar{U}_{ll'}$, and $V_{ll'}$ for $(l\leq2,4\leq l'\leq6)$ that are required for calculating the additional higher kernels $K^I_{ll'}$ for $(l\leq1,3\leq l'\leq5)$ via~\eqref{sskernelfunction2} needed to incorporate higher resonances in the $s$-channel integrals.
From~\eqref{Udef} we obtain
\begin{align}
U_{04}&=\frac{5}{2}\beta\Big\{7\alpha^2+7\beta^2-3\Big\}\ec \qquad
U_{14}=\frac{1}{2}\alpha\Big\{7\alpha^2+35\beta^2-5\Big\}\ec \qquad
U_{24}=7\alpha^2\beta\ec\\
U_{05}&=\frac{1}{8}\Big\{15-70\big(\alpha^2+3\beta^2\big)+63\big(\alpha^4+5\beta^4+10\alpha^2\beta^2\big)\Big\}\ec \qquad
U_{15}=\frac{7}{2}\alpha\beta\Big\{9\alpha^2+15\beta^2-5\Big\}\ec\nt\\
U_{25}&=\frac{1}{2}\alpha^2\Big\{9\alpha^2+63\beta^2-7\Big\}\ec \qquad
U_{06}=\frac{21}{8}\beta\Big\{5-30\big(\alpha^2+\beta^2\big)+11\big(3\alpha^4+3\beta^4+10\alpha^2\beta^2\big)\Big\}\ec\nt\\
U_{16}&=\frac{1}{8}\alpha\Big\{35-126\big(\alpha^2+5\beta^2\big)+33\big(3\alpha^4+35\beta^4+42\alpha^2\beta^2\big)\Big\}\ec \qquad
U_{26}=\frac{3}{2}\alpha^2\beta\Big\{33\alpha^2+77\beta^2-21\Big\}\ec\nt
\end{align}
and~\eqref{Ubardef} yields
{\allowdisplaybreaks
\begin{align}
\bar{U}_{04}&=-5\Big\{7b_3-14b_2+9b_1+\bar\omega\big[4-14\alpha(1-\alpha)+7\beta(1-\alpha+\beta)\big]\Big\}\ec\nt\\
\bar{U}_{14}&=21b_3-35b_2+15b_1-35\bar\omega\alpha(1-\alpha+\beta)\ec \qquad
\bar{U}_{24}=-7\Big\{b_3-b_2+2\bar\omega\alpha^2\Big\}\ec\nt\\
\bar{U}_{05}&=7\Big\{18b_4-5(9b_3-8b_2+3b_1)-\frac{15}{4}\bar\omega(1-\alpha+\beta)\big[(1-3\alpha)^2+3\beta^2\big]\Big\}\ec\nt\\
\bar{U}_{15}&=-7\Big\{12b_4-27b_3+20b_2-5b_1+\bar\omega\alpha\big[2(5-15\alpha+12\alpha^2)+15\beta(1-\alpha+\beta)\big]\Big\}\ec\nt\\
\bar{U}_{25}&=36b_4-63b_3+28b_2-63\bar\omega\alpha^2(1-\alpha+\beta)\ec\nt\\
\bar{U}_{06}&=-21\Big\{22(b_5-3b_4)+5(15b_3-8b_2+2b_1)+\bar\omega\big[2(1-9\alpha+31\alpha^2-22\alpha^3(2-\alpha))\nt\\
&\qquad\quad+\frac{\beta}{4}(1-\alpha+\beta)(3-11\alpha(6-13\alpha)+33\beta^2)\big]\Big\}\ec\nt\\
\bar{U}_{16}&=66(5b_5-14b_4)+35(27b_3-12b_2+2b_1)-\frac{21}{4}\bar\omega\alpha(1-\alpha+\beta)\big[(5-11\alpha)^2+55\beta^2\big]\ec\nt\\
\bar{U}_{26}&=-3\Big\{11(5b_5-12b_4)+7(15b_3-4b_2)+\bar\omega\alpha^2\big[56-22\alpha(7-5\alpha)+77\beta(1-\alpha+\beta)\big]\Big\}\ep
\end{align}}\noindent
From~\eqref{Vbardef} it follows that
\begin{align}
\bar{V}_{04}&=\frac{5}{6}\alpha\Big\{21(\alpha x_s)^2+63\beta(\alpha x_s)+7\alpha^2+63\beta^2-9\Big\}\ec \qquad
\bar{V}_{14}=\frac{35}{6}\alpha^2\Big\{(\alpha x_s)+3\beta\Big\}\ec \qquad
\bar{V}_{24}=\frac{7}{3}\alpha^3\ec\nt\\
\bar{V}_{05}&=\frac{105}{8}\alpha\Big\{3(\alpha x_s)^3+12\beta(\alpha x_s)^2+\big(\alpha^2+18\beta^2-2\big)(\alpha x_s)+4\beta\big(\alpha^2+3\beta^2-1\big)\Big\}\ec\nt\\
\bar{V}_{15}&=\frac{7}{8}\alpha^2\Big\{15(\alpha x_s)^2+60\beta(\alpha x_s)+9\alpha^2+90\beta^2-10\Big\}\ec \qquad
\bar{V}_{25}=\frac{21}{4}\alpha^3\Big\{(\alpha x_s)+4\beta\Big\}\ec\nt\\
\bar{V}_{06}&=\frac{21}{40}\alpha\Big\{165(\alpha x_s)^4+825\beta(\alpha x_s)^3+5\big(11\alpha^2+330\beta^2-30\big)(\alpha x_s)^2\nt\\
&\qquad+25\beta\big(11\alpha^2+66\beta^2-18\big)(\alpha x_s)+25-50\alpha^2+33\alpha^4+50\big(11\alpha^2-9\big)\beta^2+825\beta^4\Big\}\ec\nt\\
\bar{V}_{16}&=\frac{21}{40}\alpha^2\Big\{55(\alpha x_s)^3+275\beta(\alpha x_s)^2+\big(33\alpha^2+550\beta^2-50\big)(\alpha x_s)+5\beta\big(33\alpha^2+110\beta^2-30\big)\Big\}\ec\nt\\
\bar{V}_{26}&=\frac{3}{20}\alpha^3\Big\{77(\alpha x_s)^2+385\beta(\alpha x_s)+2\big(33\alpha^2+385\beta^2-35\big)\Big\}\ep
\end{align}
We refrain from explicitly spelling out the form of $b_\mu(W,W')$ for higher values of $\mu$, as these functions follow directly from their definition~\eqref{bmudef}.

\subsection{Subtracted kernel functions}
\label{subsec:spwp:subtractions}

Finally, we summarize the changes that are necessary if the subtracted versions of the HDRs are used for the $s$-channel projection.

The modified pole terms are given by
\begin{align}
\nsub{N_{l+}^I}(W)&=\bar N_{l+}^I(W)+\nsub{\Delta\bar N_{l+}^I}(W)=-\nsub{N_{(l+1)-}^I}(-W)\ec\nt\\
\twosub{\Delta\bar N_{l+}^+}(W)&=\frac{\delta_{l0}}{16\pi W}
\Bigg\{(E+m)\bigg[2\bigg(\frac{g^2}{m}+d_{00}^+-2q^2d_{01}^+\bigg)+(W-m)\big(s-s_0-q^2\big)\frac{b_{00}^+}{m}\bigg]\nt\\
&\quad-(E-m)\frac{q^2}{3}\bigg(4d_{01}^+-(W+m)\frac{b_{00}^+}{m}\bigg)\Bigg\}
+\frac{\delta_{l1}}{16\pi W}(E+m)\frac{q^2}{3}\bigg(4d_{01}^++(W-m)\frac{b_{00}^+}{m}\bigg)\nt\\
&\redonesub\frac{\delta_{l0}}{8\pi W}(E+m)\bigg(\frac{g^2}{m}+d_{00}^+\bigg)\ec\nt\\
\twosub{\Delta\bar N_{l+}^-}(W)&=\frac{\delta_{l0}}{16\pi W}
\Bigg\{(E+m)\bigg[\big(s-s_0-q^2\big)\frac{a_{00}^-}{m}+2(W-m)\bigg(-\frac{g^2}{2m^2}+b_{00}^--2q^2b_{01}^-\bigg)\bigg]\nt\\
&\quad-(E-m)\frac{q^2}{3}\bigg(\frac{a_{00}^-}{m}-4(W+m)b_{01}^-\bigg)\Bigg\}
+\frac{\delta_{l1}}{16\pi W}(E+m)\frac{q^2}{3}\bigg(\frac{a_{00}^-}{m}+4(W-m)b_{01}^-\bigg)\nt\\
&\redonesub\frac{\delta_{l0}}{8\pi W}(E+m)(W-m)\bigg(-\frac{g^2}{2m^2}+b_{00}^-\bigg)\ec
\end{align}
where for convenience we have defined non-vanishing corrections also for the unsubtracted case according to (cf.~\eqref{NIpmexpl})
\begin{equation}
\unsub{\Delta\bar N_{l+}^I}(W)=-\tilde\epsilon_-\frac{g^2}{4\pi}\frac{(E+m)(W-m)}{2W}\frac{\delta_{l0}}{m^2-a}=-\unsub{\Delta\bar N_{(l+1)-}^I}(-W)\ep
\end{equation}

The additional contributions to the $s$-channel kernels that fulfill the MacDowell symmetry relation~\eqref{macdowell} in both $(W,l)$ and $(W',l')$ can be written for all $(l\geq0,l'\geq0)$ in the symmetric form
{\allowdisplaybreaks
\begin{align}
\Delta K_{ll'}^I(W,W')
&=\widehat{\Delta K}_{ll'}^I(W,W')-\widehat{\Delta K}_{l,l'-1}^I(W,-W')+\widetilde{\Delta K}_{ll'}^I(W,W')-\widetilde{\Delta K}_{l,l'-1}^I(W,-W')\nt\\
&\quad+\frac{1}{3}\bigg\{\widetilde{\Delta K}_{ll'}^I(-W,W')-\widetilde{\Delta K}_{l,l'-1}^I(-W,-W')\nt\\
&\qquad-\widetilde{\Delta K}_{l-1,l'}^I(W,W')+\widetilde{\Delta K}_{l-1,l'-1}^I(W,-W')\bigg\}\ec\nt\\
\twosub{\widehat{\Delta K}_{ll'}^I}(W,W')
&=-\frac{W'}{W}\bigg\{\varkappa^I(W,W')h_0(s')+\varkappa^{-I}(W,W')\frac{2(s-s_0)}{(s'-s_0)^2}\bigg\}P'_{l'+1}\Big(\ste{z_s'}\Big)\delta_{l0}\nt\\
&=-\frac{W'}{W}\bigg\{\frac{\delta(W,W')(s'+s-2s_0)+\epsilon^I\rho(W,W')(s'-s)}{(s'-s_0)^2}\nt\\
&\qquad-\frac{\varkappa^I(W,W')}{s'-a}\bigg\}P'_{l'+1}\Big(\ste{z_s'}\Big)\delta_{l0}\nt\\
&\redonesub-\frac{W'}{W}\varkappa^I(W,W')h_0(s')P'_{l'+1}\Big(\ste{z_s'}\Big)\delta_{l0}\ec\nt\\
\twosub{\widetilde{\Delta K}_{ll'}^I}(W,W')
&=-\frac{W'}{W}2q^2\bigg\{\frac{\epsilon^I\rho(W,W')}{(s'-s_0)^2}P'_{l'+1}\Big(\ste{z_s'}\Big)\nt\\
&\qquad-\varkappa^{I}(W,W')h_0(s')\ste{\partial_tz_s'}P''_{l'+1}\Big(\ste{z_s'}\Big)\bigg\}\delta_{l0}\redonesub0\ec
\end{align}}\noindent
where we have used that $\epsilon^{\pm I}=\pm\epsilon^I$.
Note that for $l'=0$ the term proportional to $(s'-a)^{-1}$ cancels against the corresponding term in $K^I_{l0}(W,W')$ of~\eqref{KIexplicit} as for the nucleon pole terms (cf.\ the relation~\eqref{NIpmcheck}).

The additional contributions to $G_{lJ}$ and $H_{lJ}$ may be written as
\begin{align}
\Delta G_{lJ}(W,t')&=\widehat{\Delta G}_{lJ}(W,t')-\widehat{\Delta G}_{l+1,J}(-W,t')\qquad\forall\;(l\geq0,J\geq0)\ec\nt\\
\Delta H_{lJ}(W,t')&=\widehat{\Delta H}_{lJ}(W,t')-\widehat{\Delta H}_{l+1,J}(-W,t')\qquad\forall\;(l\geq0,J\geq1)\ec
\end{align}
where for even $J$
\begin{align}
\twosub{\widehat{\Delta G}_{lJ}}(W,t')&=\frac{E+m}{2W}(2J+1)\frac{(p_t'q_t')^J}{t'p_t'^2}\Bigg\{\ste{P_J(z_t')}\delta_{l0}\nt\\
&\qquad-2q^2\bigg(\frac{1}{t'}\ste{P_J(z_t')}+\ste{\partial_tP_J(z_t')}\bigg)\bigg(\delta_{l0}-\frac{\delta_{l1}}{3}\bigg)\Bigg\}\nt\\
&\redonesub\frac{E+m}{2W}(2J+1)\frac{(p_t'q_t')^J}{t'p_t'^2}\ste{P_J(z_t')}\delta_{l0}\ec\nt\\
\twosub{\widehat{\Delta H}_{lJ}}(W,t')&=-\frac{E+m}{2W}\frac{2J+1}{\sqrt{J(J+1)}}\frac{(p_t'q_t')^J}{t'p_t'^2}\Bigg\{\bigg(\frac{W-m}{2q_t'^2}(s-s_0)\ste{\frac{P'_J(z_t')}{z_t'}}\nt\\
&\qquad\quad+m\ste{z_t'P'_J(z_t')}\bigg)\delta_{l0}-2q^2\bigg(\frac{W-m}{4q_t'^2}\ste{\frac{P'_J(z_t')}{z_t'}}\nt\\
&\qquad\quad+m\bigg[\frac{1}{t'}\ste{z_t'P'_J(z_t')}+\ste{\partial_t(z_t'P'_J(z_t'))}\bigg]\bigg)\bigg(\delta_{l0}-\frac{\delta_{l1}}{3}\bigg)\Bigg\}\nt\\
&\redonesub-\frac{E+m}{2W}\frac{2J+1}{\sqrt{J(J+1)}}\frac{(p_t'q_t')^J}{t'p_t'^2}m\ste{z_t'P'_J(z_t')}\delta_{l0}\ec
\end{align}
and for odd $J$
\begin{align}
\twosub{\widehat{\Delta G}_{lJ}}(W,t')&=\frac{E+m}{2W}(2J+1)\frac{(p_t'q_t')^{J-1}}{t'p_t'^2}\frac{1}{2}\ste{\frac{P_J(z_t')}{z_t'}}\Bigg\{\big(s-s_0-q^2\big)\delta_{l0}+q^2\frac{\delta_{l1}}{3}\Bigg\}\redonesub0\ec\nt\\
\twosub{\widehat{\Delta H}_{lJ}}(W,t')&=-\frac{E+m}{2W}\frac{2J+1}{\sqrt{J(J+1)}}\frac{(p_t'q_t')^{J-1}}{t'p_t'^2}\Bigg\{\bigg(p_t'^2(W-m)+\frac{m}{2}(s-s_0)\bigg)\ste{P'_J(z_t')}\delta_{l0}\nt\\
&\qquad-2q^2\bigg(p_t'^2(W-m)\bigg[\frac{1}{t'}\ste{P'_J(z_t')}+\ste{\partial_tP'_J(z_t')}\bigg]\nt\\
&\qquad\quad+\frac{m}{4}\ste{P'_J(z_t')}\bigg)\bigg(\delta_{l0}-\frac{\delta_{l1}}{3}\bigg)\Bigg\}\nt\\
&\redonesub-\frac{E+m}{2W}\frac{2J+1}{\sqrt{J(J+1)}}\frac{(p_t'q_t')^{J-1}}{t'}(W-m)\ste{P'_J(z_t')}\delta_{l0}\ep
\end{align}
Note that again only even powers of momenta and $z_t'$ occur.

\section{Partial-wave projection for the \boldmath{$t$}-channel amplitudes}
\label{sec:tpwp}

In the following, we will discuss the different contributions to the $t$-channel part~\eqref{tpwhdr} of the RS system.

\subsection{Nucleon exchange}
\label{subsec:tpwp:n}

In order to carry out the projection integrals~\eqref{tprojform} we rewrite $s$ and $u$ as functions of $t$ and $z_t$ via
\begin{equation}
\label{suoftzt}
s(t,z_t)=\frac{1}{2}(\Sigma-t+4p_tq_tz_t)\ec \qquad u(t,z_t)=\frac{1}{2}(\Sigma-t-4p_tq_tz_t)\ec
\end{equation}
which allows us to cast the nucleon pole terms of the HDRs~\eqref{hdr} into the form
\begin{equation}
\bigg\{\frac{1}{m^2-s}\pm\frac{1}{m^2-u}-\frac{1\pm1}{2(m^2-a)}\bigg\}\bigg|_{[t;z_t]}=\frac{1}{2p_tq_t}\bigg\{\frac{1}{\tilde y-z_t}\mp\frac{1}{(-\tilde y)-z_t}\bigg\}-\frac{1\pm1}{2(m^2-a)}\ec
\end{equation}
where the upper/lower sign corresponds to even/odd $J$ (i.e.~to $I=+/-$) and we have defined, in analogy to~\eqref{ydef},
\begin{equation}
\tilde y(t)=\frac{t-2\mpi^2}{4p_tq_t}=\frac{m\nu_B}{p_tq_t}=z_t(s=m^2,t)=\tilde x_t(t,s'=m^2) 
\end{equation}
 ($\tilde x_t(t,s')$ will be defined in~\eqref{tildextdef}).
By noting that the orthonormality of the Legendre polynomials yields
\begin{equation}
\frac{1\pm1}{2}\int\limits_0^1\diff z\;P_J(z)P_{l=2m}(z)=\frac{\delta_{Jl}}{2l+1}=
\frac{1\mp1}{2}\int\limits_0^1\diff z\;P_J(z)P_{l=2n+1}(z) \quad\forall\;J,l(m,n\in\mathbb{N}_0)\ec
\end{equation}
the nucleon pole terms of the PWDRs~\eqref{tpwhdr} can be written as (in analogy to~\eqref{NIpm})
\begin{align}
\label{tildeNJpm}
\tilde N^J_+(t)&=\frac{g^2}{4\pi}m\bigg\{\frac{\tilde yQ_J(\tilde y)}{(p_tq_t)^{J}}-\delta_{J0}-\frac{1}{3}\frac{\delta_{J1}}{m^2-a}\bigg\}
 & &\forall\;J\geq0\ec\nt\\
\tilde N^J_-(t)&=\frac{g^2}{4\pi}\frac{\sqrt{J(J+1)}}{2J+1}\bigg\{\frac{Q_{J-1}(\tilde y)-Q_{J+1}(\tilde y)}{(p_tq_t)^{J}}-\frac{\delta_{J1}}{m^2-a}\bigg\}
 & &\forall\;J\geq1\ec
\end{align}
which for later convenience may be expressed as (in analogy to~\eqref{barNIpm})
\begin{align}
\label{hatNJpm}
\tilde N^J_+(t)&=\hat N^J_+(t)-\frac{g^2}{4\pi}\frac{m}{3}\frac{\delta_{J1}}{m^2-a}\ec
 & &\hat N^J_+(t)=\frac{g^2}{4\pi}m\bigg\{\frac{\tilde yQ_J(\tilde y)}{(p_tq_t)^{J}}-\delta_{J0}\bigg\}\ec
 & &\forall\;J\geq0\ec\nt\\
\tilde N^J_-(t)&=\hat N^J_-(t)-\frac{g^2}{4\pi}\frac{\sqrt{2}}{3}\frac{\delta_{J1}}{m^2-a}\ec
 & &\hat N^J_-(t)=\frac{g^2}{4\pi}\frac{\sqrt{J(J+1)}}{2J+1}\frac{Q_{J-1}(\tilde y)-Q_{J+1}(\tilde y)}{(p_tq_t)^{J}}\ec
 & &\forall\;J\geq1\ep
\end{align}
Note that for $t\in(\tpi,\tN)$ due to $p_t\in i\mathbb{R}$ also $\tilde y\in i\mathbb{R}$ and hence we need the analytic continuations of $Q_l(z)$ as discussed in Appendix~\ref{subsec:spwp:n}. However, the pole-term projections~\eqref{tildeNJpm},~\eqref{hatNJpm} are real for all $t$ above the logarithmic branch point singularity at $\tpi-(\mpi^2/m)^2\approx3.98\mpi^2$ of the nucleon cut (which is the left-hand cut for $\tilde y(t)^2\leq1$ along the real axis due to the $z_t$-projection of the nucleon pole terms), since $\tilde y/(p_tq_t)$ and the squares $p_t^2$ and $\tilde y^2$ are always real, and thus we can rewrite the projections solely in terms of real quantities due to the defined parity~\eqref{qlneumann} of the $Q_J(\tilde y)$.
Finally, we comment on the asymptotic behavior for $p_tq_t\to 0$, particularly including the vicinity of the aforementioned logarithmic singularity.
The ostensible poles in~\eqref{tildeNJpm} are canceled by the asymptotics of 
$Q_J(\tilde y)$ for $\tilde y\to\infty$.
In this limit, we may abort the series representation of $Q_l(z)$ valid for $|z|>1$~\cite{Bateman}
\begin{align}
Q_l(z)=\frac{2^l(l!)^2}{(2l+1)!}\bigg\{z^{-(l+1)}+\frac{(l+1)(l+2)}{2(2l+3)}z^{-(l+3)}+\frac{(l+1)(l+2)}{2(2l+3)}\frac{(l+3)(l+4)}{4(2l+5)}z^{-(l+5)}+\dots\bigg\}
\end{align}
after the first term and obtain the leading contributions
\begin{align}
\label{tchannel_pole_asym}
\tilde N^J_+(t)&=
\frac{g^2}{4\pi}\frac{J!}{(2J+1)!!}m\bigg\{\left(\frac{4}{t-2\mpi^2}\right)^J-\delta_{J0}-\frac{\delta_{J1}}{m^2-a}\bigg\}+\Ord(p_t^2q_t^2)
 & &\forall\;J\geq0\ec\nt\\
\tilde N^J_-(t)&=
\frac{g^2}{4\pi}\frac{J!}{(2J+1)!!}\sqrt{\frac{J+1}{J}}\bigg\{\left(\frac{4}{t-2\mpi^2}\right)^J-\frac{\delta_{J1}}{m^2-a}\bigg\}+\Ord(p_t^2q_t^2)
 & &\forall\;J\geq1\ep
\end{align}
In particular, it follows that the leading contribution to $\tilde N^0_+(t)$ vanishes, such that $\tilde N^0_+(t)$ even involves zeros for $p_tq_t\to 0$.
However, higher orders need to be taken into account in the approximations~\eqref{tchannel_pole_asym} in order to obtain precise numerical results in particular for $q_t\to0$, since the pole terms vary rapidly in the vicinity of $\tpi$.
Note that~\eqref{tildeNJpm} and~\eqref{tchannel_pole_asym} reduce to the results given in~\cite{Hoehler} and~\cite{Pietarinen} if the terms containing the hyperbola parameter $a$ (that only contribute for $J=1$ anyway) are dropped.

\subsection[$s$- and $u$-channel exchange]{\boldmath{$s$}- and \boldmath{$u$}-channel exchange}
\label{subsec:tpwp:s}

We may rewrite the $t$-channel partial-wave projection~\eqref{tprojform} in matrix form as
\begin{equation}
\label{tproj}
\mathbf{f}^J(t)=\int\limits^1_0\diff z_t\;\mathbf{\tilde T}^J(t,z_t)\mathbf{A}^I(s,t)\Big|_{s=s(t,z_t)}\ec
\end{equation}
where the projection kernel is given by
\begin{align}
\mathbf{\tilde T}^J(t,z_t)&=\tilde\zeta_J\begin{pmatrix}\tilde u_J&\tilde v_J\\0&\tilde w_J\end{pmatrix}\ec \qquad \tilde\zeta_J(t)=\frac{1}{4\pi(p_tq_t)^{J-1}}\ec\\
\tilde u_J(t,z_t)&=-\frac{p_t}{q_t}P_J(z_t),
 \qquad \tilde v_J(t,z_t)=mz_tP_J(z_t),
 \qquad \tilde w_J(t,z_t)=\frac{\sqrt{J(J+1)}}{2J+1}\Big[P_{J-1}(z_t)-P_{J+1}(z_t)\Big].\nt
\end{align}
For the following, we need the matrix form of both $s$- and $u$-channel HDR terms~\eqref{hdr} according to
\begin{equation}
\label{tshdr}
\mathbf{A}^I(s,t)\Big|^{s+u}_{s=s(t,z_t)}=\frac{1}{\pi}\int\limits_{s_+}^\infty\diff s'\;\mathbf{h}^I_s[t,s';z_t]\,\Im\mathbf{A}^I(s',t')\Big|_{t'=t'(s',z_s')}\ec
\end{equation}
where the kernel matrix $\mathbf{h}^I_s$ is given in~\eqref{xsdef}, and $[t,s';z_t]$ indicates that the whole integrand is to be understood as a function of these variables, which can be done by using~\eqref{suoftzt} and thereby
\begin{equation}
\Big\{h_1\mp h_2\Big\}\Big|_{[t,s';z_t]}=\bigg\{\frac{1}{s'-s}\pm\frac{1}{s'-u}-\frac{1\pm1}{2(s'-a)}\bigg\}\bigg|_{[t,s';z_t]}=\frac{1}{2p_tq_t}\bigg\{\frac{1}{\tilde x_t-z_t}\mp\frac{1}{(-\tilde x_t)-z_t}\bigg\}-\frac{1\pm1}{2(s'-a)}\ep
\end{equation}
The upper/lower sign corresponds to even/odd $J$ and we have defined in analogy to~\eqref{xsdef}
\begin{equation}
\label{tildextdef}
\tilde x_t(t,s')=\frac{t+2s'-\Sigma}{4p_tq_t}=z_t(s',t)\ep
\end{equation}
According to~\eqref{ztprimetozs}, the relation between $z_s'$ and $z_t$ in~\eqref{tshdr} is given  by
\begin{align}
\label{zsprimetozt}
z_s'(t,s';z_t)=\frac{z_t^2-\tilde\delta}{\tilde\gamma}\ec \qquad
 \tilde\gamma(t,s')&=\frac{q'^2(s'-a)}{2p_t^2q_t^2}=\gamma(s',t)\ec\nt\\
\tilde\delta(t,s')&=\frac{(t-\Sigma+2a)^2-4(s'-a)(2q'^2+\Sigma-s'-a)}{16p_t^2q_t^2}=\delta(s',t)\ep
\end{align}
Expanding the absorptive parts of~\eqref{tshdr} into $s$-channel partial waves via~\eqref{sexp} and projecting onto $t$-channel partial waves by means of~\eqref{tproj} leads us to the PWDRs for the $t$-channel partial waves
\begin{equation}
\label{tspwhdr}
\mathbf{f}^J(t)\Big|^{s+u}=\frac{1}{\pi}\int\limits^\infty_{W_+}\diff W'\;\sum\limits_{l=0}^\infty\mathbf{\tilde G}^{J l}(t,W')\,\Im\mathbf{f}^I_l(W')\ec
\end{equation}
with the kernel matrix
\begin{equation}
\mathbf{\tilde G}^{J l}(t,W')=2W'\int\limits_0^1\diff z_t\;\mathbf{\tilde T}^J(t,z_t)\mathbf{h}_s^I[t,W';z_s']\mathbf{S}^l(W',z_s')\ep
\end{equation}
As a remnant of the MacDowell symmetry, \eqref{sskernelmacdowell} induces the symmetry property
\begin{equation}
\mathbf{\tilde G}^{J l}(t,-W')=\mathbf{\tilde G}^{J l}(t,W')\boldsymbol{\sigma}_1\ec
\end{equation}
such that the parameterization with two kernel functions
\begin{equation}
\mathbf{\tilde G}^{J l}(t,W')=\begin{pmatrix}\tilde G_{Jl}(t,W')&\tilde G_{Jl}(t,-W')\\\tilde H_{Jl}(t,W')&\tilde H_{Jl}(t,-W')\end{pmatrix}
\end{equation}
is justified, where again according to $f^0_-\equiv0$ we set $\tilde H_{0l}\equiv0$ for the matrix notation. This reproduces the $s$- and $u$-channel part of~\eqref{tpwhdr}
\begin{align}
f^J_+(t)\Big|^{s+u}&=\frac{1}{\pi}\int\limits^{\infty}_{W_+}\diff W'\sum\limits^\infty_{l=0}\Big\{
\tilde G_{J l}(t,W')\,\Im f^I_{l+}(W')+\tilde G_{J l}(t,-W')\,\Im f^I_{(l+1)-}(W')\Big\}
& &\forall\;J\geq0\ec\nt\\
f^J_-(t)\Big|^{s+u}&=\frac{1}{\pi}\int\limits^{\infty}_{W_+}\diff W'\sum\limits^\infty_{l=0}\Big\{
\tilde H_{J l}(t,W')\,\Im f^I_{l+}(W')+\tilde H_{J l}(t,-W')\,\Im f^I_{(l+1)-}(W')\Big\}
& &\forall\;J\geq1\ep
\end{align}

If we introduce the abbreviations (cf.~\eqref{psidef})
\begin{align}
\tilde\psi\big[a_{kn}\big|d(W')\big]=d(W')a_{k,n+1}+d(-W')a_{kn}\ec \qquad \tilde\eta_J(t,W')=\frac{2W'}{(p_tq_t)^{J-1}}\ec
\end{align}
we find for the kernel functions
\begin{align}
\label{tskernelfunctions}
\tilde G_{Jl}(t,W')&=\tilde\eta_J\left\{-\frac{p_t}{q_t}\tilde\psi\bigg[\tilde A_{Jl}\bigg|\frac{W'+m}{E'+m}\bigg]
+m\tilde\psi\bigg[\tilde B_{Jl}\bigg|\frac{1}{E'+m}\bigg]\right\}
& &\forall\;J\geq0\ec\nt\\
\tilde H_{Jl}(t,W')&=\tilde\eta_J\frac{\sqrt{J(J+1)}}{2J+1}\tilde\psi\bigg[\tilde C_{Jl}\bigg|\frac{1}{E'+m}\bigg]
& &\forall\;J\geq1\ec
\end{align}
where the angular kernels are given by
\begin{align}
\tilde A_{Jl}(t,s')&=\int\limits_0^1\diff z_t\;P_J(z_t)\Big\{h_1\mp h_2\Big\}P_l'(z_s')\Big|_{[t,s';z_t]}\ec\nt\\
\tilde B_{Jl}(t,s')&=\int\limits_0^1\diff z_t\;P_J(z_t)z_t\Big\{h_1\pm h_2\Big\}P_l'(z_s')\Big|_{[t,s';z_t]}\ec\nt\\
\tilde C_{Jl}(t,s')&=\int\limits_0^1\diff z_t\;\Big[P_{J-1}(z_t)-P_{J+1}(z_t)\Big]\Big\{h_1\pm h_2\Big\}P_l'(z_s')\Big|_{[t,s';z_t]}=\tilde A_{J-1,l}-\tilde A_{J+1,l}\ep
\end{align}
Decomposing these kernels according to
\begin{align}
\tilde A_{Jl}(t,s')&=\frac{1}{p_tq_t}P_{l}'(\tilde z_s)Q_J(\tilde x_t)-\bar A_{Jl}(t,s')\ec\qquad
\tilde B_{Jl}(t,s')=\frac{1}{p_tq_t}P_{l}'(\tilde z_s)\tilde x_tQ_J(\tilde x_t)-\bar B_{Jl}(t,s')\ec\nt\\
\tilde C_{Jl}(t,s')&=\frac{1}{p_tq_t}P_{l}'(\tilde z_s)\Big[Q_{J-1}(\tilde x_t)-Q_{J+1}(\tilde x_t)\Big]-\bar C_{Jl}(t,s')\ec
\end{align}
with the real quantity
\begin{equation}
\tilde z_s(t,s')=\frac{\tilde x_t^2-\tdelta}{\tgamma}=1+\frac{t}{2q'^2}=z_s(s',t)
\end{equation}
and polynomial parts defined by
\begin{align}
\label{ABCbardef}
\bar A_{Jl}(t,s')&=\frac{1}{2}\int\limits^1_{-1}\diff z_t\,P_J(z_t)\left\{\frac{1}{p_tq_t}\frac{P_l'(\tilde z_s)-P_l'(z_s')}{\tilde x_t-z_t}+\frac{1\pm1}{2(s'-a)}P_l'(z_s')\right\}\ec\nt\\
\bar B_{Jl}(t,s')&=\frac{1}{2}\int\limits^1_{-1}\diff z_t\,P_J(z_t)\left\{\frac{1}{p_tq_t}\frac{\tilde x_tP_l'(\tilde z_s)-z_tP_l'(z_s')}{\tilde x_t-z_t}+\frac{1\mp1}{2(s'-a)}z_tP_l'(z_s')\right\}\ec\nt\\
\bar C_{Jl}(t,s')&=\frac{1}{2}\int\limits^1_{-1}\diff z_t\,\Big[P_{J-1}(z_t)-P_{J+1}(z_t)\Big]\left\{\frac{1}{p_tq_t}\frac{P_l'(\tilde z_s)-P_l'(z_s')}{\tilde x_t-z_t}+\frac{1\mp1}{2(s'-a)}P_l'(z_s')\right\}\nt\\
&=\bar A_{J-1,l}-\bar A_{J+1,l}\ec
\end{align}
the kernels $\tilde G_{Jl}$ and $\tilde H_{Jl}$ may be written in a recursive fashion
\begin{align}
\label{GHbardef}
\tilde G_{Jl}(t,W')&=\bar G_{Jl}(t,W')-\bar G_{J,l-1}(t,-W')\ec \qquad \bar G_{J,-1}=0\ec \qquad \forall\;J\geq0\ec\nt\\
\tilde H_{Jl}(t,W')&=\bar H_{Jl}(t,W')-\bar H_{J,l-1}(t,-W')\ec \qquad \bar H_{J,-1}=0\ec \qquad \forall\;J\geq1\ec\nt\\
\bar G_{Jl}(t,W')&=\frac{\tilde\eta_J}{E'+m}\bigg\{\frac{P_{l+1}'(\tilde z_s)}{p_tq_t}\left[-\frac{p_t}{q_t}(W'+m)+m\tilde x_t\right]Q_J(\tilde x_t)
+\frac{p_t}{q_t}(W'+m)\bar A_{J,l+1}-m\bar B_{J,l+1}\bigg\}\ec\nt\\
\bar H_{Jl}(t,W')&=\frac{\tilde\eta_J}{E'+m}\frac{\sqrt{J(J+1)}}{2J+1}\left\{\frac{P_{l+1}'(\tilde z_s)}{p_tq_t}\Big[Q_{J-1}(\tilde x_t)-Q_{J+1}(\tilde x_t)\Big]-\bar C_{J,l+1}\right\}\ec
\end{align}
keeping $\bar C_{J,l}$ just for convenience.
Note that since $\tilde x_t/(p_tq_t)$ and the squares $p_t^2$ and $\tilde x_t^2$ are always real, $\bar A_{Jl}$ is real/imaginary for $J$ even/odd and the other way around for $\bar B_{Jl}$ and $\bar C_{Jl}$. Therefore, we can conclude that the functions $\bar G_{Jl}$, $\bar H_{Jl}$ and hence the kernels $\tilde G_{Jl}$, $\tilde H_{Jl}$ are real for $t>\tpi-(\mpi^2/m)^2$, cf.\ the discussion following~\eqref{hatNJpm}.
The kernels for all combinations $(J\geq0,l\leq 2)$ explicitly read
\begin{align}
\label{expltildeGH}
\tilde G_{J0}(t,W')&=\frac{\tilde\eta_J}{E'+m}\bigg\{\frac{1}{p_tq_t}\bigg(\Big[-\frac{p_t}{q_t}(W'+m)+m\tilde x_t\Big]Q_J(\tilde x_t)-m\delta_{J0}\bigg)+\frac{p_t}{q_t}\frac{W'+m}{s'-a}\delta_{J0}-\frac{m}{3}\frac{\delta_{J1}}{s'-a}\bigg\}\ec\nt\\
\tilde H_{J0}(t,W')&=\frac{\tilde\eta_J}{E'+m}\frac{\sqrt{J(J+1)}}{2J+1}\bigg\{\frac{1}{p_tq_t}\Big[Q_{J-1}(\tilde x_t)-Q_{J+1}(\tilde x_t)\Big]-\frac{\delta_{J1}}{s'-a}\bigg\}\ec\nt\\
\tilde G_{J1}(t,W')&=-\tilde G_{J0}(t,-W')+\frac{\tilde\eta_J}{E'+m}\Bigg\{\frac{3\tilde z_s}{p_tq_t}\bigg(\Big[-\frac{p_t}{q_t}(W'+m)+m\tilde x_t\Big]Q_J(\tilde x_t)-m\delta_{J0}\bigg)\nt\\
&\qquad+\frac{W'+m}{\tgamma}\frac{p_t}{q_t}\bigg[\frac{1}{p_tq_t}\Big\{\delta_{J1}+3\tilde x_t\delta_{J0}\Big\}+\frac{1}{s'-a}\Big\{\frac{2}{5}\delta_{J2}+\Big(1-3\tdelta\Big)\delta_{J0}\Big\}\bigg]\nt\\
&\qquad-\frac{m}{\tgamma}\bigg[\frac{1}{p_tq_t}\Big\{\frac{2}{5}\delta_{J2}+\tilde x_t\delta_{J1}+\delta_{J0}\Big\}+\frac{1}{s'-a}\Big\{\frac{6}{35}\delta_{J3}+\Big(\frac{3}{5}-\tdelta\Big)\delta_{J1}\Big\}\bigg]\Bigg\}\nt\\
&=\bar{G}_{J1}(t,W')-\tilde G_{J0}(t,-W')\ec\nt\\
\tilde H_{J1}(t,W')&=-\tilde H_{J0}(t,-W')+\frac{\tilde\eta_J}{E'+m}\frac{\sqrt{J(J+1)}}{2J+1}\Bigg\{\frac{3\tilde z_s}{p_tq_t}\Big[Q_{J-1}(\tilde x_t)-Q_{J+1}(\tilde x_t)\Big]\nt\\
&\qquad-\frac{1}{\tgamma}\bigg[\frac{1}{p_tq_t}\Big\{\delta_{J2}+3\tilde x_t\delta_{J1}\Big\}+\frac{1}{s'-a}\Big\{\frac{2}{5}\delta_{J3}+3\Big(\frac{1}{5}-\tdelta\Big)\delta_{J1}\Big\}\bigg]\Bigg\}\nt\\
&=\bar{H}_{J1}(t,W')-\tilde H_{J0}(t,-W')\ec\nt\\
\tilde G_{J2}(t,W')&=-\bar{G}_{J1}(t,-W')+\frac{\tilde\eta_J}{E'+m}\Bigg\{\frac{P'_3(\tilde z_s)}{p_tq_t}\bigg(\Big[-\frac{p_t}{q_t}(W'+m)+m\tilde x_t\Big]Q_J(\tilde x_t)-m\delta_{J0}\bigg)\nt\\
&\qquad+\frac{W'+m}{\tgamma^2}\frac{p_t}{q_t}\bigg[\frac{1}{p_tq_t}\bigg\{\frac{3}{7}\delta_{J3}+\tilde x_t\delta_{J2}+\frac{5}{2}\Big(\frac{3}{5}+\tilde x_t^2-2\tdelta\Big)\delta_{J1}+\frac{15}{2}\tilde x_t\Big(\frac{1}{3}+\tilde x_t^2-2\tdelta\Big)\delta_{J0}\bigg\}\nt\\
&\qquad\quad+\frac{1}{s'-a}\bigg\{\frac{4}{21}\delta_{J4}+2\Big(\frac{3}{7}-\tdelta\Big)\delta_{J2}+\frac{15}{2}\Big(\frac{1-\tgamma^2}{5}-\frac{2}{3}\tdelta+\tdelta^2\Big)\delta_{J0}\bigg\}\bigg]\nt\\
&\qquad-\frac{m}{\tgamma^2}\bigg[\frac{1}{p_tq_t}\bigg\{\frac{4}{21}\delta_{J4}+\frac{3}{7}\tilde x_t\delta_{J3}+\Big(\frac{6}{7}+\tilde x_t^2-2\tdelta\Big)\delta_{J2}+\frac{5}{2}\Big(\frac{3}{5}+\tilde x_t^2-2\tdelta\Big)\Big(\tilde x_t\delta_{J1}+\delta_{J0}\Big)\bigg\}\nt\\
&\qquad\quad+\frac{1}{s'-a}\bigg\{\frac{20}{231}\delta_{J5}+\frac{3}{7}\Big(\frac{10}{9}-2\tdelta\Big)\delta_{J3}+\frac{5}{2}\Big(\frac{3}{7}-\frac{6}{5}\tdelta+\tdelta^2-\frac{\tgamma^2}{5}\Big)\delta_{J1}\bigg\}\bigg]\Bigg\}\nt\\
&=\bar{G}_{J2}(t,W')-\bar{G}_{J1}(t,-W')\ec\nt\\
\tilde H_{J2}(t,W')&=-\bar{H}_{J1}(t,-W')+\frac{\tilde\eta_J}{E'+m}\frac{\sqrt{J(J+1)}}{2J+1}\Bigg\{\frac{P'_3(\tilde z_s)}{p_tq_t}\Big[Q_{J-1}(\tilde x_t)-Q_{J+1}(\tilde x_t)\Big]\nt\\
&\qquad-\frac{1}{\tgamma^2}\bigg[\frac{1}{p_tq_t}\bigg\{\frac{3}{7}\delta_{J4}+\tilde x_t\delta_{J3}+\frac{5}{2}\Big(\frac{3}{7}+\tilde x_t^2-2\tdelta\Big)\delta_{J2}+\frac{15}{2}\tilde x_t\Big(\frac{1}{5}+\tilde x_t^2-2\tdelta\Big)\delta_{J1}\Big\}\nt\\
&\qquad\quad+\frac{1}{s'-a}\Big\{\frac{4}{21}\delta_{J5}+2\Big(\frac{1}{3}-\tdelta\Big)\delta_{J3}+\frac{3}{2}\Big(\frac{3}{7}-2\tdelta+5\tdelta^2-\tgamma^2\Big)\delta_{J1}\Big\}\bigg]\Bigg\}\nt\\
&=\bar{H}_{J2}(t,W')-\bar{H}_{J1}(t,-W')\ep
\end{align}
The explicit formulae for the polynomial parts $\bar A_{Jl}$, $\bar B_{Jl}$, and $\bar C_{J,l}$ for $(J\leq2,l\leq6)$ needed for calculating these kernels and furthermore the additional kernels $\tilde G_{Jl}$ and $\tilde H_{Jl}$ for $(J\leq2,3\leq l\leq5)$ via~\eqref{GHbardef} are given in Appendix~\ref{subsec:tpwp:higherkernels}.\footnote{Note that for $|a|\to\infty$, of all polynomial parts only $\bar B_{0l}$ does not vanish completely and hence $f^0_+$ receives polynomial contributions from the kernels $\tilde G_{0l}$. These remaining contributions, however, are just those that cancel with the leading terms of the $S$-wave pole terms~\eqref{tildeNJpm}, cf.\ the discussion following~\eqref{tchannel_pole_asym} as well as the explicit kernels~\eqref{expltildeGH}.}
As a check of our calculation we can reproduce the nucleon pole terms~\eqref{tildeNJpm} by (cf.~\eqref{NIpmcheck})
\begin{equation}
\label{tildeNJpmcheck}
\tilde N^J_+(t)=-f^2\tilde G_{J0}(t,-W'=m) \quad\forall\;J\geq0\ec \qquad \tilde N^J_-(t)=-f^2\tilde H_{J0}(t,-W'=m) \quad\forall\;J\geq1\ep
\end{equation}
The asymptotic behavior of the general kernel functions~\eqref{tskernelfunctions} can be deduced to be
\begin{align}
\label{tildeGHasymptotics}
&\text{for }\;p_tq_t\rightarrow 0 & \tilde G_{Jl}(t,W')&\sim \tilde H_{Jl}(t,W')\sim 1\ec\nt\\
&\text{for }\;q'\rightarrow 0 & \tilde G_{Jl}(t,W')&\sim \tilde H_{Jl}(t,W')\sim q'^{-2l}\ec & \tilde G_{Jl}(t,-W')&\sim \tilde H_{Jl}(t,-W')\sim q'^{-2l-2}\ec\nt\\
&\text{for }\;q'\rightarrow\infty & \tilde G_{Jl}(t,W')&\sim \tilde H_{Jl}(t,W')\sim q'^{-2J}\ep
\end{align}
In particular, these kernels are finite for $p_tq_t\to0$ and their precise form in this limit may be worked out in close analogy to the discussion of the pole terms in Appendix~\ref{subsec:tpwp:n} based on~\eqref{GHbardef}.
Note that both~\eqref{tildeNJpmcheck} and~\eqref{tildeGHasymptotics} obey the MacDowell symmetry relation~\eqref{macdowell}, as they should.

\subsection[$t$-channel exchange]{\boldmath{$t$}-channel exchange}
\label{subsec:tpwp:t}

We need the $t$-channel HDR terms~\eqref{sthdr} in the form
\begin{equation}
\label{tthdr}
\mathbf{A}^I(s,t)\Big|^{t}_{s=s(t,z_t)}=\frac{1}{\pi}\int\limits_{\tpi}^\infty\diff t'\;\mathbf{h}^I_t[t,t';z_t]\,\Im\mathbf{A}^I(s',t')\Big|_{s'=s'(t',z_t')}\ec
\end{equation}
where the kernel matrix $\mathbf{h}^I_t$ is given in~\eqref{xtdef}, and the integrand can be written as a function of the variables $[t,t';z_t]$ by noting that
\begin{equation}
\frac{1}{2q^2}\frac{1}{x_t-z_s}\bigg|_{[t,t';z_t]}=\frac{1}{t'-t}\ec \qquad \frac{\nu}{\nu'}\bigg|_{[t,t';z_t]}=\frac{p_tq_t}{p_t'q_t'}\frac{z_t}{z_t'}\ec
\end{equation}
and that $z_t'$ and $z_t$ are related by (cf.~\eqref{alphabetadef})
\begin{align}
\label{tildealphabetadef}
z_t'(t,t';z_t)=\sqrt{\tilde\alpha z_t^2+\tilde\beta}\ec \qquad \tilde\alpha(t,t')&=\frac{p_t^2q_t^2}{p_t'^2q_t'^2}\ec\nt\\
\tilde\beta(t,t')&=\frac{t'-t}{16p_t'^2q_t'^2}(t+t'-2\Sigma+4a)\ep
\end{align}
If we expand the absorptive part of~\eqref{tthdr} into $t$-channel partial waves by using~\eqref{texp}
and project onto $t$-channel partial waves again via~\eqref{tproj}, we obtain the following PWDRs for the $t$-channel partial waves
\begin{equation}
\label{ttpwhdr}
\mathbf{f}^J(t)\Big|^{t}=\frac{1}{\pi}\int\limits_{\tpi}^\infty\diff t'\;\sum\limits_{J'}\mathbf{\tilde K}^{J J'}(t,t')\,\Im\mathbf{f}^{J'}(t')\ec
\end{equation}
where the summation runs over even/odd values of $J'$ for even/odd values of $J$, accordingly, and the kernel matrix is defined by
\begin{equation}
\mathbf{\tilde K}^{J J'}(t,t')=\int\limits_0^1\diff z_t\;\mathbf{\tilde T}^J(t,z_t)\mathbf{h}_t^I[t,t';z_t]\mathbf{T}^{J'}(t',z_t')\ep
\end{equation}
Calculating this kernel matrix shows that it can be written with three kernel functions as
\begin{align}
\mathbf{\tilde K}^{JJ'}(t,t')&=\begin{pmatrix}\tilde K^1_{JJ'}(t,t')&\tilde K^2_{JJ'}(t,t')\\0&\tilde K^3_{JJ'}(t,t')\end{pmatrix}
 =\frac{\zeta_{JJ'}}{t'-t}\begin{pmatrix}u_{JJ'}(t,t')&v_{JJ'}(t,t')\\0&w_{JJ'}(t,t')\end{pmatrix}\ec\nt\\
\zeta_{JJ'}(t,t')&=(2J'+1)\frac{(p_t'q_t')^{J'-1}}{(p_tq_t)^{J-1}}\ec
\end{align}
where we have defined different angular kernels for even $J$ and $J'$
\begin{align}
\label{uvw_even}
u_{JJ'}&=\frac{p_tq_t'}{q_tp_t'}\int\limits_0^1\diff z_t\;P_J(z_t)P_{J'}(z_t')\ec \quad
v_{JJ'}=\frac{m}{\sqrt{J'(J'+1)}}\frac{p_t}{q_tp_t'q_t'}\int\limits^1_0\diff z_t\;P_J(z_t)\Big\{q_t^2z_t^2-q_t'^2z_t'^2\Big\}\frac{P_{J'}'(z_t')}{z_t'}\ec\nt\\
w_{JJ'}&=\frac{1}{2J+1}\sqrt{\frac{J(J+1)}{J'(J'+1)}}\frac{p_tq_t}{p_t'q_t'}\int\limits^1_0\diff z_t\;\Big\{P_{J-1}(z_t)-P_{J+1}(z_t)\Big\}z_t\frac{P_{J'}'(z_t')}{z_t'}\ec
\end{align}
and for odd $J$ and $J'$
\begin{align}
\label{uvw_odd}
u_{JJ'}&=\frac{p_t^2}{p_t'^2}\int\limits_0^1\diff z_t\;P_J(z_t)z_t\frac{P_{J'}(z_t')}{z_t'}\ec \quad
v_{JJ'}=\frac{m}{\sqrt{J'(J'+1)}}\bigg\{1-\frac{p_t^2}{p_t'^2}\bigg\}\int\limits^1_0\diff z_t\;P_J(z_t)z_tP_{J'}'(z_t')\ec\nt\\
w_{JJ'}&=\frac{1}{2J+1}\sqrt{\frac{J(J+1)}{J'(J'+1)}}\int\limits^1_0\diff z_t\;\Big\{P_{J-1}(z_t)-P_{J+1}(z_t)\Big\}P_{J'}'(z_t')\ep
\end{align}
In this way, we recover the form of the $t$-channel part given in~\eqref{tpwhdr}
\begin{align}
f^J_+(t)\Big|^{t}&=\frac{1}{\pi}\int\limits^{\infty}_{\tpi}\diff t'\sum\limits_{J'}\Big\{
\tilde K^1_{JJ'}(t,t')\,\Im f^{J'}_+(t')+\tilde K^2_{JJ'}(t,t')\,\Im f^{J'}_-(t')\Big\}
& &\forall\;J\geq0\ec\nt\\
f^J_-(t)\Big|^{t}&=\frac{1}{\pi}\int\limits^{\infty}_{\tpi}\diff t'\sum\limits_{J'}\tilde K^3_{JJ'}(t,t')\,\Im f^{J'}_-(t')
& &\forall\;J\geq1\ec
\end{align}
and according to $f^0_-\equiv0$ we set $\tilde K^3_{0J'}\equiv0\equiv\tilde K^2_{J0}$\ep

From the projection integrals~\eqref{uvw_even} and~\eqref{uvw_odd} together with the definitions~\eqref{tildealphabetadef} and
\begin{equation}
1-\frac{p_t^2}{p_t'^2}=\frac{t'-t}{4p_t'^2}\ec \qquad q_t^2z_t^2-q_t'^2z_t'^2=\frac{t'-t}{4p_t'^2}\bigg\{4q_t^2z_t^2-\frac{1}{4}(t+t'-2\Sigma+4a)\bigg\}\ec
\end{equation}
one can see that the off-diagonal term $v_{JJ'}$ is proportional to $t'-t$, as it should be.
Note also that only even powers of $z_t'$ and $z_t$ occur in the projection integrals. Therefore, the kernel functions $\tilde K^1_{JJ'}$, $\tilde K^2_{JJ'}$, and $\tilde K^3_{JJ'}$ are always real, since the prefactors contain only even powers of momenta.
The integrals can be performed with the help of~\cite{Bateman}
\begin{equation}
\label{legendreexpevenodd}
P_l(z)=\sum\limits_{\lambda=0}^{\frac{l}{2}}a^\text{ev}_{\lambda l}z^{2\lambda}\ec \qquad
P_l(z)=\sum\limits_{\lambda=0}^{\frac{l-1}{2}}a^\text{od}_{\lambda l}z^{2\lambda+1}\ec
\end{equation}
for even and odd values of $l$, respectively, where
\begin{equation}
a^\text{ev}_{\lambda l}=
\frac{(-1)^{\lambda+\frac{l}{2}}(2\lambda+l-1)!}{2^{l-1}\Big(\frac{l}{2}-\lambda\Big)!\Big(\lambda+\frac{l}{2}-1\Big)!(2\lambda)!}\ec \qquad
a^\text{od}_{\lambda l}=
\frac{(-1)^{\lambda+\frac{l-1}{2}}(2\lambda+l)!}{2^{l-1}\Big(\frac{l-1}{2}-\lambda\Big)!\Big(\lambda+\frac{l-1}{2}\Big)!(2\lambda+1)!}\ec
\end{equation}
which also follow from reordering the expansion
\begin{equation}
P_l(z)=\frac{1}{2^l}\sum\limits_{\lambda=0}^{\big\lfloor\frac{l}{2}\big\rfloor}\frac{(-1)^\lambda(2l-2\lambda)!}{\lambda!(l-\lambda)!(l-2\lambda)!}z^{l-2\lambda}\ec \qquad
\bigg\lfloor\frac{l}{2}\bigg\rfloor=\begin{cases}\frac{l}{2}&\text{for even $l$}\ec\\\frac{l-1}{2}&\text{for odd $l$}\ep\end{cases}
\end{equation}
In this way, the required non-vanishing integrals may be written for even $J$ and $J'$ as
{\allowdisplaybreaks
\begin{align}
\label{proj_even}
\int\limits_0^1\diff z_t\;P_J(z_t)P_{J'}(z_t')&=\sum\limits_{\lambda'=\frac{J}{2}}^{\frac{J'}{2}}a^\text{ev}_{\lambda'J'}
\sum\limits_{\mu=\frac{J}{2}}^{\lambda'}\begin{pmatrix}\lambda'\\\mu\end{pmatrix}
\tilde\alpha^\mu\tilde\beta^{\lambda'-\mu}\tilde a^\text{ev}_{J\mu}\ec\nt\\
\int\limits^1_0\diff z_t\;P_J(z_t)\Big\{q_t^2z_t^2-q_t'^2z_t'^2\Big\}\frac{P_{J'}'(z_t')}{z_t'}&=
\sum\limits_{\lambda'=\max\{\frac{J}{2},1\}}^{\frac{J'}{2}}2\lambda'a^\text{ev}_{\lambda'J'}\nt\\
&\hspace{-100pt}\times\Bigg\{
q_t^2\sum\limits_{\mu=\max\{\frac{J}{2}-1,0\}}^{\lambda'-1}\begin{pmatrix}\lambda'-1\\\mu\end{pmatrix}
\tilde\alpha^\mu\tilde\beta^{\lambda'-1-\mu}\tilde a^\text{ev}_{J,\mu+1}-
q_t'^2\sum\limits_{\mu=\frac{J}{2}}^{\lambda'}\begin{pmatrix}\lambda'\\\mu\end{pmatrix}
\tilde\alpha^\mu\tilde\beta^{\lambda'-\mu}\tilde a^\text{ev}_{J\mu}\Bigg\}\ec\nt\\
\int\limits^1_0\diff z_t\;\Big\{P_{J-1}(z_t)-P_{J+1}(z_t)\Big\}z_t\frac{P_{J'}'(z_t')}{z_t'}&=\nt\\
&\hspace{-100pt}\sum\limits_{\lambda'=\max\{\frac{J}{2},1\}}^{\frac{J'}{2}}2\lambda'a^\text{ev}_{\lambda'J'}
\sum\limits_{\mu=\max\{\frac{J}{2}-1,0\}}^{\lambda'-1}\begin{pmatrix}\lambda'-1\\\mu\end{pmatrix}
\tilde\alpha^\mu\tilde\beta^{\lambda'-1-\mu}\tilde a^\text{od}_{J-1,\mu+1}\nt\\
&\hspace{-100pt}-\sum\limits_{\lambda'=\frac{J}{2}+1}^{\frac{J'}{2}}2\lambda'a^\text{ev}_{\lambda'J'}
\sum\limits_{\mu=\frac{J}{2}}^{\lambda'-1}\begin{pmatrix}\lambda'-1\\\mu\end{pmatrix}
\tilde\alpha^\mu\tilde\beta^{\lambda'-1-\mu}\tilde a^\text{od}_{J+1,\mu+1}\ec
\end{align}}\noindent
and for odd $J$ and $J'$ as
\begin{align}
\label{proj_odd}
\int\limits_0^1\diff z_t\;P_J(z_t)z_t\frac{P_{J'}(z_t')}{z_t'}&=\sum\limits_{\lambda'=\frac{J-1}{2}}^{\frac{J'-1}{2}}a^\text{od}_{\lambda'J'}
\sum\limits_{\mu=\frac{J-1}{2}}^{\lambda'}\begin{pmatrix}\lambda'\\\mu\end{pmatrix}
\tilde\alpha^\mu\tilde\beta^{\lambda'-\mu}\tilde a^\text{od}_{J,\mu+1}\ec\nt\\
\int\limits^1_0\diff z_t\;P_J(z_t)z_tP_{J'}'(z_t')&=\sum\limits_{\lambda'=\frac{J-1}{2}}^{\frac{J'-1}{2}}(2\lambda'+1)a^\text{od}_{\lambda'J'}
\sum\limits_{\mu=\frac{J-1}{2}}^{\lambda'}\begin{pmatrix}\lambda'\\\mu\end{pmatrix}
\tilde\alpha^\mu\tilde\beta^{\lambda'-\mu}\tilde a^\text{od}_{J,\mu+1}\ec\nt\\
\int\limits^1_0\diff z_t\;\Big\{P_{J-1}(z_t)-P_{J+1}(z_t)\Big\}P_{J'}'(z_t')&=\nt\\
&\hspace{-90pt}\sum\limits_{\lambda'=\frac{J-1}{2}}^{\frac{J'-1}{2}}(2\lambda'+1)a^\text{od}_{\lambda'J'}
\sum\limits_{\mu=\frac{J-1}{2}}^{\lambda'}\begin{pmatrix}\lambda'\\\mu\end{pmatrix}
\tilde\alpha^\mu\tilde\beta^{\lambda'-\mu}\tilde a^\text{ev}_{J-1,\mu}\nt\\
&\hspace{-100pt}-\sum\limits_{\lambda'=\frac{J+1}{2}}^{\frac{J'-1}{2}}(2\lambda'+1)a^\text{od}_{\lambda'J'}
\sum\limits_{\mu=\frac{J+1}{2}}^{\lambda'}\begin{pmatrix}\lambda'\\\mu\end{pmatrix}
\tilde\alpha^\mu\tilde\beta^{\lambda'-\mu}\tilde a^\text{ev}_{J+1,\mu}\ec
\end{align}
with the definitions (for even and odd values of $J$, respectively)\footnote{These identities are similar to the Saalsch\"{u}tz formula~\eqref{saalschuetz} employed in~\cite{BaackeSteiner}. Note that $(-1)!!=0!!=1$.}
\begin{align}
\label{saalschuetz2}
\tilde a^\text{ev}_{J\mu}&=\sum\limits_{\lambda=0}^{\frac{J}{2}}\frac{a^\text{ev}_{\lambda J}}{2(\mu+\lambda)+1}
 =2^J\frac{\left(\mu+\frac{J}{2}\right)!(2\mu)!}{\left(\mu-\frac{J}{2}\right)!(2\mu+J+1)!}
 =\frac{(2\mu)!}{(2\mu-J)!!(2\mu+1+J)!!} \qquad \left(\mu\geq \frac{J}{2}\right)\ec\nt\\
\tilde a^\text{od}_{J\mu}&=\sum\limits_{\lambda=0}^{\frac{J-1}{2}}\frac{a^\text{od}_{\lambda J}}{2(\mu+\lambda)+1}
 =2^J\frac{\left(\mu+\frac{J-1}{2}\right)!(2\mu-1)!}{\left(\mu-\frac{J+1}{2}\right)!(2\mu+J)!}
 =\frac{(2\mu-1)!}{(2\mu-1-J)!!(2\mu+J)!!} \qquad \left(\mu\geq \frac{J-1}{2}\right)\ep
\end{align}
We can conclude that the following kernels vanish:
\begin{equation}
\label{vanishingtildeK}
\mathbf{\tilde K}^{JJ'}(t,t')=0\quad\forall\;J'<J\ec
\end{equation}
and by using the identities
\begin{align}
(2J+1)a^\text{ev}_{\frac{J}{2},J}\tilde a^\text{ev}_{J,\frac{J}{2}}&=1\ec &
J a^\text{ev}_{\frac{J}{2},J}\tilde a^\text{od}_{J-1,\frac{J}{2}}&=1\ec & &\text{for even }J\ec\nt\\
(2J+1)a^\text{od}_{\frac{J-1}{2},J}\tilde a^\text{od}_{J,\frac{J+1}{2}}&=1\ec &
J a^\text{od}_{\frac{J-1}{2},J}\tilde a^\text{ev}_{J-1,\frac{J-1}{2}}&=1\ec & &\text{for odd }J\ec
\end{align}
it follows that the non-vanishing kernels for $J'=J$ take the form
\begin{align}
\label{tildeKJJ}
\tilde K_{JJ}^1(t,t')&=\frac{p_t^2}{p_t'^2}\frac{1}{t'-t}=\frac{1}{t'-t}-\frac{1}{t'-\tN}=\frac{t}{t'}\frac{1}{t'-t}-\frac{\tN}{t'}\frac{1}{t'-\tN} & &\forall\;J\geq0\ec\nt\\
\tilde K_{JJ}^2(t,t')&=\sqrt{\frac{J}{J+1}}\frac{m}{4p_t'^2}=\sqrt{\frac{J}{J+1}}\frac{m}{t'-\tN} & &\forall\;J\geq1\ec\nt\\
\tilde K_{JJ}^3(t,t')&=\frac{1}{t'-t} & &\forall\;J\geq1\ec
\end{align}
from which one can immediately read off the relation (valid for all $J$)
\begin{equation}
\label{tildeK12rel}
\tilde K_{JJ}^2(t,t')=m\sqrt{\frac{J}{J+1}}\left\{\tilde K_{JJ}^3(t,t')-\tilde K_{JJ}^1(t,t')\right\}\ep
\end{equation}
This together with
\begin{align}
\label{tildeK02}
\tilde K_{02}^1(t,t')&=\frac{5}{16}\frac{p_t^2}{p_t'^2}\Big\{t+t'-2\Sigma+6a\Big\}\ec &
\tilde K_{02}^2(t,t')&=\frac{5m}{16\sqrt{6}}\frac{p_t^2}{p_t'^2}\Big\{4q_t^2-3(t+t'-2\Sigma+4a)\Big\}\nt\ec\\
\tilde K_{13}^1(t,t')&=\frac{7}{48}\frac{p_t^2}{p_t'^2}\Big\{t+t'-2\Sigma+10a\Big\}\ec &
\tilde K_{13}^2(t,t')&=\frac{7m}{64\sqrt{3}}\frac{1}{p_t'^2}\Big\{8p_t^2q_t^2+(t'-t)(t+t'-2\Sigma+5a)\Big\}\nt\ec\\
\tilde K_{13}^3(t,t')&=\frac{7}{8\sqrt{6}}\Big\{t+t'-2\Sigma+5a\Big\}\ec
\end{align}
completes the calculation of the $t$-channel kernels with $(J\leq3,J'\leq3)$.
Finally, from \eqref{proj_even} and \eqref{proj_odd} we may infer the asymptotic behavior of the non-vanishing kernels
\begin{align}
\label{tildeKasymptotics}
&\text{for }\;p_t \rightarrow 0 & \tilde K^1_{JJ'}(t,t')&\sim  p_t^2\ec & \tilde K^2_{JJ'}(t,t')\sim\tilde K^3_{JJ'}(t,t')&\sim 1\ec\nt\\
&\text{for }\;q_t \rightarrow 0 & \tilde K^1_{JJ'}(t,t')&\sim \tilde K^2_{JJ'}(t,t')\sim \tilde K^3_{JJ'}(t,t')\sim 1\ec\nt\\
&\text{for }\;t \rightarrow\infty & \tilde K^1_{JJ'}(t,t')&\sim \tilde K^2_{JJ'}(t,t')\sim t^{J'-J}\ec & \tilde K^3_{JJ'}(t,t')&\sim t^{J'-J-1}\ec\nt\\
&\text{for }\;p_t'\rightarrow 0 & \tilde K^1_{JJ'}(t,t')&\sim \tilde K^2_{JJ'}(t,t')\sim p_t'^{-2}\ec & \tilde K^3_{JJ'}(t,t')&\sim 1\ec\nt\\
&\text{for }\;q_t'\rightarrow 0 & \tilde K^1_{JJ'}(t,t')&\sim \tilde K^2_{JJ'}(t,t')\sim \tilde K^3_{JJ'}(t,t')\sim 1\ec\nt\\
&\text{for }\;t'\rightarrow\infty & \tilde K^1_{JJ'}(t,t')&\sim  t'^{J'-J-2}\ec & \tilde K^2_{JJ'}(t,t')\sim\tilde K^3_{JJ'}(t,t')&\sim t'^{J'-J-1}\ep
\end{align}
Note that the kernel $\tilde K^2_{02}(t,t')$ exceptionally has better convergence properties
\begin{equation}
\label{tilde2K02asymptotics}
\tilde K^2_{02}(t,t')\sim p_t^2 \quad \text{for }\;p_t \rightarrow 0\ec \qquad \tilde K^2_{02}(t,t')\sim 1 \quad \text{for }\;t'\rightarrow\infty\ep
\end{equation}

\subsection{Higher kernel functions}
\label{subsec:tpwp:higherkernels}

The explicit form of the polynomial parts $\bar A_{Jl}$, $\bar B_{Jl}$, and $\bar C_{J,l}$ from~\eqref{ABCbardef} for $(J\leq2,l\leq6)$ that are needed in order to calculate the kernels $\tilde G_{Jl}$ and $\tilde H_{Jl}$ for $(J\leq2,l\leq5)$ via~\eqref{GHbardef} explicitly read 
\begin{align}
\bar{A}_{01}&=\frac{1}{s'-a}\ec\qquad
\bar{A}_{02}=\frac{3}{\tgamma}\bigg[\frac{\tilde{x}_t}{p_tq_t}+\frac{1}{s'-a}\bigg\{\frac{1}{3}-\tdelta\bigg\}\bigg]\ec\\
\bar{A}_{03}&=\frac{15}{2\tgamma^2}\bigg[\frac{\tilde{x}_t}{p_tq_t}\bigg\{\tilde{x}_t^2+\frac{1}{3}-2\tdelta\bigg\}+\frac{1}{s'-a}\bigg\{\frac{1-\tgamma^2}{5}-\frac{2}{3}\tdelta+\tdelta^2\bigg\}\bigg]\ec\nt\\
\bar{A}_{04}&=\frac{35}{2\tgamma^3}\bigg[\frac{\tilde{x}_t}{p_tq_t}\bigg\{\tilde{x}_t^4+\tilde{x}_t^2\bigg(\frac{1}{3}-3\tdelta\bigg)+\frac{1}{5}-\frac{3}{7}\tgamma^2-\tdelta+3\tdelta^2\bigg\}\nt\\
&\qquad+\frac{1}{s'-a}\bigg\{\frac{1}{7}-\frac{\tgamma^2}{7}\bigg(1-3\tdelta\bigg)-\frac{3}{5}\tdelta+\tdelta^2-\tdelta^3\bigg\}\bigg]\ec\nt\\
\bar{A}_{05}&=\frac{315}{2\tgamma^4}\bigg[\frac{\tilde{x}_t}{p_tq_t}\bigg\{\frac{\tilde{x}_t^6}{4}+\tilde{x}_t^4\bigg(\frac{1}{12}-\tdelta\bigg)+\tilde{x}_t^2\bigg(\frac{1}{20}-\frac{\tgamma^2}{6}-\frac{\tdelta}{3}+\frac{3}{2}\tdelta^2\bigg)+\frac{1}{28}-\frac{\tgamma^2}{3}\bigg(\frac{1}{6}-\tdelta\bigg)-\frac{\tdelta}{5}\nt\\
&\qquad\quad+\frac{\tdelta^2}{2}-\tdelta^3\bigg\}
+\frac{1}{s'-a}\bigg\{\frac{1}{36}-\frac{\tgamma^2}{6}\bigg(\frac{1}{5}-\frac{\tgamma^2}{14}-\frac{2}{3}\tdelta+\tdelta^2\bigg)-\frac{\tdelta}{7}+\frac{3}{10}\tdelta^2-\frac{\tdelta^3}{3}+\frac{\tdelta^4}{4}\bigg\}\bigg]\ec\nt\\
\bar{A}_{06}&=\frac{3465}{4\tgamma^5}\bigg[\frac{\tilde{x}_t}{p_tq_t}\bigg\{\frac{\tilde{x}_t^8}{10}+\tilde{x}_t^4\bigg(\frac{1}{50}-\frac{\tgamma^2}{11}-\frac{\tdelta}{6}+\tdelta^2\bigg)+\tilde{x}_t^2\bigg(\frac{1}{70}-\frac{\tgamma^2}{11}\bigg(\frac{1}{3}-3\tdelta\bigg)-\frac{\tdelta}{10}+\frac{\tdelta^2}{3}-\tdelta^3\bigg)\nt\\
&\qquad\quad+\frac{\tilde{x}_t^6}{2}\bigg(\frac{1}{15}-\tdelta\bigg)+\frac{1}{90}-\frac{\tgamma^2}{11}\bigg(\frac{1}{5}-\frac{\tgamma^2}{6}-\tdelta+3\tdelta^2\bigg)-\frac{\tdelta}{14}+\frac{\tdelta^2}{5}-\frac{\tdelta^3}{3}+\frac{\tdelta^4}{2}\bigg\}\nt\\
&\qquad+\frac{1}{s'-a}\bigg\{\frac{1}{110}-\frac{\tgamma^2}{11}\bigg(\frac{1}{7}-\frac{\tgamma^2}{6}\bigg(\frac{1}{3}-\tdelta\bigg)-\frac{3}{5}\tdelta+\tdelta^2-\tdelta^3\bigg)-\frac{\tdelta}{18}+\frac{\tdelta^2}{7}-\frac{\tdelta^3}{5}+\frac{\tdelta^4}{6}-\frac{\tdelta^5}{10}\bigg\}\bigg]\ec\nt
\end{align}
\begin{align}
\bar{A}_{11}&=0\ec\qquad
\bar{A}_{12}=\frac{1}{p_tq_t}\frac{1}{\tgamma}\ec\qquad
\bar{A}_{13}=\frac{1}{p_tq_t}\frac{5}{2\tgamma^2}\bigg\{\tilde{x}_t^2+\frac{3}{5}-2\tdelta\bigg\}\ec\\
\bar{A}_{14}&=\frac{1}{p_tq_t}\frac{35}{2\tgamma^3}\bigg\{\frac{\tilde{x}_t^4}{3}+\tilde{x}_t^2\bigg(\frac{1}{5}-\tdelta\bigg)+\frac{1-\tgamma^2}{7}-\frac{3}{5}\tdelta+\tdelta^2\bigg\}\ec\nt\\
\bar{A}_{15}&=\frac{1}{p_tq_t}\frac{315}{2\tgamma^4}\bigg\{\frac{\tilde{x}_t^6}{12}+\tilde{x}_t^4\bigg(\frac{1}{20}-\frac{\tdelta}{3}\bigg)+\tilde{x}_t^2\bigg(\frac{1}{28}-\frac{\tgamma^2}{18}-\frac{\tdelta}{5}+\frac{\tdelta^2}{2}\bigg)\nt\\
&\qquad+\frac{1}{36}-\frac{\tgamma^2}{3}\bigg(\frac{1}{10}-\frac{\tdelta}{3}\bigg)-\frac{\tdelta}{7}+\frac{3}{10}\tdelta^2-\frac{\tdelta^3}{3}\bigg\}\ec\nt\\
\bar{A}_{16}&=\frac{1}{p_tq_t}\frac{3465}{4\tgamma^5}\bigg\{\frac{\tilde{x}_t^8}{30}+\tilde{x}_t^4\bigg(\frac{1}{70}-\frac{\tgamma^2}{33}-\frac{\tdelta}{10}+\frac{\tdelta^2}{3}\bigg)+\tilde{x}_t^2\bigg(\frac{1}{90}-\frac{\tgamma^2}{11}\bigg(\frac{1}{5}-\tdelta\bigg)-\frac{\tdelta}{14}+\frac{\tdelta^2}{5}-\frac{\tdelta^3}{3}\bigg)\nt\\
&\qquad+\frac{\tilde{x}_t^6}{2}\bigg(\frac{1}{25}-\frac{\tdelta}{3}\bigg)+\frac{1}{110}-\frac{\tgamma^2}{11}\bigg(\frac{1}{7}-\frac{\tgamma^2}{18}-\frac{3}{5}\tdelta+\tdelta^2\bigg)-\frac{\tdelta}{18}+\frac{\tdelta^2}{7}-\frac{\tdelta^3}{5}+\frac{\tdelta^4}{6}\bigg\}\ec\nt
\end{align}
\begin{align}
\bar{A}_{21}&=0\ec\qquad
\bar{A}_{22}=\frac{2}{5\tgamma}\frac{1}{s'-a}\ec\qquad
\bar{A}_{23}=\frac{1}{\tgamma^2}\bigg[\frac{\tilde{x}_t}{p_tq_t}+\frac{1}{s'-a}\bigg\{\frac{6}{7}-2\tdelta\bigg\}\bigg]\ec\\
\bar{A}_{24}&=\frac{7}{\tgamma^3}\bigg[\frac{\tilde{x}_t}{p_tq_t}\bigg\{\frac{\tilde{x}_t^2}{3}+\frac{2}{7}-\tdelta\bigg\}+\frac{1}{s'-a}\bigg\{\frac{5}{21}-\frac{\tgamma^2}{7}-\frac{6}{7}\tdelta+\tdelta^2\bigg\}\bigg]\ec\nt\\
\bar{A}_{25}&=\frac{21}{\tgamma^4}\bigg[\frac{\tilde{x}_t}{p_tq_t}\bigg\{\frac{\tilde{x}_t^4}{4}+\tilde{x}_t^2\bigg(\frac{3}{14}-\tdelta\bigg)+\frac{5}{28}-\frac{\tgamma^2}{6}-\frac{6}{7}\tdelta+\frac{3}{2}\tdelta^2\bigg\}\nt\\
&\qquad+\frac{1}{s'-a}\bigg\{\frac{5}{33}-\tgamma^2\bigg(\frac{1}{7}-\frac{\tdelta}{3}\bigg)-\frac{5}{7}\tdelta+\frac{9}{7}\tdelta^2-\tdelta^3\bigg\}\bigg]\ec\nt\\
\bar{A}_{26}&=\frac{231}{2\tgamma^5}\bigg[\frac{\tilde{x}_t}{p_tq_t}\bigg\{\frac{\tilde{x}_t^6}{10}+\tilde{x}_t^4\bigg(\frac{3}{35}-\frac{\tdelta}{2}\bigg)+\tilde{x}_t^2\bigg(\frac{1}{14}-\frac{\tgamma^2}{11}-\frac{3}{7}\tdelta+\tdelta^2\bigg)+\frac{2}{33}-\frac{\tgamma^2}{11}\bigg(\frac{6}{7}-3\tdelta\bigg)-\frac{5}{14}\tdelta\nt\\
&\qquad\quad+\frac{6}{7}\tdelta^2-\tdelta^3\bigg\}
+\frac{1}{s'-a}\bigg\{\frac{15}{286}-\frac{\tgamma^2}{11}\bigg(\frac{5}{7}-\frac{\tgamma^2}{6}-\frac{18}{7}\tdelta+3\tdelta^2\bigg)-\frac{10}{33}\tdelta+\frac{5}{7}\tdelta^2-\frac{6}{7}\tdelta^3+\frac{\tdelta^4}{2}\bigg\}\bigg]\ec\nt
\end{align}
\begin{align}
\bar{B}_{01}&=\frac{1}{p_tq_t}\ec\qquad
\bar{B}_{02}=\frac{1}{p_tq_t}\frac{3}{\tgamma}\bigg\{\tilde{x}_t^2+\frac{1}{3}-\tdelta\bigg\}\ec\\
\bar{B}_{03}&=\frac{1}{p_tq_t}\frac{15}{2\tgamma^2}\bigg\{\tilde{x}_t^4+\tilde{x}_t^2\bigg(\frac{1}{3}-2\tdelta\bigg)+\frac{1-\tgamma^2}{5}-\frac{2}{3}\tdelta+\tdelta^2\bigg\}\ec\nt\\
\bar{B}_{04}&=\frac{1}{p_tq_t}\frac{35}{2\tgamma^3}\bigg\{\tilde{x}_t^6+\tilde{x}_t^4\bigg(\frac{1}{3}-3\tdelta\bigg)+\tilde{x}_t^2\bigg(\frac{1}{5}-\frac{3}{7}\tgamma^2-\tdelta+3\tdelta^2\bigg)+\frac{1}{7}-\frac{\tgamma^2}{7}\bigg(1-3\tdelta\bigg)-\frac{3}{5}\tdelta+\tdelta^2-\tdelta^3\bigg\}\ec\nt\\
\bar{B}_{05}&=\frac{1}{p_tq_t}\frac{315}{2\tgamma^4}\bigg\{\frac{\tilde{x}_t^8}{4}+\tilde{x}_t^4\bigg(\frac{1}{20}-\frac{\tgamma^2}{6}-\frac{\tdelta}{3}+\frac{3}{2}\tdelta^2\bigg)+\tilde{x}_t^2\bigg(\frac{1}{28}-\frac{\tgamma^2}{3}\bigg(\frac{1}{6}-\tdelta\bigg)-\frac{\tdelta}{5}+\frac{\tdelta^2}{2}-\tdelta^3\bigg)\nt\\
&\qquad+\tilde{x}_t^6\bigg(\frac{1}{12}-\tdelta\bigg)+\frac{1}{36}-\frac{\tgamma^2}{6}\bigg(\frac{1}{5}-\frac{\tgamma^2}{14}-\frac{2}{3}\tdelta+\tdelta^2\bigg)-\frac{\tdelta}{7}+\frac{3}{10}\tdelta^2-\frac{\tdelta^3}{3}+\frac{\tdelta^4}{4}\bigg\}\ec\nt\\
\bar{B}_{06}&=\frac{1}{p_tq_t}\frac{3465}{4\tgamma^5}\bigg\{\frac{\tilde{x}_t^{10}}{10}+\tilde{x}_t^6\bigg(\frac{1}{50}-\frac{\tgamma^2}{11}-\frac{\tdelta}{6}+\tdelta^2\bigg)+\tilde{x}_t^4\bigg(\frac{1}{70}-\frac{\tgamma^2}{11}\bigg(\frac{1}{3}-3\tdelta\bigg)-\frac{\tdelta}{10}+\frac{\tdelta^2}{3}-\tdelta^3\bigg)\nt\\
&\qquad+\frac{\tilde{x}_t^8}{2}\bigg(\frac{1}{15}-\tdelta\bigg)+\tilde{x}_t^2\bigg(\frac{1}{90}-\frac{\tgamma^2}{11}\bigg(\frac{1}{5}-\frac{\tgamma^2}{6}-\tdelta+3\tdelta^2\bigg)-\frac{\tdelta}{14}+\frac{\tdelta^2}{5}-\frac{\tdelta^3}{3}+\frac{\tdelta^4}{2}\bigg)\nt\\
&\qquad+\frac{1}{110}-\frac{\tgamma^2}{11}\bigg(\frac{1}{7}-\frac{\tgamma^2}{6}\bigg(\frac{1}{3}-\tdelta\bigg)-\frac{3}{5}\tdelta+\tdelta^2-\tdelta^3\bigg)-\frac{\tdelta}{18}+\frac{\tdelta^2}{7}-\frac{\tdelta^3}{5}+\frac{\tdelta^4}{6}-\frac{\tdelta^5}{10}\bigg\}\ec\nt
\end{align}
\begin{align}
\bar{B}_{11}&=\frac{1}{s'-a}\frac{1}{3}\ec\qquad
\bar{B}_{12}=\frac{1}{\tgamma}\bigg[\frac{\tilde{x}_t}{p_tq_t}+\frac{1}{s'-a}\bigg\{\frac{3}{5}-\tdelta\bigg\}\bigg]\ec\\
\bar{B}_{13}&=\frac{5}{2\tgamma^2}\bigg[\frac{\tilde{x}_t}{p_tq_t}\bigg\{\tilde{x}_t^2+\frac{3}{5}-2\tdelta\bigg\}+\frac{1}{s'-a}\bigg\{\frac{3}{7}-\frac{\tgamma^2}{5}-\frac{6}{5}\tdelta+\tdelta^2\bigg\}\bigg]\ec\nt\\
\bar{B}_{14}&=\frac{35}{2\tgamma^3}\bigg[\frac{\tilde{x}_t}{p_tq_t}\bigg\{\frac{\tilde{x}_t^4}{3}+\tilde{x}_t^2\bigg(\frac{1}{5}-\tdelta\bigg)+\frac{1-\tgamma^2}{7}-\frac{3}{5}\tdelta+\tdelta^2\bigg\}\nt\\
&\qquad+\frac{1}{s'-a}\bigg\{\frac{1}{9}-\frac{\tgamma^2}{7}\bigg(\frac{3}{5}-\tdelta\bigg)-\frac{3}{7}\tdelta+\frac{3}{5}\tdelta^2-\frac{\tdelta^3}{3}\bigg\}\bigg]\ec\nt\\
\bar{B}_{15}&=\frac{105}{2\tgamma^4}\bigg[\frac{\tilde{x}_t}{p_tq_t}\bigg\{\frac{\tilde{x}_t^6}{4}+\tilde{x}_t^4\bigg(\frac{3}{20}-\tdelta\bigg)+\tilde{x}_t^2\bigg(\frac{3}{28}-\frac{\tgamma^2}{6}-\frac{3}{5}\tdelta+\frac{3}{2}\tdelta^2\bigg)+\frac{1}{12}-\tgamma^2\bigg(\frac{1}{10}-\frac{\tdelta}{3}\bigg)-\frac{3}{7}\tdelta\nt\\
&\qquad\quad+\frac{9}{10}\tdelta^2-\tdelta^3\bigg\}+\frac{1}{s'-a}\bigg\{\frac{3}{44}-\tgamma^2\bigg(\frac{1}{14}-\frac{\tgamma^2}{84}-\frac{\tdelta}{5}+\frac{\tdelta^2}{6}\bigg)-\frac{\tdelta}{3}+\frac{9}{14}\tdelta^2-\frac{3}{5}\tdelta^3+\frac{\tdelta^4}{4}\bigg\}\bigg]\ec\nt\\
\bar{B}_{16}&=\frac{3465}{4\tgamma^5}\bigg[\frac{\tilde{x}_t}{p_tq_t}\bigg\{\frac{\tilde{x}_t^8}{30}+\tilde{x}_t^4\bigg(\frac{1}{70}-\frac{\tgamma^2}{33}-\frac{\tdelta}{10}+\frac{\tdelta^2}{3}\bigg)+\tilde{x}_t^2\bigg(\frac{1}{90}-\frac{\tgamma^2}{11}\bigg(\cfrac{1}{5}-\tdelta\bigg)-\frac{\tdelta}{14}+\frac{\tdelta^2}{5}-\frac{\tdelta^3}{3}\bigg)\nt\\
&\qquad\quad+\frac{\tilde{x}_t^6}{2}\bigg(\frac{1}{25}-\frac{\tdelta}{3}\bigg)+\frac{1}{110}-\frac{\tgamma^2}{11}\bigg(\frac{1}{7}-\frac{\tgamma^2}{18}-\frac{3}{5}\tdelta+\tdelta^2\bigg)-\frac{\tdelta}{18}+\frac{\tdelta^2}{7}-\frac{\tdelta^3}{5}+\frac{\tdelta^4}{6}\bigg\}\nt\\
&\qquad+\frac{1}{s'-a}\bigg\{\frac{1}{130}-\frac{\tgamma^2}{11}\bigg(\frac{1}{9}-\frac{\tgamma^2}{6}\bigg(\frac{1}{5}-\frac{\tdelta}{3}\bigg)-\frac{3}{7}\tdelta+\frac{3}{5}\tdelta^2-\frac{\tdelta^3}{3}\bigg)-\frac{\tdelta}{22}+\frac{\tdelta^2}{9}-\frac{\tdelta^3}{7}+\frac{\tdelta^4}{10}-\frac{\tdelta^5}{30}\bigg\}\ec\nt
\end{align}
\begin{align}
\bar{B}_{21}&=0\ec\qquad
\bar{B}_{22}=\frac{1}{p_tq_t}\frac{2}{5\tgamma}\ec\qquad
\bar{B}_{23}=\frac{1}{p_tq_t}\frac{1}{\tgamma^2}\bigg\{\tilde{x}_t^2+\frac{6}{7}-2\tdelta\bigg\}\ec\\
\bar{B}_{24}&=\frac{1}{p_tq_t}\frac{7}{\tgamma^3}\bigg\{\frac{\tilde{x}_t^4}{3}+\tilde{x}_t^2\bigg(\frac{2}{7}-\tdelta\bigg)+\frac{5}{21}-\frac{\tgamma^2}{7}-\frac{6}{7}\tdelta+\tdelta^2\bigg\}\ec\nt\\
\bar{B}_{25}&=\frac{1}{p_tq_t}\frac{21}{\tgamma^4}\bigg\{\frac{\tilde{x}_t^6}{4}+\tilde{x}_t^4\bigg(\frac{3}{14}-\tdelta\bigg)+\tilde{x}_t^2\bigg(\frac{5}{28}-\frac{\tgamma^2}{6}-\frac{6}{7}\tdelta+\frac{3}{2}\tdelta^2\bigg)\nt\\
&\qquad+\frac{5}{33}-\tgamma^2\bigg(\frac{1}{7}-\frac{\tdelta}{3}\bigg)-\frac{5}{7}\tdelta+\frac{9}{7}\tdelta^2-\tdelta^3\bigg\}\ec\nt\\
\bar{B}_{26}&=\frac{1}{p_tq_t}\frac{231}{2\tgamma^5}\bigg\{\frac{\tilde{x}_t^8}{10}+\tilde{x}_t^4\bigg(\frac{1}{14}-\frac{\tgamma^2}{11}-\frac{3}{7}\tdelta+\tdelta^2\bigg)+\tilde{x}_t^2\bigg(\frac{2}{33}-\frac{\tgamma^2}{11}\bigg(\frac{6}{7}-3\tdelta\bigg)-\frac{5}{14}\tdelta+\frac{6}{7}\tdelta^2-\tdelta^3\bigg)\nt\\
&\qquad+\tilde{x}_t^6\bigg(\frac{3}{35}-\frac{\tdelta}{2}\bigg)+\frac{15}{286}-\frac{\tgamma^2}{11}\bigg(\frac{5}{7}-\frac{\tgamma^2}{6}-\frac{18}{7}\tdelta+3\tdelta^2\bigg)-\frac{10}{33}\tdelta+\frac{5}{7}\tdelta^2-\frac{6}{7}\tdelta^3+\frac{\tdelta^4}{2}\bigg\}\ec\nt
\end{align}
\begin{align}
\bar{C}_{11}&=\frac{1}{s'-a}\ec\qquad
\bar{C}_{12}=\frac{3}{\tgamma}\bigg[\frac{\tilde{x}_t}{p_tq_t}+\frac{1}{s'-a}\bigg\{\frac{1}{5}-\tdelta\bigg\}\bigg]\ec\\
\bar{C}_{13}&=\frac{15}{2\tgamma^2}\bigg[\frac{\tilde{x}_t}{p_tq_t}\bigg\{\tilde{x}_t^2+\frac{1}{5}-2\tdelta\bigg\}+\frac{1}{s'-a}\bigg\{\frac{3}{35}-\frac{\tgamma^2}{5}-\frac{2}{5}\tdelta+\tdelta^2\bigg\}\bigg]\ec\nt\\
\bar{C}_{14}&=\frac{105}{2\tgamma^3}\bigg[\frac{\tilde{x}_t}{p_tq_t}\bigg\{\frac{\tilde{x}_t^4}{3}+\tilde{x}_t^2\bigg(\frac{1}{15}-\tdelta\bigg)+\frac{1}{35}-\frac{\tgamma^2}{7}-\frac{\tdelta}{5}+\tdelta^2\bigg\}\nt\\
&\qquad+\frac{1}{s'-a}\bigg\{\frac{1}{63}-\frac{\tgamma^2}{7}\bigg(\frac{1}{5}-\tdelta\bigg)-\frac{3}{35}\tdelta+\frac{\tdelta^2}{5}-\frac{\tdelta^3}{3}\bigg\}\bigg]\ec\nt\\
\bar{C}_{15}&=\frac{315}{2\tgamma^4}\bigg[\frac{\tilde{x}_t}{p_tq_t}\bigg\{\frac{\tilde{x}_t^6}{4}+\tilde{x}_t^4\bigg(\frac{1}{20}-\tdelta\bigg)+\tilde{x}_t^2\bigg(\frac{3}{140}-\frac{\tgamma^2}{6}-\frac{\tdelta}{5}+\frac{3}{2}\tdelta^2\bigg)+\frac{1}{84}-\frac{\tgamma^2}{3}\bigg(\frac{1}{10}-\tdelta\bigg)-\frac{3}{35}\tdelta\nt\\
&\qquad\quad+\frac{3}{10}\tdelta^2-\tdelta^3\bigg\}+\frac{1}{s'-a}\bigg\{\frac{1}{132}-\tgamma^2\bigg(\frac{1}{70}-\frac{\tgamma^2}{84}-\frac{\tdelta}{15}+\frac{\tdelta^2}{6}\bigg)-\frac{\tdelta}{21}+\frac{9}{70}\tdelta^2-\frac{\tdelta^3}{5}+\frac{\tdelta^4}{4}\bigg\}\bigg]\ec\nt\\
\bar{C}_{16}&=\frac{3465}{4\tgamma^5}\bigg[\frac{\tilde{x}_t}{p_tq_t}\bigg\{\frac{\tilde{x}_t^8}{10}+\tilde{x}_t^4\bigg(\frac{3}{350}-\frac{\tgamma^2}{11}-\frac{\tdelta}{10}+\tdelta^2\bigg)+\tilde{x}_t^2\bigg(\frac{1}{210}-\frac{\tgamma^2}{11}\bigg(\frac{1}{5}-3\tdelta\bigg)-\frac{3}{70}\tdelta+\frac{\tdelta^2}{5}-\tdelta^3\bigg)\nt\\
&\qquad\quad+\frac{\tilde{x}_t^6}{2}\bigg(\frac{1}{25}-\tdelta\bigg)+\frac{1}{330}-\frac{\tgamma^2}{11}\bigg(\frac{3}{35}-\frac{\tgamma^2}{6}-\frac{3}{5}\tdelta+3\tdelta^2\bigg)-\frac{\tdelta}{42}+\frac{3}{35}\tdelta^2-\frac{\tdelta^3}{5}+\frac{\tdelta^4}{2}\bigg\}\nt\\
&
\quad+\frac{1}{s'-a}\bigg\{\frac{3}{1430}-\frac{\tgamma^2}{11}\bigg(\frac{1}{21}-\frac{\tgamma^2}{6}\bigg(\frac{1}{5}-\tdelta\bigg)-\frac{9}{35}\tdelta+\frac{3}{5}\tdelta^2-\tdelta^3\bigg)-\frac{\tdelta}{66}+\frac{\tdelta^2}{21}-\frac{3}{35}\tdelta^3+\frac{\tdelta^4}{10}-\frac{\tdelta^5}{10}\bigg\}\bigg]\ec\nt
\end{align}
\begin{align}
\bar{C}_{21}&=0\ec\qquad
\bar{C}_{22}=\frac{1}{p_tq_t}\frac{1}{\tgamma}\ec\qquad
\bar{C}_{23}=\frac{1}{p_tq_t}\frac{5}{2\tgamma^2}\bigg\{\tilde{x}_t^2+\frac{3}{7}-2\tdelta\bigg\}\ec\\
\bar{C}_{24}&=\frac{1}{p_tq_t}\frac{35}{2\tgamma^3}\bigg\{\frac{\tilde{x}_t^4}{3}+\tilde{x}_t^2\bigg(\frac{1}{7}-\tdelta\bigg)+\frac{5}{63}-\frac{\tgamma^2}{7}-\frac{3}{7}\tdelta+\tdelta^2\bigg\}\ec\nt\\
\bar{C}_{25}&=\frac{1}{p_tq_t}\frac{105}{2\tgamma^4}\bigg\{\frac{\tilde{x}_t^6}{4}+\tilde{x}_t^4\bigg(\frac{3}{28}-\tdelta\bigg)+\tilde{x}_t^2\bigg(\frac{5}{84}-\frac{\tgamma^2}{6}-\frac{3}{7}\tdelta+\frac{3}{2}\tdelta^2\bigg)\nt\\
&\qquad+\frac{5}{132}-\tgamma^2\bigg(\frac{1}{14}-\frac{\tdelta}{3}\bigg)-\frac{5}{21}\tdelta+\frac{9}{14}\tdelta^2-\tdelta^3\bigg\}\ec\nt\\
\bar{C}_{26}&=\frac{1}{p_tq_t}\frac{1155}{4\tgamma^5}\bigg\{\frac{\tilde{x}_t^8}{10}+\tilde{x}_t^4\bigg(\frac{1}{42}-\frac{\tgamma^2}{11}-\frac{3}{14}\tdelta+\tdelta^2\bigg)+\tilde{x}_t^2\bigg(\frac{1}{66}-\frac{\tgamma^2}{11}\bigg(\frac{3}{7}-3\tdelta\bigg)-\frac{5}{42}\tdelta+\frac{3}{7}\tdelta^2-\tdelta^3\bigg)\nt\\
&\qquad+\frac{\tilde{x}_t^6}{2}\bigg(\frac{3}{35}-\tdelta\bigg)+\frac{3}{286}-\frac{\tgamma^2}{11}\bigg(\frac{5}{21}-\frac{\tgamma^2}{6}-\frac{9}{7}\tdelta+3\tdelta^2\bigg)-\frac{5}{66}\tdelta+\frac{5}{21}\tdelta^2-\frac{3}{7}\tdelta^3+\frac{\tdelta^4}{2}\bigg\}\ep\nt
\end{align}

\subsection{Subtracted kernel functions}
\label{subsec:tpwp:subtractions}

Here, we give the modifications of the nucleon-pole-term projections and the kernel functions for the $t$-channel projection, that are required by the subtractions performed in Sect.~\ref{subsec:subRS:HDRs}.

To start with, the $n$-times subtracted nucleon-pole-term projections may be written as
\begin{align}
\label{DeltahatNJpmsubn}
\nsub{\tilde N^J_\pm}(t)&=\hat N^J_\pm(t)+\nsub{\Delta\hat N^J_\pm}(t)\ec\nt\\
\twosub{\Delta\hat N^J_+}(t)&=-\frac{p_t^2}{4\pi}\bigg(\frac{g^2}{m}+d_{00}^++d_{01}^+t-b_{00}^+\frac{q_t^2}{3}\bigg)\delta_{J0}
+\frac{m}{12\pi}\bigg(-\frac{g^2}{2m^2}+b_{00}^-+b_{01}^-t-a_{00}^-\frac{p_t^2}{m^2}\bigg)\delta_{J1}\nt\\
&\quad+\frac{b_{00}^+}{30\pi}\delta_{J2}\nt\\
&\redonesub-\frac{p_t^2}{4\pi}\bigg(\frac{g^2}{m}+d_{00}^+\bigg)\delta_{J0}+\frac{m}{12\pi}\bigg(-\frac{g^2}{2m^2}+b_{00}^-\bigg)\delta_{J1}\ec\nt\\
\twosub{\Delta\hat N^J_-}(t)&=\frac{\sqrt{2}}{12\pi}\bigg(-\frac{g^2}{2m^2}+b_{00}^-+b_{01}^-t\bigg)\delta_{J1}+\frac{b_{00}^+}{30\pi}\frac{\sqrt{6}}{2m}\delta_{J2}\ec\nt\\
&\redonesub\frac{\sqrt{2}}{12\pi}\bigg(-\frac{g^2}{2m^2}+b_{00}^-\bigg)\delta_{J1}\ec
\end{align}
where in analogy to the $s$-channel projection we have defined unsubtracted corrections (cf.~\eqref{hatNJpm})
\begin{equation}
\label{DeltahatNJpmunsub}
\unsub{\Delta\hat N^J_+}(t)=-\frac{g^2}{4\pi}\frac{m}{3}\frac{\delta_{J1}}{m^2-a}\ec\qquad
\unsub{\Delta\hat N^J_-}(t)=-\frac{g^2}{4\pi}\frac{\sqrt{2}}{3}\frac{\delta_{J1}}{m^2-a}\ec
\end{equation}
which are constant and non-zero only for $J=1$, in order to split off all terms that are either constant or contain subthreshold parameters.
Note that for both one and two subtractions the full nucleon-pole-term projections fulfill the threshold relations~\eqref{threshold_tildeN} for $p_t\to0$, but no longer for $q_t\to0$.
However, the subtraction-independent parts of the pole terms $\hat N^J_\pm$ still fulfill the relations~\eqref{threshold_tildeN} for $p_tq_t\to0$.

The necessary update of the $s$-channel kernels $\tilde G_{Jl}(t,W')$ and $\tilde H_{Jl}(t,W')$ may be achieved by adding
\begin{align}
\twosub{\Delta\bar A_{Jl}}(t,s')&=\bigg\{\bigg(h_0(s')-\frac{t}{(s'-s_0)^2}\bigg)P_l'\Big(\ste{z_s'}\Big)
+h_0(s')\,t\ste{\partial_tz_s'}P_l''\Big(\ste{z_s'}\Big)\bigg\}\delta_{J0}\nt\\
&\quad+\frac{4}{3}\frac{p_tq_t}{(s'-s_0)^2}P_l'\Big(\ste{z_s'}\Big)\delta_{J1}\nt\\
&\redonesub h_0(s')P_l'\Big(\ste{z_s'}\Big)\delta_{J0}\ec\nt\\
\twosub{\Delta\bar B_{Jl}}(t,s')&=\bigg\{\bigg(h_0(s')-\frac{t}{(s'-s_0)^2}\bigg)P_l'\Big(\ste{z_s'}\Big)
+h_0(s')\,t\ste{\partial_tz_s'}P_l''\Big(\ste{z_s'}\Big)\bigg\}\frac{\delta_{J1}}{3}\nt\\
&\quad+\frac{4}{3}\frac{p_tq_t}{(s'-s_0)^2}P_l'\Big(\ste{z_s'}\Big)\Big(\delta_{J0}+\frac{2}{5}\delta_{J2}\Big)\nt\\
&\redonesub h_0(s')P_l'\Big(\ste{z_s'}\Big)\frac{\delta_{J1}}{3}\ec\nt\\
\twosub{\Delta\bar C_{Jl}}(t,s')&=\bigg\{\bigg(h_0(s')-\frac{t}{(s'-s_0)^2}\bigg)P_l'\Big(\ste{z_s'}\Big)
+h_0(s')\,t\ste{\partial_tz_s'}P_l''\Big(\ste{z_s'}\Big)\bigg\}\delta_{J1}\nt\\
&\quad+\frac{4}{3}\frac{p_tq_t}{(s'-s_0)^2}P_l'\Big(\ste{z_s'}\Big)\delta_{J2}\nt\\
&\redonesub h_0(s')P_l'\Big(\ste{z_s'}\Big)\delta_{J1}\ec
\end{align}
respectively, to $\bar A_{Jl}$, $\bar B_{Jl}$, and $\bar C_{Jl}$ at the pertinent places in~\eqref{GHbardef}, leading to corresponding $\Delta\tilde G_{Jl}$ and $\Delta\tilde H_{Jl}$.
Note that also in both the once- and twice-subtracted case $\Delta\bar C_{Jl}=\Delta\bar A_{J-1,l}-\Delta\bar A_{J+1,l}$ is still valid (for $J\geq1$, here actually $\Delta\bar C_{Jl}=\Delta\bar A_{J-1,l}$).

The additional contributions to the $t$-channel kernels $\mathbf{\tilde K}^{JJ'}(t,t')$ amount, for even $J$ and $J'$, to
\begin{align}
\twosub{\Delta\tilde K^1_{JJ'}}(t,t')&=-(2J'+1)(p_t'q_t')^{J'}\frac{p_t^2}{p_t'^2}\frac{1}{t'}
\bigg\{\bigg(1+\frac{t}{t'}\bigg)\ste{P_{J'}(z_t')}+t\ste{\partial_tP_{J'}(z_t')}\bigg\}\delta_{J0}\nt\\
&\redonesub -(2J'+1)(p_t'q_t')^{J'}\frac{p_t^2}{p_t'^2}\frac{1}{t'}\ste{P_{J'}(z_t')}\delta_{J0}\ec\nt\\
\twosub{\Delta\tilde K^2_{JJ'}}(t,t')&=\frac{2J'+1}{\sqrt{J'(J'+1)}}(p_t'q_t')^{J'}\frac{p_t^2}{p_t'^2}\frac{m}{t'}
\bigg\{\bigg[\bigg(1+\frac{t}{t'}\bigg)\ste{z_t'P_{J'}'(z_t')}+t\ste{\partial_t(z_t'P_{J'}'(z_t'))}\bigg]\delta_{J0}\nt\\
&\qquad-\frac{1}{3}\frac{q_t^2}{q_t'^2}\ste{\frac{P_{J'}'(z_t')}{z_t'}}\bigg(\delta_{J0}+\frac{2}{5}\frac{\delta_{J2}}{p_t^2q_t^2}\bigg)\bigg\}\nt\\
&\redonesub \frac{2J'+1}{\sqrt{J'(J'+1)}}(p_t'q_t')^{J'}\frac{p_t^2}{p_t'^2}\frac{m}{t'}\ste{z_t'P_{J'}'(z_t')}\delta_{J0}\ec\nt\\
\twosub{\Delta\tilde K^3_{JJ'}}(t,t')&=-\frac{2J'+1}{\sqrt{J'(J'+1)}}(p_t'q_t')^{J'-2}\frac{\sqrt{6}}{15}\frac{1}{t'}\ste{\frac{P_{J'}'(z_t')}{z_t'}}\delta_{J2}
\redonesub 0\ec
\end{align}
while for odd $J$ and $J'$ one finds
\begin{align}
\twosub{\Delta\tilde K^1_{JJ'}}(t,t')&=-(2J'+1)(p_t'q_t')^{J'-1}\frac{p_t^2}{p_t'^2}\frac{1}{t'}
\ste{\frac{P_{J'}(z_t')}{z_t'}}\frac{\delta_{J1}}{3}
\redonesub 0\ec\nt\\
\twosub{\Delta\tilde K^2_{JJ'}}(t,t')&=-\frac{2J'+1}{\sqrt{J'(J'+1)}}(p_t'q_t')^{J'-1}\frac{m}{t'}
\bigg\{\bigg(1-\frac{p_t^2}{p_t'^2}+\frac{t}{t'}\bigg)\ste{P_{J'}'(z_t')}
+t\ste{\partial_tP_{J'}'(z_t')}\bigg\}\frac{\delta_{J1}}{3}\nt\\
&\redonesub -\frac{2J'+1}{\sqrt{J'(J'+1)}}(p_t'q_t')^{J'-1}\frac{m}{t'}\ste{P_{J'}'(z_t')}\frac{\delta_{J1}}{3}\ec\nt\\
\twosub{\Delta\tilde K^3_{JJ'}}(t,t')&=-\frac{2J'+1}{\sqrt{J'(J'+1)}}(p_t'q_t')^{J'-1}\frac{\sqrt{2}}{t'}
\bigg\{\bigg(1+\frac{t}{t'}\bigg)\ste{P_{J'}'(z_t')}+t\ste{\partial_tP_{J'}'(z_t')}\bigg\}\frac{\delta_{J1}}{3}\nt\\
&\redonesub -\frac{2J'+1}{\sqrt{J'(J'+1)}}(p_t'q_t')^{J'-1}\frac{\sqrt{2}}{t'}\ste{P_{J'}'(z_t')}\frac{\delta_{J1}}{3}\ep
\end{align}
Furthermore, $\mathbf{\Delta}\mathbf{\tilde K}^{JJ'}=0$ for $J>2$ or $J'<J$, the latter being in agreement with~\eqref{vanishingtildeK}.
In all cases only even powers of $z_t'$ and the primed momenta occur, and hence the additional kernel terms are always real.
Here, we refrain from explicitly expanding the Legendre polynomials using~\eqref{legendreexpevenodd} as in Appendix~\ref{subsec:tpwp:t}, but only give one example to demonstrate this point (for even $J$ and $J'$)
\begin{equation}
\ste{\partial_t(z_t'P_{J'}'(z_t'))}=2\ste{\partial_tz_t'^2}\sum\limits_{\lambda=1}^{\frac{J'}{2}}a^\text{ev}_{\lambda J'}\lambda^2\ste{z_t'^2}^{\lambda-1}\ep
\end{equation}
For later convenience, we explicitly state all those subtracted kernels with $0\leq J'\leq3$ that differ from their unsubtracted form
{\allowdisplaybreaks
\begin{align}
\label{subtractedtildeKkernels}
\twosub{\tilde K^1_{00}}(t,t')&=\frac{t^2}{t'^2}\tilde K^1_{00}(t,t')
\redonesub\frac{t}{t'}\tilde K^1_{00}(t,t')\ec
\qquad
\twosub{\tilde K^1_{11}}(t,t')=\frac{t}{t'}\tilde K^1_{11}(t,t')
\redonesub\tilde K^1_{11}(t,t')\ec\nt\\
\twosub{\tilde K^1_{02}}(t,t')&=\frac{5}{8}\frac{p_t^2}{p_t'^2}\bigg\{\bigg(1+\frac{t}{t'}\bigg)\frac{\tN\tpi}{4t'}-\frac{t}{t'}s_0\bigg\}
\redonesub\frac{5}{16}\frac{p_t^2}{p_t'^2}\bigg\{t+\frac{\tN\tpi}{2t'}\bigg\}\ec\nt\\
\twosub{\tilde K^1_{13}}(t,t')&=\frac{7}{48}\frac{p_t^2}{p_t'^2}\bigg\{t+\frac{3\tN\tpi}{2t'}\bigg\}
\redonesub\tilde K^1_{13}(t,t')\ec\nt\\
\twosub{\tilde K^2_{11}}(t,t')&=\frac{t\tN}{t'^2}\tilde K^2_{11}(t,t')
\redonesub\frac{\tN}{t'}\tilde K^2_{11}(t,t')\ec
\qquad
\twosub{\tilde K^2_{22}}(t,t')=\frac{\tN}{t'}\tilde K^2_{22}(t,t')
\redonesub\tilde K^2_{22}(t,t')\ec\nt\\
\twosub{\tilde K^2_{02}}(t,t')&=\frac{5m}{4\sqrt{6}}\frac{p_t^2q_t^2}{p_t'^2}\frac{\tN}{t'}
\redonesub-\frac{5m}{8\sqrt{6}}\frac{p_t^2}{p_t'^2}\bigg\{t+\frac{\tpi}{2}\bigg\}\ec\nt\\
\twosub{\tilde K^2_{13}}(t,t')&=\frac{7m}{16\sqrt{3}}\bigg\{\frac{p_t^2}{p_t'^2}\bigg[2q_t^2-t-\frac{\tN\tpi}{4t'}\bigg]+\bigg[\bigg(1+\frac{t}{t'}\bigg)\frac{\tN\tpi}{4t'}-\frac{t}{t'}s_0\bigg]\bigg\}\nt\\
&\redonesub\frac{7m}{16\sqrt{3}}\bigg\{\frac{p_t^2}{p_t'^2}\Big[2q_t^2-\Big(t+t'-4s_0+5a\Big)\Big]+\bigg[t+\frac{\tN\tpi}{4t'}\bigg]\bigg\}\ec\nt\\
\twosub{\tilde K^3_{11}}(t,t')&=\frac{t^2}{t'^2}\tilde K^3_{11}(t,t')
\redonesub\frac{t}{t'}\tilde K^3_{11}(t,t')\ec
\qquad
\twosub{\tilde K^3_{22}}(t,t')=\frac{t}{t'}\tilde K^3_{22}(t,t')
\redonesub\tilde K^3_{22}(t,t')\ec\nt\\
\twosub{\tilde K^3_{13}}(t,t')&=\frac{7}{8\sqrt{6}}\bigg\{\bigg(1+\frac{t}{t'}\bigg)\frac{\tN\tpi}{4t'}-\frac{t}{t'}s_0\bigg\}
\redonesub\frac{7}{8\sqrt{6}}\bigg\{t+\frac{\tN\tpi}{4t'}\bigg\}\ec
\end{align}}\noindent
still obeying the threshold-behavior relation~\eqref{threshold_tildeK12}.
Note that at the level of two subtractions all these kernels are independent of $a$ (which is, however, not true for only one subtraction and $J\geq3$ or without subtracting), and that the exceptionally safe behavior of $\tilde K^2_{02}(t,t')$ at $\tN$ is preserved (cf.~\eqref{tilde2K02asymptotics}):
\begin{equation}
\label{tilde2K02asymptoticsnsub}
\nsub{\tilde K^2_{02}}(t\to\tN,t')=\Ord(p_t^2) \qquad\forall\;n\geq0\ep
\end{equation}

\section{Ranges of convergence}
\label{sec:convergence}

In this Appendix, we will analyze both the convergence of the partial-wave expansion of the imaginary parts inside the integrals and the convergence of the partial-wave projection of the full HDR equations.
For the rest of this section we may work as if no subtractions were necessary.

\subsection{Boundaries of the double spectral regions}
\label{subsec:convergence:doublespectralregions}

The following analysis is performed in the spirit of~\cite{Hoehler,piK:RS,ggpipi}.\footnote{Note that the authors of~\cite{piK:RS} corrected their discussion of the boundaries of the double spectral regions in~\cite{piK:K0star}.}
The basic {\it assumption} is that the $T$-matrix element (and hence the scattering amplitudes $\{A(s,t),B(s,t)\}\propto T(s,t)/(16\pi)$) fulfills Mandelstam analyticity~\cite{Mandelstam:analyticity}, i.e.\ that it can be written in terms of double spectral density functions $\rho_{su}$, $\rho_{tu}$, and $\rho_{st}$ according to\footnote{Mandelstam analyticity can at any rate be justified in the framework of perturbation theory~\cite{Mandelstam:analyticity,Mandelstam:boxintegral,Mandelstam:boundaries}. While for $\pi\pi$ scattering the validity of the Mandelstam representation can even be shown rigorously in a finite region~\cite{Martin1,Martin2}, for $\pi N$ scattering (involving unequal masses and spin) at least the uniqueness of amplitudes satisfying this representation is ensured by the MacDowell symmetry~\cite{MacDowell:continuation,CheungChen-Cheung}.}
\begin{equation}
T(s,t)=\frac{1}{\pi^2}\iint\diff s'\diff u'\frac{\rho_{su}(s',u')}{(s'-s)(u'-u)}+ 
\frac{1}{\pi^2}\iint\diff t'\diff u'\frac{\rho_{tu}(t',u')}{(t'-t)(u'-u)}+
\frac{1}{\pi^2}\iint\diff s'\diff t'\frac{\rho_{st}(s',t')}{(s'-s)(t'-t)}\ep
\end{equation}
The integration ranges are determined by those regions in the Mandelstam plane where the corresponding double spectral densities have support. The boundaries of these so-called double spectral regions will be the central objects of the following discussion.

\begin{figure}
\centering
\includegraphics[width=3.5cm,angle=-90]{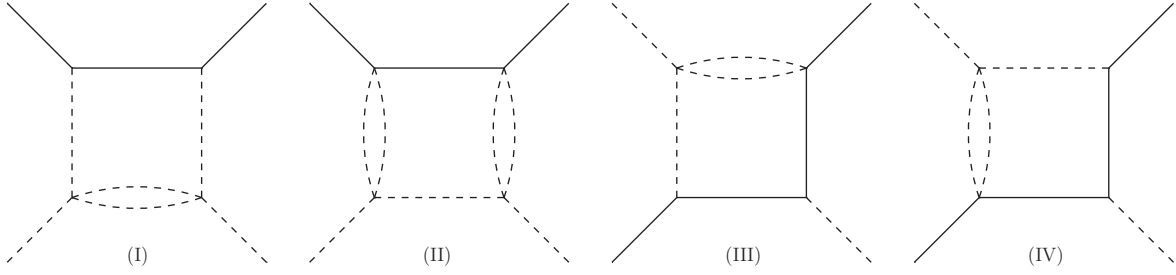}
\caption[Box graphs constraining the boundaries of the double spectral regions.]{Box graphs constraining the boundaries of the double spectral regions. Solid lines denote nucleons and dashed lines denote pions.}
\label{fig:boundary}
\end{figure}
\begin{figure}
\centering
\includegraphics[scale=0.8]{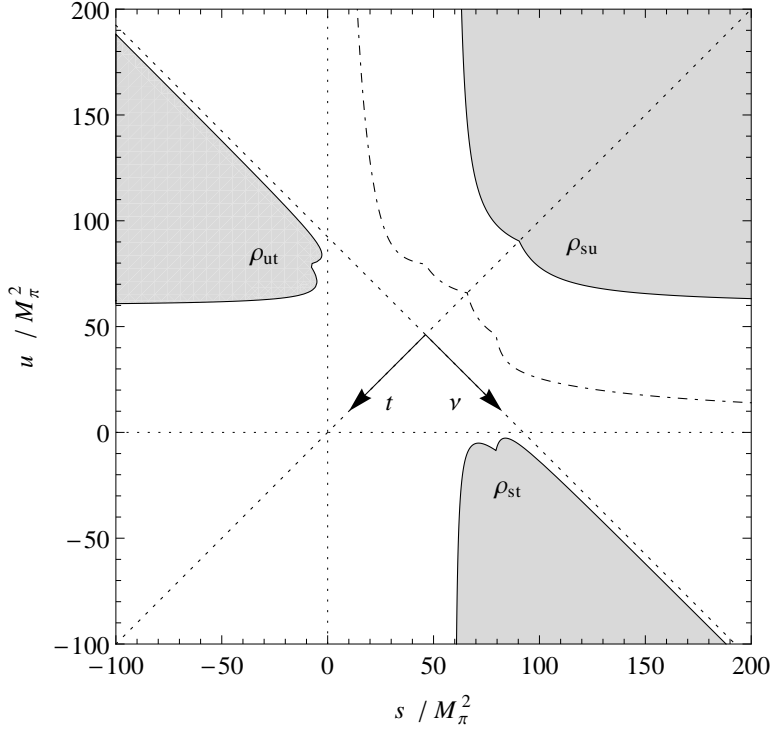}
\caption[Double spectral regions for $\pi N$ scattering.]{Double spectral regions for $\pi N$ scattering (shaded) and boundaries of $\rho_{st}$ and $\rho_{ut}$ reflected in the corresponding scattering angle (dot-dashed).}
\label{fig:rhobounds}
\end{figure}

The three double spectral densities can be derived by studying the consequences of unitarity in the 2-intermediate-particle approximation. We consider the corresponding lowest-lying intermediate states as depicted in Fig.~\ref{fig:boundary} (as unitarity diagrams, i.e.\ with on-shell intermediate particles), where the inelastic (referring to the intermediate state of the $s$-channel process) diagram (I) and the elastic diagram (II) yield the boundary of the support of $\rho_{st}$ (from which, due to $s\leftrightarrow u$ crossing symmetry, the result for $\rho_{ut}$ directly follows), while (III) and (IV) are relevant for calculating the boundary of the support of $\rho_{su}$. This leads to boundary functions (cf.~\cite{Hoehler})
\begin{align}
\label{bIbII}
b_\text{I}(s,t)&=\big(t-4\mpi^2\big)\lambda\big(s,m^2,4\mpi^2\big)-16\mpi^4\big(s+3\Sigma_-\big)\ec\nt\\
b_\text{II}(s,t)&=\big(t-16\mpi^2\big)\lambda_s-64\mpi^4s 
\end{align}
for the boundary of $\rho_{st}$ and thus $b_\text{I}(u,t)$ and $b_\text{II}(u,t)$ for the boundary of $\rho_{ut}$, as well as
\begin{align}
b_\text{III}(s,u)&=\lambda_u\lambda\big(s,m^2,4\mpi^2\big)-16\mpi^2\Big[m^2su-\Sigma_-^2\big(m^2-t(s,u)\big)\Big]\ec\nt\\
b_\text{IV}(s,u)&=\lambda_s\lambda\big(u,m^2,4\mpi^2\big)-16\mpi^2\Big[m^2su-\Sigma_-^2\big(m^2-t(s,u)\big)\Big]
\end{align}
for the boundary of $\rho_{su}$, where we only need to consider $b_\text{III}(s,u)=b_\text{IV}(u,s)$ due to $s\leftrightarrow u$ symmetry.
The whole support of all three double spectral densities is then given by the union of the regions allowed by the non-trivial constraints that the corresponding boundary functions be non-negative.
Furthermore, trivial constraints arise from the lower kinematical bounds of the corresponding physical regions that are given by the asymptotes of the boundary functions in question, e.g.\ for the inelastic diagram (I) we find the asymptotes $s=(m+2\mpi)^2$ and $t=(2\mpi)^2=\tpi$, and for the elastic diagram (II) we obtain $s=(m+\mpi)^2=s_+$ and $t=(4\mpi)^2$.
Therefore, by defining the following abbreviations for the solutions of the implicit equations
\begin{equation}
b_\text{I/II}(s,t)\overset{!}{=}0 \qquad\Rightarrow\qquad t=T_\text{I/II}(s)\ec \quad s=S_\text{I/II}(t)\ec
\end{equation}
the boundary of the support of e.g.\ $\rho_{st}$ is described by
\begin{equation}
\label{Tstdef}
T_{st}(s)=\min\{T_\text{I}(s),T_\text{II}(s)\}=\begin{cases}T_\text{II}(s)\hspace{6.3em}\text{for }s_+<s<(m+2\mpi)^2\ec\\\min\{T_\text{I}(s),T_\text{II}(s)\}\quad\text{for }(m+2\mpi)^2<s\ec\end{cases}
\end{equation}
with the functions
\begin{align}
T_\text{I}(s)&=\frac{4\mpi^2\big(s-m^2-2\mpi^2\big)^2}{\lambda\big(s,m^2,4\mpi^2\big)}>4\mpi^2 \qquad \forall\;s>(m+2\mpi)^2\ec\nt\\
T_\text{II}(s)&=\frac{16\mpi^2\big(s-\Sigma_-\big)^2}{\lambda_s}>16\mpi^2 \qquad \forall\;s>s_+\ec
\end{align}
again limited by the physical constraints, such that by definition $T_{st}(s)>4\mpi^2$ for $s>s_+$.
The boundaries of all three double spectral regions are shown in Fig.~\ref{fig:rhobounds}.
The asymptotes of $\rho_{st}$ are $s=s_+$ and $t=\tpi$ and hence those of $\rho_{ut}$ are $u=s_+$ and $t=\tpi$, while the symmetric asymptotes of $\rho_{su}$ are $s=s_+$ and $u=s_+$.

\subsection{Lehmann ellipse constraints}
\label{subsec:convergence:lehmannellipse}

The boundaries of the double spectral regions limit the range of validity of the HDRs in two ways:
\begin{enumerate}
\item The partial-wave expansions of the imaginary parts inside the HDR integrals (internal/primed kinematics) in the unphysical regions for both $s$- and $t$-channel partial waves converge only for CMS scattering angle cosines $z'$ within the corresponding large Lehmann ellipses~\cite{Lehmann}. These ellipses are the largest ellipses in the complex $z'$ plane, centered at the origin with foci at $z'=\pm1$, that do not reach into any double spectral region.
\item For a given value of the parameter $a$, the hyperbolae $(s-a)(u-a)=b$ with asymptotes $s=a$ and $u=a$ must not enter any double spectral region for all values of the parameter $b$ that are necessary for the partial-wave projections of the full HDR equations (external/unprimed kinematics) in given kinematical ranges. Trivial geometrical constraints on $a$ arise already from the asymptotes of the double spectral regions.
\end{enumerate}
In this section we will show how the (large) Lehmann ellipse constraint can be translated for a given $a$ into a constraint on $b$, each for both the expansions in $s$- and $t$-channel partial waves.
For any allowed fixed $a$, the allowed values of $b$ are those fulfilling both of the above requirements, and the (limited) freedom in the choice of $a$ in the construction of the HDRs can be used in order to optimize the convergence properties of the PWHDRs.
In the two subsequent sections we will investigate numerically how these limits on $b$ (for given $a$) yield the ranges of convergence of the full RS system via the restrictions that are necessary for both the projections onto $s$- and $t$-channel partial waves to converge.

For the partial-wave expansion of the $s$-channel contributions, the Lehmann ellipse constraint states that the expansion converges for angles $z_s'(s',t')=1+2s't'/\lambda_{s'}$ (cf.~\eqref{primedkinematics}) inside the ellipse
\begin{equation}
\label{sLehmannellipse}
\frac{(\Re z_s')^2}{A_s^2}+\frac{(\Im z_s')^2}{B_s^2}=1\ec
\end{equation}
with foci at $z_s'=\pm1$ (corresponding to the physical constraint $-1\leq z_s'\leq1$), i.e.\ semimajor and semiminor axis $A_s$ and $B_s$ are related by
\begin{equation}
A_s^2-B_s^2=1\ep
\end{equation}
Since for given $t'$ the angle $z_s'$ is always real in the integration range $s'>s_+$, the maximal value of $z_s'$ for given $s'$ not entering the support of $\rho_{st}$ follows from the corresponding maximally allowed value of $t'$ (according to~\eqref{Tstdef} for the internal (primed) variables) and thus reads 
\begin{equation}
z_{s'}^\text{max}(s')=1+\frac{2s'}{\lambda_{s'}}T_{st}(s')=A_s \qquad \forall\;s'>s_+\ep
\end{equation}
From the geometrical condition $-A_s\leq z_s'\leq A_s$ then follows
\begin{equation}
\label{szreflectionbound}
-z_{s'}^\text{max}\leq z_s'\leq z_{s'}^\text{max} \qquad \forall\;s'>s_+\ec
\end{equation}
and the lower bound due to this reflection in $z_s'$ is actually stronger than the restrictions imposed by $\rho_{su}$ as shown by the dot-dashed line in Fig.~\ref{fig:rhobounds}, where the $z_s'$-reflected boundary of the support of $\rho_{st}$ for $\nu>0$ is given by
\begin{equation}
\label{szreflectionboundexplicitly}
u\Big(s,t\Big(s,-z_s\big(s,t=T_{st}(s)\big)\Big)\Big)=\frac{\Sigma_-^2}{s}+T_{st}(s)\ec
\end{equation}
with the asymptote $u=4\mpi^2$ for $s\to\infty$ due to $T_{st}(s)$.
Furthermore, due to $s\leftrightarrow u$ symmetry $\rho_{ut}$ yields exactly the same constraints as $\rho_{st}$ (including the $z_u'$-reflected boundary for $\nu<0$), and hence we only need to consider the latter.\footnote{Note that both the $s$- and $u$-channel physical regions fit well in between $\rho_{st}$, $\rho_{ut}$, and their reflected boundaries.}
The possible values of $t'$ for given $s'$ are then restricted by (cf.~\cite{Hoehler})
\begin{equation}
-\frac{\lambda_{s'}}{s'}-T_{st}(s')\leq t'\leq T_{st}(s') \qquad \forall\;s'>s_+\ep
\end{equation}
Via the linear relation~\eqref{bofsta} for the internal kinematics this range for $t'$ can be translated into a range of allowed values of $b(s',t';a)$ for given $a$ according to (cf.~\eqref{szreflectionboundexplicitly})
\begin{align}
\label{bsminusplus}
b_s^-(s',a)&\leq b\leq b_s^+(s',a) \qquad \forall\;s'>s_+>a\ec\nt\\
b_s^-(s',a)&=(s'-a)\big(\Sigma-s'-T_{st}(s')-a\big)\ec\nt\\
b_s^+(s',a)&=(s'-a)\Big(\Sigma-s'+\frac{\lambda_{s'}}{s'}+T_{st}(s')-a\Big)=(s'-a)\bigg\{\frac{\Sigma_-^2}{s'}+T_{st}(s')-a\bigg\}\ec
\end{align}
where we have used that from the asymptotes $s=s_+$ and $u=s_+$ of the double spectral regions it is geometrically clear from Fig.~\ref{fig:rhobounds} that the allowed values of the hyperbola's asymptotic parameter $a$ are trivially limited to $a<s_+$ (independent of $b$), and hence we have $s'>a$ for all $s'>s_+$.
By invoking the asymptotes $s=\tpi$ and $u=\tpi$ of the $z'$-reflected boundaries of $\rho_{st}$ and $\rho_{ut}$ (cf.~\eqref{szreflectionboundexplicitly}) we can deduce that the allowed range of $a$ is actually geometrically limited by $a<\tpi$, which is the reason why the ``fixed-$t$ limit'' $|a|\to\infty$ actually reduces to $a\to-\infty$.
Now, we may define the highest lower and the lowest upper bound 
\begin{equation}
\tilde b_s^-(a)=\max_{s'>s_+}b_s^-(s',a)\ec \qquad \tilde b_s^+(a)=\min_{s'>s_+}b_s^+(s',a)\ec 
\end{equation}
as the maximum/minimum value of $b_s^{-/+}(s',a)$ within the integration range $s'>s_+$, which then finally determines the allowed values of $b$ for given $a$ by
\begin{equation}
\tilde b_s^-(a)\leq b\leq\tilde b_s^+(a) \qquad \forall\;s'>s_+>a\ec
\end{equation}
for the $s$-channel parts of the HDRs.

The Lehmann ellipse constraint for the partial-wave expansion of the $t$-channel contributions limits the convergence of the expansion to angles $z_t'(s',t')=m\nu'/(p_t'q_t')$ (cf.~\eqref{primedkinematics}) inside an ellipse similar to~\eqref{sLehmannellipse} centered at the origin with foci at $z_t'=\pm1$
\begin{equation}
\label{tLehmannellipse}
\frac{(\Re z_t')^2}{A_t^2}+\frac{(\Im z_t')^2}{B_t^2}=1\ec \qquad A_t^2-B_t^2=1\ep
\end{equation}
The argument for the $t$-channel contributions is more intricate, since inside the integration range $t'>\tpi$ the angle $z_t'$ becomes purely imaginary for $\tpi<t'<\tN$, and hence no relations similar to~\eqref{szreflectionbound} are possible.
However, as the relation between $z_t'$ and $b$ is non-linear anyway (cf.~\eqref{internalkinematicsofsta})
\begin{equation}
z_t'^2=\frac{(t'-\Sigma+2a)^2-4b(s',t';a)}{16p_t'^2q_t'^2}\ec
\end{equation}
where all squares are real but not necessarily positive, we are interested in the resulting Lehmann ellipse constraint for $z_t'^2$.
By squaring equation~\eqref{tLehmannellipse} for general complex $z_t'$ we arrive at
\begin{equation}
\label{tLehmannellipsesquared}
\frac{\big(\Re\{z_t'^2\}-\frac{1}{2}\big)^2}{\tilde A_t^2}+\frac{\big(\Im\{z_t'^2\}\big)^2}{\tilde B_t^2}=1\ec
\end{equation}
which corresponds to an ellipse in the complex $z_t'^2$ plane shifted to the right by $(A_t^2-B_t^2)/2=1/2$. Hence, it is centered at $(1/2,0)$ with the semimajor and semiminor axes given by
\begin{equation}
\tilde A_t=\frac{A_t^2+B_t^2}{2}=A_t^2-\frac{1}{2}\ec \qquad \tilde B_t=A_tB_t=A_t\sqrt{A_t^2-1}\ec
\end{equation}
such that the foci are at $1/2\mp\sqrt{\tilde A_t^2-\tilde B_t^2}=1/2\mp1/2$ (corresponding to the physical constraint $0\leq z_t'^2\leq1$).
Since for $t'>\tpi$ we have $z_t'^2=\Re\{z_t'^2\}$, the geometrical condition $1/2-\tilde A_t\leq z_t'^2\leq1/2+\tilde A_t$ leads to the analog of~\eqref{szreflectionbound}
\begin{equation}
\label{tz2reflectionbound}
1-A_t^2=-B_t^2\leq z_t'^2\leq A_t^2\ec
\end{equation}
where it is important to note that on the right-hand side the relation between $z_t'$ and $A_t$ is not fixed due to the squares, while the reflection bound on the left-hand side is again more restrictive than the corresponding bound due to $\rho_{su}$, and hence we only have to look at the boundaries of the support of $\rho_{st}$.
For the following it turns out to be advantageous to rewrite the boundary functions $b_\text{I,II}(s,t)$ of~\eqref{bIbII} in terms of $(\nu,t)$, since the quantity $\nu(z_t,t)=p_tq_tz_t/m$ is always real
\begin{align}
b_\text{I}(\nu,t)&=\big(t-4\mpi^2\big)\Big\{\frac{1}{4}\big(t-4m\nu+6\mpi^2\big)^2-16m^2\mpi^2\Big\}+8\mpi^4\Big\{t-4m\nu-\Sigma-6\Sigma_-\Big\}\overset{!}{=}0\ec\nt\\
b_\text{II}(\nu,t)&=\big(t-16\mpi^2\big)\Big\{\frac{1}{4}\big(t-4m\nu\big)^2-4m^2\mpi^2\Big\}+32\mpi^4\Big\{t-4m\nu-\Sigma\Big\}\overset{!}{=}0\ep
\end{align}
Solving these implicit quadratic equations for $\nu(t)$ yields the physical solutions (cf.~\cite{Hoehler})
\begin{align}
\nu_\text{I}(t)&=\frac{\big(t-2\mpi^2\big)\big(t+4\mpi^2\big)+8\mpi\sqrt{t}\sqrt{\big(t-4\mpi^2\big)m^2+\mpi^4}}{4m\big(t-4\mpi^2\big)}>0
& &\forall\;t>4\mpi^2=\tpi\ec\nt\\
\nu_\text{II}(t)&=\frac{\big(t-8\mpi\big)^2+4\mpi\sqrt{t}\sqrt{\big(t-16\mpi^2\big)m^2+16\mpi^4}}{4m\big(t-16\mpi^2\big)}>0
& &\forall\;t>16\mpi^2=4\tpi\ec
\end{align}
again limited by the physical constraints, where each sign of the root is fixed by $z_t(\nu,t)=m\nu/(p_tq_t)\propto+\nu$ and hence $z_t^\text{max}=+m\nu^\text{max}/(p_tq_t)$ in the physical $t$-channel region $t>4m^2=\tN$.
Defining the (positive) combined upper bound on $\nu$ according to
\begin{equation}
\label{Nstdef}
N_{st}(t)=\min\{\nu_\text{I}(t),\nu_\text{II}(t)\}=\begin{cases}\nu_\text{I}(t)\hspace{6.4em}\text{for }\tpi<t<4\tpi\ec\\\min\{\nu_\text{I}(t),\nu_\text{II}(t)\}\quad\text{for }4\tpi<t\ec\end{cases}
\end{equation}
and resorting to the geometrical constraints of the original $t$-channel Lehmann ellipse~\eqref{tLehmannellipse} for $z_t'$, the maximally allowed value of the real angle $z_t'=\Re z_t'$ for given $t'>\tN$ not entering the support of $\rho_{st}$ is given by
\begin{equation}
z_{t'}^\text{max}(t')=\frac{m}{p_t'q_t'}N_{st}(t')=A_t \qquad \forall\;t'>\tN\ec
\end{equation}
and thus~\eqref{tz2reflectionbound} in this case leads to
\begin{equation}
\label{tz2boundhigh}
1-\frac{m^2}{p_t'^2q_t'^2}N_{st}(t')^2\leq z_t'^2\leq\frac{m^2}{p_t'^2q_t'^2}N_{st}(t')^2 \qquad \forall\;t'>\tN\ep
\end{equation}
In contrast, for $\tpi<t'<\tN$ we have $p_t'=ip_-'$ with real $p_-'$. Accordingly, for the purely imaginary angle $z_t'=i\,\Im z_t'$ it follows from~\eqref{tLehmannellipse} that
\begin{equation}
\left|\Im z_t'(t')\right|=\left|-\frac{m\nu'}{p_-'q_t'}\right|\leq\frac{m}{p_-'q_t'}N_{st}(t')=B_t \quad\Rightarrow\quad
B_t^2=-\frac{m^2}{p_t'^2q_t'^2}N_{st}(t')^2 \qquad \forall\;\tpi<t'<\tN\ec
\end{equation}
which plugged into~\eqref{tz2reflectionbound} yields
\begin{equation}
\label{tz2boundlow}
\frac{m^2}{p_t'^2q_t'^2}N_{st}(t')^2\leq z_t'^2\leq1-\frac{m^2}{p_t'^2q_t'^2}N_{st}(t')^2 \qquad \forall\;\tpi<t'<\tN\ep
\end{equation}
However, from both~\eqref{tz2boundhigh} with $p_t'^2>0$ for all $t'>\tN$ and~\eqref{tz2boundlow} with $p_t'^2<0$ for all $\tpi<t'<\tN$ we arrive at the same constraints on $\nu'^2$ for given $t'>\tpi$ (cf.~\cite{Hoehler})
\begin{equation}
\label{nubound}
\frac{p_t'^2q_t'^2}{m^2}-N_{st}(t')^2\leq\nu'^2\leq N_{st}(t')^2 \qquad \forall\;t'>\tpi\ep
\end{equation}
By virtue of the linear relation (cf.~\eqref{sunuoftab})
\begin{equation}
16m^2\nu'^2=(t'-\Sigma+2a)^2-4b\ec
\end{equation}
this range for $\nu'^2$ can then be translated into a range for $b(\nu'^2,t';a)$ according to
\begin{align}
\label{btminusplus}
b_t^-(t',a)&\leq b\leq b_t^+(t',a) \qquad \forall\;t'>\tpi>a\ec\nt\\
b_t^-(t',a)&=\frac{1}{4}(t'-\Sigma+2a)^2-4m^2N_{st}(t')^2\ec\nt\\
b_t^+(t',a)&=\frac{1}{4}(t'-\Sigma+2a)^2-4p_t'^2q_t'^2+4m^2N_{st}(t')^2=(t'-\Sigma)a+a^2+\Sigma_-^2+4m^2N_{st}(t')^2\ec
\end{align}
where we have included the geometrical constraint on $a$ as discussed below equation~\eqref{bsminusplus}.
Defining again the highest lower and the lowest upper bound 
\begin{equation}
\tilde b_t^-(a)=\max_{t'>\tpi}b_t^-(t',a)\ec \qquad \tilde b_t^+(a)=\min_{t'>\tpi}b_t^+(t',a)
\end{equation}
as the maximum/minimum value of $b_t^{-/+}(s',a)$ within the integration range $t'>\tpi$, we can finally give the range of allowed values of $b$ for given $a$ by
\begin{equation}
\tilde b_t^-(a)\leq b\leq\tilde b_t^+(a) \qquad \forall\;t'>\tpi>a\ec
\end{equation}
for the $t$-channel parts of the HDRs.

\subsection[$s$-channel partial-wave projection]{\boldmath{$s$}-channel partial-wave projection}
\label{subsec:convergence:sprojection}

As mentioned before, it turns out that the constraints due to $\rho_{ut}$ and $\rho_{su}$ are equal to or weaker than the restrictions due to $\rho_{st}$.
Therefore, we only need to consider the corresponding constraints for the $s$-channel partial-wave projection of both the $s$-channel partial-wave expanded and the $t$-channel partial-wave expanded HDR parts.
However, the strategy to find the optimal value of $a$ and the corresponding range of convergence in $s$ is the same in both cases:
from the Lehmann ellipse constraint it follows that all allowed values of $b$ must obey\footnote{Note that the lower bounds coincide: $\tilde b_s^-(a)=\tilde b_t^-(a)$ for all $a<s_+$.}
\begin{equation}
\label{tildebstconstraint}
\tilde b_{s,t}^-(a)\leq b\leq\tilde b_{s,t}^+(a)\ec
\end{equation}
for all $s'>s_+$ and $t'>\tpi$, i.e.\ within the corresponding integration ranges, respectively.
The limits $-1\leq z_s\leq1$ of the scattering angle for the physical $s$-channel reaction translate into
\begin{equation}
-4q^2=-\frac{\lambda_s}{s}\leq t\leq0 \qquad \forall\;s>s_+\ec
\end{equation}
and hence for given $a<s_+<s$ the bounds on $b$ due to the $s$-channel partial-wave projection are given by (cf.~\eqref{bsminusplus})
\begin{align}
b_s^\text{min}(s,a)&\leq b\leq b_s^\text{max}(s,a) \qquad \forall\;s>s_+>a\ec\\
b_s^\text{min}(s,a)&=(s-a)(\Sigma-s-a)\ec \qquad
b_s^\text{max}(s,a)=(s-a)\Big(\Sigma-s+\frac{\lambda_s}{s}-a\Big)=(s-a)\bigg\{\frac{\Sigma_-^2}{s}-a\bigg\}\ep\nt
\end{align}
The maximally allowed value of $s$ for given $a$, $s_{s,t}^\text{max}(a)$, is then the largest value of $s$ such that for given $a$ both $b_s^\text{min}(s,a)$ and $b_s^\text{max}(s,a)$ lie within the ranges $\big[\tilde b_{s,t}^-(a),\tilde b_{s,t}^+(a)\big]$, respectively.
Equating the boundary values of $b$ from both the $s$- and $t$-channel partial-wave expansions and the $s$-channel partial-wave projection yields
\begin{align}
b_s^\text{min}(s,a)\overset{!}{=}\tilde b_{s,t}^-(a) \quad\Rightarrow\quad s=s_{s,t}^-(a)\ec\nt\\
b_s^\text{max}(s,a)\overset{!}{=}\tilde b_{s,t}^+(a) \quad\Rightarrow\quad s=s_{s,t}^+(a)\ec
\end{align}
where $s_{s,t}^-$ and $s_{s,t}^+$ denote the corresponding maximal solutions for given $a$, leads to two equations for the two wanted unknowns $\tilde s_{s,t}^\text{max}$ and $\tilde a_{s,t}^s$ defined by
\begin{equation}
\tilde s_{s,t}^\text{max}=\max_{a<s_+} s_{s,t}^\text{max}(a)=s_{s,t}^\text{max}(\tilde a_{s,t}^s)\ep
\end{equation}
Explicitly, they follow from equating the maximal solutions
\begin{align}
s_{s,t}^-(a)&\overset{!}{=}s_{s,t}^+(a) \quad\Rightarrow\quad
 a=\tilde a_{s,t}^s\ec \quad s_{s,t}^-(\tilde a_{s,t}^s)=s_{s,t}^+(\tilde a_{s,t}^s)=\tilde s_{s,t}^\text{max}\ec\nt\\
s_{s,t}^\pm(a)&=\max\Big\{s_{s,t}^{\pm(-)}(a),s_{s,t}^{\pm(+)}(a)\Big\}\ec\nt\\
s_{s,t}^{-(\pm)}(a)&=\frac{\Sigma}{2}\pm\sqrt{\left(\frac{\Sigma}{2}-a\right)^2-\tilde b_{s,t}^-(a)}\ec\nt\\
s_{s,t}^{+(\pm)}(a)&=\frac{1}{2a}\Bigg\{\Big[a^2+\Sigma_-^2-\tilde b_{s,t}^+(a)\Big]\pm\sqrt{\Big[a^2+\Sigma_-^2-\tilde b_{s,t}^+(a)\Big]^2-4a^2\Sigma_-^2}\Bigg\}\ec
\end{align}
where for $s>s_+>\Sigma/2$ we have $s_{s,t}^-=s_{s,t}^{-(+)}$ and for in addition e.g.\ $a<0$ we have $s_{s,t}^+=s_{s,t}^{+(-)}$.
The maximum value of the two other (i.e.\ minimal) solutions for $\tilde a_{s,t}$ then yields the highest lower bound on $s$ and thus we can write
\begin{equation}
\tilde s_{s,t}^\text{min}=\max\Big\{s_+,s_{s,t}^{-(-)}(\tilde a_{s,t}^s),s_{s,t}^{+(+)}(\tilde a_{s,t}^s)\Big\} \qquad\text{for }s>s_+\text{ and }a<0\ep
\end{equation}

For the $s$-channel parts, solving the equations numerically for all allowed $a<s_+$ leads to the following optimal value of $a$ and corresponding range of convergence in $s>s_+=59.64\,\mpi^2=(1.08\GeV)^2$ 
\begin{equation}
\tilde a_s^s=-128.30\,\mpi^2\ec \quad s_+<s<\tilde s_s^\text{max}=106.09\,\mpi^2\ec \quad \tilde b_s^-(\tilde a_s^s)=26860\,\mpi^4\ec \quad \tilde b_s^+(\tilde a_s^s)=34388\,\mpi^4\ec
\end{equation}
in agreement with the unpublished Appendix~E of~\cite{HiteSteiner}.\footnote{Appendix~E of~\cite{HiteSteiner} deals with finding the optimal values for the $s$-channel partial-wave projection of the $s$-channel partial-wave expanded absorptive parts of the HDRs only and follows a similar scheme.
The quoted results are $\tilde s_s^\text{max}\gtrsim105\,\mpi^2$ for $\tilde a_s^s\approx-117\,\mpi^2$.}

For the $t$-channel parts, this procedure results in
\begin{equation}
\label{schannelrange}
\tilde a_t^s=-23.19\,\mpi^2\ec \quad s_+<s<\tilde s_t^\text{max}=97.30\,\mpi^2\ec \quad \tilde b_t^-(\tilde a_t^s)=2202\,\mpi^4\ec \quad \tilde b_t^+(\tilde a_t^s)=5212\,\mpi^4\ep
\end{equation}

\begin{figure}[t!]
\centering
\includegraphics[width=0.49\linewidth,clip]{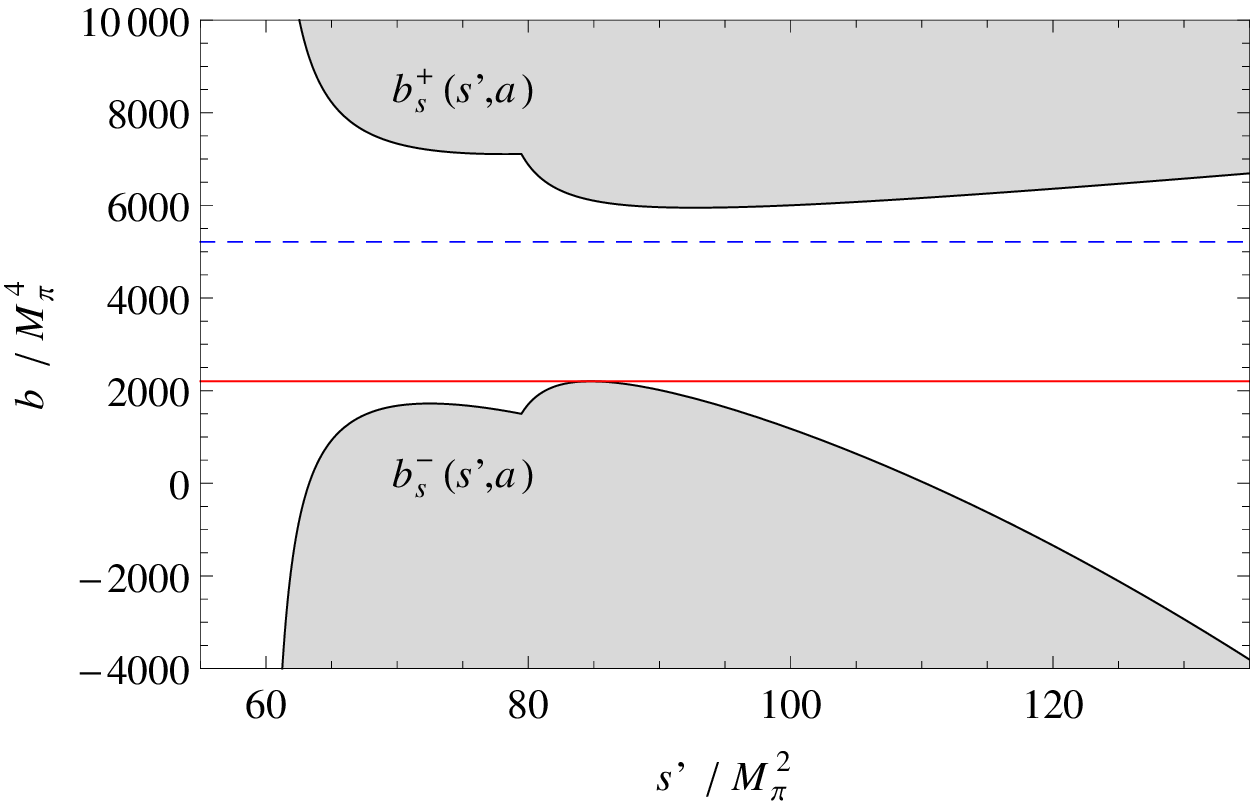}
\includegraphics[width=0.49\linewidth,clip]{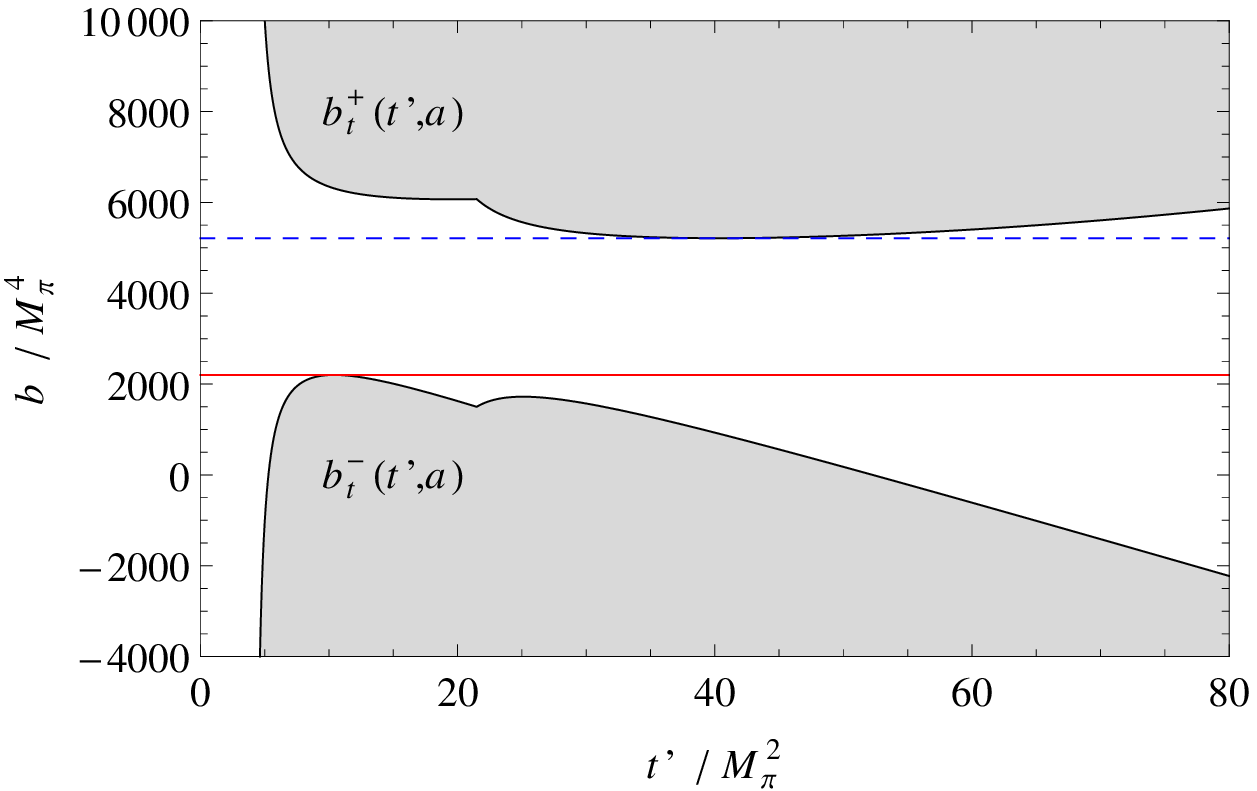}
\caption[Allowed ranges of $b$ for $s$-channel partial-wave projection.]{Allowed ranges of $b$ for $s$-channel partial-wave projection with $a=\tilde a_t^s=-23.19\,\mpi^2$ for $s$-channel (left) and $t$-channel (right) partial-wave expansion. Horizontal lines correspond to $\tilde b_t^-(a)=2202\,\mpi^4$ (solid) and $\tilde b_t^+(a)=5212\,\mpi^4$ (dashed).}
\label{fig:bstplusminus}
\end{figure}

In conclusion, the $s$-channel constraints are weaker than the $t$-channel ones, which can also be deduced from Fig.~\ref{fig:bstplusminus}, where the situation for $\tilde a_t^s=-23.19\,\mpi^2$ is shown:
for this $a$ the range of $b$ limited by $\tilde b_t^-(\tilde a_t^s)$ and $\tilde b_t^+(\tilde a_t^s)$ for the $t$-channel partial-wave expansion also lies within the allowed range of $b$ for the $s$-channel partial-wave expansion, and hence for the interval of $s$ given in~\eqref{schannelrange} this range of $b$ covers the interval $\big[b_s^\text{min}(s,\tilde a_t^s),b_s^\text{max}(s,\tilde a_t^s)\big]$ that is needed for the $s$-channel partial-wave projection.
By construction, the resulting family of hyperbolae does cross neither any double spectral region nor their $z'$-reflected boundaries as depicted in Fig.~\ref{fig:hyperbolae}(left), and thus~\eqref{schannelrange} corresponding to $\sqrt{\tilde s^\text{max}}=\sqrt{97.30}\,\mpi=1.38\GeV$ constitutes the result for the $s$-channel partial-wave projection, in agreement with~\cite{Hoehler}.\footnote{As combined result for both $s$- and $t$-channel contributions, the numbers $\tilde s^\text{max}=97\,\mpi^2$ for $\tilde a^s=-23\,\mpi^2$ are quoted in~\cite{Hoehler} without further explanation and giving only a vague reference for these numerical values, which is most probably meant to be~\cite{Stahov:Diss}. However, roughly the same numbers are also given more recently in~\cite{Stahov:2002}.}

\subsection[$t$-channel partial-wave projection]{\boldmath{$t$}-channel partial-wave projection}
\label{subsec:convergence:tprojection}

\begin{figure}[t!]
\centering
\includegraphics[width=0.49\linewidth,clip]{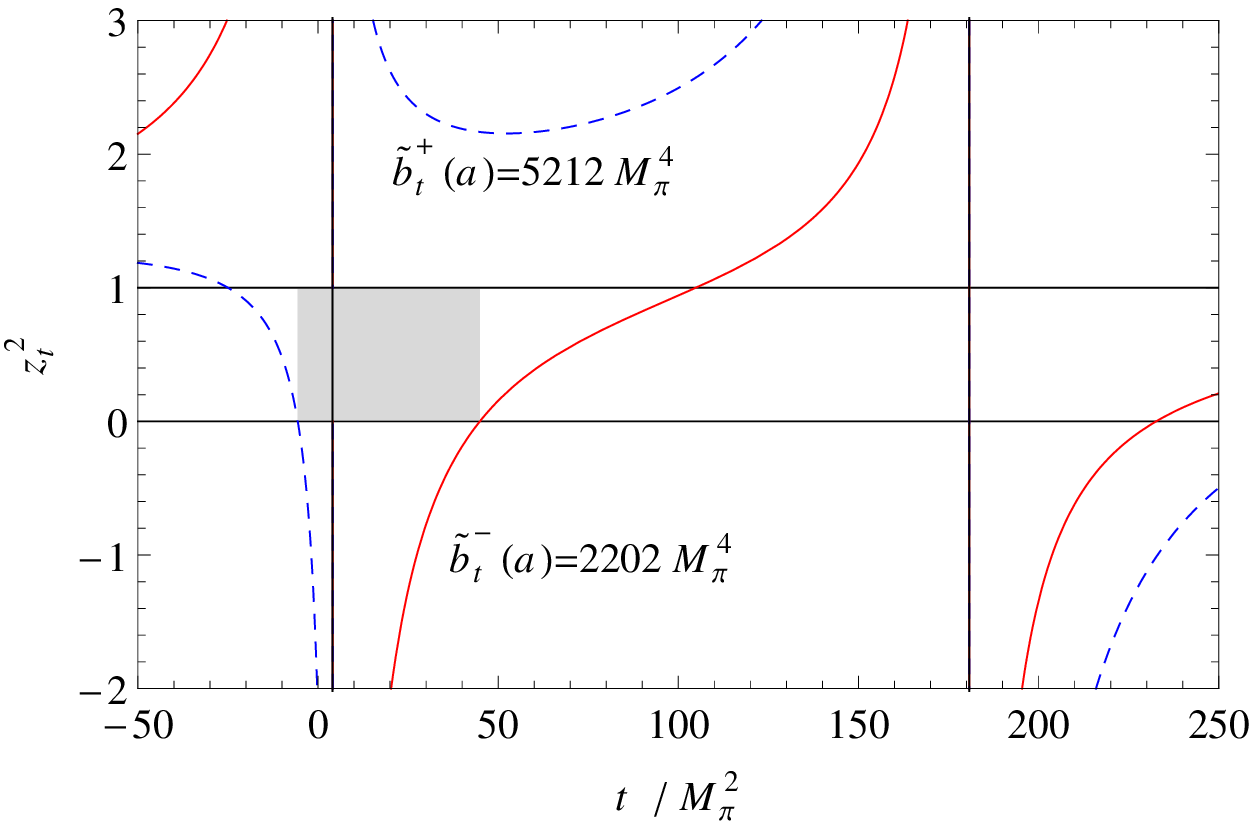}
\includegraphics[width=0.49\linewidth,clip]{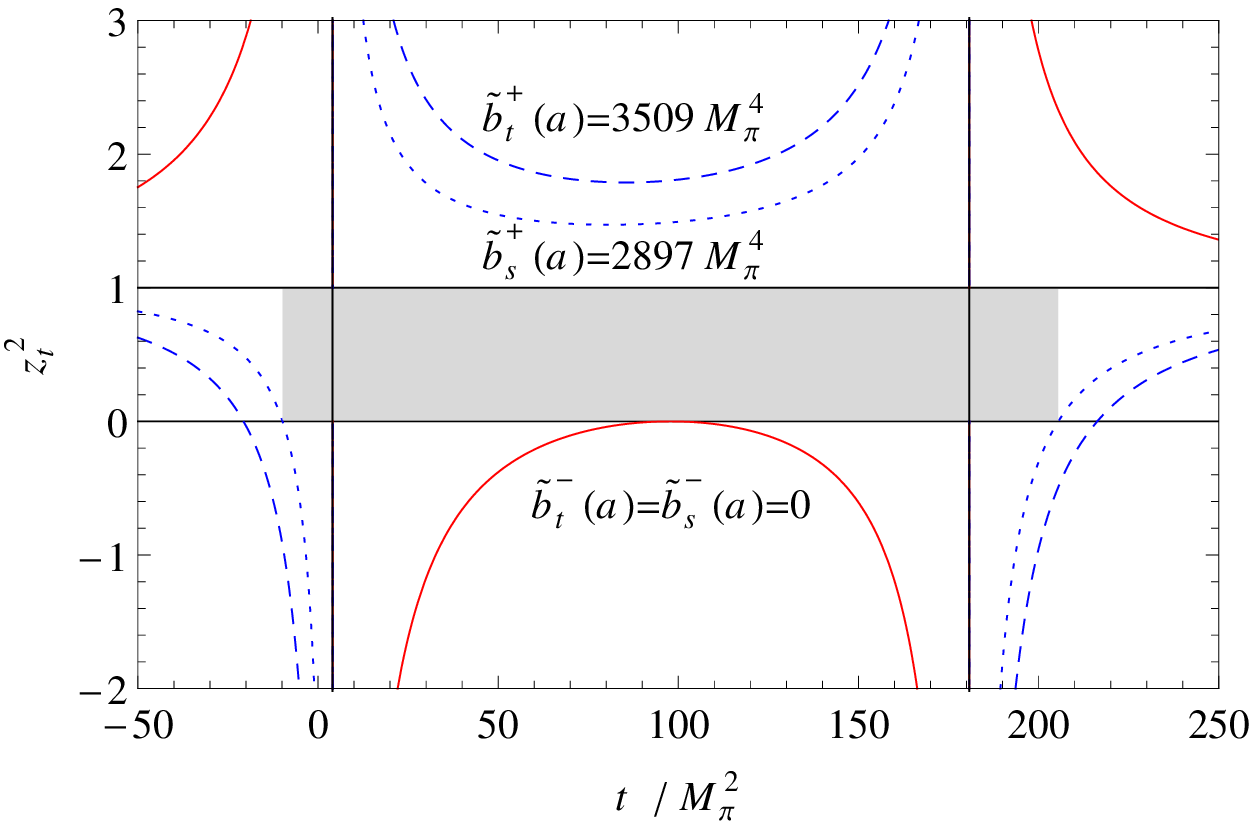}
\caption[Ranges of convergence in $t$ for $t$-channel partial-wave projection.]{Ranges of convergence in $t$ for $t$-channel partial-wave projection from full coverage (shaded area) of the physical band $0\leq z_t^2\leq1$ for $a=\tilde a_t^s=-23.19\,\mpi^2$ (left) and $a=\tilde a_{s,t}^t=-2.71\,\mpi^2$ (right). Vertical lines indicate thresholds $\tpi$ and $\tN$.}
\label{fig:zt2st}
\end{figure}

\begin{figure}[t!]
\centering
\includegraphics[width=0.49\linewidth,clip]{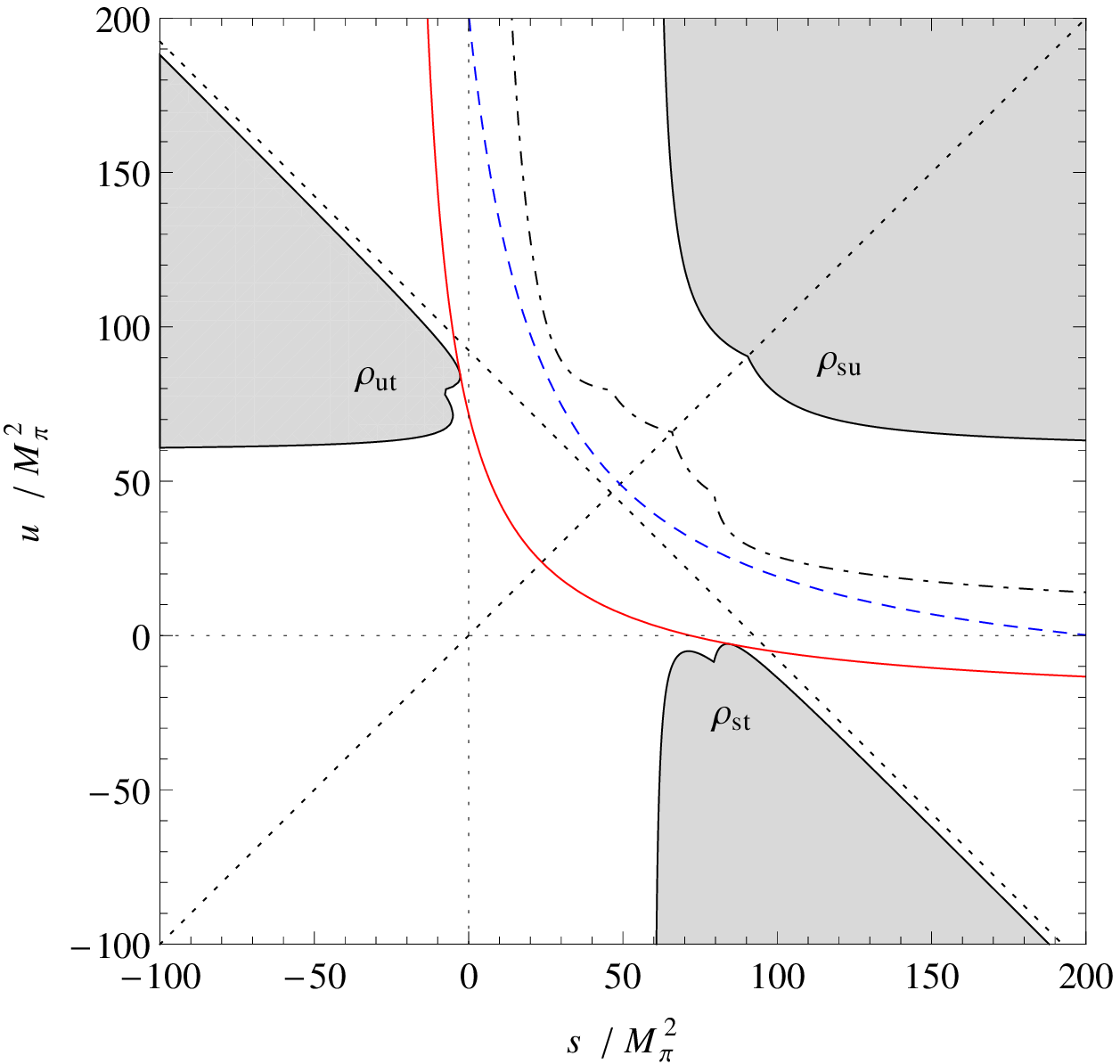}
\includegraphics[width=0.49\linewidth,clip]{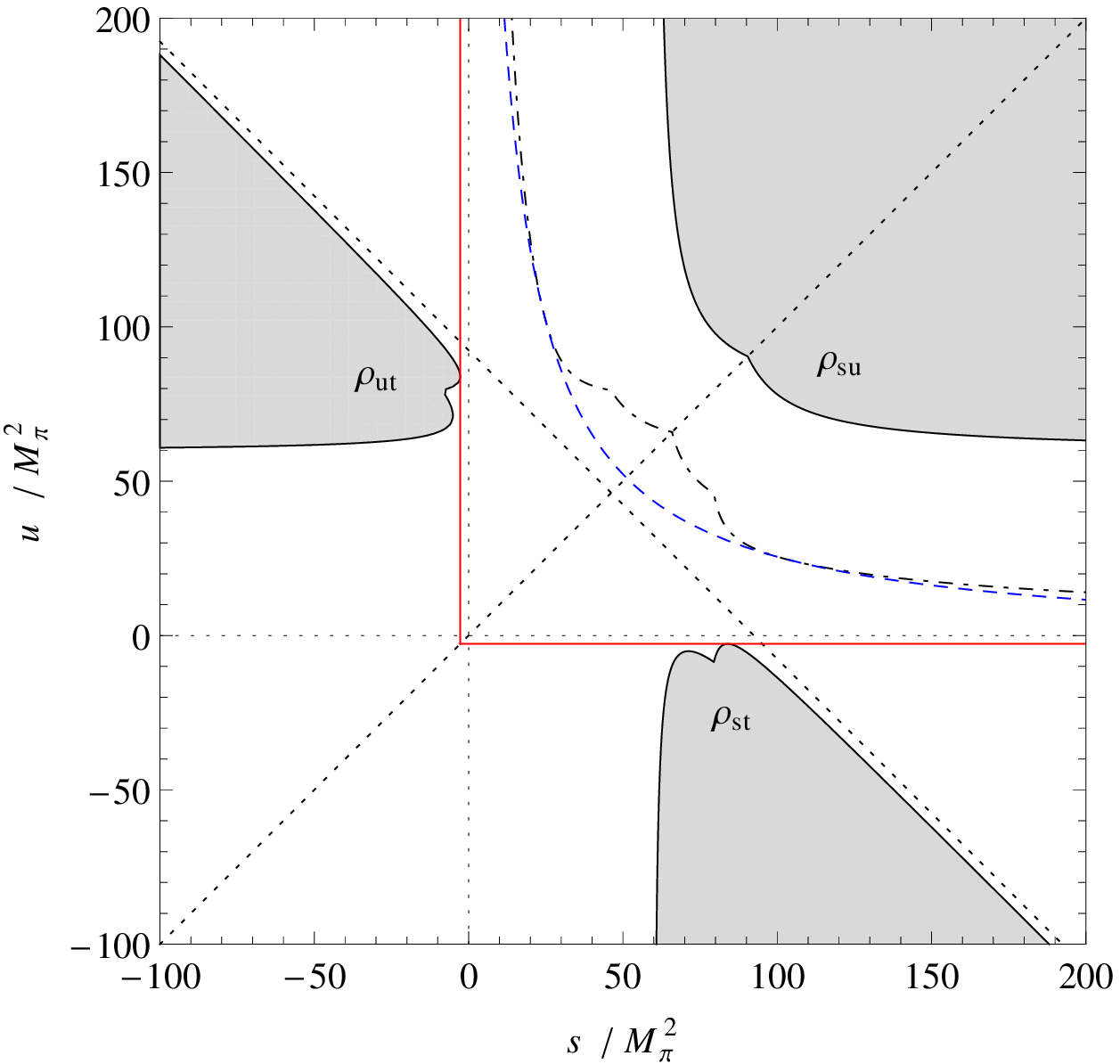}
\caption[Double spectral regions and limiting hyperbolae for $s$- and $t$-channel partial-wave projection]{Double spectral regions and limiting hyperbolae for $s$- and $t$-channel partial-wave projection. Left: for $a=\tilde a_t^s=-23.19\,\mpi^2$ with $\tilde b_t^-(a)=2202\,\mpi^4$ (solid) and $\tilde b_t^+(a)=5212\,\mpi^4$ (dashed). Right: for $a=\tilde a_{s,t}^t=-2.71\,\mpi^2$ with $\tilde b_{s,t}^-(a)=0$ (solid) and $\tilde b_s^+(a)=2897\,\mpi^4$ (dashed).}
\label{fig:hyperbolae}
\end{figure}

The relation between the range of $b$ permitted by the Lehmann ellipse constraint~\eqref{tildebstconstraint} and the corresponding range of convergence in $t$ for the projection of the HDR equations onto $t$-channel partial waves for given $a$ is most easily established on the basis of the squared $t$-channel scattering angle $z_t^2$, which must cover the range 
\begin{equation}
\label{zt2constraint}
0\leq z_t^2(t,a,b)=\frac{(t-\Sigma+2a)^2-4b}{16p_t^2q_t^2}=\frac{(t-\Sigma+2a)^2-4b}{(t-\tpi)(t-\tN)}\leq1
\end{equation}
for both the $s$-channel and $t$-channel partial-wave expanded parts, since (as discussed after the $t$-channel partial-wave projection formulae~\eqref{tprojform}) the integrands are always functions of the real square $z_t^2$ even between the thresholds $\tpi$ and $\tN$. Furthermore, $0\leq z_t^2\leq1$ is not only a necessary condition for $0\leq z_t\leq1$ but also equivalent to $-1\leq z_t\leq1$, which in turn is already sufficient to perform the partial-wave projections in our case (cf.\ the discussion in Appendix~\ref{sec:tpwp}).
Therefore, the range~\eqref{zt2constraint} of $z_t^2$ constitutes the necessary and sufficient condition not only for the physical region $t>\tN$, but for all kinematical regions.
Obviously, for $\tpi<t<\tN$ and given $a$, $z_t^2$ can only be non-negative for $b$ non-negative and large enough.
Translating~\eqref{zt2constraint} into ranges for $b$ while taking care of the signs of $p_t^2$ and $q_t^2$ in the different kinematical regions yields (cf.~\eqref{btminusplus})
\begin{align}
b_t^\text{min}(t,a)&\leq b\leq b_t^\text{max}(t,a) \qquad \forall\;\tpi<t<\tN\ec\\
b_t^\text{max}(t,a)&\leq b\leq b_t^\text{min}(t,a) \qquad \,\forall\;t>\tN\text{ (or }t<\tpi)\ec\nt\\
b_t^\text{min}(t,a)&=\frac{1}{4}(t-\Sigma+2a)^2\geq0\ec \qquad
b_t^\text{max}(t,a)=\frac{1}{4}(t-\Sigma+2a)^2-4p_t^2q_t^2=(t-\Sigma)a+a^2+\Sigma_-^2\ec\nt
\end{align}
where the superscripts min/max refer to both the (at least partially) unphysical kinematical range $t>\tpi$ needed in our RS system as well as the corresponding min/max values $0/1$ of $z_t^2$.
Solving these equations for $t$ yields (cf.\ $t_{(\pm)}(\nu=0;a,b)$ of~\eqref{sunuoftab})
\begin{equation}
\label{t0zt2pmt1zt2}
t_0^{(\pm)}(a,b_t^\text{min})=\Sigma-2a\pm2\sqrt{b_t^\text{min}}\ec \qquad
t_1(a,b_t^\text{max})=\Sigma-a+\frac{1}{a}\Big[b_t^\text{max}-\Sigma_-^2\Big]\ec
\end{equation}
and the range of convergence in $t$ for given $a$ is the kinematical range in which all values between $b_t^\text{min}(t,a)$ and $b_t^\text{max}(t,a)$ are covered by both intervals $[\tilde b_{s,t}^-(a),\tilde b_{s,t}^+(a)]$.
Between the thresholds (i.e.\ for $\tpi<t<\tN$) this amounts to the conditions $\tilde b_{s,t}^-(a)\leq b_t^\text{min}(t,a)$ and $b_t^\text{max}(t,a)\leq\tilde b_{s,t}^+(a)$, while below or above the thresholds (i.e.\ for $t<\tpi$ or $\tN<t$) we have $\tilde b_{s,t}^-(a)\leq b_t^\text{max}(t,a)$ and $b_t^\text{min}(t,a)\leq\tilde b_{s,t}^+(a)$.\footnote{Accordingly, at the thresholds the respective min/max values are identical: $b_t^\text{min}(\tpi,a)=b_t^\text{max}(\tpi,a)=\big(a-\Sigma_-\big)^2\geq0$ and $b_t^\text{min}(\tN,a)=b_t^\text{max}(\tN,a)=\big(a+\Sigma_-\big)^2\geq0$.}
Equivalently, we can demand that for given $a$ the band $0\leq z_t^2(t,a,b)\leq1$ must be fully covered by the area between $z_t^2(t,a,\tilde b_{s,t}^-(a))$ and $z_t^2(t,a,\tilde b_{s,t}^+(a))$ in order to determine the range of validity in $t$.
The situation that results from using the set~\eqref{schannelrange} of optimal parameters for the $s$-channel partial-wave projection derived in the previous section is shown in Fig.~\ref{fig:zt2st}(left):
the $t$-channel projection is then valid for $-5.63\,\mpi^2<t<44.92\,\mpi^2$ (denoted by the shaded area of coverage), and the reason for this rather low upper bound on $t$ is that the curve for $\tilde b_t^-=2202\,\mpi^4$ changes sign between the thresholds and thus enters the critical band $0\leq z_t^2\leq1$, which is hence no longer fully covered by the allowed area.
Indeed, the range of convergence can be significantly improved if $z_t^2(t,a,\tilde b_{s,t}^-(a))\leq0$ (and of course also $z_t^2(t,a,\tilde b_{s,t}^+(a))\geq1$) for all $t$ between the thresholds.
From~\eqref{zt2constraint} it is clear that for $t\in(\tpi,\tN)$ we have $z_t^2(t,a,b)\leq0$ if and only if $b\leq(t-\Sigma+2a)^2/4$, such that the curves for the lower limits $\tilde b_{s,t}^-(a)$ of $b$ will be tangent to the zero axis provided that $\tilde b_{s,t}^-(a)=0$.
Solving this numerically yields
\begin{equation}
\tilde b_s^-(a)=\tilde b_t^-(a)\overset{!}{=}0 \quad\Rightarrow\quad a=\tilde a_{s,t}^t=-2.71\,\mpi^2\ec
\end{equation}
which is unambiguous since it turns out that $\tilde b_{s,t}^-(a)>0$ for $a<\tilde a_{s,t}^t$ as well as $\tilde b_{s,t}^-(a)<0$ for $\tilde a_{s,t}^t<a<s_+$ (where we have used the numerical equality of the lower bounds for both $s$- and $t$-channel partial-wave expansion).
Furthermore, the curves for $\tilde b_{s,t}^+(a)$ start to enter the critical band due to change of sign at $\tN$ for $a>2.58\,\mpi^2$ and $a>9.17\,\mpi^2$, respectively (however, the geometrical constraint $a<\tpi$ is partially tighter anyway).
Thus, $\tilde a_{s,t}^t$ is the smallest value of $a$ such that the critical band is fully covered between the thresholds, which is shown in Fig.~\ref{fig:zt2st}(right).
From this figure and equation~\eqref{t0zt2pmt1zt2} it is clear that in this case we can deduce the corresponding upper and lower bounds $t_{s,t}^\text{min}(a)$ and $t_{s,t}^\text{max}(a)$ on $t$ by the intercepts $t_0^{(\pm)}(a,\tilde b_{s,t}^+(a))$ of $z_t^2(t,a,\tilde b_{s,t}^+(a))$ with the zero axis below and above the thresholds, respectively.
Since moreover both $t_{s,t}^\text{max}(a)=t_0^{(+)}(a,\tilde b_{s,t}^+(a))$ are strictly decreasing in the allowed ranges of $a$, the minimal allowed value $a=\tilde a_{s,t}^t$ is also the optimal one yielding $\tilde t_{s,t}^\text{min}=t_0^{(-)}(\tilde a_{s,t}^t,\tilde b_{s,t}^+(\tilde a_{s,t}^t))$ and  $\tilde t_{s,t}^\text{max}=t_0^{(+)}(\tilde a_{s,t}^t,\tilde b_{s,t}^+(\tilde a_{s,t}^t))$.
This procedure results in
\begin{align}
\tilde b_s^+(\tilde a_{s,t}^t)&=2897\,\mpi^4 \quad\Rightarrow\quad \ \,-9.84\,\mpi^2\leq t\leq205.45\,\mpi^2\ec\nt\\
\tilde b_t^+(\tilde a_{s,t}^t)&=3509\,\mpi^4 \quad\Rightarrow\quad -20.67\,\mpi^2\leq t\leq216.28\,\mpi^2\ec
\end{align}
where the $s$-channel Lehmann ellipse constraint proves slightly more restrictive,
and thus the final result for the $t$-channel partial-wave projection reads
\begin{equation}
\label{tchannelrange}
\tilde a_{s,t}^t=-2.71\,\mpi^2\ec \quad \tpi<t<\tilde t_s^\text{max}=205.45\,\mpi^2\ec \quad \tilde b_{s,t}^-(\tilde a_{s,t}^t)=0\ec \quad \tilde b_s^+(\tilde a_{s,t}^t)=2897\,\mpi^4\ec
\end{equation}
which corresponds to $\sqrt{\tilde t^\text{max}}=\sqrt{205.45}\,\mpi=2.00\GeV$.
Again, ascertaining that the resulting family of hyperbolae does enter neither any double spectral region nor their $z'$-reflected boundaries, which is shown in Fig.~\ref{fig:hyperbolae}(right), completes the derivation of the final result~\eqref{tchannelrange} for the $t$-channel partial-wave projection.
It is interesting to note that the domain of validity in $t$ is much bigger as the one in $s$, which is reflected by the possibility to use only the positive half $0\leq z_t\leq1$ of the range of the scattering angle due to Bose symmetry in the $t$-channel; in particular the range of convergence connects the physical regions for the $s$- and $u$-channel reactions, where $t\leq0$, with the $t$-channel physical region $t\geq\tN$.
                                                                                                                                                                                                                                                                                      
The complicated interplay between $a$, $\tilde b_{s,t}^\pm(a)$, and $z_t^2(t,a,b)$ in the different kinematical regions is the reason why it is not possible to treat the $t$-channel projection in analogy to the $s$-channel projection in the previous section:
equating again the corresponding boundary values of $b$ from both the $s$- and $t$-channel partial-wave expansions and the $t$-channel partial-wave projection, and subsequently equating the corresponding maximal solutions in order to obtain $\tilde t_{s,t}^\text{max}$ as the maximal upper limit on $t$ for $t>\tN$ leads to entering or even crossing the critical band between the thresholds.

\section{Asymptotic regions and Regge theory}
\label{sec:regge}

The asymptotic $s$- and $t$-channel contributions of the HDRs~\eqref{hdr} to the invariant amplitudes are defined by splitting the corresponding integration ranges $s_+\leq s'\leq\infty$ and $\tpi\leq t'\leq\infty$ at some appropriate values $\sa=\Wa^2$ and $\ta$, respectively, which yields the following asymptotic contributions
\begin{align}
\label{hdrasym}
\asym{A^+}(s,t)&=\frac{1}{\pi}\int\limits^\infty_{\sa}\diff s'\bigg[\frac{1}{s'-s}+\frac{1}{s'-u}-\frac{1}{s'-a}\bigg]\Im A^+(s',z_s')
+\frac{1}{\pi}\int\limits_{\ta}^\infty\diff t'\frac{\Im A^+(t',z_t')}{t'-t}\ec\nt\\
\asym{A^-}(s,t)&=\frac{1}{\pi}\int\limits^\infty_{\sa}\diff s'\bigg[\frac{1}{s'-s}-\frac{1}{s'-u}\bigg]\Im A^-(s',z_s')
+\frac{1}{\pi}\int\limits_{\ta}^\infty\diff t'\frac{s-u}{s'-u'}\frac{\Im A^-(t',z_t')}{t'-t}\ec\nt\\
\asym{B^+}(s,t)&=\frac{1}{\pi}\int\limits^\infty_{\sa}\diff s'\bigg[\frac{1}{s'-s}-\frac{1}{s'-u}\bigg]\Im B^+(s',z_s')
+\frac{1}{\pi}\int\limits_{\ta}^\infty\diff t'\frac{s-u}{s'-u'}\frac{\Im B^+(t',z_t')}{t'-t}\ec\nt\\
\asym{B^-}(s,t)&=\frac{1}{\pi}\int\limits^\infty_{\sa}\diff s'\bigg[\frac{1}{s'-s}+\frac{1}{s'-u}-\frac{1}{s'-a}\bigg]\Im B^-(s',z_s')
+\frac{1}{\pi}\int\limits_{\ta}^\infty\diff t'\frac{\Im B^-(t',z_t')}{t'-t}\ec
\end{align}
and the remaining non-asymptotic parts are given by the corresponding integrals over $s_+\leq s'\leq\sa$ and $\tpi\leq t'\leq\ta$, respectively, plus the nucleon pole terms $N^I(s,t)$ for the amplitudes $B^I(s,t)$.
The internal (primed) kinematics are given by (cf.\ Sect.~\ref{subsec:preliminaries:hdrs} and especially~\eqref{sunuoftab})
\begin{align}
s'(t';a,b)&=\frac{1}{2}\Big(\Sigma-t'+\sqrt{(t'-\Sigma+2a)^2-4b}\Big)\ec \qquad t'(s';a,b)=-\frac{b}{s'-a}+\Sigma-s'-a\ec\nt\\
u'(t';a,b)&=\frac{1}{2}\Big(\Sigma-t'-\sqrt{(t'-\Sigma+2a)^2-4b}\Big)\ec
\end{align}
where the parameter $b$ is fixed by the external (unprimed) kinematics as
\begin{equation}
(s-a)(\Sigma-s-t-a)=b=(s'-a)(u'-a)\ec
\end{equation}
such that
\begin{equation}
s'(u';a,b)=\frac{b}{u'-a}+a\ec \qquad u'(s';a,b)=\frac{b}{s'-a}+a\ep
\end{equation}
Thus (for given $a$ and finite $b$), for the $s$-channel integrals we need the asymptotic behavior in the limit
\begin{equation}
\label{sasym}
s'\rightarrow\infty \quad\Rightarrow\quad t'\rightarrow-\infty\ec \quad u'\rightarrow a\ec
\end{equation}
while in the $t$-channel integrals the asymptotic behavior is determined by 
\begin{equation}
\label{tasym}
t'\rightarrow\infty \quad\Rightarrow\quad u'\rightarrow-\infty\ec \quad s'\rightarrow a\ep
\end{equation}

From~\eqref{hdrasym} the asymptotic parts of the $s$- and $t$-channel partial waves may then be deduced by the projection formulae~\eqref{sproj} and~\eqref{tproj} as
\begin{align}
\label{pwprojasym}
\asym{f_{l+}^I}(W)&=\int\limits^1_{-1}\diff z_s\Big\{R_{l,l+1}^1(W,z_s)\asym{A^I}(W,z_s)+R_{l,l+1}^2(W,z_s)\asym{B^I}(W,z_s)\Big\}\ec\nt\\
\asym{f_{(l+1)-}^I}(W)&=\int\limits^1_{-1}\diff z_s\Big\{R_{l+1,l}^1(W,z_s)\asym{A^I}(W,z_s)+R_{l+1,l}^2(W,z_s)\asym{B^I}(W,z_s)\Big\}\ec\nt\\
\asym{f_+^J}(t)&=\tilde\zeta_J(t)\int\limits_0^1\diff z_t\Big\{\tilde u_J(t,z_t)\asym{A^I}(t,z_t)+\tilde v_J(t,z_t)\asym{B^I}(t,z_t)\Big\}\ec\nt\\
\asym{f_-^J}(t)&=\tilde\zeta_J(t)\int\limits_0^1\diff z_t\,\tilde w_J(t,z_t)\asym{B^I}(t,z_t)\ec
\end{align}
where for the $t$-channel partial waves we have again $I=+/-$ for even/odd $J$.
Note that for these asymptotic contributions we do not expand the absorptive parts inside the integrals in order to take into account the high-energy behavior of the full invariant amplitudes as given by Regge theory~\cite{Regge}.
Therefore, also for the so-called driving terms (i.e.\ the sums of all higher partial waves that are not taken into account explicitly~\cite{ACGL,piK:RS} as well as the asymptotic contributions of the lower partial waves treated dynamically) the integration ranges are limited by $\sa$ and $\ta$ in order to avoid double counting of the asymptotic regions.
This procedure follows~\cite{ACGL,piK:RS}, motivated by the observations that,
first, for higher and higher energies one would be forced to explicitly use higher and higher partial waves as well in order to ensure the validity of the partial-wave expansion, and second, no available information in the asymptotic regime is lost without need.
In Sect.~\ref{subsec:piNMO:input} we explicitly demonstrate the matching of the Regge model to truncated sums of the lowest partial waves with $l\leq l_\text{max}$ for $l_\text{max}\in\{3,4,5\}$.

In the following, the contributions from the asymptotic regions in both channels will be examined in the framework of Regge theory. For a general introduction see e.g.~\cite{Collins}.

\subsection[$s$-channel asymptotics]{\boldmath{$s$}-channel asymptotics}
\label{subsec:regge:s}

First of all, contributions from $t$-channel Regge trajectories, i.e.\ the leading Pomeron ($I_t=0$) trajectory $\alpha_P(t')\approx\alpha^{(0)}_P=1$ (roughly independent of $t'$ but with exponential residue function $\beta_P(t')=\sigma_P\exp\frac{b_Pt'}{2}$, where $\sigma_P$ represents the asymptotic total-cross-section value for $\pi\pi$ scattering and $b_P$ is the width of the diffraction peak, cf.~\cite{ACGL,piK:RS}) as well as the $\rho$ ($I_t=1$) and $f$ ($I_t=0$) trajectories $\alpha_\rho(t')=\alpha^{(0)}_\rho+\alpha^{(1)}_\rho t'$ (and $\alpha_f(t')$ in analogy) should be negligible, since due to~\eqref{sasym} they will behave as
\begin{equation}
\Im\Amp(s',t')\sim\beta_P(t')s'^{\alpha_P(t')}\sim e^{\frac{b_Pt'}{2}}s'\sim e^{-s'}s'\ec \qquad 
\Im\Amp(s',t')\sim\beta_\rho(t')s'^{\alpha_\rho(t')}\sim s'^{\alpha^{(0)}_\rho+t'\alpha^{(1)}_\rho}\sim s'^{-s'}
\end{equation}
for $s'\to\infty$, leading to an exponential suppression.

Let us briefly review the $u$-channel-exchange contributions to the $s$-channel reactions of backward $\pi N$ scattering as discussed in~\cite{piNuRegge}.
The invariant amplitudes can be parameterized according to (cf.\ also~\cite{Hoehler})
\begin{align}
A(s',u')=\sum\limits_i\frac{\beta^A_i(u')\zeta_i(u')}{\Gamma\big(\alpha_i(u')-\frac{1}{2}\big)}\bigg(\frac{s'}{s_R}\bigg)^{\alpha_i(u')-\frac{1}{2}}\ec \qquad
B(s',u')=\sum\limits_i\frac{\beta^B_i(u')\zeta_i(u')}{\Gamma\big(\alpha_i(u')-\frac{1}{2}\big)}\bigg(\frac{s'}{s_R}\bigg)^{\alpha_i(u')-\frac{1}{2}}\ec
\end{align}
where both sums run over the four trajectories $i\in\{N_\alpha,N_\gamma,\Delta_\delta,\Delta_\beta\}$, and the Regge propagators $\zeta_i(u')$ are given by
\begin{equation}
\zeta_i(u')=\frac{1+\mathcal{S}_i\,\exp\big(-i\pi\big[\alpha_i(u')-\frac{1}{2}\big]\big)}{\sin\big(\pi\big[\alpha_i(u')-\frac{1}{2}\big]\big)}\ep
\end{equation}
Besides the scaling factor $s_R=1\GeV^2$, the following Regge residues $\beta_i^{A/B}(u')$ and Regge trajectories $\alpha_i(u')$ are employed:
\begin{equation}
\beta_i^A(u')=a_i+b_iu'\ec \qquad \beta_i^B(u')=c_i+d_iu'\ec \qquad \alpha_i(u')=\alpha^{(0)}_i+\alpha'u'\ec
\end{equation}
i.e.\ both the residues and the trajectories are linearly parameterized, and for the latter an identical slope $\alpha^{(1)}_i=\alpha'$ is used for all $i$.
The signature $\mathcal{S}_i=(-1)^{J_i-\frac{1}{2}}$ of the trajectory $i$ is positive for $N_\alpha$ and $\Delta_\beta$ and negative for $N_\gamma$ and $\Delta_\delta$.
Since $\Im\zeta_i(u')=-\mathcal{S}_i$, we may conclude that the imaginary parts of the invariant amplitudes in the $u$-channel isospin basis $I_u\in\{1/2=N,3/2=\Delta\}$ can be written as
\begin{equation}
\Im A^N(s',u')=\sum\limits_{i\in\{N_\alpha,N_\gamma\}}\tilde\beta_i^A(u')\bigg(\frac{s'}{s_R}\bigg)^{\alpha_i(u')-\frac{1}{2}}\ec \qquad 
\Im A^\Delta(s',u')=\sum\limits_{i\in\{\Delta_\delta,\Delta_\beta\}}\tilde\beta_i^A(u')\bigg(\frac{s'}{s_R}\bigg)^{\alpha_i(u')-\frac{1}{2}}\ec
\end{equation}
with the abbreviations
\begin{equation}
\tilde\beta_i^A(u')=-\frac{\mathcal{S}_i\beta_i^A(u')}{\Gamma\big(\alpha_i(u')-\frac{1}{2}\big)}\ec
\end{equation}
and analogously for the $B$ amplitudes.
Using now the isospin crossing relations~\eqref{sanduchannelcrossing}, we finally obtain the absorptive parts
{\allowdisplaybreaks
\begin{align}
\Im A^+(s',u'(s',t'))&=+\frac{1}{3}\sum\limits_{i\in\{N_\alpha,N_\gamma\}}\tilde\beta_i^A(u')\bigg(\frac{s'}{s_R}\bigg)^{\alpha_i(u')-\frac{1}{2}}
+\frac{2}{3}\sum\limits_{i\in\{\Delta_\delta,\Delta_\beta\}}\tilde\beta_i^A(u')\bigg(\frac{s'}{s_R}\bigg)^{\alpha_i(u')-\frac{1}{2}}\ec\nt\\
\Im A^-(s',u'(s',t'))&=-\frac{1}{3}\sum\limits_{i\in\{N_\alpha,N_\gamma\}}\tilde\beta_i^A(u')\bigg(\frac{s'}{s_R}\bigg)^{\alpha_i(u')-\frac{1}{2}}
+\frac{1}{3}\sum\limits_{i\in\{\Delta_\delta,\Delta_\beta\}}\tilde\beta_i^A(u')\bigg(\frac{s'}{s_R}\bigg)^{\alpha_i(u')-\frac{1}{2}}\ec\nt\\
\Im B^+(s',u'(s',t'))&=+\frac{1}{3}\sum\limits_{i\in\{N_\alpha,N_\gamma\}}\tilde\beta_i^B(u')\bigg(\frac{s'}{s_R}\bigg)^{\alpha_i(u')-\frac{1}{2}}
+\frac{2}{3}\sum\limits_{i\in\{\Delta_\delta,\Delta_\beta\}}\tilde\beta_i^B(u')\bigg(\frac{s'}{s_R}\bigg)^{\alpha_i(u')-\frac{1}{2}}\ec\nt\\
\Im B^-(s',u'(s',t'))&=-\frac{1}{3}\sum\limits_{i\in\{N_\alpha,N_\gamma\}}\tilde\beta_i^B(u')\bigg(\frac{s'}{s_R}\bigg)^{\alpha_i(u')-\frac{1}{2}}
+\frac{1}{3}\sum\limits_{i\in\{\Delta_\delta,\Delta_\beta\}}\tilde\beta_i^B(u')\bigg(\frac{s'}{s_R}\bigg)^{\alpha_i(u')-\frac{1}{2}}\ec
\end{align}}\noindent
where the dependence on $(s',t')$ can be translated into dependencies on $(s',z_s')$ for the $s$-channel integrals and $(t',z_t')$ for the $t$-channel integrals via~\eqref{internalkinematicsofsta}.
For convenience we also give the numerical values of~\cite{piNuRegge} for the 21 real parameters in Table~\ref{tab:uregge}.
\begin{table}
\centering
\renewcommand{\arraystretch}{1.3}
\begin{tabular}{c r r r r}
\toprule
 & $N_\alpha\quad$ & $N_\gamma\quad$ & $\Delta_\delta\quad$  & $\Delta_\beta\quad$ \\
\midrule
$a~\big[\GeV^{-1}\big]$\! & $-60.68$ & $  47.22$ & $ -75.15$ & $1419.99$ \\
$b~\big[\GeV^{-3}\big]$\! & $326.52$ & $-215.84$ & $-138.75$ & $3052.84$ \\
$c~\big[\GeV^{-2}\big]$\! & $546.40$ & $-101.11$ & $  64.16$ & $-192.64$ \\
$d~\big[\GeV^{-4}\big]$\! & $307.42$ & $-128.04$ & $  86.77$ & $-695.81$ \\
\midrule
$\alpha^{(0)}$ & $-0.36$ & $-0.62$ & $0.03$ & $-2.65$ \\
\midrule
$\alpha'~\big[\GeV^{-2}\big]$\! & \multicolumn{4}{c}{$0.908$} \\
\bottomrule
\end{tabular}
\renewcommand{\arraystretch}{1.0}
\caption[Regge-model parameter values.]{Regge-model parameter values for backward $\pi N$ scattering as given in~\cite{piNuRegge}.}
\label{tab:uregge}
\end{table}

As a byproduct, we can use these relations to infer the high-energy behavior of the HDR $s$-channel integrals:
from the trajectory parameters given in Table~\ref{tab:uregge} it follows that the high-energy tail of the integrals will be governed by the $\Delta_\delta$ trajectory.
Explicitly, the integrands for $A^+$ and $B^-$ will behave as
\begin{equation}
s'^{-1}s'^{\alpha_{\Delta_\delta}(a)-\frac{1}{2}}=s'^{\alpha'a-1.47}=
\begin{cases}
s'^{-1.88}\quad\text{for }a=-23.19\,\mpi^2\ec\\
s'^{-1.52}\quad\text{for }a=-2.71\,\mpi^2\ec
\end{cases}
\end{equation}
for $u'\to a$,
whereas the integrands for $A^-$ and $B^+$ fall off faster by one power in $s'$ (cf.~\eqref{hdrasym}).
We thus conclude that the $s$-channel part of the (unsubtracted) HDRs~\eqref{hdr} converges in principle for $a<26.57\,\mpi^2$.
Note that in order to investigate the behavior of these asymptotic contributions in the ``fixed-$t$ limit'' $a\to-\infty$ (as discussed in Appendix~\ref{subsec:convergence:lehmannellipse}) it is important to take the limits in the correct order, since $u'\to a$ only after $s'\to\infty$.
Since $\alpha_i(u')-\frac{1}{2}<-1$ for sufficiently large and negative $a$, the $s$-channel Regge contributions vanish in the limit $s'\to\infty$ for such values of $a$.
As shown in Sect.~\ref{subsec:piNMO:results}, these asymptotic contributions are numerically small for the optimal value of $a$ (and a reasonable choice of $\sa$), and thus they can be safely neglected for $a\to-\infty$, regardless of the pathological behavior of the Regge model due to the Gamma function in this case.

\subsection[$t$-channel asymptotics]{\boldmath{$t$}-channel asymptotics}
\label{subsec:regge:t}

Similarly to the previous section one could use Regge theory to describe the $t$-channel asymptotic region.
However, the significance of these contributions in view of the corresponding low-energy region differs strongly from the $s$-channel: while contributing crucially to the dispersive integrals, the pseudophysical region $\tpi\leq t\leq\tN$ cannot be constrained from experiment, but requires an analytic continuation.
Within our system of RS equations this task naturally takes the form of a MO problem, as explained in Sects.~\ref{sec:RS} and \ref{sec:piNMO}.
The solution of these equations becomes rather involved once intermediate states other than $\pi\pi$ are energetically allowed, which happens around $1\GeV$ (especially $\bar KK$ above $2\mK$).
In view of the ensuing uncertainty of the $t$-channel partial waves even below the $\bar NN$ threshold it is clear that the inclusion of phase-shift solutions above $\tN$~\cite{Anisovich}, and even more so the modeling of the high-energy region, will be of little practical relevance.
Moreover, as shown explicitly in Sect.~\ref{subsec:piNMO:results}, already the $s$-channel Regge contributions are numerically immaterial, in particular if subtractions are performed, which provides evidence that also the high-energy region in the $t$-channel can be safely ignored.
For these reasons, we will not consider the $t$-channel asymptotic region any further.

\subsection{Subtracted asymptotics}
\label{subsec:regge:asymptotics}

Here, we show how to incorporate the effects due to subtractions into the Regge description of the asymptotic parts of the corresponding subtracted HDRs~\eqref{hdr1sub} and~\eqref{hdr2sub}.
However, according to Appendix~\ref{subsec:regge:t} all asymptotic $t$-channel contributions will be neglected.

For the high-energy tail $s'>\sa$ of the $s$-channel integrals, according to Appendix~\ref{subsec:regge:s} the absorptive parts may generically be written as sums of Regge-trajectory contributions 
\begin{equation}
\Im X^{I_u}(s',u'(s',t'))=\sum\limits_{i}\tilde\beta_i^X(u')\left(\frac{s'}{s_R}\right)^{\alpha_i(u')-\frac{1}{2}} \quad\text{for }X\in\{A,B\}\ec
\end{equation}
with summands of the generic form (i.e.\ dropping the indices $X$ and $i$ for the time being)
\begin{equation}
\tilde\beta(u')=-\frac{\mathcal{S}\beta(u')}{\Gamma\big(\alpha(u')-\frac{1}{2}\big)}\ec \qquad
\beta(u')=\beta^{(0)}+\beta^{(1)}u'\ec \qquad
\alpha(u')=\alpha^{(0)}+\alpha'u'\ep
\end{equation}
While the evaluation of the Regge contributions is straightforward in the un- and once-subtracted case, for two subtractions one furthermore needs the derivative
\begin{align}
\ste{\partial_t\left\{\tilde\beta(u')\left(\frac{s'}{s_R}\right)^{\alpha(u')-\frac{1}{2}}\right\}}
&=\ste{\partial_tt'}\ste{\frac{\mathcal{S}}{\Gamma\Big(\alpha\big(u'(s',t')\big)-\frac{1}{2}\Big)}\left(\frac{s'}{s_R}\right)^{\alpha\big(u'(s',t')\big)-\frac{1}{2}}}\\
&\quad\times\ste{\beta^{(1)}+\alpha'\beta\big(u'(s',t')\big)\left\{\log\frac{s'}{s_R}-\Psi\Big(\alpha\big(u'(s',t')\big)-\tfrac{1}{2}\Big)\right\}}\ec\nt
\end{align}
where $\Psi(z)$ denotes the digamma function defined as the logarithmic derivative of the gamma function
\begin{equation}
\Psi(z)=\frac{\diff}{\diff z}\log\Gamma(z)=\frac{\Gamma'(z)}{\Gamma(z)}\ep
\end{equation}
To this end, one may use $u'(s',t')=\Sigma-s'-t'$ and (cf.~\eqref{zsdeltzs})
\begin{equation}
\ste{t'}=-\frac{(s'-s_0)^2}{s'-a}\ec \qquad \ste{\partial_tt'}=\frac{s_0-a}{s'-a}\ep
\end{equation}
After utilizing the crossing relations in order to rewrite the Regge contributions in the $I\in\{+,-\}$ isospin basis and expressing $t'$ as well as the corresponding kernel functions in terms of $(s',z_s')$, we can perform the partial-wave projections of the $s$-channel contributions onto both $s$- and $t$-channel partial waves according to~\eqref{pwprojasym}, where again the implicit kinematical dependencies have to be taken into account accordingly.

Finally, we demonstrate the projection onto the lowest $t$-channel partial waves with $J\leq2$ explicitly.
The $n$-times subtracted versions of~\eqref{pwprojasym} immediately lead to
\begin{align}
\label{pwprojasymtexpl}
\nsubasym{f^0_+}(t)&=\frac{1}{4\pi}\int\limits_0^1\diff z_t\;p_t^2
\Bigg\{-\!\nsubasym{A^+}(t,z_t)+4mq_t^2z_t^2\frac{\nsubasym{B^+}(t,z_t)}{4p_tq_tz_t}\Bigg\}\ec\nt\\
\nsubasym{f^1_+}(t)&=\frac{1}{4\pi}\int\limits_0^1\diff z_t\;z_t^2
\Bigg\{-4p_t^2\frac{\nsubasym{A^-}(t,z_t)}{4p_tq_tz_t}+m\!\nsubasym{B^-}(t,z_t)\Bigg\}\ec\nt\\
\nsubasym{f^1_-}(t)&=\frac{1}{4\pi}\int\limits_0^1\diff z_t\;\frac{1-z_t^2}{\sqrt{2}}\nsubasym{B^-}(t,z_t)\ec\nt\\
\nsubasym{f^2_+}(t)&=\frac{1}{4\pi}\int\limits_0^1\diff z_t\;\frac{3z_t^2-1}{2q_t^2}
\Bigg\{-\!\nsubasym{A^+}(t,z_t)+4mq_t^2z_t^2\frac{\nsubasym{B^+}(t,z_t)}{4p_tq_tz_t}\Bigg\}\ec\nt\\
\nsubasym{f^2_-}(t)&=\frac{1}{4\pi}\int\limits_0^1\diff z_t\;2\sqrt{6}\;z_t^2\big(1-z_t^2\big)\frac{\nsubasym{B^+}(t,z_t)}{4p_tq_tz_t}\ec
\end{align}
again written in terms of quantities that are always real since $4p_tq_tz_t=4m\nu$.
Here, the asymptotic $s$-channel contributions to the invariant amplitudes for e.g.\ the twice-subtracted case read (i.e.\ as functions of $(t,z_t)$, cf.~\eqref{zsprimetozt} for $z_s'(t,s';z_t)$)
\begin{align}
\label{sasymA2sub}
\twosub{A^+}_\text{s-asym}(t,z_t)&=\frac{1}{\pi}\int\limits_{\sa}^\infty\diff s'
\Bigg\{\bigg[\frac{2(s'-s_0)+t}{(s'-s_0+\frac{t}{2})^2-4p_t^2q_t^2z_t^2}-\frac{1}{s'-a}\bigg]\Im A^+(s',z_s')\nt\\
&\qquad-\bigg(h_0(s')-\frac{t}{(s'-s_0)^2}\bigg)\ste{\Im A^+(s',z_s')}-h_0(s')\,t\ste{\partial_t\Im A^+(s',z_s')}\Bigg\}\ec\nt\\
\frac{\twosub{A^-}_\text{s-asym}(t,z_t)}{4p_tq_tz_t}&=\frac{1}{\pi}\int\limits_{\sa}^\infty\diff s'
\Bigg\{\frac{\Im A^-(s',z_s')}{(s'-s_0+\frac{t}{2})^2-4p_t^2q_t^2z_t^2}-\frac{\ste{\Im A^-(s',z_s')}}{(s'-s_0)^2}\Bigg\}\ec
\end{align}
and analogously for $B^-(t,z_t)$ and $B^+(t,z_t)/(4p_tq_tz_t)$.
Note that again only real squares of momenta and $z_t$ occur and hence these formulae are valid in all kinematical regions.
Furthermore, by rewriting the general $t$-channel partial-wave projections~\eqref{tprojform} for both even and odd $J$ in terms of real quantities (i.e.\ $\nu$-even amplitudes and squares of momenta as well as squares of $z_t$) as above, the partial waves exhibit ostensible poles at $\tpi$ for all $J\geq2$ and in addition at $\tN$ for all $J\geq3$, while from the discussion of their threshold behavior in Sect.~\ref{subsubsec:RS:thrfJpm} we know that these poles are immaterial.
The reason for this behavior can be understood by first noting that for $p_tq_t\to0$ the asymptotic ($s$-channel) contributions~\eqref{sasymA2sub} no longer depend on $z_t$. The orthogonality of the Legendre polynomials $P_J(z_t)$ for even $J\geq2$ and odd $J\geq3$ then balances the poles and leads to the expected finite (but non-vanishing) values of the partial waves at both the pseudothreshold $\tpi$ and the threshold $\tN$ (cf.\ the explicit case for $f^2_+(t)$ in~\eqref{pwprojasymtexpl}).

\end{document}